# A Catalog of Sloan Magnitudes for the Brightest Stars – Version 2

Anthony Mallama, 14012 Lancaster Lane, Bowie, MD, 20715, anthony.mallama@gmail.com

2018 May 23

Abstract

A new version of the Catalog containing Sloan magnitudes for the brightest stars is presented. The accuracy of the data indicates that the Catalog is a reliable source of comparison star magnitudes for astronomical photometry. Version 2 complements the APASS database of fainter stars.

1. Introduction

This paper presents a revised version of the Catalog of Sloan magnitudes for the brightest stars (Mallama, 2014). The revision corrects a small discrepancy in one of the equations. Specifically, the coefficient, 0.953, of the term $((R-I) – C_1)$ in equation (3) in section A.1 of the Appendix should have been 0.935. While this discrepancy makes only a very slight difference in the results, a corrected Catalog is provided here as Version 2.

The new version of the Catalog is described and its validation is discussed in Section 2. The Catalog is recommended as a bright star extension to the APASS database in Section 3. Version 2 of the Catalog is included after the Reference section.

2. The new version

The changes to the Catalog from the original to Version 2 prompted an examination of the differences in the colors and magnitudes between the two datasets. Table 1 indicates that the mean difference is 0.00 for all magnitudes and colors. Therefore there is no discernible systematic bias between the original and Version 2. The corresponding standard deviations are all 0.00 or 0.01 magnitude. Meanwhile, the estimated Catalog uncertainties of Johnson-Cousins colors are 0.02 magnitude and the Sloan uncertainties are all 0.03 magnitude except for the u' band for which it is larger (Mallama 2014). Thus, the errors due to the slight inconsistency in the equation are insignificant.

Table 1. Differences between the original Catalog and Version 2 along with the uncertainties

|        | Mean Difference | Standard Deviation | Uncertainty |
|--------|-----------------|--------------------|-------------|
| V-Rc   | 0.00            | 0.00               | 0.02        |
| R-Ic   | 0.00            | 0.01               | 0.02        |
| u'     | 0.00            | 0.00               | 0.08        |
| g'     | 0.00            | 0.00               | 0.03        |
| r'     | 0.00            | 0.00               | 0.03        |
| i'     | 0.00            | 0.01               | 0.03        |
| z'     | 0.00            | 0.01               | 0.03        |

The overall quality of the Sloan magnitudes in the original Catalog was established by Mallama and Kroubsek (2015) using two independent techniques. One was to compare the Catalog magnitudes with those synthesized from Hubble Space Telescope (HST) spectra. The other was to compare the Catalog magnitudes with those directly observed using CCD photometry. The uncertainties determined by the two techniques supported those quoted by Mallama for the original Catalog. The coefficient change only affected 1 of the 23 stellar comparisons. The differences for that star were 0.00 for the u', g' and r' bands, 0.01 for i' and 0.02 for z'. So, the statistical results are not appreciably changed and Version 2 of the Catalog is validated.

3. A bright star extension to APASS

The APASS database Data Release 9 (Henden et al. 2016) contains B, V, g', r' and i' values for stars between magnitudes 10 and 17. Meanwhile, the Catalog of bright star magnitudes contains values for all of those bands in addition to U, R, I, u', z' and the color indices $V – R_C$ and $R_C – I_C$ for stars between magnitudes -2 and +7. The U, B, V, R and I magnitudes in the Catalog are those of the primary standard stars observed by Johnson et al. (1966). Therefore, the Catalog is proposed as an extension of APASS to brighter magnitudes.

4. Conclusions

This paper has presented Version 2 of a Catalog of Sloan magnitudes for the brightest stars. The data values have been validated by comparison with CCD photometry and with synthetic magnitudes derived from HST spectra. The new version of the Catalog is proposed as a bright star extension to APASS. Version 2 of the Catalog follows the reference section. Each Sloan magnitude is accompanied by its estimated uncertainty in units of 0.01 magnitude.

```
           <--------- Johnson --------->  < Cousins >   <------------------ Sloan ------------------>
  HD         V     U-B    B-V   V-R   R-I    V-Rc  Rc-Ic    <- u'->    <- g'->    <- r'->    <- i'->    <- z'->
   496     3.88   0.84   1.03  0.75  0.52    0.52  0.47    6.62  5    4.35  4    3.55  4    3.30  3    3.18  3
   571     5.04   0.26   0.40  0.42  0.29    0.27  0.28    6.65  9    5.17  3    4.96  2    4.91  2    4.90  3
  1013     4.80   1.93   1.57  1.34  1.13    0.90  1.03    9.31 14    5.57  5    4.19  5    3.39  3    2.91  3
  1280     4.61   0.05   0.06  0.08  0.01    0.03  0.03    5.75 13    4.55  3    4.71  2    4.90  2    5.05  3
  1404     4.52   0.07   0.05  0.08  0.00    0.03  0.01    5.68 14    4.45  3    4.62  2    4.83  2    4.99  3
  1522     3.55   1.17   1.22  0.85  0.59    0.59  0.52    6.84  7    4.13  4    3.15  5    2.85  3    2.69  3
  1581     4.23   0.02   0.58  0.49  0.34    0.32  0.33    5.61  4    4.45  3    4.09  2    3.97  2    3.93  2
  2151     2.80   0.11   0.62  0.50  0.34    0.33  0.34    4.33  3    3.05  3    2.64  2    2.53  2    2.49  2
  2261     2.40   0.88   1.09  0.81  0.59    0.56  0.52    5.23  4    2.90  4    2.04  4    1.74  3    1.58  3
  2262     3.94   0.10   0.17  0.14  0.08    0.07  0.09    5.20 11    3.94  3    3.99  2    4.12  2    4.24  3
  2772     4.73  -0.36  -0.10  0.00 -0.12   -0.02 -0.07    5.22  8    4.58  2    4.88  3    5.18  3    5.39  3
  3360     3.66  -0.89  -0.19 -0.08 -0.21   -0.08 -0.13    3.38  4    3.46  2    3.86  3    4.21  3    4.46  4
  3369     4.36  -0.55  -0.16 -0.04 -0.12   -0.05 -0.07    4.56  6    4.18  2    4.54  3    4.83  3    5.04  4
  3546     4.38   0.47   0.87  0.68  0.51    0.47  0.46    6.53  3    4.76  4    4.11  3    3.87  2    3.75  2
  3627     3.28   1.48   1.28  0.92  0.66    0.64  0.59    7.02 12    3.89  4    2.84  5    2.48  3    2.28  3
  3651     5.86   0.58   0.85  0.65  0.39    0.44  0.37    8.15  4    6.23  4    5.60  3    5.46  2    5.39  2
  3817     5.33   0.60   0.89  0.71  0.46    0.49  0.42    7.67  4    5.72  4    5.04  3    4.84  2    4.75  2
  3901     4.81  -0.66  -0.10 -0.01 -0.12   -0.03 -0.06    4.90  3    4.66  2    4.97  3    5.25  3    5.46  3
  3919     4.59   0.72   0.97  0.75  0.52    0.52  0.47    7.13  4    5.03  4    4.27  3    4.03  2    3.90  2
  4128     2.02   0.88   1.01  0.72  0.51    0.50  0.46    4.80  6    2.48  4    1.70  4    1.47  3    1.34  3
  4188     4.75   0.84   1.00  0.74  0.51    0.51  0.46    7.47  6    5.21  4    4.43  4    4.19  3    4.07  2
  4628     5.76   0.58   0.88  0.77  0.47    0.53  0.45    8.06  4    6.15  4    5.46  3    5.24  2    5.12  2
  4656     4.44   1.88   1.51  1.17  0.87    0.80  0.78    8.85 15    5.18  5    3.88  5    3.32  4    3.00  3
  4727     4.53  -0.58  -0.15 -0.03 -0.15   -0.04 -0.09    4.70  5    4.35  2    4.70  3    5.02  3    5.24  4
  4813     5.20  -0.01   0.50  0.46  0.28    0.30  0.29    6.50  3    5.38  3    5.08  2    5.02  2    5.01  2
  5015     4.82   0.12   0.53  0.48  0.30    0.31  0.30    6.31  3    5.02  3    4.69  2    4.61  2    4.59  2
  5112     4.78   1.92   1.57  1.23  0.91    0.84  0.82    9.28 14    5.55  5    4.19  5    3.60  4    3.25  3
  5234     4.84   1.26   1.22  0.92  0.61    0.64  0.54    8.25  9    5.42  4    4.42  4    4.10  3    3.93  3
  5395     4.64   0.70   0.96  0.77  0.50    0.53  0.46    7.15  4    5.07  4    4.32  3    4.09  2    3.97  2
  5448     3.87   0.15   0.12  0.15  0.08    0.08  0.09    5.17 14    3.84  3    3.93  2    4.06  2    4.17  3
  5516     4.42   0.69   0.94  0.73  0.48    0.50  0.44    6.91  4    4.84  4    4.12  3    3.90  2    3.79  2
  5612     6.32   0.58   0.90  0.67  0.46    0.45  0.42    8.64  3    6.72  4    6.05  3    5.85  2    5.75  2
  5848     4.26   1.33   1.21  0.89  0.60    0.62  0.53    7.76 11    4.83  4    3.85  4    3.54  3    3.38  3
  6116     5.98   0.09   0.16  0.18  0.05    0.10  0.06    7.22 11    5.97  3    6.02  2    6.18  2    6.31  3
  6186     4.28   0.71   0.96  0.78  0.52    0.54  0.47    6.81  4    4.71  4    3.96  3    3.71  2    3.59  2
  6482     6.12   0.85   1.01  0.73  0.53    0.50  0.48    8.86  6    6.58  4    5.80  4    5.55  3    5.42  3
  6582     5.18   0.09   0.69  0.63  0.41    0.43  0.39    6.72  5    5.46  3    4.97  2    4.80  2    4.73  2
  6595     3.31   0.56   0.90  0.71  0.52    0.49  0.47    5.60  3    3.71  4    3.02  3    2.78  2    2.65  2
  6763     5.53  -0.01   0.34  0.35  0.18    0.22  0.19    6.74  5    5.62  3    5.48  2    5.51  2    5.56  3
```

```
 6805        3.45  1.19  1.16  0.83  0.58    0.57  0.52    6.74  9   3.99 4   3.07 4   2.78 3   2.62 3
 6811        4.25 -0.34 -0.07  0.03 -0.05   -0.01 -0.01    4.79  8   4.12 2   4.39 2   4.62 3   4.79 3
 6860        2.05  1.96  1.57  1.24  1.00    0.84  0.91    6.60 15   2.82 5   1.46 5   0.78 4   0.38 3
 6961        4.34  0.13  0.17  0.18  0.07    0.10  0.08    5.64 12   4.34 3   4.38 2   4.52 2   4.64 3
 7087        4.66  0.81  1.03  0.76  0.54    0.52  0.49    7.36  4   5.13 4   4.33 4   4.07 3   3.93 3
 7106        4.51  1.00  1.10  0.82  0.58    0.57  0.52    7.51  6   5.02 4   4.14 4   3.85 3   3.70 3
 7318        4.66  0.84  1.03  0.75  0.53    0.52  0.48    7.40  5   5.13 4   4.33 4   4.08 3   3.95 3
 7570        4.96  0.10  0.58  0.49  0.36    0.32  0.35    6.45  3   5.18 3   4.82 2   4.69 2   4.64 2
 7804        5.17  0.08  0.07  0.11  0.03    0.05  0.05    6.35 13   5.11 3   5.26 2   5.43 2   5.57 3
 7964        4.76  0.10  0.03  0.08  0.05    0.03  0.06    5.95 15   4.68 3   4.86 2   5.02 2   5.15 3
 8207        4.90  0.97  1.08  0.81  0.53    0.56  0.48    7.84  6   5.40 4   4.54 4   4.29 3   4.16 3
 8491        4.74  0.93  1.04  0.78  0.52    0.54  0.47    7.61  7   5.22 4   4.40 4   4.15 3   4.03 2
 8512        3.59  0.92  1.06  0.76  0.56    0.52  0.50    6.45  6   4.08 4   3.25 4   2.97 3   2.83 3
 8799        4.83  0.00  0.42  0.41  0.23    0.26  0.24    6.10  3   4.97 3   4.75 2   4.73 2   4.75 2
 9138        4.84  1.52  1.38  1.06  0.74    0.74  0.66    8.69 10   5.50 4   4.34 5   3.90 3   3.65 3
 9270        3.62  0.74  0.97  0.72  0.50    0.50  0.46    6.19  5   4.06 4   3.31 4   3.08 3   2.97 2
 9362        3.95  0.70  0.99  0.75  0.51    0.52  0.46    6.48  4   4.40 4   3.63 4   3.39 2   3.27 2
 9408        4.72  0.76  1.00  0.77  0.53    0.53  0.48    7.33  4   5.18 4   4.39 3   4.14 2   4.01 2
 9826        4.10  0.06  0.54  0.46  0.29    0.30  0.30    5.51  3   4.30 3   3.97 2   3.90 2   3.88 2
 9900        5.56  1.43  1.38  0.97  0.71    0.67  0.63    9.29  8   6.22 4   5.08 5   4.68 4   4.45 3
 9927        3.57  1.44  1.28  0.96  0.65    0.67  0.58    7.26 11   4.18 4   3.12 4   2.77 3   2.57 3
10205        4.94 -0.41 -0.09 -0.01 -0.07   -0.03 -0.03    5.37  7   4.80 2   5.10 2   5.34 3   5.53 3
10307        4.96  0.11  0.62  0.53  0.33    0.35  0.33    6.49  3   5.21 3   4.79 2   4.69 2   4.65 2
10348        5.99  0.88  1.01  0.77  0.46    0.53  0.42    8.77  6   6.45 4   5.66 4   5.46 2   5.36 2
10380        4.44  1.56  1.36  1.06  0.71    0.74  0.63    8.34 12   5.09 4   3.94 4   3.53 3   3.30 3
10476        5.24  0.49  0.84  0.69  0.43    0.47  0.41    7.40  3   5.61 4   4.97 3   4.79 2   4.70 2
10700        3.50  0.21  0.72  0.62  0.47    0.42  0.45    5.22  3   3.80 3   3.28 2   3.06 2   2.95 2
10761        4.26  0.68  0.96  0.74  0.48    0.51  0.44    6.74  4   4.69 4   3.95 3   3.73 2   3.63 2
11171        4.68  0.03  0.32  0.29  0.16    0.18  0.18    5.93  6   4.76 3   4.65 2   4.70 2   4.76 2
11257        5.94 -0.02  0.30  0.33  0.17    0.21  0.19    7.11  5   6.01 3   5.91 2   5.94 2   6.00 3
11353        3.72  1.06  1.14  0.80  0.55    0.55  0.49    6.82  7   4.25 4   3.35 4   3.08 3   2.94 3
11415        3.38 -0.62 -0.15 -0.04 -0.12   -0.05 -0.06    3.49  4   3.20 2   3.56 3   3.84 3   4.05 4
11559        4.63  0.70  0.94  0.73  0.47    0.50  0.43    7.13  4   5.05 4   4.33 3   4.12 2   4.02 2
11636        2.65  0.10  0.13  0.14  0.08    0.07  0.09    3.89 12   2.63 3   2.71 2   2.85 2   2.96 3
11753        5.11 -0.14 -0.06  0.02 -0.06   -0.01 -0.04    5.92 12   4.98 2   5.25 2   5.51 3   5.71 3
11973        4.79  0.09  0.28  0.28  0.16    0.17  0.17    6.10  8   4.85 3   4.78 2   4.82 2   4.89 3
12055        4.83  0.51  0.88  0.70  0.45    0.48  0.42    7.04  3   5.22 4   4.55 3   4.36 2   4.26 2
12111        4.54  0.06  0.16  0.18  0.08    0.10  0.09    5.74 11   4.53 3   4.58 2   4.71 2   4.83 3
12216        3.98  0.04 -0.01  0.06  0.00    0.01  0.01    5.06 15   3.88 3   4.10 2   4.31 3   4.47 3
12274        4.01  1.89  1.56  1.26  1.04    0.85  0.94    8.46 14   4.77 5   3.42 5   2.70 4   2.28 3
12303        5.04 -0.32 -0.08  0.03 -0.05   -0.01 -0.01    5.60  9   4.90 2   5.18 3   5.41 3   5.59 3
```

```
12471     5.50   0.06   0.03   0.06   0.03    0.01   0.05    6.63 14    5.42  3    5.61  2    5.78  2    5.92  3
12524     5.14   1.82   1.49   1.22   0.89    0.83   0.80    9.46 14    5.86  4    4.57  5    4.00  3    3.66  3
12533     2.10   0.92   1.21   0.94   0.68    0.65   0.61    5.05  3    2.67  4    1.67  4    1.29  3    1.08  3
12929     2.00   1.13   1.15   0.84   0.62    0.58   0.55    5.20  8    2.54  4    1.62  4    1.29  3    1.11  3
13041     4.83   0.17   0.12   0.14   0.07    0.07   0.08    6.16 14    4.80  3    4.90  2    5.04  2    5.16  3
13161     3.00   0.12   0.14   0.14   0.08    0.07   0.09    4.27 12    2.98  3    3.06  2    3.19  2    3.30  3
13267     6.35  -0.43   0.33   0.34   0.26    0.21   0.26    6.98  6    6.44  3    6.31  2    6.27  2    6.27  3
13476     6.43   0.24   0.60   0.55   0.51    0.36   0.43    8.12  4    6.67  3    6.26  2    6.05  2    5.95  2
13594     6.06  -0.07   0.40   0.38   0.25    0.24   0.26    7.22  3    6.19  3    5.99  2    5.96  2    5.97  2
13596     5.70   1.95   1.55   1.24   0.97    0.84   0.88   10.23 15    6.46  5    5.11  5    4.47  4    4.08  3
13611     4.37   0.61   0.88   0.67   0.49    0.45   0.45    6.72  4    4.76  4    4.10  3    3.88  2    3.77  2
14055     4.01   0.02   0.02   0.03   0.00   -0.01   0.01    5.08 13    3.93  3    4.13  2    4.34  2    4.50  3
14872     4.70   1.90   1.53   1.18   0.87    0.81   0.78    9.15 15    5.45  5    4.13  5    3.58  4    3.25  3
15089     4.53   0.06   0.13   0.16   0.06    0.08   0.07    5.72 11    4.51  3    4.59  2    4.74  2    4.86  3
15130     4.89  -0.05  -0.02  -0.02  -0.03   -0.04  -0.01    5.85 13    4.78  2    5.03  2    5.27  2    5.44  3
15318     4.29  -0.11  -0.06   0.02  -0.05   -0.01  -0.03    5.14 13    4.16  2    4.43  2    4.68  3    4.86  3
15371     4.25  -0.50  -0.14  -0.03  -0.13   -0.04  -0.08    4.53  6    4.08  2    4.42  3    4.72  3    4.95  3
15798     4.75  -0.01   0.45   0.41   0.27    0.26   0.27    6.02  3    4.90  3    4.66  2    4.61  2    4.60  2
16046     4.89  -0.14  -0.06   0.02  -0.07   -0.01  -0.05    5.70 12    4.76  2    5.03  2    5.30  3    5.50  3
16161     4.86   0.53   0.86   0.65   0.49    0.44   0.47    7.09  3    5.24  4    4.60  3    4.36  2    4.23  2
16739     4.92   0.14   0.59   0.50   0.30    0.33   0.30    6.47  3    5.15  3    4.77  2    4.69  2    4.67  2
16754     4.75   0.06   0.06   0.10   0.01    0.04   0.03    5.90 13    4.69  3    4.84  2    5.03  2    5.18  3
16815     4.11   0.74   1.02   0.79   0.56    0.55   0.50    6.71  4    4.58  4    3.77  3    3.49  2    3.35  2
16861     6.30   0.06   0.06   0.05   0.02    0.01   0.04    7.45 13    6.24  3    6.41  2    6.59  2    6.73  3
16895     4.13   0.00   0.49   0.46   0.30    0.30   0.30    5.43  3    4.30  3    4.01  2    3.94  2    3.92  2
16908     4.67  -0.63  -0.13  -0.02  -0.13   -0.04  -0.07    4.78  4    4.50  2    4.84  3    5.13  3    5.35  3
16970     3.47   0.07   0.09   0.11   0.04    0.05   0.05    4.65 13    3.42  3    3.55  2    3.72  2    3.85  3
17081     4.25  -0.45  -0.14  -0.02  -0.14   -0.04  -0.09    4.60  7    4.08  2    4.42  3    4.73  3    4.96  4
17094     4.27   0.07   0.31   0.30   0.19    0.19   0.19    5.57  7    4.35  3    4.24  2    4.27  2    4.32  3
17206     4.46   0.00   0.48   0.43   0.27    0.28   0.28    5.76  3    4.63  3    4.36  2    4.30  2    4.30  2
17361     4.52   1.03   1.11   0.80   0.58    0.55   0.52    7.56  7    5.04  4    4.16  4    3.87  3    3.71  3
17573     3.63  -0.38  -0.10  -0.02  -0.11   -0.04  -0.07    4.10  8    3.48  2    3.79  2    4.08  3    4.29  3
17584     4.23   0.08   0.34   0.30   0.23    0.19   0.23    5.56  6    4.32  3    4.20  2    4.19  2    4.22  2
17652     4.46   0.69   0.99   0.76   0.54    0.52   0.49    6.97  3    4.91  4    4.14  3    3.88  2    3.74  2
17709     4.53   1.92   1.56   1.21   0.95    0.83   0.86    9.02 14    5.29  5    3.95  5    3.32  4    2.95  3
17824     4.77   0.61   0.90   0.70   0.47    0.48   0.43    7.13  4    5.17  4    4.48  3    4.28  2    4.17  2
18322     3.87   0.98   1.12   0.79   0.58    0.55   0.52    6.85  6    4.39  4    3.51  4    3.22  3    3.06  3
18331     5.17   0.05   0.08   0.11   0.05    0.05   0.06    6.32 12    5.12  3    5.25  2    5.41  2    5.54  3
18411     4.70   0.12   0.06   0.10   0.04    0.04   0.05    5.93 15    4.64  3    4.79  2    4.96  2    5.09  3
18449     4.94   1.28   1.25   0.89   0.64    0.62   0.57    8.40  9    5.53  4    4.52  5    4.17  3    3.99  3
18519     4.63   0.08   0.04   0.05   0.02    0.01   0.04    5.80 14    4.56  3    4.74  2    4.92  2    5.06  3
```

```
18604     4.70 -0.45 -0.12 -0.03 -0.11    -0.04 -0.06    5.06  7    4.54  2    4.87  3    5.15  3    5.36  3
18622     2.91  0.12  0.12  0.14  0.08     0.07  0.08    4.17 13    2.88  3    2.98  2    3.11  2    3.23  3
18778     5.95  0.09  0.15  0.13  0.07     0.06  0.09    7.19 11    5.94  3    6.01  2    6.14  2    6.25  3
18978     4.09  0.08  0.16  0.13  0.09     0.06  0.10    5.32 11    4.08  3    4.15  2    4.28  2    4.39  3
19275     4.88  0.05  0.02  0.08 -0.01     0.03  0.00    5.99 14    4.80  3    4.99  2    5.20  3    5.37  3
19476     3.81  0.83  0.98  0.74  0.50     0.51  0.46    6.51  6    4.25  4    3.49  3    3.26  2    3.15  2
19656     4.64  1.02  1.11  0.83  0.57     0.57  0.51    7.67  7    5.16  4    4.27  4    3.98  3    3.83  3
19787     4.37  0.86  1.03  0.77  0.51     0.53  0.46    7.14  5    4.84  4    4.03  4    3.80  3    3.67  2
20010     3.85  0.05  0.51  0.46  0.31     0.30  0.31    5.23  3    4.04  3    3.73  2    3.65  2    3.62  2
20150     4.89 -0.01 -0.02  0.06 -0.03     0.01 -0.01    5.90 14    4.78  2    5.01  2    5.25  3    5.43  3
20320     4.80  0.09  0.23  0.22  0.11     0.13  0.11    6.08  9    4.83  3    4.82  2    4.92  2    5.02  3
20468     4.82  1.55  1.49  1.07  0.76     0.75  0.68    8.77  8    5.54  4    4.29  5    3.83  4    3.58  4
20644     4.47  1.77  1.54  1.20  0.88     0.82  0.79    8.75 12    5.22  5    3.90  5    3.33  4    3.00  3
20677     4.95  0.07  0.04  0.08  0.00     0.03  0.01    6.10 14    4.88  3    5.05  2    5.26  2    5.42  3
20794     4.27  0.22  0.71  0.62  0.40     0.42  0.38    5.99  3    4.57  3    4.05  2    3.90  2    3.83  2
20894     5.52  0.59  0.88  0.66  0.47     0.45  0.43    7.84  4    5.91  4    5.25  3    5.05  2    4.94  2
21120     3.60  0.62  0.89  0.68  0.45     0.47  0.42    5.96  4    3.99  4    3.32  3    3.13  2    3.04  2
21447     5.10  0.05  0.04  0.09 -0.01     0.03  0.00    6.22 14    5.03  3    5.20  2    5.42  2    5.58  3
21754     4.10  1.02  1.13  0.77  0.54     0.53  0.49    7.14  6    4.63  4    3.74  4    3.48  3    3.34  3
21770     5.32 -0.02  0.41  0.41  0.22     0.26  0.23    6.55  3    5.45  3    5.24  2    5.23  2    5.26  3
21790     4.73 -0.27 -0.09 -0.01 -0.08    -0.03 -0.05    5.35 10    4.59  2    4.89  2    5.16  3    5.36  3
22203     4.28 -0.35 -0.12  0.00 -0.12    -0.02 -0.07    4.77  9    4.12  2    4.44  3    4.73  3    4.95  4
22484     4.28  0.08  0.57  0.49  0.32     0.32  0.32    5.74  3    4.50  3    4.14  2    4.04  2    4.01  2
22663     4.58  0.77  1.04  0.84  0.57     0.58  0.51    7.23  4    5.06  4    4.22  3    3.94  2    3.79  2
23227     5.00 -0.60 -0.16 -0.05 -0.15    -0.06 -0.09    5.13  5    4.82  2    5.18  3    5.50  3    5.73  3
23401     4.66  0.07  0.03  0.13  0.04     0.06  0.05    5.81 14    4.58  3    4.75  2    4.92  3    5.05  3
23754     4.23  0.01  0.42  0.39  0.22     0.25  0.23    5.51  4    4.37  3    4.15  2    4.14  2    4.17  2
24160     4.17  0.68  0.95  0.71  0.50     0.49  0.46    6.65  4    4.60  4    3.87  3    3.64  2    3.52  2
24398     2.85 -0.78  0.12  0.14  0.09     0.07  0.14    2.89  8    2.82  3    2.92  2    2.99  2    3.07  3
24479     5.04 -0.16 -0.10  0.07 -0.08     0.02 -0.05    5.80 13    4.89  2    5.18  3    5.45  3    5.65  4
24555     4.45  0.42  0.68  0.59  0.40     0.39  0.38    6.43  5    4.73  3    4.25  2    4.10  2    4.03  2
25490     3.91  0.06  0.03  0.08  0.00     0.03  0.01    5.04 14    3.83  3    4.01  2    4.22  2    4.38  3
25604     4.37  0.95  1.07  0.79  0.53     0.55  0.48    7.28  6    4.86  4    4.02  4    3.77  3    3.63  3
25642     4.29 -0.04  0.02  0.08  0.02     0.03  0.04    5.28 12    4.21  3    4.40  2    4.58  2    4.72  3
26612     4.93  0.07  0.33  0.35  0.21     0.22  0.22    6.24  6    5.02  3    4.88  2    4.89  2    4.92  3
26630     4.15  0.64  0.95  0.79  0.54     0.55  0.48    6.57  3    4.58  4    3.83  3    3.57  2    3.44  2
26673     4.71  0.64  1.01  0.80  0.57     0.55  0.51    7.17  3    5.17  4    4.37  3    4.09  2    3.94  2
26690     5.29  0.00  0.36  0.35  0.19     0.22  0.20    6.52  4    5.39  3    5.24  2    5.26  2    5.30  3
26722     4.84  0.51  0.80  0.66  0.45     0.45  0.42    7.01  4    5.19  3    4.59  3    4.40  2    4.31  2
26846     4.86  1.14  1.18  0.85  0.57     0.59  0.51    8.09  7    5.41  4    4.47  4    4.18  3    4.03  3
26912     4.30 -0.51 -0.05  0.05 -0.06     0.01 -0.01    4.62  4    4.18  2    4.43  2    4.66  3    4.83  3
```

```
26965     4.43  0.44  0.82  0.69  0.45    0.47  0.43   6.51  3   4.79  4   4.17  3   3.96  2   3.86  2
26967     3.86  1.01  1.10  0.86  0.59    0.60  0.52   6.87  7   4.37  4   3.48  4   3.18  2   3.02  2
27022     5.27  0.47  0.81  0.64  0.43    0.43  0.40   7.39  3   5.62  4   5.03  3   4.85  2   4.76  2
27045     4.94  0.11  0.26  0.26  0.11    0.16  0.12   6.27  9   4.99  3   4.94  2   5.04  2   5.13  3
27376     3.56 -0.37 -0.12 -0.01 -0.11   -0.03 -0.06   4.03  9   3.40  2   3.72  3   4.00  3   4.21  3
27442     4.44  1.07  1.08  0.83  0.55    0.57  0.49   7.52  8   4.94  4   4.07  4   3.81  3   3.67  2
27820     5.12  0.11  0.07  0.10  0.05    0.04  0.06   6.34 14   5.06  3   5.21  2   5.37  2   5.50  3
27861     5.17  0.08  0.08  0.10  0.03    0.04  0.05   6.36 13   5.12  3   5.26  2   5.43  2   5.57  3
28028     3.96  1.80  1.49  1.17  0.83    0.80  0.75   8.25 14   4.68  4   3.40  5   2.89  3   2.59  3
28100     4.69  0.72  0.98  0.72  0.51    0.50  0.46   7.24  4   5.13  4   4.38  4   4.14  3   4.02  2
28749     4.91  1.43  1.31  0.94  0.67    0.65  0.60   8.60 10   5.54  4   4.46  5   4.09  3   3.88  3
28978     5.68  0.12  0.05  0.12  0.04    0.06  0.05   6.91 15   5.61  3   5.77  2   5.94  3   6.07  3
29085     4.51  0.71  0.98  0.75  0.54    0.52  0.49   7.05  4   4.95  4   4.19  3   3.93  2   3.80  2
29094     4.27  0.79  1.22  0.98  0.69    0.68  0.62   7.05  4   4.85  4   3.83  4   3.44  3   3.22  3
29140     4.26  0.12  0.18  0.19  0.10    0.11  0.11   5.56 11   4.26  3   4.30  2   4.41  2   4.51  3
29291     3.82  0.72  0.98  0.75  0.49    0.52  0.45   6.37  4   4.26  4   3.50  3   3.28  2   3.17  2
29503     3.87  1.03  1.09  0.84  0.56    0.58  0.50   6.90  7   4.37  4   3.50  4   3.22  3   3.08  2
29763     4.29 -0.57 -0.14 -0.03 -0.14   -0.04 -0.08   4.48  5   4.12  2   4.46  3   4.76  3   4.98  3
29875     4.45  0.00  0.34  0.34  0.21    0.21  0.22   5.67  5   4.54  3   4.41  2   4.41  2   4.44  3
30614     4.29 -0.88  0.03  0.11  0.00    0.05  0.07   4.15  8   4.21  3   4.39  2   4.53  3   4.66  3
30652     3.19 -0.01  0.46  0.42  0.26    0.27  0.27   4.46  3   3.35  3   3.09  2   3.05  2   3.05  2
30739     4.35  0.03  0.01  0.06  0.00    0.01  0.01   5.43 14   4.26  3   4.46  2   4.67  2   4.84  3
30834     4.77  1.58  1.41  1.09  0.78    0.76  0.70   8.72 11   5.45  4   4.25  5   3.78  3   3.51  3
30836     3.68 -0.81 -0.16 -0.05 -0.16   -0.06 -0.08   3.53  3   3.50  2   3.86  3   4.16  3   4.38  4
31109     4.40  0.18  0.23  0.32  0.17    0.20  0.17   5.81 11   4.43  3   4.39  2   4.43  3   4.49  3
31278     4.47 -0.02 -0.02  0.09 -0.01    0.03  0.01   5.47 14   4.36  2   4.58  2   4.79  3   4.96  3
31295     4.67  0.09  0.08  0.11  0.03    0.05  0.05   5.87 13   4.62  3   4.75  2   4.93  2   5.07  3
31398     2.69  1.78  1.53  1.06  0.82    0.74  0.74   6.98 12   3.44  5   2.15  6   1.64  4   1.35  4
31421     4.06  1.10  1.15  0.88  0.63    0.61  0.56   7.22  7   4.60  4   3.66  4   3.33  3   3.15  3
31647     4.95  0.01  0.05  0.05  0.03    0.01  0.05   6.03 12   4.88  3   5.06  2   5.23  2   5.37  3
31767     4.49  1.56  1.40  1.05  0.70    0.73  0.62   8.41 11   5.17  4   3.99  5   3.59  3   3.36  3
31910     4.03  0.63  0.93  0.70  0.45    0.48  0.41   6.43  4   4.45  4   3.74  3   3.55  2   3.46  2
32147     6.21  1.00  1.06  0.85  0.49    0.59  0.47   9.18  7   6.70  4   5.84  3   5.60  2   5.48  2
32249     4.81 -0.75 -0.19 -0.08 -0.19   -0.08 -0.12   4.72  3   4.61  2   5.01  3   5.35  3   5.59  4
32309     4.92 -0.14 -0.05  0.02 -0.05   -0.01 -0.03   5.74 12   4.80  2   5.06  2   5.31  3   5.49  3
32630     3.18 -0.67 -0.18 -0.05 -0.17   -0.06 -0.11   3.21  4   2.99  2   3.37  3   3.70  3   3.93  4
32831     4.55  1.19  1.20  0.91  0.65    0.63  0.58   7.86  8   5.12  4   4.13  4   3.78  3   3.59  3
32887     3.19  1.77  1.46  1.10  0.81    0.77  0.73   7.43 14   3.90  4   2.66  5   2.16  3   1.87  3
33111     2.79  0.10  0.13  0.14  0.08    0.07  0.09   4.03 12   2.77  3   2.85  2   2.99  2   3.10  3
33262     4.72 -0.04  0.52  0.48  0.30    0.31  0.30   5.99  4   4.91  3   4.59  2   4.51  2   4.49  2
33276     4.82  0.18  0.32  0.32  0.19    0.20  0.19   6.27  9   4.90  3   4.79  2   4.81  2   4.86  3
```

```
33641      4.88   0.10   0.18   0.19   0.10      0.11   0.11    6.15 11   4.88  3   4.92  2   5.03  2   5.13  3
33802      4.44  -0.40  -0.09  -0.02  -0.09     -0.04  -0.04    4.88  7   4.30  2   4.60  2   4.86  3   5.06  3
33856      4.45   1.14   1.19   0.85   0.59      0.59   0.52    7.68  7   5.01  4   4.05  4   3.75  3   3.59  3
33949      4.36  -0.34  -0.10  -0.01  -0.09     -0.03  -0.05    4.88  9   4.21  2   4.52  3   4.78  3   4.98  3
34029      0.08   0.45   0.80   0.60   0.44      0.40   0.41    2.16  3   0.43  3  -0.15  3  -0.33  2  -0.42  2
34317      6.42   0.03  -0.02   0.05  -0.03      0.01  -0.01    7.48 15   6.31  2   6.54  2   6.78  3   6.96  3
34334      4.54   1.27   1.27   0.98   0.71      0.68   0.63    7.99  8   5.14  4   4.09  4   3.68  3   3.45  3
34411      4.71   0.13   0.62   0.53   0.32      0.35   0.32    6.26  3   4.96  3   4.54  2   4.44  2   4.41  2
34503      3.59  -0.46  -0.12  -0.02  -0.11     -0.04  -0.06    3.94  7   3.43  2   3.75  3   4.04  3   4.25  3
34642      4.83   0.79   1.00   0.74   0.55      0.51   0.49    7.49  5   5.29  4   4.51  4   4.24  3   4.10  2
34748      6.35  -0.74  -0.11   0.01  -0.14     -0.02  -0.06    6.32  3   6.19  2   6.51  3   6.79  3   7.00  3
34816      4.29  -1.01  -0.25  -0.12  -0.28     -0.11  -0.18    3.82  4   4.06  2   4.51  3   4.92  3   5.20  4
34968      4.71  -0.10  -0.05   0.05  -0.05      0.01  -0.03    5.58 13   4.59  2   4.84  2   5.09  3   5.28  3
34989      5.80  -0.88  -0.13   0.01  -0.13     -0.02  -0.05    5.57  4   5.63  2   5.96  3   6.23  3   6.44  4
35007      5.69  -0.66  -0.11  -0.02  -0.15     -0.04  -0.08    5.77  3   5.53  2   5.85  3   6.15  3   6.38  3
35039      4.74  -0.79  -0.16  -0.05  -0.18     -0.06  -0.11    4.62  3   4.56  2   4.92  3   5.25  3   5.49  3
35149      5.00  -0.87  -0.15  -0.05  -0.18     -0.06  -0.10    4.77  4   4.82  2   5.18  3   5.50  3   5.73  3
35299      5.70  -0.88  -0.21  -0.08  -0.19     -0.08  -0.11    5.43  3   5.49  2   5.90  3   6.23  3   6.47  4
35369      4.12   0.70   0.96   0.72   0.50      0.50   0.46    6.63  4   4.55  4   3.82  3   3.58  2   3.47  2
35407      6.32  -0.63  -0.15  -0.05  -0.16     -0.06  -0.10    6.42  4   6.14  2   6.50  3   6.82  3   7.06  3
35468      1.64  -0.88  -0.22  -0.09  -0.22     -0.09  -0.14    1.36  3   1.42  2   1.85  3   2.20  3   2.46  4
35497      1.65  -0.49  -0.13  -0.01  -0.10     -0.03  -0.05    1.95  6   1.48  2   1.81  3   2.09  3   2.29  4
35588      6.16  -0.75  -0.18  -0.06  -0.19     -0.06  -0.12    6.08  3   5.97  2   6.35  3   6.69  3   6.93  4
35640      6.23  -0.23  -0.05   0.01  -0.05     -0.02  -0.03    6.93 10   6.11  2   6.37  2   6.62  3   6.81  3
35708      4.89  -0.77  -0.14  -0.05  -0.15     -0.06  -0.08    4.80  3   4.72  2   5.07  3   5.37  3   5.58  3
35912      6.44  -0.75  -0.17  -0.07  -0.18     -0.07  -0.11    6.37  3   6.25  2   6.63  3   6.96  3   7.20  3
36079      2.84   0.47   0.82   0.65   0.44      0.44   0.41    4.96  3   3.20  4   2.59  3   2.41  2   2.32  2
36134      5.78   0.99   1.15   0.83   0.57      0.57   0.51    8.79  5   6.32  4   5.40  4   5.12  3   4.96  3
36166      5.78  -0.85  -0.20  -0.06  -0.20     -0.06  -0.13    5.55  3   5.57  2   5.98  3   6.32  3   6.57  4
36267      4.20  -0.56  -0.13  -0.06  -0.14     -0.06  -0.08    4.40  5   4.03  2   4.38  2   4.68  3   4.90  3
36285      6.32  -0.84  -0.19  -0.08  -0.17     -0.08  -0.10    6.11  3   6.12  2   6.52  3   6.84  3   7.07  4
36351      5.46  -0.82  -0.17  -0.04  -0.22     -0.05  -0.15    5.29  3   5.27  2   5.64  3   6.01  3   6.27  4
36371      4.77  -0.44   0.35   0.37   0.27      0.24   0.27    5.40  7   4.87  3   4.71  2   4.67  2   4.67  3
36430      6.21  -0.74  -0.18  -0.06  -0.17     -0.06  -0.10    6.14  3   6.02  2   6.40  3   6.72  3   6.95  4
36512      4.62  -1.07  -0.26  -0.12  -0.26     -0.11  -0.16    4.06  5   4.38  2   4.85  3   5.23  3   5.50  4
36597      3.87   1.08   1.14   0.82   0.60      0.57   0.53    7.00  7   4.40  4   3.49  4   3.19  3   3.02  3
36673      2.57   0.25   0.20   0.22   0.21      0.13   0.19    4.05 14   2.59  3   2.59  2   2.62  2   2.67  3
36741      6.58  -0.80  -0.16  -0.07  -0.21     -0.07  -0.14    6.44  3   6.40  2   6.77  3   7.12  3   7.38  3
36777      5.36   0.07   0.05   0.08   0.04      0.03   0.05    6.52 14   5.29  3   5.46  2   5.63  2   5.76  3
36822      4.41  -0.97  -0.15  -0.01  -0.17     -0.03  -0.08    4.05  6   4.23  2   4.58  3   4.88  3   5.10  4
37160      4.09   0.66   0.95   0.76   0.55      0.52   0.49    6.54  4   4.52  4   3.78  3   3.51  2   3.37  2
```

```
37171     5.90   1.93   1.60   1.19   0.94    0.81   0.85   10.43 13   6.69  5    5.31  6    4.69  4    4.33  4
37468     3.80  -1.01  -0.24  -0.08  -0.24   -0.08  -0.14    3.33  4   3.57  2    4.01  3    4.37  3    4.63  4
37507     4.81   0.11   0.13   0.15   0.06    0.08   0.07    6.07 12   4.79  3    4.87  2    5.02  2    5.14  3
37711     4.86  -0.63  -0.12  -0.01  -0.12   -0.03  -0.06    4.98  4   4.70  2    5.02  3    5.30  3    5.51  3
37742     1.77  -1.06  -0.21  -0.08  -0.20   -0.08  -0.10    1.25  6   1.56  2    1.97  3    2.30  3    2.53  4
37795     2.64  -0.47  -0.12  -0.02  -0.10   -0.04  -0.07    2.97  6   2.48  2    2.80  3    3.10  3    3.31  3
37984     4.90   1.06   1.17   0.88   0.61    0.61   0.54    8.01  6   5.45  4    4.50  4    4.18  3    4.01  3
38104     5.47   0.07   0.03   0.08   0.01    0.03   0.03    6.62 14   5.39  3    5.57  2    5.76  2    5.91  3
38393     3.60   0.01   0.47   0.45   0.26    0.29   0.27    4.91  3   3.76  3    3.49  2    3.45  2    3.45  2
38656     4.53   0.69   0.94   0.73   0.49    0.50   0.45    7.02  4   4.95  4    4.23  3    4.00  2    3.89  2
38678     3.55   0.06   0.10   0.12   0.03    0.06   0.05    4.72 12   3.51  3    3.63  2    3.80  2    3.94  3
38771     2.05  -1.02  -0.18  -0.02  -0.18   -0.04  -0.09    1.60  6   1.86  2    2.23  3    2.54  3    2.77  4
38899     4.91  -0.16  -0.07   0.02  -0.08   -0.01  -0.05    5.69 12   4.78  2    5.05  2    5.33  3    5.53  3
38944     4.74   1.93   1.62   1.36   1.07    0.91   0.97    9.28 13   5.54  5    4.11  5    3.37  4    2.93  3
39003     3.97   1.09   1.14   0.82   0.56    0.57   0.50    7.11  7   4.50  4    3.59  4    3.32  3    3.17  3
39283     5.00   0.12   0.05   0.09   0.02    0.03   0.04    6.23 15   4.93  3    5.10  2    5.28  2    5.42  3
39357     4.59   0.03  -0.02   0.04   0.00    0.00   0.01    5.65 15   4.48  2    4.72  2    4.93  2    5.09  3
39364     3.85   0.71   0.98   0.86   0.56    0.60   0.50    6.39  4   4.29  4    3.50  3    3.22  2    3.08  2
39425     3.12   1.21   1.16   0.85   0.58    0.59   0.52    6.43  9   3.66  4    2.73  4    2.44  3    2.28  3
40035     3.72   0.91   0.99   0.77   0.50    0.53   0.46    6.53  7   4.17  4    3.39  3    3.16  2    3.05  2
40111     4.82  -0.94  -0.06   0.06  -0.11    0.01  -0.02    4.55  7   4.69  2    4.95  3    5.20  3    5.38  3
40136     3.72   0.00   0.33   0.33   0.16    0.21   0.18    4.94  5   3.81  3    3.68  2    3.72  2    3.78  3
40494     4.36  -0.66  -0.18  -0.07  -0.17   -0.07  -0.10    4.40  4   4.17  2    4.55  3    4.87  3    5.11  4
40657     4.52   1.21   1.22   0.93   0.67    0.65   0.60    7.87  8   5.10  4    4.09  4    3.72  3    3.52  3
40808     3.96   1.08   1.14   0.82   0.58    0.57   0.52    7.09  7   4.49  4    3.58  4    3.29  3    3.14  3
40967     4.94  -0.58  -0.12  -0.03  -0.09   -0.04  -0.03    5.12  4   4.78  2    5.11  3    5.36  3    5.54  3
41116     4.15   0.52   0.87   0.68   0.45    0.47   0.42    6.37  3   4.53  4    3.88  3    3.69  2    3.59  2
41695     4.67   0.00   0.05   0.09   0.00    0.03   0.01    5.73 12   4.60  3    4.77  2    4.97  2    5.14  3
41753     4.42  -0.67  -0.15  -0.06  -0.16   -0.06  -0.10    4.46  4   4.24  2    4.60  3    4.92  3    5.16  3
42545     4.92  -0.59  -0.12  -0.02  -0.14   -0.04  -0.08    5.09  4   4.76  2    5.08  3    5.39  3    5.61  3
42560     4.48  -0.66  -0.17  -0.05  -0.16   -0.06  -0.10    4.53  4   4.29  2    4.67  3    4.98  3    5.21  4
42818     4.80   0.00   0.03   0.07  -0.01    0.02   0.00    5.85 13   4.72  3    4.91  2    5.12  2    5.29  3
43039     4.35   0.81   1.01   0.80   0.54    0.55   0.49    7.04  5   4.81  4    4.01  3    3.75  2    3.61  2
43232     3.96   1.42   1.31   0.97   0.64    0.67   0.57    7.64 10   4.59  4    3.50  5    3.16  3    2.97  3
43244     6.52   0.09   0.27   0.27   0.14    0.16   0.14    7.83  8   6.57  3    6.51  2    6.59  2    6.68  3
43261     6.09   0.62   0.90   0.69   0.49    0.47   0.45    8.46  4   6.49  4    5.81  3    5.58  2    5.47  2
43445     5.01  -0.24  -0.09   0.02  -0.05   -0.01  -0.03    5.67 11   4.87  2    5.16  3    5.41  3    5.59  3
43785     4.37   0.83   1.00   0.73   0.51    0.50   0.46    7.08  6   4.83  4    4.05  4    3.81  3    3.69  2
43834     5.09   0.33   0.72   0.61   0.32    0.41   0.32    6.97  3   5.39  3    4.88  3    4.78  2    4.75  2
44402     3.02  -0.72  -0.18  -0.05  -0.18   -0.06  -0.11    2.98  3   2.83  2    3.21  3    3.54  3    3.78  4
44762     3.85   0.52   0.88   0.67   0.47    0.45   0.43    6.07  3   4.24  4    3.58  3    3.37  2    3.27  2
```

```
44769     4.31  0.14  0.20  0.19  0.10     0.11  0.12    5.64 11    4.33  3    4.34  2    4.44  2    4.54  3
45542     4.14 -0.49 -0.14 -0.03 -0.11    -0.04 -0.07    4.43  7    3.97  2    4.31  3    4.60  3    4.81  3
45813     4.48 -0.61 -0.17 -0.06 -0.16    -0.06 -0.10    4.59  5    4.29  2    4.67  3    4.99  3    5.22  4
46300     4.50 -0.18  0.00  0.10  0.04     0.04  0.07    5.29 10    4.41  3    4.61  2    4.75  3    4.88  3
46933     4.54 -0.01 -0.06  0.04 -0.02     0.00 -0.01    5.53 15    4.41  2    4.68  2    4.90  3    5.08  3
47105     1.92  0.05  0.00  0.06 -0.01     0.01  0.00    3.02 15    1.83  3    2.04  2    2.25  2    2.42  3
47174     4.79  1.30  1.23  0.90  0.60     0.62  0.53    8.26  9    5.37  4    4.37  4    4.06  3    3.90  3
47205     3.92  0.99  1.05  0.79  0.51     0.55  0.46    6.87  7    4.40  4    3.57  4    3.34  3    3.21  2
47442     4.42  1.04  1.15  0.86  0.60     0.60  0.53    7.50  6    4.96  4    4.03  4    3.72  3    3.56  3
47667     4.81  1.68  1.49  1.03  0.73     0.72  0.65    8.94 11    5.53  4    4.29  6    3.86  4    3.62  4
47670     3.17 -0.41 -0.11  0.00 -0.07    -0.02 -0.03    3.59  7    3.01  2    3.33  3    3.57  3    3.76  3
48099     6.37 -0.94 -0.05  0.07 -0.05     0.02  0.02    6.10  7    6.25  2    6.50  3    6.69  3    6.85  3
48329     2.98  1.46  1.40  0.96  0.61     0.67  0.54    6.76  9    3.66  4    2.50  5    2.19  4    2.02  4
48433     4.49  1.17  1.16  0.86  0.60     0.60  0.53    7.75  8    5.03  4    4.10  4    3.79  3    3.63  3
48737     3.36  0.06  0.43  0.39  0.23     0.25  0.24    4.71  4    3.50  3    3.28  2    3.26  2    3.28  2
48915    -1.46 -0.05  0.00  0.00 -0.03    -0.02 -0.01   -0.49 12   -1.55  3   -1.33  2   -1.09  2   -0.91  3
49293     4.46  1.04  1.11  0.79  0.55     0.55  0.49    7.51  7    4.98  4    4.10  4    3.83  3    3.69  3
49878     4.55  1.66  1.36  1.02  0.71     0.71  0.63    8.58 14    5.20  4    4.06  5    3.66  3    3.43  3
50019     3.60  0.14  0.10  0.10  0.08     0.04  0.08    4.88 14    3.56  3    3.68  2    3.82  2    3.94  3
50522     4.35  0.51  0.85  0.65  0.44     0.44  0.40    6.54  3    4.72  4    4.09  3    3.91  2    3.83  2
50635     4.65  0.08  0.30  0.32  0.20     0.20  0.23    5.96  7    4.72  3    4.62  2    4.62  2    4.64  3
50778     4.08  1.69  1.43  1.13  0.78     0.78  0.70    8.19 13    4.77  4    3.55  5    3.08  3    2.80  3
50973     4.90  0.05  0.03  0.05  0.01     0.01  0.03    6.02 14    4.82  3    5.01  2    5.20  2    5.35  3
51199     4.69  0.06  0.36  0.38  0.19     0.24  0.21    6.00  6    4.79  3    4.63  2    4.64  2    4.68  3
51799     4.95  1.92  1.69  1.48  1.18     0.97  1.07    9.52 11    5.78  5    4.28  7    3.44  4    2.93  4
52089     1.50 -0.93 -0.21 -0.09 -0.21    -0.09 -0.13    1.16  4    1.29  2    1.71  3    2.05  3    2.30  4
54719     4.42  1.40  1.26  0.96  0.63     0.67  0.56    8.05 11    5.02  4    3.97  4    3.64  3    3.46  3
54810     4.92  0.78  1.03  0.79  0.52     0.55  0.47    7.58  4    5.39  4    4.58  4    4.33  2    4.21  2
54893     4.83 -0.69 -0.18 -0.07 -0.15    -0.07 -0.09    4.83  4    4.64  2    5.02  3    5.33  3    5.56  4
55185     4.15  0.04  0.00  0.06  0.01     0.01  0.03    5.24 14    4.06  3    4.27  2    4.46  2    4.61  3
56169     5.05  0.09  0.08  0.15  0.06     0.08  0.07    6.25 13    5.00  3    5.12  2    5.27  3    5.40  3
56456     4.76 -0.29 -0.10 -0.01 -0.08    -0.03 -0.04    5.35 10    4.61  2    4.92  3    5.18  3    5.38  3
56537     3.58  0.09  0.12  0.12  0.05     0.06  0.07    4.80 12    3.55  3    3.65  2    3.81  2    3.93  3
56577     4.78  1.87  1.70  1.32  0.96     0.89  0.87    9.28 10    5.62  5    4.14  6    3.50  4    3.13  4
56986     3.53  0.04  0.34  0.35  0.19     0.22  0.20    4.81  6    3.62  3    3.48  2    3.50  2    3.55  3
57821     4.96 -0.38 -0.05  0.06 -0.04     0.01 -0.01    5.45  7    4.84  2    5.09  2    5.32  3    5.49  3
58142     4.64 -0.02 -0.02 -0.01 -0.05    -0.03 -0.03    5.64 14    4.53  2    4.78  2    5.03  2    5.22  3
58207     3.79  0.84  1.04  0.77  0.50     0.53  0.46    6.54  5    4.27  4    3.45  4    3.22  3    3.10  3
58367     4.99  0.78  1.01  0.73  0.50     0.50  0.46    7.64  4    5.45  4    4.67  4    4.44  3    4.32  3
58946     4.18 -0.02  0.32  0.32  0.19     0.20  0.20    5.36  5    4.26  3    4.15  2    4.17  2    4.21  3
58972     4.30  1.53  1.43  1.11  0.79     0.77  0.71    8.19  9    4.99  4    3.77  5    3.29  3    3.01  3
```

```
59294     4.55  1.37  1.29  0.90  0.64   0.62  0.57   8.15  9   5.16 4   4.12 5   3.77 4   3.58 3
59612     4.84  0.19  0.24  0.27  0.27   0.16  0.23   6.26 11   4.88 3   4.84 2   4.83 2   4.85 3
60522     4.06  1.94  1.54  1.24  0.91   0.84  0.82   8.57 15   4.81 5   3.48 5   2.88 3   2.54 3
60532     4.45  0.07  0.51  0.51  0.27   0.34  0.28   5.86  3   4.64 3   4.32 2   4.26 2   4.26 2
61110     4.91  0.11  0.40  0.39  0.22   0.25  0.23   6.31  6   5.04 3   4.84 2   4.83 2   4.86 2
61330     4.53 -0.31 -0.09 -0.02 -0.07  -0.04 -0.03   5.10  9   4.39 2   4.69 2   4.94 3   5.13 3
61421     0.37  0.03  0.42  0.42  0.23   0.27  0.24   1.68  4   0.51 3   0.28 2   0.27 2   0.28 3
61497     4.99  0.09  0.08  0.18  0.04   0.10  0.05   6.19 13   4.94 3   5.05 2   5.22 3   5.36 3
61935     3.93  0.88  1.02  0.77  0.52   0.53  0.47   6.72  6   4.40 4   3.60 4   3.35 3   3.22 2
62345     3.57  0.70  0.92  0.71  0.45   0.49  0.42   6.06  5   3.98 4   3.28 3   3.08 2   2.99 2
62509     1.14  0.86  1.00  0.75  0.50   0.52  0.46   3.89  6   1.60 4   0.82 4   0.59 3   0.47 2
62576     4.59  1.98  1.63  1.31  0.97   0.88  0.88   9.20 14   5.39 5   3.97 5   3.32 4   2.94 3
62623     3.95 -0.05  0.18  0.26  0.22   0.16  0.21   5.02  8   3.95 3   3.97 2   3.97 2   4.01 3
62721     4.87  1.76  1.46  1.15  0.83   0.79  0.75   9.09 13   5.58 4   4.33 5   3.81 3   3.51 3
63700     3.35  1.18  1.25  0.88  0.55   0.61  0.49   6.67  6   3.94 4   2.93 5   2.66 3   2.52 3
63744     4.71  0.92  1.06  0.76  0.57   0.52  0.51   7.57  6   5.20 4   4.37 4   4.09 3   3.94 3
63922     4.11 -1.01 -0.18 -0.08 -0.17  -0.08 -0.08   3.68  6   3.92 2   4.31 3   4.61 3   4.83 3
64096     5.16  0.06  0.60  0.52  0.36   0.34  0.35   6.61  4   5.40 3   5.00 2   4.87 2   4.82 2
64145     4.98  0.11  0.09  0.13  0.05   0.06  0.07   6.21 14   4.93 3   5.06 2   5.21 2   5.33 3
64440     3.73  0.75  1.05  0.81  0.56   0.56  0.45   6.36  3   4.21 4   3.38 4   3.15 2   3.04 3
64740     4.63 -0.92 -0.23 -0.13 -0.21  -0.11 -0.14   4.29  3   4.41 2   4.85 3   5.21 3   5.46 4
64760     4.24 -0.99 -0.14 -0.02 -0.12  -0.04 -0.03   3.86  6   4.07 2   4.41 3   4.66 3   4.86 4
65456     4.79  0.16  0.15  0.20  0.15   0.11  0.14   6.12 13   4.78 3   4.83 2   4.91 2   4.99 3
65810     4.61  0.08  0.08  0.15  0.06   0.08  0.07   5.80 13   4.56 3   4.68 2   4.83 3   4.96 3
66141     4.38  1.28  1.25  0.99  0.67   0.69  0.60   7.84  9   4.97 4   3.93 4   3.56 3   3.35 3
67006     4.84  0.00  0.05  0.11  0.00   0.05  0.01   5.90 12   4.77 3   4.93 2   5.14 3   5.30 3
67797     4.40 -0.60 -0.15 -0.03 -0.14  -0.04 -0.08   4.54  5   4.22 2   4.57 3   4.88 3   5.10 4
68290     4.72  0.77  0.96  0.73  0.48   0.50  0.44   7.33  5   5.15 4   4.41 3   4.20 2   4.09 2
69142     4.44  1.09  1.17  0.84  0.64   0.58  0.57   7.60  7   4.99 4   4.05 4   3.71 3   3.52 3
69267     3.53  1.77  1.48  1.12  0.78   0.78  0.70   7.78 13   4.25 4   2.99 5   2.51 3   2.24 3
70060     4.45  0.11  0.22  0.21  0.13   0.12  0.14   5.76 10   4.48 3   4.47 2   4.55 2   4.63 3
70442     5.58  0.50  0.77  0.64  0.45   0.43  0.42   7.72  5   5.91 3   5.34 3   5.15 2   5.06 2
70555     4.83  1.60  1.45  1.00  0.75   0.70  0.67   8.83 10   5.53 4   4.33 6   3.88 4   3.63 4
70930     4.82 -0.84 -0.15 -0.04 -0.13  -0.05 -0.05   4.63  4   4.64 2   5.00 3   5.27 3   5.48 4
71115     5.13  0.64  0.94  1.25 -0.06   0.85  0.02   7.55  4   5.55 4   4.68 2   4.89 5   5.04 3
71155     3.90 -0.03 -0.02  0.03 -0.05  -0.01 -0.03   4.88 13   3.79 2   4.03 2   4.28 2   4.47 3
71369     3.36  0.52  0.85  0.69  0.42   0.47  0.39   5.57  3   3.73 4   3.09 3   2.92 2   2.84 2
73108     4.61  1.16  1.17  0.89  0.63   0.62  0.56   7.86  8   5.16 4   4.21 4   3.87 3   3.69 3
73155     5.01  1.38  1.33  0.97  0.68   0.67  0.61   8.65  9   5.65 4   4.55 5   4.17 3   3.95 3
73262     4.17  0.02  0.00  0.04  0.01   0.00  0.03   5.23 14   4.08 3   4.29 2   4.48 2   4.63 3
73471     4.43  1.28  1.20  0.89  0.56   0.62  0.50   7.86 10   5.00 4   4.02 4   3.74 3   3.60 3
```

```
74006    3.98  0.65  0.93  0.66  0.48   0.45  0.44   6.41  4   4.40 4   3.70 4   3.49 3   3.38 2
74137    4.88  0.92  1.07  0.82  0.53   0.57  0.48   7.75  6   5.37 4   4.52 4   4.27 3   4.14 2
74198    4.66  0.01  0.02  0.06  0.00   0.01  0.01   5.72 13   4.58 3   4.77 2   4.98 2   5.14 3
74272    4.77  0.12  0.12  0.21  0.19   0.12  0.17   6.03 13   4.74 3   4.81 2   4.87 2   4.93 3
74395    4.61  0.51  0.84  0.65  0.44   0.44  0.40   6.80  3   4.98 4   4.36 3   4.18 2   4.09 2
74442    3.94  0.99  1.08  0.78  0.54   0.54  0.49   6.91  7   4.44 4   3.59 4   3.33 3   3.19 3
74485    6.13  0.63  0.94  0.70  0.49   0.48  0.45   8.54  3   6.55 4   5.84 3   5.61 2   5.50 2
74575    3.69 -0.87 -0.18 -0.08 -0.16  -0.08 -0.08   3.45  3   3.50 2   3.89 3   4.19 3   4.41 3
74739    4.02  0.78  1.03  0.75  0.49   0.52  0.45   6.68  4   4.49 4   3.69 4   3.47 3   3.36 3
74918    4.32  0.64  0.90  0.68  0.46   0.47  0.42   6.72  4   4.72 4   4.04 3   3.84 2   3.74 2
75063    3.91 -0.05  0.00  0.05  0.13   0.01  0.14   4.88 12   3.82 3   4.03 2   4.11 2   4.19 3
75137    4.37 -0.05 -0.05  0.05 -0.07   0.01 -0.04   5.31 14   4.25 2   4.50 2   4.77 3   4.96 3
75691    4.01  1.40  1.26  0.95  0.67   0.66  0.60   7.64 11   4.61 4   3.57 4   3.20 3   2.99 3
76294    3.10  0.82  1.00  0.71  0.49   0.49  0.45   5.80  5   3.56 4   2.79 4   2.57 3   2.45 3
76483    4.89  0.13  0.10  0.13  0.05   0.06  0.07   6.16 14   4.85 3   4.96 2   5.12 2   5.24 3
76644    3.14  0.07  0.19  0.22  0.07   0.13  0.09   4.37 10   3.15 3   3.17 2   3.30 2   3.41 3
76756    4.26  0.15  0.13  0.14  0.04   0.07  0.06   5.57 13   4.24 3   4.32 2   4.48 2   4.61 3
76827    4.76  1.88  1.53  1.47  1.26   0.97  1.15   9.18 14   5.51 5   4.13 6   3.21 3   2.66 3
76943    3.97  0.06  0.43  0.40  0.22   0.26  0.23   5.32  4   4.11 3   3.89 2   3.88 2   3.90 2
77258    4.45  0.38  0.65  0.55  0.41   0.36  0.35   6.36  5   4.71 3   4.27 2   4.14 2   4.09 2
77327    3.60  0.01  0.00  0.07  0.01   0.02  0.03   4.65 14   3.51 3   3.71 2   3.90 3   4.05 3
77350    5.46 -0.10 -0.03  0.02 -0.04  -0.01 -0.02   6.34 12   5.35 2   5.59 2   5.83 2   6.01 3
77912    4.56  0.81  1.04  0.75  0.49   0.52  0.45   7.26  4   5.04 4   4.23 4   4.00 3   3.89 3
78004    3.75  1.21  1.20  0.82  0.60   0.57  0.53   7.08  8   4.32 4   3.36 5   3.05 3   2.89 3
78154    4.81  0.02  0.49  0.45  0.28   0.29  0.29   6.14  3   4.98 3   4.70 2   4.63 2   4.62 2
78209    4.48  0.12  0.27  0.27  0.11   0.16  0.13   5.83  9   4.53 3   4.47 2   4.56 2   4.65 3
78362    4.67  0.15  0.35  0.32  0.13   0.20  0.15   6.10  8   4.77 3   4.63 2   4.70 2   4.77 2
78541    4.56  1.87  1.61  1.24  0.90   0.84  0.81   9.01 12   5.35 5   3.96 6   3.38 4   3.04 4
79469    3.88 -0.12 -0.07 -0.01 -0.05  -0.03 -0.03   4.71 13   3.75 2   4.03 2   4.28 3   4.47 3
79917    4.94  1.06  1.11  0.79  0.57   0.55  0.55   8.02  7   5.46 4   4.58 4   4.26 3   4.09 3
79940    4.62  0.10  0.45  0.38  0.28   0.24  0.27   6.04  4   4.77 3   4.54 2   4.49 2   4.49 2
80007    1.68  0.03  0.00  0.07  0.02   0.02  0.04   2.76 14   1.59 3   1.79 2   1.98 2   2.12 3
80081    3.82  0.05  0.06  0.12  0.03   0.06  0.05   4.96 13   3.76 3   3.91 2   4.08 2   4.22 3
80493    3.13  1.95  1.55  1.23  0.90   0.84  0.81   7.66 15   3.89 5   2.55 5   1.96 3   1.62 3
80499    4.81  0.61  0.93  0.72  0.47   0.50  0.43   7.18  3   5.23 4   4.51 3   4.30 2   4.20 2
80586    4.79  0.67  0.94  0.70  0.44   0.48  0.41   7.25  4   5.21 4   4.50 3   4.31 2   4.22 2
80874    4.72  2.03  1.61  1.40  1.08   0.93  0.98   9.39 15   5.51 5   4.08 6   3.33 4   2.89 4
81146    4.46  1.31  1.23  0.91  0.63   0.63  0.56   7.95 10   5.04 4   4.04 4   3.70 3   3.52 3
81169    4.68  0.63  0.91  0.71  0.47   0.49  0.43   7.07  4   5.09 4   4.39 3   4.18 2   4.08 2
81188    2.50 -0.75 -0.18 -0.04 -0.20  -0.05 -0.12   2.42  3   2.31 2   2.68 3   3.03 3   3.27 4
81797    1.97  1.72  1.45  1.04  0.77   0.72  0.69   6.13 13   2.67 4   1.46 5   0.99 4   0.73 3
```

```
81799   4.68  1.17  1.13  0.88  0.56    0.61  0.50    7.92  9    5.21  4    4.29  4    4.01  2    3.87  3
81817   4.30  1.72  1.48  1.13  0.74    0.78  0.66    8.48 12    5.02  4    3.76  5    3.32  3    3.07  3
81937   3.67  0.10  0.33  0.34  0.18    0.21  0.19    5.02  7    3.76  3    3.63  2    3.66  2    3.71  3
81997   4.61  0.01  0.45  0.43  0.22    0.28  0.24    5.91  3    4.76  3    4.51  2    4.50  2    4.52  2
82150   4.51  1.68  1.44  1.01  0.76    0.70  0.73    8.61 12    5.21  4    4.01  5    3.50  4    3.21  3
82308   4.31  1.89  1.54  1.23  0.90    0.84  0.81    8.75 14    5.06  5    3.73  5    3.15  3    2.80  3
82328   3.18  0.03  0.46  0.44  0.27    0.29  0.28    4.51  3    3.34  3    3.08  2    3.02  2    3.02  2
82434   3.60 -0.07  0.36  0.36  0.22    0.23  0.23    4.74  3    3.70  3    3.54  2    3.54  2    3.57  3
82446   4.57  0.10  0.10  0.16  0.06    0.08  0.07    5.80 13    4.53  3    4.63  2    4.79  2    4.91  3
82621   4.51  0.04  0.00  0.09  0.01    0.03  0.03    5.60 14    4.42  3    4.62  2    4.81  3    4.96  3
82741   4.81  0.76  0.99  0.76  0.53    0.52  0.48    7.42  4    5.26  4    4.49  3    4.23  2    4.10  2
83425   4.68  1.45  1.32  1.06  0.71    0.74  0.63    8.41 10    5.31  4    4.19  4    3.78  3    3.55  3
83446   4.35  0.12  0.17  0.15  0.10    0.08  0.11    5.64 12    4.35  3    4.40  2    4.51  2    4.61  3
83618   3.91  1.45  1.32  0.99  0.67    0.69  0.60    7.64 10    4.54  4    3.44  5    3.07  3    2.87  3
83754   5.05 -0.58 -0.15 -0.07 -0.15   -0.07 -0.09    5.22  5    4.87  2    5.24  3    5.55  3    5.78  3
83808   3.52  0.21  0.49  0.41  0.23    0.26  0.25    5.11  5    3.69  3    3.42  2    3.39  2    3.41  2
83953   4.78 -0.58 -0.12  0.02 -0.09   -0.01 -0.04    4.96  4    4.62  2    4.94  3    5.19  3    5.39  4
84367   4.79  0.36  0.50  0.47  0.32    0.31  0.32    6.59  8    4.97  3    4.67  2    4.57  2    4.54  2
84737   5.10  0.08  0.62  0.53  0.34    0.35  0.33    6.58  4    5.35  3    4.93  2    4.82  2    4.79  2
85235   4.60  0.08  0.03  0.10  0.00    0.04  0.01    5.76 14    4.52  3    4.70  2    4.91  3    5.07  3
85444   4.11  0.65  0.92  0.69  0.47    0.47  0.43    6.53  4    4.52  4    3.82  3    3.62  2    3.51  2
85503   3.88  1.40  1.22  0.91  0.58    0.63  0.52    7.48 12    4.46  4    3.46  4    3.17  3    3.01  3
85622   4.58  0.99  1.20  0.83  0.63    0.57  0.56    7.62  4    5.15  4    4.19  5    3.85  3    3.67  3
86663   4.70  1.92  1.60  1.43  1.08    0.95  0.98    9.22 13    5.49  5    4.06  6    3.31  4    2.86  4
87504   4.59 -0.27 -0.10 -0.01 -0.09   -0.03 -0.05    5.20 10    4.44  2    4.75  3    5.01  3    5.21  3
87696   4.49  0.07  0.18  0.18  0.07    0.10  0.09    5.72 10    4.49  3    4.53  2    4.66  2    4.78  3
87737   3.53 -0.21 -0.04  0.09  0.02    0.03  0.06    4.26 10    3.41  2    3.65  3    3.81  3    3.94  4
87837   4.37  1.75  1.45  1.13  0.77    0.78  0.69    8.57 14    5.07  4    3.83  5    3.37  3    3.11  3
87887   4.50 -0.06 -0.04  0.06 -0.05    0.01 -0.03    5.43 13    4.38  2    4.63  2    4.87  3    5.06  3
87901   1.35 -0.36 -0.11 -0.02 -0.10   -0.04 -0.06    1.84  9    1.19  2    1.51  3    1.79  3    2.00  3
88284   3.61  0.92  1.00  0.77  0.48    0.53  0.44    6.44  7    4.07  4    3.28  3    3.07  2    2.96  2
88355   6.44  0.02  0.46  0.41  0.27    0.26  0.28    7.76  3    6.60  3    6.35  2    6.29  2    6.29  2
88955   3.85  0.08  0.05  0.02  0.02   -0.01  0.04    5.02 14    3.78  3    3.97  2    4.15  2    4.29  3
89021   3.45  0.06  0.03  0.08 -0.01    0.03  0.00    4.58 14    3.37  3    3.55  2    3.77  2    3.94  3
89025   3.44  0.19  0.31  0.31  0.19    0.19  0.19    4.90  9    3.52  3    3.41  2    3.44  2    3.49  3
89125   5.82 -0.05  0.50  0.48  0.27    0.31  0.28    7.06  3    6.00  3    5.70  2    5.64  2    5.64  2
89449   4.80  0.01  0.45  0.45  0.23    0.29  0.25    6.10  3    4.95  3    4.70  2    4.67  2    4.69  2
89758   3.05  1.89  1.59  1.28  0.96    0.86  0.87    7.52 13    3.83  5    2.45  5    1.81  4    1.43  3
89998   4.83  1.09  1.12  0.80  0.58    0.55  0.52    7.96  8    5.35  4    4.46  4    4.17  3    4.02  3
90277   4.74  0.18  0.25  0.26  0.14    0.16  0.15    6.16 11    4.78  3    4.74  2    4.81  2    4.88  3
90362   5.56  1.86  1.52  1.25  0.95    0.85  0.86    9.95 14    6.30  5    4.98  5    4.35  3    3.98  3
```

```
90537    4.21  0.65  0.90  0.69  0.46   0.47  0.42   6.62  4   4.61 4   3.93 3   3.73 2   3.63 2
90610    4.25  1.63  1.45  1.10  0.79   0.77  0.71   8.29 11   4.95 4   3.72 5   3.24 3   2.96 3
91312    4.75  0.08  0.23  0.20  0.09   0.11  0.10   6.02  9   4.78 3   4.77 2   4.89 2   4.99 2
91480    5.16 -0.02  0.34  0.33  0.17   0.21  0.19   6.35  5   5.25 3   5.12 2   5.15 2   5.21 3
92139    3.84  0.07  0.30  0.25  0.16   0.15  0.17   5.13  7   3.91 3   3.83 2   3.88 2   3.94 2
93497    2.69  0.57  0.90  0.68  0.49   0.47  0.45   4.99  3   3.09 4   2.41 3   2.19 2   2.08 2
93813    3.11  1.27  1.24  0.93  0.64   0.65  0.57   6.55  9   3.70 4   2.68 4   2.34 3   2.15 3
94264    3.83  0.92  1.04  0.83  0.54   0.57  0.49   6.68  6   4.31 4   3.47 3   3.21 2   3.08 2
94890    4.60  0.84  1.03  0.75  0.53   0.52  0.51   7.34  5   5.07 4   4.27 4   3.99 3   3.84 2
95129    6.00  1.92  1.59  1.49  1.29   0.98  1.18  10.51 14   6.78 5   5.35 6   4.40 4   3.83 4
95272    4.07  0.98  1.09  0.80  0.55   0.55  0.49   7.03  6   4.57 4   3.71 4   3.44 3   3.30 3
95370    4.39  0.12  0.11  0.13  0.07   0.06  0.08   5.65 13   4.36 3   4.46 2   4.60 2   4.72 3
95418    2.37  0.00 -0.02  0.06 -0.04   0.01 -0.02   3.39 14   2.26 2   2.49 2   2.74 3   2.92 3
95578    4.75  1.92  1.62  1.33  0.96   0.89  0.87   9.28 13   5.55 5   4.13 5   3.49 3   3.11 3
95608    4.42  0.05  0.05  0.06 -0.02   0.01 -0.01   5.55 13   4.35 3   4.53 2   4.75 2   4.93 3
95689    1.79  0.92  1.07  0.81  0.58   0.56  0.52   4.66  6   2.28 4   1.43 4   1.14 3   0.99 2
96097    4.63  0.08  0.33  0.33  0.17   0.21  0.18   5.95  7   4.72 3   4.59 2   4.63 2   4.69 3
96833    3.01  1.12  1.14  0.84  0.57   0.58  0.51   6.19  8   3.54 4   2.63 4   2.34 3   2.19 3
97277    4.48  0.05  0.03  0.07  0.02   0.02  0.04   5.60 14   4.40 3   4.59 2   4.77 2   4.91 3
97603    2.56  0.11  0.12  0.13  0.03   0.06  0.05   3.81 13   2.53 3   2.63 2   2.80 2   2.93 3
97633    3.35  0.08 -0.02  0.05 -0.03   0.01 -0.01   4.48 16   3.24 2   3.47 2   3.71 3   3.89 3
98058    4.47  0.15  0.21  0.27  0.11   0.16  0.12   5.82 11   4.49 3   4.48 2   4.58 2   4.68 3
98118    5.18  1.84  1.52  1.27  0.94   0.86  0.85   9.54 14   5.92 5   4.59 5   3.97 3   3.61 3
98262    3.49  1.55  1.40  1.06  0.70   0.74  0.62   7.40 10   4.17 4   2.98 5   2.58 3   2.36 3
98353    4.79  0.03  0.12  0.11  0.03   0.05  0.05   5.93 11   4.76 3   4.86 2   5.04 2   5.18 3
98430    3.56  0.99  1.11  0.83  0.60   0.57  0.53   6.55  6   4.08 4   3.19 4   2.88 3   2.72 3
98664    4.05 -0.13 -0.06  0.02 -0.07  -0.01 -0.04   4.88 12   3.92 2   4.19 2   4.45 3   4.65 3
98839    5.00  0.80  0.98  0.75  0.46   0.52  0.42   7.66  5   5.44 4   4.68 3   4.48 2   4.38 2
99028    3.94  0.07  0.41  0.39  0.21   0.25  0.22   5.30  5   4.07 3   3.86 2   3.86 2   3.89 2
99211    4.08  0.10  0.21  0.23  0.11   0.13  0.12   5.37 10   4.10 3   4.10 2   4.20 2   4.30 3
99648    4.95  0.80  1.00  0.75  0.50   0.52  0.46   7.62  5   5.41 4   4.63 4   4.40 3   4.28 2
100029   3.85  1.97  1.62  1.31  0.99   0.88  0.90   8.45 14   4.65 5   3.23 5   2.57 4   2.17 3
100407   3.54  0.70  0.93  0.70  0.48   0.48  0.44   6.04  5   3.96 4   3.25 3   3.03 2   2.93 2
100600   5.95 -0.65 -0.16 -0.06 -0.18  -0.06 -0.12   6.02  4   5.77 2   6.14 3   6.47 3   6.71 3
100889   4.70 -0.16 -0.08  0.01 -0.07  -0.02 -0.04   5.47 12   4.56 2   4.85 2   5.11 3   5.31 3
100920   4.30  0.74  1.01  0.72  0.52   0.50  0.47   6.89  4   4.76 4   3.98 4   3.74 3   3.61 3
101431   4.70 -0.22 -0.07  0.01 -0.05  -0.02 -0.03   5.40 10   4.57 2   4.85 2   5.09 3   5.28 3
101501   5.34  0.23  0.74  0.61  0.36   0.41  0.35   7.09  4   5.65 3   5.12 3   5.00 2   4.95 2
102070   4.72  0.74  0.97  0.73  0.48   0.50  0.44   7.29  5   5.16 4   4.41 3   4.19 2   4.09 2
102224   3.72  1.16  1.18  0.88  0.60   0.61  0.53   6.98  8   4.27 4   3.32 4   3.01 3   2.85 3
102870   3.60  0.11  0.55  0.48  0.28   0.31  0.29   5.09  3   3.81 3   3.47 2   3.40 2   3.39 2
```

```
102964      4.46   1.46   1.30   0.94   0.67     0.65   0.60    8.19 11    5.08  4    4.01  5    3.64  3    3.43  3
103095      6.45   0.17   0.75   0.66   0.45     0.45   0.43    8.13  5    6.77  3    6.21  2    6.01  2    5.91  2
103287      2.44   0.03   0.00   0.00  -0.03    -0.02  -0.01    3.52 14    2.35  3    2.57  2    2.81  2    2.98  3
104321      4.67   0.10   0.13   0.16   0.04     0.08   0.06    5.91 12    4.65  3    4.73  2    4.89  2    5.02  3
104979      4.12   0.64   0.99   0.74   0.49     0.51   0.45    6.57  3    4.57  4    3.80  4    3.58  3    3.47  2
105452      4.02  -0.02   0.32   0.30   0.18     0.19   0.19    5.20  5    4.10  3    3.99  2    4.02  2    4.07  3
105707      2.98   1.47   1.34   0.93   0.64     0.65   0.57    6.74 10    3.62  4    2.52  5    2.18  4    1.99  3
105937      3.96  -0.61  -0.15  -0.10  -0.16    -0.09  -0.10    4.09  5    3.78  2    4.16  2    4.47  3    4.70  3
106591      3.31   0.07   0.08   0.06   0.00     0.01   0.01    4.48 13    3.26  3    3.41  2    3.62  2    3.78  3
106625      2.58  -0.35  -0.11  -0.04  -0.09    -0.05  -0.05    3.08  9    2.42  2    2.75  2    3.01  3    3.21  3
107259      3.90   0.07   0.02   0.08   0.00     0.03   0.01    5.04 14    3.82  3    4.01  2    4.21  2    4.38  3
107328      4.96   1.15   1.16   0.89   0.61     0.62   0.54    8.19  8    5.50  4    4.56  4    4.24  3    4.07  3
107383      4.74   0.79   1.01   0.79   0.52     0.55   0.47    7.40  5    5.20  4    4.40  3    4.16  2    4.03  2
107705      6.40   0.08   0.60   0.43   0.28     0.29   0.29    7.87  3    6.64  3    6.26  3    6.20  2    6.19  2
107950      4.80   0.62   0.87   0.70   0.44     0.49   0.41    7.15  5    5.18  4    4.52  3    4.33  2    4.24  2
108483      3.91  -0.77  -0.19  -0.11  -0.20    -0.09  -0.13    3.80  3    3.71  2    4.11  3    4.46  3    4.71  3
108767      2.94  -0.09  -0.05  -0.05  -0.04    -0.04  -0.02    3.83 13    2.82  2    3.09  2    3.34  2    3.52  3
109085      4.32   0.02   0.37   0.38   0.18     0.26   0.19    5.59  5    4.43  3    4.25  2    4.28  2    4.33  3
109358      4.27   0.05   0.59   0.54   0.31     0.37   0.31    5.70  4    4.50  3    4.10  2    4.01  2    3.98  2
109379      2.64   0.65   0.88   0.61   0.44     0.42   0.41    5.04  5    3.03  4    2.38  3    2.20  3    2.11  2
109485      4.81  -0.01   0.00   0.06  -0.02     0.03   0.00    5.83 13    4.72  3    4.92  2    5.15  3    5.32  3
109787      3.86   0.03   0.05   0.05   0.04     0.02   0.05    4.96 13    3.79  3    3.96  2    4.13  2    4.27  3
109799      5.45   0.03   0.32   0.31   0.19     0.21   0.20    6.70  6    5.53  3    5.41  2    5.43  2    5.48  3
110014      4.67   1.36   1.22   0.90   0.61     0.64   0.54    8.22 11    5.25  4    4.25  4    3.93  3    3.76  3
110073      4.64  -0.40  -0.08  -0.02  -0.09    -0.02  -0.05    5.09  7    4.50  2    4.79  2    5.06  3    5.26  3
110304      2.17  -0.01  -0.01   0.03   0.00     0.00   0.01    3.19 14    2.07  3    2.29  2    2.50  2    2.66  3
110379      2.74  -0.03   0.36   0.29   0.19     0.19   0.20    3.93  4    2.84  3    2.70  2    2.72  2    2.76  2
110458      4.66   1.01   1.10   0.75   0.57     0.53   0.51    7.67  7    5.17  4    4.31  4    4.03  3    3.87  3
110646      5.93   0.47   0.86   0.67   0.49     0.47   0.45    8.07  3    6.31  4    5.66  3    5.43  2    5.32  2
110897      5.95  -0.03   0.55   0.54   0.29     0.37   0.30    7.25  4    6.16  3    5.79  2    5.72  2    5.70  2
110956      4.65  -0.63  -0.16  -0.11  -0.21    -0.09  -0.14    4.74  4    4.47  2    4.85  2    5.21  3    5.47  3
111397      5.70   0.07   0.02   0.04  -0.01     0.01   0.00    6.84 14    5.62  3    5.81  2    6.03  2    6.20  3
111456      5.85  -0.04   0.46   0.44   0.28     0.30   0.29    7.08  3    6.01  3    5.74  2    5.68  2    5.67  3
111915      4.33   1.58   1.37   0.98   0.75     0.69   0.67    8.26 12    4.99  4    3.85  5    3.40  4    3.15  3
111968      4.27   0.12   0.21   0.20   0.14     0.13   0.14    5.58 11    4.29  3    4.29  2    4.37  2    4.45  3
112092      4.03  -0.75  -0.17  -0.14  -0.27    -0.11  -0.18    3.96  3    3.84  2    4.24  2    4.64  3    4.92  3
112097      6.25   0.05   0.27   0.25   0.15     0.16   0.17    7.50  8    6.30  3    6.24  2    6.30  2    6.36  3
112300      3.38   1.79   1.59   1.53   1.33     1.01   1.21    7.71 11    4.16  5    2.74  8    1.74  6    1.17  5
112412      5.60  -0.03   0.34   0.24   0.23     0.15   0.23    6.78  4    5.69  3    5.58  2    5.57  2    5.59  2
112769      4.79   1.96   1.56   1.35   0.95     0.91   0.86    9.34 15    5.55  5    4.17  5    3.54  3    3.17  3
113139      4.93   0.01   0.36   0.37   0.21     0.25   0.22    6.18  5    5.03  3    4.87  2    4.87  2    4.90  3
```

```
113226      2.84   0.74   0.94   0.64   0.45      0.45  0.42     5.39  5    3.26  4    2.56  4    2.37  3    2.27  3
113703      4.71  -0.57  -0.14  -0.06  -0.13     -0.05 -0.04     4.90  5    4.54  2    4.89  3    5.15  3    5.35  3
113791      4.27  -0.76  -0.19  -0.09  -0.19     -0.07 -0.10     4.17  3    4.07  2    4.47  3    4.79  3    5.02  4
113996      4.82   1.87   1.45   1.18   0.81      0.82  0.73     9.19 16    5.52  4    4.27  4    3.77  3    3.48  3
114038      5.19   1.12   1.14   0.82   0.58      0.58  0.52     8.37  8    5.72  4    4.81  4    4.52  3    4.36  3
114330      4.38   0.01  -0.01   0.05   0.00      0.02  0.01     5.42 14    4.28  3    4.50  2    4.70  3    4.87  3
114613      4.85   0.31   0.70   0.57   0.36      0.39  0.35     6.69  3    5.14  3    4.65  3    4.52  2    4.47  2
114710      4.26   0.08   0.58   0.49   0.30      0.33  0.31     5.72  3    4.48  3    4.11  2    4.03  2    4.00  2
115617      4.74   0.26   0.71   0.58   0.36      0.40  0.35     6.52  3    5.04  3    4.53  3    4.41  2    4.36  2
115659      3.00   0.66   0.92   0.60   0.47      0.41  0.43     5.44  4    3.41  4    2.74  4    2.53  3    2.43  3
115892      2.73   0.01   0.03   0.05  -0.01      0.02  0.00     3.80 13    2.65  3    2.84  2    3.05  2    3.22  3
116656      2.06   0.03   0.02  -0.04  -0.02     -0.04 -0.01     3.15 14    1.98  3    2.19  2    2.42  2    2.59  3
116842      4.02   0.08   0.16   0.17   0.07      0.10  0.09     5.25 11    4.01  3    4.06  2    4.20  2    4.31  3
116976      4.75   1.06   1.10   0.79   0.53      0.56  0.48     7.83  8    5.26  4    4.39  4    4.13  3    4.00  3
117176      4.98   0.26   0.71   0.61   0.39      0.42  0.37     6.76  3    5.28  3    4.76  2    4.61  2    4.55  2
117440      3.88   1.04   1.16   0.84   0.59      0.59  0.52     6.96  6    4.42  4    3.49  4    3.19  3    3.03  3
117675      4.69   1.96   1.60   1.46   1.16      0.97  1.05     9.26 14    5.48  5    4.04  6    3.22  4    2.72  4
118098      3.38   0.08   0.12   0.07   0.06      0.03  0.07     4.59 12    3.35  3    3.46  2    3.61  2    3.74  3
118232      4.70   0.11   0.12   0.11   0.02      0.06  0.05     5.95 13    4.67  3    4.77  2    4.94  2    5.08  3
118623      4.83   0.09   0.23   0.27   0.13      0.18  0.14     6.11  9    4.86  3    4.83  2    4.91  2    4.99  3
119605      5.60   0.41   0.81   0.64   0.45      0.45  0.42     7.64  3    5.95  4    5.35  3    5.16  2    5.06  2
119756      4.23   0.00   0.38   0.33   0.21      0.22  0.22     5.47  4    4.34  3    4.17  2    4.18  2    4.21  2
120136      4.50   0.05   0.48   0.41   0.24      0.28  0.25     5.87  3    4.67  3    4.40  2    4.37  2    4.38  2
120315      1.86  -0.68  -0.19  -0.12  -0.18     -0.09 -0.12     1.87  4    1.66  2    2.07  3    2.40  3    2.65  3
120477      4.07   1.87   1.52   1.20   0.87      0.83  0.78     8.47 14    4.81  5    3.50  5    2.94  3    2.62  3
120818      6.65   0.07   0.12   0.08   0.01      0.04  0.03     7.85 12    6.62  3    6.73  2    6.91  2    7.06  3
120955      4.73  -0.56  -0.14  -0.04  -0.13     -0.04 -0.07     4.93  5    4.56  2    4.90  3    5.19  3    5.40  4
121263      2.55  -0.91  -0.22  -0.14  -0.20     -0.11 -0.12     2.23  3    2.33  2    2.77  3    3.11  3    3.36  3
121325      3.41   1.84   1.57   1.26   0.98      0.88  0.95     7.80 12    4.18  5    2.81  5    2.09  3    1.66  3
121370      2.68   0.20   0.58   0.44   0.29      0.30  0.30     4.31  4    2.90  3    2.54  2    2.47  2    2.45  2
121743      3.83  -0.83  -0.21  -0.13  -0.22     -0.10 -0.15     3.62  3    3.62  2    4.04  3    4.41  3    4.67  3
121790      3.87  -0.80  -0.20  -0.14  -0.21     -0.11 -0.14     3.71  3    3.66  2    4.08  3    4.44  3    4.69  3
122408      4.26   0.13   0.10   0.15   0.06      0.09  0.07     5.53 14    4.22  3    4.32  2    4.47  3    4.60  3
123123      3.28   1.04   1.12   0.87   0.55      0.62  0.49     6.34  7    3.80  4    2.89  4    2.62  2    2.48  2
123139      2.06   0.90   0.99   0.76   0.53      0.54  0.48     4.86  7    2.51  4    1.73  3    1.48  2    1.35  2
123299      3.65  -0.08  -0.05  -0.03  -0.07     -0.03 -0.05     4.55 13    3.53  2    3.80  2    4.07  2    4.27  3
123999      4.83   0.07   0.54   0.44   0.29      0.30  0.29     6.26  3    5.03  3    4.70  2    4.63  2    4.62  2
124294      4.21   1.47   1.32   1.07   0.75      0.76  0.67     7.96 11    4.84  4    3.71  4    3.27  3    3.02  3
124850      4.09   0.03   0.51   0.50   0.27      0.34  0.27     5.45  3    4.28  3    3.95  2    3.90  2    3.90  3
124897     -0.05   1.28   1.23   0.97   0.65      0.69  0.58     3.40  9    0.53  4   -0.50  4   -0.85  2   -1.04  3
125162      4.18   0.05   0.08   0.02   0.03      0.00  0.05     5.33 12    4.13  3    4.29  2    4.46  2    4.60  3
```

```
125238      3.55 -0.72 -0.18 -0.10 -0.14    -0.08 -0.07    3.51  3    3.36  2    3.75  3    4.04  3    4.25  3
125337      4.52  0.09  0.13  0.10  0.04     0.05  0.06    5.75 12    4.50  3    4.59  2    4.75  2    4.88  3
125351      4.81  0.92  1.06  0.76  0.53     0.54  0.48    7.67  6    5.30  4    4.46  4    4.21  3    4.08  3
125451      5.41 -0.03  0.38  0.34  0.19     0.23  0.21    6.61  4    5.52  3    5.35  2    5.36  2    5.40  2
125473      4.05 -0.11 -0.03 -0.01 -0.02    -0.02  0.00    4.92 12    3.94  2    4.19  2    4.41  2    4.58  3
125642      6.33  0.05  0.05  0.06 -0.03     0.03 -0.01    7.46 13    6.26  3    6.43  2    6.66  2    6.84  3
125932      4.75  1.53  1.31  0.97  0.67     0.69  0.60    8.58 12    5.38  4    4.28  5    3.91  3    3.71  3
126053      6.27  0.08  0.63  0.56  0.32     0.38  0.32    7.76  4    6.52  3    6.09  2    5.99  2    5.95  2
126660      4.06  0.01  0.50  0.42  0.25     0.28  0.27    5.38  3    4.24  3    3.95  2    3.90  2    3.90  2
126868      4.82  0.21  0.74  0.58  0.37     0.40  0.36    6.55  4    5.13  3    4.61  3    4.46  2    4.40  2
127665      3.59  1.44  1.30  0.92  0.65     0.65  0.59    7.29 11    4.21  4    3.14  5    2.78  3    2.58  3
127700      4.25  1.70  1.44  1.05  0.75     0.74  0.68    8.38 13    4.95  4    3.73  5    3.28  3    3.02  3
128167      4.47 -0.08  0.37  0.34  0.19     0.23  0.22    5.60  3    4.58  3    4.41  2    4.42  2    4.45  3
128345      4.05 -0.56 -0.15 -0.09 -0.11    -0.07 -0.05    4.24  5    3.87  2    4.24  3    4.50  3    4.70  3
129116      4.00 -0.69 -0.17 -0.11 -0.15    -0.09 -0.07    4.01  4    3.81  2    4.20  3    4.49  3    4.70  3
129174      4.54 -0.32 -0.04  0.07 -0.02     0.03  0.02    5.12  8    4.42  2    4.66  3    4.86  3    5.02  4
129456      4.05  1.53  1.35  1.01  0.73     0.72  0.66    7.90 11    4.70  4    3.56  5    3.13  3    2.88  3
129502      3.88 -0.01  0.38  0.40  0.22     0.27  0.23    5.11  4    3.99  3    3.80  2    3.79  2    3.82  3
129685      4.92 -0.03  0.01  0.06  0.02     0.03  0.05    5.92 13    4.83  3    5.03  2    5.20  2    5.34  3
129972      4.60  0.75  0.98  0.73  0.48     0.52  0.45    7.19  4    5.04  4    4.28  3    4.06  2    3.95  2
130109      3.73 -0.03 -0.01  0.07 -0.02     0.03  0.00    4.72 13    3.63  3    3.84  2    4.06  3    4.23  3
130694      4.41  1.47  1.40  1.07  0.76     0.76  0.69    8.21  9    5.09  4    3.89  5    3.43  3    3.17  3
130807      4.33 -0.61 -0.14 -0.05 -0.11    -0.04 -0.05    4.46  4    4.16  2    4.50  3    4.77  3    4.97  3
130819      5.16 -0.04  0.41  0.38  0.19     0.26  0.22    6.37  3    5.29  3    5.08  2    5.09  2    5.12  2
130841      2.75  0.09  0.15  0.14  0.04     0.08  0.06    3.99 11    2.74  3    2.80  2    2.96  2    3.09  3
131873      2.08  1.77  1.47  1.11  0.76     0.79  0.69    6.32 13    2.79  4    1.54  5    1.07  3    0.81  3
132052      4.49  0.04  0.32  0.32  0.17     0.21  0.18    5.76  6    4.57  3    4.45  2    4.49  2    4.54  3
132058      2.68 -0.85 -0.22 -0.10 -0.17    -0.08 -0.09    2.44  3    2.46  2    2.89  3    3.19  3    3.42  4
132200      3.13 -0.78 -0.20 -0.08 -0.16    -0.07 -0.08    3.00  3    2.92  2    3.33  3    3.62  3    3.84  4
133124      4.82  1.83  1.50  1.19  0.79     0.82  0.72    9.16 14    5.55  5    4.25  5    3.76  3    3.48  3
133165      4.40  0.87  1.04  0.81  0.54     0.57  0.49    7.19  5    4.88  4    4.04  3    3.78  2    3.64  2
133208      3.50  0.71  0.97  0.65  0.44     0.45  0.42    6.03  4    3.94  4    3.21  4    3.02  3    2.93  3
133242      3.89 -0.58 -0.14 -0.10 -0.11    -0.08 -0.05    4.06  5    3.72  2    4.08  2    4.34  3    4.54  3
133582      4.55  1.34  1.23  0.93  0.65     0.66  0.59    8.08 10    5.13  4    4.11  4    3.75  3    3.56  3
133955      4.05 -0.68 -0.18 -0.08 -0.15    -0.07 -0.08    4.06  4    3.86  2    4.24  3    4.54  3    4.76  4
134083      4.93 -0.02  0.43  0.40  0.21     0.27  0.24    6.17  3    5.07  3    4.84  2    4.83  2    4.85  2
134481      3.70 -0.12 -0.03 -0.05 -0.01    -0.04  0.03    4.56 12    3.59  2    3.85  2    4.04  2    4.20  3
134687      4.82 -0.67 -0.17 -0.10 -0.16    -0.08 -0.09    4.85  4    4.63  2    5.01  3    5.32  3    5.55  3
134759      4.54 -0.38 -0.08 -0.04 -0.09    -0.04 -0.05    5.02  7    4.40  2    4.70  2    4.97  3    5.18  3
135153      4.91  0.29  0.37  0.36  0.31     0.24  0.28    6.54 10    5.02  3    4.85  2    4.79  2    4.79  3
135734      4.27 -0.37 -0.09 -0.02 -0.04    -0.02  0.00    4.75  8    4.13  2    4.42  2    4.64  3    4.81  3
```

```
135742      2.61 -0.37 -0.11 -0.04 -0.10    -0.04 -0.05    3.08  8   2.45 2   2.77 3   3.04 3   3.24 3
135758      4.33  1.07  1.10  0.81  0.56     0.57  0.51    7.42  8   4.84 4   3.96 4   3.68 3   3.53 3
136422      3.56  1.87  1.54  1.19  0.87     0.82  0.79    7.98 14   4.31 5   2.98 5   2.42 4   2.09 3
136504      3.37 -0.73 -0.18 -0.11 -0.16    -0.09 -0.08    3.32  3   3.18 2   3.57 3   3.87 3   4.09 3
136664      4.54 -0.62 -0.15 -0.06 -0.14    -0.05 -0.07    4.65  4   4.36 2   4.72 3   5.01 3   5.22 3
137006      6.12  0.05  0.26  0.24  0.12     0.15  0.14    7.37  8   6.17 3   6.12 2   6.20 2   6.27 3
137107      4.98  0.04  0.58  0.48  0.28     0.33  0.30    6.39  4   5.20 3   4.83 2   4.76 2   4.74 2
137391      4.32  0.06  0.31  0.30  0.15     0.20  0.17    5.61  7   4.40 3   4.29 2   4.34 2   4.41 3
137759      3.29  1.23  1.16  0.78  0.60     0.55  0.54    6.63 10   3.83 4   2.92 5   2.60 3   2.43 3
138481      5.02  1.91  1.59  1.27  0.93     0.87  0.85    9.52 13   5.80 5   4.41 5   3.80 3   3.43 3
138629      5.02  0.11  0.07  0.15  0.04     0.09  0.06    6.24 14   4.96 3   5.09 2   5.25 3   5.38 3
138716      4.62  0.85  1.01  0.77  0.52     0.54  0.48    7.36  6   5.08 4   4.28 3   4.03 2   3.90 2
138905      3.91  0.74  1.02  0.71  0.55     0.50  0.50    6.51  4   4.38 4   3.59 4   3.31 3   3.17 3
139063      3.57  1.59  1.39  1.00  0.71     0.71  0.64    7.52 12   4.24 4   3.08 5   2.66 3   2.43 3
139127      4.33  1.72  1.43  1.03  0.73     0.73  0.66    8.48 13   5.02 4   3.82 5   3.38 4   3.14 3
139365      3.65 -0.69 -0.18 -0.11 -0.19    -0.09 -0.11    3.65  4   3.46 2   3.85 3   4.18 3   4.42 3
139521      4.67  0.73  1.00  0.68  0.49     0.48  0.47    7.24  4   5.13 4   4.36 4   4.12 3   4.00 3
139663      4.96  1.51  1.33  0.96  0.68     0.68  0.61    8.77 11   5.60 4   4.49 5   4.10 3   3.89 3
139664      4.64 -0.02  0.40  0.35  0.20     0.23  0.22    5.87  3   4.77 3   4.57 2   4.57 2   4.60 2
139798      5.75 -0.02  0.36  0.33  0.18     0.22  0.20    6.96  4   5.85 3   5.70 2   5.72 2   5.76 3
139997      4.72  1.94  1.58  1.28  0.94     0.87  0.86    9.25 14   5.49 5   4.11 5   3.49 3   3.12 3
140008      4.75 -0.54 -0.14 -0.10 -0.16    -0.08 -0.09    4.98  6   4.58 2   4.94 2   5.25 3   5.48 3
140027      6.01  0.61  0.90  0.69  0.46     0.49  0.43    8.37  4   6.41 4   5.72 3   5.52 2   5.41 2
140159      4.53  0.02  0.05  0.04 -0.01     0.01  0.01    5.62 13   4.46 3   4.64 2   4.84 2   5.01 3
140573      2.64  1.25  1.17  0.81  0.56     0.57  0.51    6.01 10   3.19 4   2.25 4   1.97 3   1.82 3
140775      5.58  0.03  0.04  0.04 -0.01     0.01  0.01    6.68 13   5.51 3   5.69 2   5.90 2   6.06 3
141004      4.43  0.10  0.60  0.51  0.32     0.35  0.32    5.93  3   4.67 3   4.27 2   4.17 2   4.13 2
141477      4.09  1.95  1.62  1.25  0.98     0.86  0.89    8.66 13   4.89 5   3.48 6   2.82 4   2.42 3
141513      3.53 -0.10 -0.04 -0.01 -0.05    -0.02 -0.02    4.41 12   3.41 2   3.67 2   3.91 2   4.09 3
141556      3.95 -0.14 -0.04 -0.02 -0.07    -0.02 -0.04    4.77 11   3.83 2   4.09 2   4.35 2   4.54 3
141637      4.68 -0.73 -0.06  0.01 -0.10    -0.01 -0.03    4.69  3   4.55 2   4.82 2   5.07 3   5.26 3
141680      5.23  0.82  1.02  0.77  0.53     0.54  0.48    7.94  5   5.70 4   4.89 4   4.63 2   4.50 2
141795      3.70  0.09  0.16  0.08  0.05     0.04  0.08    4.94 11   3.69 3   3.77 2   3.91 2   4.04 2
141891      2.85  0.05  0.29  0.32  0.15     0.21  0.18    4.11  7   2.91 3   2.82 2   2.86 2   2.92 3
141992      4.78  1.88  1.54  1.22  0.85     0.84  0.77    9.21 14   5.53 5   4.20 5   3.65 3   3.34 3
142091      4.82  0.87  1.00  0.76  0.49     0.54  0.45    7.58  6   5.28 4   4.49 3   4.26 2   4.14 2
142105      4.32  0.05  0.04  0.06  0.01     0.03  0.04    5.44 14   4.25 3   4.42 2   4.60 2   4.75 3
142114      4.59 -0.67 -0.06 -0.02 -0.09    -0.02 -0.03    4.68  3   4.46 2   4.74 2   4.98 3   5.17 3
142198      4.16  0.82  1.01  0.72  0.52     0.51  0.48    6.86  5   4.62 4   3.84 4   3.59 3   3.46 2
142373      4.62  0.00  0.57  0.48  0.32     0.33  0.33    5.97  4   4.84 3   4.48 2   4.37 2   4.33 2
142669      3.86 -0.82 -0.20 -0.11 -0.20    -0.09 -0.12    3.67  3   3.65 2   4.06 3   4.40 3   4.64 4
```

```
142780      5.38   1.97   1.64   1.49   1.25      0.99   1.15      9.99 13    6.19   5    4.71   7    3.80   4    3.24   4
142860      3.86  -0.03   0.48   0.49   0.24      0.33   0.26      5.12  3    4.03   3    3.73   2    3.69   2    3.70   3
143107      4.15   1.28   1.23   0.89   0.62      0.63   0.56      7.60  9    4.73   4    3.73   4    3.39   3    3.21   3
143118      3.41  -0.84  -0.22  -0.14  -0.21     -0.11  -0.17      3.19  3    3.19   2    3.63   3    4.01   3    4.29   3
143546      4.65   0.63   0.92   0.67   0.47      0.47   0.44      7.04  4    5.06   4    4.36   3    4.15   2    4.04   2
143699      4.89  -0.57  -0.14  -0.07  -0.14     -0.06  -0.08      5.08  5    4.72   2    5.07   3    5.36   3    5.58   3
143807      4.98  -0.20  -0.07   0.01  -0.08     -0.01  -0.05      5.71 11    4.85   2    5.12   2    5.39   3    5.59   3
143894      4.83   0.05   0.07   0.10   0.01      0.05   0.04      5.97 13    4.77   3    4.91   2    5.10   2    5.24   3
144197      4.72   0.15   0.23   0.20   0.13      0.13   0.13      6.08 11    4.75   3    4.74   2    4.83   2    4.92   3
144206      4.76  -0.32  -0.11   0.02  -0.10      0.00  -0.07      5.30  9    4.60   2    4.91   3    5.19   3    5.41   4
144217      2.59  -0.86  -0.08  -0.02  -0.09     -0.02  -0.01      2.42  5    2.45   2    2.74   2    2.97   3    3.15   3
144284      4.03   0.10   0.52   0.45   0.25      0.31   0.27      5.49  3    4.22   3    3.91   2    3.86   2    3.85   2
144294      4.23  -0.68  -0.18  -0.10  -0.17     -0.08  -0.09      4.24  4    4.04   2    4.43   3    4.73   3    4.96   3
144470      3.97  -0.82  -0.05   0.06  -0.08      0.03   0.00      3.87  5    3.85   2    4.09   3    4.31   3    4.49   3
144608      4.33   0.51   0.84   0.65   0.43      0.45   0.41      6.52  3    4.70   4    4.07   3    3.89   2    3.80   2
145328      4.76   0.87   1.01   0.75   0.54      0.53   0.49      7.53  6    5.22   4    4.43   4    4.16   3    4.02   2
145482      4.59  -0.74  -0.16  -0.05  -0.17     -0.04  -0.09      4.53  3    4.41   2    4.77   3    5.08   3    5.31   4
145502      4.01  -0.64   0.03   0.11   0.03      0.06   0.09      4.19  4    3.93   3    4.10   2    4.23   3    4.35   3
145502      4.01  -0.64   0.03   0.11   0.03      0.06   0.09      4.19  4    3.93   3    4.10   2    4.23   3    4.35   3
145570      4.93   0.10   0.09   0.12   0.02      0.07   0.05      6.15 13    4.88   3    5.00   2    5.18   2    5.32   3
146051      2.75   1.96   1.59   1.29   1.03      0.88   0.94      7.32 14    3.53   5    2.14   5    1.43   4    1.00   3
146624      4.77  -0.01   0.02   0.08  -0.05      0.04  -0.02      5.80 13    4.69   3    4.87   2    5.12   3    5.30   3
146686      4.02   1.17   1.08   0.78   0.58      0.55   0.52      7.23 11    4.52   4    3.66   4    3.37   3    3.21   3
146738      5.78   0.10   0.07   0.07   0.00      0.03   0.02      6.99 14    5.72   3    5.87   2    6.07   2    6.22   3
146791      3.23   0.75   0.98   0.70   0.49      0.49   0.45      5.82  4    3.67   4    2.92   4    2.69   3    2.58   2
146834      6.28   0.77   1.07   0.83   0.62      0.59   0.56      8.95  3    6.77   4    5.91   4    5.58   2    5.40   2
147084      4.57   0.66   0.84   0.87   0.84      0.62   0.76      6.96  6    4.94   4    4.24   2    3.71   2    3.40   3
147449      4.82   0.02   0.34   0.31   0.15      0.21   0.17      6.07  5    4.91   3    4.78   2    4.83   2    4.89   2
147584      4.91   0.03   0.55   0.54   0.28      0.37   0.30      6.29  3    5.12   3    4.75   2    4.68   2    4.66   2
147677      4.85   0.80   0.97   0.73   0.46      0.52   0.43      7.50  6    5.29   4    4.53   3    4.33   2    4.23   2
147700      4.50   0.84   1.03   0.78   0.50      0.55   0.46      7.24  5    4.97   4    4.16   4    3.92   2    3.80   2
147933      4.63  -0.55   0.22   0.36   0.31      0.24   0.30      5.04  6    4.66   3    4.60   2    4.52   3    4.51   4
147971      4.47  -0.53  -0.07   0.04  -0.04      0.01   0.03      4.75  4    4.34   2    4.60   3    4.80   3    4.95   4
148367      4.63   0.06   0.16   0.16   0.07      0.10   0.09      5.83 11    4.62   3    4.68   2    4.80   2    4.92   3
148387      2.74   0.70   0.91   0.61   0.46      0.42   0.43      5.22  5    3.15   4    2.48   4    2.27   3    2.17   2
148605      4.79  -0.79  -0.07  -0.06  -0.14     -0.05  -0.07      4.72  4    4.66   2    4.95   2    5.23   2    5.45   3
148703      4.23  -0.80  -0.16  -0.06  -0.15     -0.05  -0.07      4.09  3    4.05   2    4.41   3    4.70   3    4.92   4
148786      4.27   0.71   0.92   0.65   0.44      0.45   0.42      6.77  5    4.68   4    3.99   4    3.80   3    3.71   2
148856      2.77   0.69   0.94   0.64   0.47      0.45   0.44      5.26  4    3.19   4    2.49   4    2.28   3    2.17   3
148857      3.83   0.02   0.01  -0.01  -0.01     -0.02   0.01      4.90 14    3.74   3    3.96   2    4.17   2    4.33   3
149161      4.85   1.82   1.49   1.20   0.86      0.83   0.78      9.17 14    5.57   4    4.28   5    3.73   3    3.41   3
```

```
149212     5.01 -0.11 -0.06  0.05 -0.02   0.02  0.01   5.86 13   4.88 2   5.14 3   5.35 3   5.52 4
149438     2.81 -1.01 -0.25 -0.12 -0.25  -0.09 -0.15   2.34  4   2.58 2   3.03 3   3.40 3   3.66 4
149447     4.16  1.94  1.57  1.20  1.00   0.84  0.97   8.69 15   4.93 5   3.57 5   2.84 4   2.40 3
149630     4.20 -0.10 -0.01  0.03 -0.01   0.00  0.02   5.09 12   4.10 3   4.32 2   4.52 2   4.68 3
150680     2.81  0.21  0.65  0.51  0.32   0.35  0.33   4.49  3   3.07 3   2.64 2   2.53 2   2.50 2
150997     3.50  0.61  0.92  0.67  0.48   0.47  0.45   5.87  4   3.91 4   3.21 3   2.99 2   2.88 2
151613     4.84 -0.06  0.39  0.35  0.21   0.23  0.22   6.01  3   4.96 3   4.78 2   4.77 2   4.80 2
151680     2.29  1.16  1.16  0.86  0.60   0.61  0.54   5.53  8   2.83 4   1.89 4   1.58 3   1.41 3
151769     4.66  0.07  0.47  0.45  0.25   0.31  0.27   6.05  4   4.82 3   4.55 2   4.50 2   4.50 2
151985     3.57 -0.84 -0.21 -0.13 -0.22  -0.10 -0.14   3.35  3   3.36 2   3.78 3   4.14 3   4.40 3
152614     4.38 -0.32 -0.08 -0.09 -0.08  -0.07 -0.03   4.94  9   4.24 2   4.55 2   4.80 2   4.99 3
152863     6.08  0.62  0.92  0.70  0.48   0.49  0.45   8.46  4   6.49 4   5.78 3   5.56 2   5.45 2
153597     4.90 -0.03  0.48  0.45  0.27   0.31  0.29   6.16  3   5.07 3   4.78 2   4.72 2   4.71 2
153808     3.92 -0.11 -0.01 -0.01 -0.04  -0.02 -0.02   4.80 11   3.82 3   4.05 2   4.29 2   4.47 3
154143     4.98  1.93  1.60  1.44  1.21   0.96  1.11   9.51 13   5.77 5   4.33 6   3.46 4   2.93 4
154494     4.91  0.06  0.12  0.12  0.02   0.07  0.05   6.09 12   4.88 3   4.98 2   5.15 2   5.29 3
155125     2.42  0.09  0.05  0.03  0.01   0.00  0.04   3.60 14   2.35 3   2.53 2   2.71 2   2.86 3
155203     3.34  0.09  0.40  0.36  0.20   0.24  0.22   4.72  5   3.47 3   3.27 2   3.27 2   3.30 2
155763     3.17 -0.43 -0.11 -0.06 -0.12  -0.05 -0.06   3.56  7   3.01 2   3.34 2   3.62 3   3.83 3
156164     3.13  0.08  0.08  0.05  0.03   0.02  0.05   4.32 13   3.08 3   3.23 2   3.39 2   3.53 3
156266     4.72  1.11  1.15  0.80  0.59   0.57  0.53   7.89  7   5.26 4   4.34 4   4.04 3   3.87 3
156274     5.48  0.38  0.80  0.72  0.46   0.51  0.44   7.47  3   5.83 3   5.21 2   4.99 2   4.89 2
156283     3.16  1.66  1.44  0.96  0.72   0.68  0.65   7.24 12   3.86 4   2.67 6   2.24 4   2.00 4
156384     5.91  0.82  1.04  0.94  0.59   0.67  0.57   8.63  4   6.39 4   5.52 3   5.18 2   4.99 2
156729     4.66 -0.03  0.05  0.09  0.00   0.05  0.03   5.68 11   4.59 3   4.75 2   4.94 2   5.09 3
156897     4.38 -0.06  0.41  0.37  0.22   0.25  0.23   5.56  3   4.51 3   4.30 2   4.30 2   4.32 2
156928     4.31  0.04  0.03  0.06  0.00   0.03  0.03   5.42 14   4.23 3   4.41 2   4.60 2   4.76 3
157792     4.16  0.12  0.28  0.29  0.12   0.19  0.14   5.51  9   4.22 3   4.14 2   4.22 2   4.29 3
157950     4.54 -0.08  0.41  0.38  0.21   0.26  0.23   5.69  3   4.67 3   4.46 2   4.45 2   4.48 2
157999     4.33  1.58  1.50  1.10  0.77   0.78  0.70   8.33  8   5.06 5   3.78 5   3.31 4   3.04 3
158408     2.68 -0.81 -0.23 -0.14 -0.23  -0.11 -0.13   2.49  3   2.46 2   2.90 3   3.25 3   3.50 4
158899     4.41  1.68  1.44  1.04  0.75   0.74  0.68   8.51 12   5.11 4   3.89 5   3.44 4   3.18 3
159217     4.59 -0.10 -0.02 -0.01 -0.05  -0.02 -0.02   5.48 12   4.48 2   4.73 2   4.97 2   5.16 3
159433     4.29  0.90  1.09  0.75  0.55   0.53  0.50   7.14  5   4.79 4   3.94 4   3.67 3   3.52 3
159532     1.87  0.22  0.40  0.35  0.20   0.23  0.21   3.42  8   2.00 3   1.80 2   1.82 2   1.85 2
159541     4.88  0.03  0.26  0.24  0.13   0.15  0.16   6.10  7   4.93 3   4.88 2   4.94 2   5.01 3
159560     4.88  0.04  0.27  0.24  0.13   0.15  0.15   6.12  7   4.93 3   4.88 2   4.94 2   5.02 3
159561     2.07  0.10  0.15  0.14  0.08   0.08  0.10   3.32 12   2.06 3   2.12 2   2.25 2   2.36 3
159876     3.52  0.12  0.24  0.20  0.13   0.13  0.15   4.85 10   3.56 3   3.53 2   3.61 2   3.68 3
159975     4.63 -0.18  0.11  0.18  0.11   0.11  0.13   5.48  7   4.60 3   4.68 2   4.77 3   4.86 3
160032     4.77 -0.04  0.40  0.29  0.22   0.19  0.24   5.97  3   4.90 3   4.72 2   4.70 2   4.72 2
```

```
160835      6.36  1.18  1.20  0.92  0.59    0.65  0.53     9.65  8    6.93  4    5.93  4    5.63  3    5.47  3
160915      4.87 -0.03  0.47  0.41  0.24    0.28  0.26     6.12  3    5.03  3    4.77  2    4.73  2    4.74  2
160922      4.80 -0.01  0.43  0.41  0.22    0.28  0.24     6.06  3    4.94  3    4.71  2    4.69  2    4.71  3
161096      2.77  1.24  1.17  0.82  0.57    0.58  0.52     6.13 10    3.32  4    2.38  4    2.09  3    1.94  3
161868      3.75  0.04  0.04  0.04  0.00    0.01  0.03     4.86 13    3.68  3    3.86  2    4.05  2    4.20  3
162003      4.58  0.01  0.43  0.38  0.23    0.26  0.25     5.87  3    4.72  3    4.50  2    4.47  2    4.49  2
163588      3.75  1.21  1.18  0.83  0.59    0.59  0.53     7.07  9    4.30  4    3.36  4    3.05  3    2.89  3
163770      3.87  1.46  1.35  0.90  0.63    0.64  0.57     7.63 10    4.52  4    3.42  5    3.07  4    2.89  4
163917      3.34  0.87  0.99  0.71  0.48    0.50  0.45     6.10  7    3.79  4    3.03  4    2.80  3    2.69  2
163955      4.75 -0.03 -0.05  0.05 -0.02    0.02  0.01     5.72 14    4.63  2    4.88  3    5.09  3    5.26  3
164058      2.22  1.88  1.52  1.14  0.85    0.80  0.77     6.64 15    2.96  5    1.66  5    1.11  4    0.80  3
164259      4.62 -0.01  0.39  0.33  0.19    0.22  0.21     5.86  4    4.74  3    4.56  2    4.57  2    4.61  2
164349      4.68  1.21  1.27  0.87  0.59    0.62  0.53     8.05  7    5.28  4    4.25  5    3.95  3    3.78  3
164353      3.97 -0.65  0.02  0.10  0.00    0.05  0.07     4.14  4    3.89  3    4.07  2    4.22  3    4.34  3
164577      4.42  0.02  0.04  0.07  0.01    0.03  0.04     5.50 13    4.35  3    4.52  2    4.70  2    4.85  3
165135      2.99  0.77  1.01  0.73  0.51    0.52  0.47     5.62  4    3.45  4    2.67  4    2.42  3    2.29  2
165634      4.55  0.76  0.95  0.66  0.52    0.47  0.48     7.14  5    4.98  4    4.26  4    4.01  3    3.88  2
165760      4.63  0.73  0.97  0.69  0.50    0.49  0.46     7.19  4    5.07  4    4.33  4    4.09  3    3.97  2
165777      3.73  0.09  0.12  0.14  0.05    0.08  0.08     4.95 12    3.70  3    3.79  2    3.94  2    4.06  3
166063      4.53  0.78  1.01  0.70  0.48    0.49  0.45     7.18  4    4.99  4    4.21  4    3.99  3    3.88  3
166182      4.35 -0.81 -0.15 -0.06 -0.18   -0.05 -0.13     4.20  3    4.17  2    4.53  3    4.88  3    5.13  3
166205      4.36  0.03  0.02  0.05 -0.01    0.02  0.01     5.45 14    4.28  3    4.47  2    4.68  2    4.84  3
166620      6.40  0.59  0.87  0.77  0.49    0.54  0.47     8.71  4    6.78  4    6.10  3    5.85  2    5.73  2
167818      4.63  1.80  1.66  1.21  0.93    0.84  0.85     9.02  9    5.45  5    4.02  6    3.40  4    3.04  4
168454      2.70  1.55  1.38  1.00  0.68    0.71  0.61     6.59 11    3.36  4    2.21  5    1.82  3    1.60  3
168656      4.84  0.61  0.91  0.69  0.45    0.49  0.42     7.20  4    5.25  4    4.55  3    4.35  2    4.25  2
168720      4.96  2.00  1.58  1.34  1.00    0.91  0.91     9.57 16    5.73  5    4.34  5    3.66  3    3.25  3
168723      3.25  0.65  0.94  0.70  0.50    0.49  0.46     5.68  4    3.67  4    2.95  3    2.71  2    2.59  2
168775      4.34  1.19  1.17  0.86  0.55    0.61  0.50     7.63  9    4.89  4    3.94  4    3.66  3    3.52  3
169022      1.85 -0.13 -0.03  0.00 -0.01   -0.01  0.02     2.69 11    1.74  2    1.98  2    2.18  2    2.34  3
169156      4.68  0.72  0.94  0.72  0.47    0.51  0.44     7.21  5    5.10  4    4.37  3    4.16  2    4.05  2
169414      3.84  1.16  1.18  0.85  0.60    0.60  0.54     7.10  8    4.39  4    3.44  4    3.13  3    2.96  3
169420      4.81  0.93  1.30  1.10  0.84    0.78  0.76     7.82  3    5.43  4    4.31  4    3.77  2    3.46  2
169467      3.51 -0.65 -0.17 -0.13 -0.21   -0.10 -0.14     3.57  4    3.32  2    3.71  2    4.07  3    4.33  3
169916      2.81  0.90  1.04  0.75  0.56    0.53  0.51     5.64  6    3.29  4    2.47  4    2.19  3    2.04  3
169981      5.83  0.09  0.07  0.08  0.04    0.04  0.06     7.03 14    5.77  3    5.92  2    6.08  2    6.21  3
170073      4.99  0.05  0.08  0.05  0.02    0.02  0.05     6.14 12    4.94  3    5.09  2    5.26  2    5.40  3
170153      3.58 -0.06  0.49  0.44  0.31    0.30  0.31     4.80  3    3.75  3    3.47  2    3.38  2    3.36  2
170296      4.71  0.04  0.07  0.08  0.05    0.04  0.07     5.84 12    4.65  3    4.80  2    4.95  2    5.08  3
170693      4.82  1.11  1.19  0.86  0.63    0.61  0.57     8.01  7    5.38  4    4.41  4    4.07  3    3.89  3
170845      4.64  0.76  1.02  0.69  0.50    0.49  0.46     7.27  4    5.11  4    4.32  4    4.09  3    3.97  3
```

```
171443      3.83   1.53   1.34   0.97   0.68      0.69   0.61      7.68 12    4.47  4    3.36  5    2.97  3    2.75  3
171635      4.80   0.44   0.61   0.53   0.31      0.36   0.29      6.77  7    5.04  3    4.63  2    4.56  2    4.55  2
171978      5.74   0.06   0.06   0.07   0.04      0.03   0.06      6.89 13    5.68  3    5.83  2    6.00  2    6.13  3
172167      0.03   0.00   0.00  -0.04  -0.03     -0.04  -0.01      1.06 13   -0.06  3    0.17  2    0.39  2    0.57  3
172910      4.87  -0.71  -0.18  -0.14  -0.16     -0.11  -0.07      4.84  4    4.68  2    5.08  2    5.37  3    5.59  3
173300      3.16  -0.36  -0.11  -0.01  -0.11     -0.02  -0.06      3.65  9    3.00  2    3.32  3    3.59  3    3.80  3
173648      4.36   0.16   0.19   0.15   0.08      0.09   0.10      5.72 12    4.37  3    4.40  2    4.52  2    4.62  3
173654      5.64   0.08   0.14   0.13   0.07      0.08   0.09      6.86 12    5.62  3    5.70  2    5.83  2    5.94  3
173667      4.19   0.02   0.46   0.39   0.26      0.26   0.27      5.51  3    4.35  3    4.10  2    4.05  2    4.04  2
173764      4.22   0.84   1.09   0.79   0.57      0.56   0.51      6.99  4    4.72  4    3.86  4    3.58  3    3.43  3
173780      4.84   1.22   1.20   0.88   0.61      0.62   0.55      8.19  9    5.41  4    4.43  4    4.10  3    3.93  3
173880      4.36   0.08   0.12   0.09   0.01      0.05   0.04      5.57 12    4.33  3    4.44  2    4.61  2    4.76  3
174974      4.83   1.28   1.40   1.01   0.69      0.72   0.62      8.37  5    5.51  4    4.33  5    3.93  3    3.71  3
175190      4.98   1.50   1.32   0.94   0.66      0.67   0.60      8.77 11    5.61  4    4.52  5    4.15  3    3.95  3
175191      2.03  -0.75  -0.22  -0.11  -0.21     -0.09  -0.13      1.93  4    1.81  2    2.24  3    2.59  3    2.84  4
175492      4.60   0.50   0.78   0.64   0.46      0.45   0.43      6.74  4    4.93  3    4.36  3    4.15  2    4.05  2
175535      4.93   0.57   0.90   0.68   0.45      0.48   0.42      7.23  3    5.33  4    4.65  3    4.45  2    4.35  2
175775      3.51   1.13   1.18   0.80   0.59      0.57   0.53      6.73  7    4.06  4    3.12  5    2.82  3    2.66  3
176303      5.22   0.07   0.53   0.42   0.30      0.28   0.30      6.64  3    5.42  3    5.10  2    5.02  2    5.00  2
176411      4.02   1.04   1.08   0.76   0.52      0.54   0.48      7.06  8    4.52  4    3.67  4    3.42  3    3.29  3
176437      3.24  -0.08  -0.05  -0.03  -0.01     -0.03   0.01      4.14 13    3.12  2    3.39  2    3.60  2    3.76  3
176524      4.82   1.10   1.15   0.85   0.56      0.60   0.51      7.98  7    5.36  4    4.43  4    4.14  3    3.99  3
176678      4.02   1.04   1.09   0.79   0.54      0.56   0.49      7.06  8    4.52  4    3.66  4    3.39  3    3.25  3
176687      2.59   0.05   0.08   0.05   0.01      0.02   0.04      3.74 12    2.54  3    2.69  2    2.87  2    3.01  3
177196      5.02   0.09   0.19   0.21   0.09      0.13   0.10      6.28 10    5.03  3    5.04  2    5.16  2    5.27  3
177241      3.77   0.86   1.00   0.72   0.53      0.51   0.48      6.52  6    4.23  4    3.45  4    3.20  3    3.07  2
177716      3.31   1.14   1.20   0.87   0.59      0.62   0.52      6.55  7    3.88  4    2.90  4    2.60  3    2.44  3
177724      2.99  -0.01   0.01   0.01   0.00     -0.01   0.01      4.02 13    2.90  3    3.11  2    3.32  2    3.48  3
177756      3.43  -0.27  -0.09  -0.03  -0.09     -0.03  -0.06      4.05 10    3.29  2    3.59  2    3.87  3    4.08  3
178253      4.11   0.09   0.04   0.04   0.00      0.01   0.01      5.29 14    4.04  3    4.22  2    4.43  2    4.59  3
178345      4.11   1.07   1.20   0.82   0.61      0.58   0.58      7.25  5    4.68  4    3.71  5    3.36  3    3.16  3
179950      4.83   0.32   0.55   0.49   0.34      0.33   0.30      6.60  6    5.04  3    4.69  2    4.61  2    4.59  2
180163      4.38  -0.67  -0.14  -0.10  -0.15     -0.08  -0.08      4.43  3    4.21  2    4.57  2    4.87  3    5.09  3
180554      4.77  -0.55  -0.04   0.01  -0.07     -0.01  -0.03      5.04  3    4.65  2    4.91  2    5.15  3    5.34  3
180711      3.07   0.78   1.00   0.70   0.51      0.49   0.46      5.71  5    3.53  4    2.76  4    2.52  3    2.40  3
180777      5.14   0.00   0.31   0.31   0.17      0.21   0.18      6.35  6    5.22  3    5.11  2    5.15  2    5.20  3
180809      4.37   1.23   1.25   0.87   0.59      0.62   0.52      7.76  7    4.96  4    3.95  5    3.65  3    3.49  3
181276      3.76   0.74   0.97   0.63   0.47      0.44   0.43      6.33  5    4.20  4    3.48  4    3.27  3    3.17  3
181984      4.45   1.45   1.25   0.90   0.58      0.64   0.52      8.14 12    5.04  4    4.02  4    3.73  3    3.57  3
182564      4.59   0.06   0.02   0.04  -0.03      0.01  -0.01      5.72 14    4.51  3    4.70  2    4.94  2    5.12  3
182568      4.97  -0.72  -0.09   0.02  -0.14      0.00  -0.07      4.98  3    4.83  2    5.11  3    5.40  3    5.62  3
```

```
182640      3.36   0.04   0.32   0.25   0.16    0.16   0.17    4.63  6   3.44  3   3.34  2   3.39  2   3.45  2
182835      4.67   0.51   0.60   0.51   0.46    0.35   0.38    6.73  9   4.91  3   4.51  2   4.35  2   4.28  2
183439      4.45   1.82   1.50   1.21   0.97    0.84   0.88    8.78 14   5.18  5   3.88  5   3.23  3   2.85  3
183912      3.08   0.62   1.13   0.87   0.66    0.62   0.59    5.58  5   3.61  4   2.69  4   2.32  3   2.12  3
184006      3.79   0.11   0.14   0.11   0.07    0.06   0.08    5.05 12   3.77  3   3.85  2   3.99  2   4.11  3
184171      4.74  -0.67  -0.13   0.01  -0.16   -0.01  -0.10    4.80  3   4.57  2   4.90  3   5.22  3   5.46  4
184406      4.45   1.24   1.18   0.91   0.61    0.64   0.54    7.81  9   5.00  4   4.03  4   3.72  3   3.55  3
184606      4.99  -0.43  -0.07  -0.01  -0.09   -0.02  -0.05    5.40  6   4.86  2   5.14  2   5.40  3   5.60  3
184707      4.60  -0.15  -0.06  -0.03  -0.06   -0.03  -0.04    5.40 12   4.47  2   4.75  2   5.01  3   5.21  3
184915      4.96  -0.87   0.00   0.06  -0.04    0.03   0.03    4.82  7   4.87  3   5.07  2   5.26  3   5.41  3
184930      4.36  -0.44  -0.09   0.02  -0.08    0.00  -0.04    4.75  6   4.22  2   4.50  3   4.76  3   4.95  3
184960      5.74   0.00   0.48   0.46   0.28    0.31   0.28    7.04  3   5.91  3   5.62  2   5.56  2   5.55  3
185144      4.69   0.37   0.80   0.65   0.41    0.45   0.39    6.67  3   5.04  3   4.44  3   4.27  2   4.20  2
185395      4.47  -0.02   0.39   0.35   0.21    0.23   0.22    5.69  4   4.59  3   4.41  2   4.40  2   4.43  2
185734      4.70   0.79   0.95   0.68   0.47    0.48   0.43    7.33  6   5.13  4   4.40  4   4.20  3   4.09  2
185958      4.37   0.89   1.05   0.71   0.50    0.50   0.46    7.19  5   4.85  4   4.04  4   3.81  3   3.69  3
186408      5.95   0.20   0.64   0.45   0.33    0.31   0.33    7.61  3   6.21  3   5.80  3   5.69  2   5.65  2
186427      6.20   0.21   0.66   0.44   0.34    0.30   0.33    7.88  3   6.47  3   6.05  3   5.94  2   5.90  2
186791      2.72   1.68   1.52   1.07   0.75    0.76   0.67    6.87 10   3.46  5   2.18  6   1.73  4   1.48  4
186882      2.87  -0.10  -0.02  -0.01  -0.02   -0.02  -0.01    3.76 12   2.76  2   3.01  2   3.23  2   3.41  3
187362      5.00   0.06   0.10   0.14   0.06    0.08   0.07    6.17 12   4.96  3   5.06  2   5.22  2   5.34  3
188119      3.82   0.52   0.89   0.64   0.48    0.45   0.44    6.05  3   4.21  4   3.55  3   3.34  3   3.23  2
188209      5.62  -0.97  -0.07   0.10  -0.12    0.05  -0.03    5.30  8   5.49  2   5.74  3   5.99  3   6.18  4
188260      4.58  -0.13  -0.06   0.03  -0.05    0.00  -0.03    5.41 12   4.45  2   4.71  2   4.96  3   5.15  3
188310      4.68   0.89   1.05   0.76   0.57    0.54   0.51    7.50  5   5.16  4   4.34  4   4.05  3   3.90  3
188376      4.70   0.32   0.75   0.59   0.37    0.41   0.36    6.58  3   5.02  3   4.48  3   4.35  2   4.29  2
188512      3.72   0.49   0.86   0.66   0.49    0.47   0.47    5.89  3   4.10  4   3.45  3   3.21  2   3.08  2
188603      4.50   1.56   1.46   1.02   0.73    0.72   0.65    8.45  9   5.21  4   3.98  5   3.56  4   3.32  4
188892      4.95  -0.52  -0.09   0.01  -0.10   -0.01  -0.06    5.23  5   4.81  2   5.10  3   5.37  3   5.58  3
188947      3.93   0.88   1.03   0.74   0.52    0.52   0.47    6.72  6   4.40  4   3.60  4   3.35  3   3.23  3
189005      4.82   0.54   0.89   0.65   0.47    0.45   0.43    7.08  3   5.21  4   4.55  3   4.34  2   4.24  2
189037      4.92   0.06   0.12   0.10   0.06    0.05   0.08    6.10 12   4.89  3   4.99  2   5.13  2   5.25  3
189319      3.47   1.93   1.57   1.20   0.92    0.83   0.83    7.98 14   4.24  5   2.89  5   2.28  4   1.93  3
190248      3.56   0.46   0.76   0.61   0.34    0.42   0.33    5.64  4   3.88  3   3.33  3   3.22  2   3.18  2
190940      4.50   1.50   1.31   0.93   0.65    0.66   0.58    8.29 12   5.13  4   4.05  5   3.69  3   3.50  3
191692      3.22  -0.12  -0.07  -0.07  -0.05   -0.06  -0.03    4.05 13   3.09  2   3.38  2   3.63  2   3.82  3
192310      5.73   0.64   0.88   0.71   0.45    0.50   0.43    8.12  5   6.12  4   5.44  3   5.24  2   5.14  2
192425      4.95   0.01   0.09   0.10   0.00    0.05   0.01    6.05 11   4.90  3   5.03  2   5.24  2   5.40  3
192514      4.83   0.16   0.09   0.16   0.08    0.10   0.08    6.13 15   4.78  3   4.89  2   5.03  3   5.14  3
192696      4.30   0.08   0.11   0.12   0.06    0.07   0.07    5.50 12   4.27  3   4.37  2   4.52  2   4.65  3
192806      4.52   1.11   1.26   0.96   0.71    0.68   0.63    7.75  5   5.12  4   4.07  4   3.66  3   3.43  3
```

```
192836     6.12   0.92   1.04   0.77   0.53      0.54   0.48      8.97  6    6.60  4    5.78  4    5.52  3    5.39  2
192876     4.26   0.81   1.08   0.79   0.53      0.56   0.48      6.99  4    4.76  4    3.90  4    3.65  3    3.52  3
192907     4.39  -0.09  -0.05  -0.02  -0.06     -0.02  -0.03      5.28 13    4.27  2    4.53  2    4.78  2    4.97  3
192947     3.58   0.69   0.95   0.69   0.47      0.49   0.43      6.07  4    4.01  4    3.28  3    3.07  2    2.97  2
193322     5.84  -0.77   0.10   0.15   0.01      0.09   0.08      5.89  8    5.80  3    5.90  2    6.04  2    6.16  3
193432     4.76  -0.11  -0.04   0.01  -0.06     -0.01  -0.04      5.62 12    4.64  2    4.90  2    5.15  3    5.34  3
193495     3.08   0.27   0.79   0.55   0.50      0.38   0.46      4.92  4    3.42  3    2.86  3    2.63  3    2.52  2
193592     5.76   0.02   0.12   0.13   0.06      0.08   0.07      6.89 11    5.73  3    5.82  2    5.97  2    6.10  3
193702     6.23   0.03   0.06   0.06   0.03      0.03   0.05      7.34 13    6.17  3    6.33  2    6.50  2    6.64  3
193924     1.94  -0.71  -0.20  -0.09  -0.16     -0.07  -0.10      1.90  4    1.73  2    2.14  3    2.46  3    2.69  4
194093     2.23   0.54   0.67   0.49   0.34      0.33   0.31      4.37  8    2.50  3    2.06  3    1.97  2    1.95  2
194317     4.44   1.50   1.33   1.01   0.67      0.72   0.60      8.24 11    5.08  4    3.96  4    3.59  3    3.38  3
194943     4.81   0.05   0.37   0.35   0.19      0.23   0.20      6.12  5    4.92  3    4.75  2    4.77  2    4.81  3
195295     4.02   0.30   0.40   0.36   0.23      0.24   0.23      5.68  9    4.15  3    3.95  2    3.94  2    3.97  2
195593     6.18   0.74   1.01   0.85   0.69      0.60   0.62      8.77  4    6.64  4    5.82  3    5.43  2    5.21  2
195725     4.22   0.16   0.20   0.17   0.09      0.10   0.10      5.58 12    4.24  3    4.25  2    4.37  2    4.48  3
195810     4.04  -0.48  -0.12  -0.02  -0.11     -0.02  -0.06      4.36  6    3.88  2    4.20  3    4.48  3    4.69  4
196180     4.69   0.11   0.11   0.13   0.05      0.08   0.06      5.93 13    4.66  3    4.76  2    4.91  2    5.05  3
196524     3.63   0.08   0.44   0.40   0.24      0.27   0.25      5.02  4    3.78  3    3.54  2    3.51  2    3.53  2
196574     4.33   0.69   0.96   0.67   0.46      0.47   0.42      6.83  4    4.76  4    4.03  4    3.83  3    3.74  2
196724     4.82  -0.07  -0.02   0.03  -0.04      0.00  -0.02      5.75 12    4.71  2    4.95  2    5.19  3    5.37  3
196867     3.77  -0.21  -0.06   0.00  -0.04     -0.01  -0.02      4.49 10    3.64  2    3.91  2    4.15  3    4.33  3
197692     4.13   0.04   0.43   0.36   0.20      0.24   0.22      5.46  4    4.27  3    4.05  2    4.05  2    4.08  2
197752     4.91   1.18   1.19   0.85   0.58      0.60   0.52      8.20  8    5.47  4    4.51  4    4.22  3    4.06  3
197912     4.23   0.88   1.06   0.77   0.53      0.54   0.48      7.04  5    4.72  4    3.88  4    3.63  3    3.50  3
197989     2.46   0.87   1.03   0.73   0.54      0.52   0.49      5.24  6    2.93  4    2.13  4    1.87  3    1.73  3
198001     3.77   0.02   0.00   0.07   0.00      0.03   0.02      4.83 14    3.68  3    3.88  2    4.08  3    4.24  3
198084     4.52   0.10   0.54   0.47   0.28      0.32   0.29      5.99  3    4.72  3    4.38  2    4.32  2    4.31  2
198149     3.43   0.61   0.92   0.67   0.49      0.47   0.45      5.80  4    3.84  4    3.14  3    2.92  2    2.81  2
198542     4.12   1.91   1.63   1.25   0.94      0.86   0.85      8.64 12    4.92  5    3.51  6    2.89  4    2.52  4
198743     4.73   0.10   0.32   0.28   0.14      0.18   0.16      6.08  7    4.81  3    4.70  2    4.77  2    4.84  2
198809     4.61   0.46   0.82   0.68   0.46      0.48   0.42      6.72  3    4.97  4    4.34  3    4.14  2    4.05  2
199081     4.77  -0.58  -0.14  -0.07  -0.13     -0.06  -0.08      4.94  5    4.60  2    4.95  3    5.24  3    5.46  3
199629     3.94   0.00   0.02   0.05   0.00      0.02   0.01      4.99 13    3.86  3    4.05  2    4.26  2    4.42  3
199766     5.23   0.02   0.46   0.32   0.28      0.21   0.28      6.55  3    5.39  3    5.16  2    5.10  2    5.09  2
200499     4.84   0.05   0.18   0.16   0.07      0.10   0.08      6.04 10    4.84  3    4.88  2    5.02  2    5.14  3
200761     4.07   0.01  -0.01   0.01  -0.02     -0.01  -0.01      5.11 14    3.97  3    4.20  2    4.42  2    4.60  3
200905     3.70   1.80   1.65   1.20   0.90      0.83   0.81      8.08  9    4.51  5    3.10  6    2.51  4    2.17  4
200914     4.49   1.90   1.60   1.28   0.98      0.87   0.89      8.98 13    5.28  5    3.88  5    3.22  4    2.83  3
201092     6.03   1.23   1.37   1.17   0.83      0.82   0.80      9.49  5    6.69  4    5.50  4    4.92  2    4.59  2
201251     4.56   1.73   1.56   1.17   0.82      0.81   0.74      8.80 10    5.32  5    3.98  5    3.47  4    3.18  4
```

```
201381      4.52  0.67  0.94  0.69  0.46      0.49  0.42      6.98  4    4.94  4    4.22  3    4.02  2    3.93  2
202109      3.20  0.76  0.99  0.70  0.48      0.49  0.44      5.81  4    3.65  4    2.89  4    2.67  3    2.57  2
202275      4.49 -0.01  0.50  0.43  0.28      0.29  0.29      5.79  3    4.67  3    4.38  2    4.31  2    4.30  2
202850      4.23 -0.39  0.13  0.15  0.14      0.09  0.17      4.81  3    4.21  3    4.28  2    4.33  2    4.40  3
203280      2.45  0.11  0.22  0.21  0.11      0.13  0.13      3.76 10    2.48  3    2.47  2    2.56  2    2.65  3
203454      6.40  0.00  0.53  0.47  0.34      0.32  0.33      7.73  3    6.60  3    6.27  2    6.16  2    6.12  2
203608      4.22 -0.13  0.48  0.47  0.30      0.32  0.30      5.34  4    4.39  3    4.10  2    4.02  2    4.00  3
204075      3.74  0.60  1.00  0.64  0.43      0.45  0.40      6.14  3    4.20  4    3.45  4    3.27  3    3.19  3
204139      5.78  1.74  1.44  1.05  0.76      0.74  0.68      9.96 14    6.48  4    5.26  5    4.81  3    4.55  3
204381      4.51  0.60  0.91  0.68  0.47      0.48  0.43      6.86  4    4.92  4    4.22  3    4.02  2    3.91  2
204724      4.57  1.93  1.62  1.25  1.09      0.86  0.99      9.11 13    5.37  5    3.96  6    3.20  4    2.75  3
204867      2.87  0.57  0.84  0.61  0.41      0.42  0.38      5.14  4    3.24  4    2.62  3    2.47  2    2.40  2
205435      4.02  0.56  0.89  0.71  0.50      0.50  0.46      6.30  3    4.41  4    3.73  3    3.50  2    3.38  2
205512      4.91  1.00  1.08  0.82  0.54      0.58  0.49      7.89  7    5.41  4    4.54  4    4.28  2    4.15  2
205765      6.25  0.05  0.06  0.08  0.02      0.04  0.04      7.39 13    6.19  3    6.34  2    6.52  2    6.67  3
205767      4.69  0.13  0.19  0.17  0.10      0.10  0.11      6.01 11    4.70  3    4.73  2    4.84  2    4.94  3
206453      4.73  0.51  0.88  0.67  0.49      0.47  0.45      6.94  3    5.12  4    4.45  3    4.23  2    4.12  2
206538      6.11  0.08  0.07  0.11  0.05      0.06  0.06      7.29 13    6.05  3    6.19  2    6.35  2    6.48  3
206672      4.67 -0.69 -0.12 -0.02 -0.13     -0.02 -0.07      4.70  3    4.51  2    4.83  3    5.12  3    5.33  4
206742      4.34 -0.11 -0.05 -0.01 -0.03     -0.02 -0.01      5.20 12    4.22  2    4.48  2    4.72  2    4.90  3
206826      4.51  0.01  0.49  0.38  0.28      0.26  0.28      5.83  3    4.68  3    4.41  2    4.35  2    4.34  2
206859      4.31  0.96  1.18  0.80  0.56      0.57  0.50      7.29  4    4.86  4    3.92  5    3.65  3    3.50  3
206901      4.12  0.03  0.44  0.37  0.25      0.25  0.26      5.44  3    4.27  3    4.04  2    4.00  2    4.01  2
206952      4.57  1.09  1.10  0.84  0.54      0.59  0.49      7.69  8    5.08  4    4.19  4    3.93  2    3.80  2
207198      5.94 -0.64  0.31  0.28  0.17      0.18  0.20      6.28 10    6.02  3    5.91  2    5.93  2    5.98  3
207330      4.24 -0.72 -0.12 -0.05 -0.12     -0.04 -0.06      4.23  3    4.08  2    4.41  3    4.68  3    4.89  3
207971      3.01 -0.37 -0.12 -0.05 -0.06     -0.04 -0.02      3.48  9    2.85  2    3.18  3    3.42  3    3.60  3
208501      5.80 -0.02  0.72  0.68  0.60      0.48  0.51      7.20  8    6.10  3    5.56  2    5.28  2    5.13  3
209100      4.69  0.99  1.06  0.88  0.56      0.62  0.54      7.65  7    5.18  4    4.31  3    4.00  2    3.83  2
209625      5.31  0.15  0.23  0.18  0.10      0.11  0.12      6.67 11    5.34  3    5.33  2    5.44  2    5.53  2
209688      4.46  1.66  1.37  1.00  0.78      0.71  0.70      8.50 14    5.12  4    3.97  5    3.50  3    3.23  3
209747      4.84  1.80  1.44  1.07  0.76      0.76  0.68      9.11 15    5.54  4    4.31  5    3.86  3    3.60  3
209790      4.29  0.09  0.34  0.32  0.13      0.21  0.14      5.63  7    4.38  3    4.25  2    4.33  2    4.41  3
209819      4.25 -0.27 -0.07 -0.05 -0.09     -0.04 -0.05      4.88  9    4.12  2    4.41  2    4.68  2    4.88  3
209952      1.74 -0.47 -0.13 -0.08 -0.06     -0.07 -0.01      2.07  7    1.57  2    1.92  2    2.15  3    2.33  3
209975      5.11 -0.83  0.08  0.16  0.03      0.10  0.10      5.06  8    5.06  3    5.17  2    5.29  3    5.40  3
210027      3.76 -0.03  0.44  0.40  0.25      0.27  0.26      5.00  3    3.91  3    3.67  2    3.63  2    3.64  2
210049      4.50  0.06  0.05  0.07  0.05      0.03  0.06      5.64 13    4.43  3    4.60  2    4.76  2    4.89  3
210418      3.55  0.10  0.07  0.05  0.04      0.02  0.05      4.76 14    3.49  3    3.65  2    3.81  2    3.95  3
210702      5.95  0.73  0.95  0.72  0.49      0.51  0.45      8.50  5    6.38  4    5.64  3    5.42  2    5.31  2
210807      4.79  0.61  0.92  0.69  0.48      0.49  0.44      7.16  4    5.20  4    4.50  3    4.28  2    4.18  2
```

```
211073      4.49  1.44  1.39  1.01  0.74   0.72  0.66   8.24  8   5.16 4   3.99 5   3.56 3   3.31 3
211388      4.13  1.62  1.46  1.00  0.72   0.71  0.64   8.16 10   4.84 4   3.62 6   3.20 4   2.97 4
211391      4.15  0.80  0.99  0.70  0.48   0.49  0.44   6.81  5   4.60 4   3.84 4   3.62 3   3.52 2
212061      3.84 -0.13 -0.06  0.04 -0.04   0.01 -0.02   4.67 12   3.71 2   3.97 3   4.22 3   4.40 3
212097      4.81 -0.19  0.00  0.09 -0.01   0.05  0.01   5.59  9   4.72 3   4.91 2   5.12 3   5.29 3
212120      4.56 -0.50 -0.10 -0.02 -0.11  -0.02 -0.06   4.86  5   4.41 2   4.72 3   4.99 3   5.20 3
212496      4.44  0.77  1.02  0.75  0.57   0.53  0.51   7.08  4   4.91 4   4.11 4   3.82 3   3.67 2
212593      4.58 -0.34  0.09  0.13  0.10   0.08  0.13   5.20  4   4.53 3   4.65 2   4.74 2   4.83 3
212943      4.80  0.88  1.06  0.78  0.56   0.55  0.50   7.61  5   5.29 4   4.45 4   4.17 3   4.03 3
213009      3.97  0.81  1.03  0.73  0.61   0.52  0.54   6.67  4   4.44 4   3.64 4   3.33 3   3.16 3
213320      4.81 -0.14 -0.08  0.00 -0.04  -0.01 -0.02   5.61 13   4.67 2   4.96 2   5.20 3   5.38 3
213398      4.29  0.02  0.01  0.02  0.02   0.00  0.04   5.36 14   4.20 3   4.41 2   4.59 2   4.74 3
213420      4.51 -0.74 -0.09 -0.01 -0.11  -0.02 -0.04   4.49  3   4.37 2   4.66 3   4.92 3   5.11 3
213558      3.77  0.00  0.01  0.00 -0.03  -0.01 -0.01   4.81 13   3.68 3   3.90 2   4.13 2   4.31 3
213998      4.00 -0.28 -0.10 -0.06 -0.07  -0.05 -0.04   4.60 10   3.85 2   4.17 2   4.43 3   4.63 3
214454      4.63  0.11  0.24  0.22  0.14   0.14  0.15   5.95 10   4.67 3   4.64 2   4.71 2   4.78 3
214680      4.88 -1.05 -0.20 -0.09 -0.21  -0.07 -0.11   4.38  6   4.67 2   5.08 3   5.41 3   5.65 4
214748      4.16 -0.34 -0.12 -0.06 -0.10  -0.05 -0.06   4.67  9   4.00 2   4.33 2   4.61 3   4.82 3
214868      4.46  1.36  1.33  0.92  0.68   0.65  0.61   8.07  8   5.10 4   4.00 5   3.62 4   3.41 3
214923      3.40 -0.24 -0.09 -0.04 -0.07  -0.04 -0.04   4.06 11   3.26 2   3.56 2   3.82 3   4.02 3
214994      4.79 -0.01 -0.01  0.04 -0.02   0.01 -0.01   5.81 14   4.69 3   4.91 2   5.14 3   5.31 3
215104      4.85  0.81  1.03  0.76  0.65   0.54  0.58   7.55  4   5.32 4   4.51 4   4.16 3   3.97 2
215167      4.69  1.56  1.36  0.97  0.72   0.69  0.64   8.59 12   5.34 4   4.21 5   3.80 3   3.56 3
215182      2.95  0.57  0.86  0.64  0.48   0.45  0.39   5.23  4   3.33 4   2.69 3   2.52 2   2.45 2
215648      4.19 -0.02  0.50  0.43  0.31   0.29  0.31   5.47  3   4.37 3   4.08 2   3.99 2   3.97 2
215665      3.94  0.91  1.08  0.76  0.51   0.54  0.46   6.80  5   4.44 4   3.59 4   3.35 3   3.23 3
216032      3.98  1.94  1.59  1.19  0.95   0.82  0.86   8.52 14   4.76 5   3.39 6   2.76 4   2.39 4
216131      3.48  0.68  0.94  0.68  0.47   0.48  0.43   5.95  4   3.90 4   3.19 3   2.98 2   2.88 2
216228      3.53  0.90  1.05  0.83  0.51   0.59  0.46   6.36  6   4.01 4   3.17 3   2.93 2   2.81 2
216336      4.46 -0.14 -0.04  0.02  0.02   0.00  0.04   5.28 11   4.34 2   4.59 2   4.78 3   4.92 3
216446      4.74  1.35  1.26  0.94  0.67   0.67  0.60   8.30 10   5.34 4   4.29 4   3.92 3   3.72 3
216627      3.28  0.08  0.05  0.07  0.04   0.03  0.06   4.45 14   3.21 3   3.38 2   3.54 2   3.67 3
216735      4.90  0.00  0.00  0.04 -0.02   0.01 -0.01   5.93 13   4.81 3   5.02 2   5.24 2   5.42 3
216763      4.21  0.70  0.97  0.74  0.61   0.52  0.54   6.73  4   4.65 4   3.89 3   3.58 2   3.41 2
216956      1.16  0.06  0.09  0.06  0.02   0.03  0.04   2.33 12   1.11 3   1.25 2   1.43 2   1.57 3
217014      5.50  0.20  0.67  0.54  0.34   0.37  0.34   7.17  3   5.77 3   5.31 2   5.20 2   5.16 2
217166      6.43  0.16  0.64  0.53  0.36   0.36  0.35   8.03  3   6.69 3   6.25 2   6.13 2   6.08 2
217382      4.72  1.70  1.43  1.09  0.73   0.77  0.65   8.85 13   5.41 4   4.19 5   3.76 3   3.52 3
217891      4.52 -0.50 -0.12 -0.02 -0.13  -0.02 -0.08   4.81  6   4.36 2   4.68 3   4.98 3   5.20 3
218031      4.66  0.88  1.06  0.74  0.57   0.52  0.51   7.47  5   5.15 4   4.32 4   4.04 3   3.89 3
218045      2.48 -0.06 -0.04  0.01 -0.03  -0.01 -0.01   3.41 13   2.36 2   2.62 2   2.85 3   3.03 3
```

```
218240      4.48  0.59  0.90  0.71  0.49    0.50  0.45    6.81  4   4.88  4   4.19  3   3.96  2   3.85  2
218329      4.51  1.88  1.58  1.26  1.02    0.86  0.92    8.96 13   5.28  5   3.91  5   3.21  4   2.80  3
218356      4.77  1.15  1.35  0.97  0.68    0.69  0.61    8.11  4   5.42  4   4.30  5   3.92  3   3.70  3
218376      4.86 -0.87 -0.03  0.10 -0.11    0.05 -0.03    4.70  7   4.75  2   4.97  3   5.22  3   5.41  4
218452      5.33  1.72  1.41  1.02  0.75    0.72  0.67    9.47 14   6.01  4   4.83  5   4.38  3   4.13  3
218594      3.64  1.22  1.23  0.84  0.60    0.59  0.53    7.00  8   4.22  4   3.23  5   2.93  3   2.76  3
218640      4.68  0.38  0.64  0.54  0.36    0.37  0.35    6.58  5   4.94  3   4.50  2   4.38  2   4.33  2
218658      4.42  0.45  0.82  0.65  0.43    0.45  0.40    6.52  3   4.78  4   4.17  3   3.99  2   3.90  2
218670      3.90  0.86  1.02  0.75  0.56    0.53  0.50    6.66  6   4.37  4   3.57  4   3.29  3   3.15  2
219080      4.53  0.04  0.30  0.29  0.16    0.19  0.18    5.78  7   4.60  3   4.50  2   4.55  2   4.61  3
219134      5.57  0.89  1.00  0.83  0.53    0.59  0.51    8.36  7   6.03  4   5.22  3   4.94  2   4.79  2
219215      4.22  1.87  1.55  1.28  1.08    0.87  0.98    8.64 13   4.98  5   3.62  5   2.87  3   2.42  3
219449      4.25  1.01  1.11  0.79  0.56    0.56  0.50    7.26  6   4.77  4   3.88  4   3.61  3   3.46  3
219615      3.69  0.57  0.91  0.72  0.51    0.51  0.46    6.00  3   4.10  4   3.39  3   3.15  2   3.03  2
219734      4.86  1.97  1.67  1.46  1.26    0.97  1.15    9.48 13   5.68  5   4.19  7   3.28  4   2.72  4
219784      4.41  1.06  1.13  0.84  0.63    0.59  0.56    7.50  7   4.94  4   4.02  4   3.69  3   3.51  3
219916      4.76  0.49  0.84  0.65  0.45    0.45  0.42    6.92  3   5.13  4   4.50  3   4.31  2   4.21  2
220321      3.98  0.95  1.10  0.82  0.60    0.58  0.53    6.91  5   4.49  4   3.61  4   3.30  3   3.14  3
220657      4.41  0.14  0.61  0.54  0.32    0.37  0.32    5.97  3   4.65  3   4.24  2   4.14  2   4.12  2
220704      4.40  1.80  1.47  1.13  0.81    0.79  0.73    8.68 14   5.11  4   3.85  5   3.35  3   3.07  3
220954      4.30  1.03  1.08  0.80  0.53    0.57  0.48    7.32  8   4.80  4   3.94  4   3.69  3   3.55  3
221115      4.56  0.74  0.94  0.74  0.45    0.52  0.42    7.11  5   4.98  4   4.25  3   4.06  2   3.96  2
221507      4.37 -0.36 -0.09  0.00 -0.07   -0.01 -0.05    4.87  8   4.23  2   4.52  3   4.79  3   4.98  3
221565      4.72  0.01  0.02  0.07  0.03    0.03  0.05    5.78 13   4.64  3   4.82  2   5.00  2   5.14  3
222095      4.74  0.09  0.08  0.11  0.05    0.06  0.06    5.94 13   4.69  3   4.82  2   4.98  2   5.11  3
222173      4.29 -0.28 -0.11  0.00 -0.09   -0.01 -0.05    4.89 10   4.13  2   4.44  3   4.71  3   4.91  4
222368      4.13  0.00  0.51  0.44  0.31    0.30  0.31    5.45  3   4.32  3   4.01  2   3.93  2   3.90  2
222404      3.21  0.95  1.03  0.75  0.51    0.53  0.46    6.10  7   3.68  4   2.87  4   2.64  3   2.52  2
222439      4.14 -0.24 -0.08 -0.01 -0.07   -0.02 -0.04    4.81 10   4.00  2   4.29  2   4.55  3   4.75  3
222574      4.84  0.49  0.81  0.62  0.42    0.43  0.39    6.98  4   5.19  4   4.60  3   4.43  2   4.36  2
222603      4.51  0.07  0.21  0.18  0.10    0.11  0.11    5.76  9   4.53  3   4.54  2   4.65  2   4.75  3
222661      4.51 -0.13 -0.04  0.02 -0.05    0.00 -0.02    5.35 12   4.39  2   4.64  2   4.89  3   5.07  3
222842      4.93  0.63  0.96  0.71  0.48    0.50  0.44    7.35  3   5.36  4   4.62  3   4.41  2   4.30  2
223352      4.57 -0.01  0.00  0.04 -0.02    0.01 -0.01    5.59 13   4.48  3   4.69  2   4.91  2   5.09  3
224427      4.66  1.67  1.59  1.46  1.34    0.97  1.22    8.83  8   5.44  5   4.01  6   2.99  4   2.42  4
224572      4.88 -0.83 -0.06  0.04 -0.11    0.01 -0.04    4.76  5   4.75  2   5.01  3   5.27  3   5.47  3
224617      4.01  0.07  0.42  0.38  0.24    0.26  0.25    5.37  4   4.15  3   3.93  2   3.90  2   3.92  2
225132      4.56 -0.12 -0.04  0.03 -0.04    0.00 -0.02    5.41 12   4.44  2   4.69  2   4.93  3   5.11  3
 16160      5.82  0.79  0.97  0.83  0.53    0.57  0.51    8.46  5   6.26  4   5.48  3   5.20  2   5.05  2
 17506      3.79  1.90  1.69  1.23  0.89    0.84  0.80    8.33 10   4.62  5   3.17  6   2.60  5   2.27  4
 60863      4.61 -0.43 -0.12 -0.03 -0.11   -0.04 -0.07    5.00  7   4.45  2   4.78  3   5.06  3   5.28  3
```

```
 74180          3.83  0.28  0.70  0.63  0.60    0.43  0.49    5.63  3   4.12  3   3.61  2   3.35  2   3.21  2
 90839          4.84 -0.01  0.52  0.48  0.28    0.31  0.29    6.15  3   5.03  3   4.71  2   4.65  2   4.64  2
135722          3.49  0.68  0.95  0.73  0.51    0.52  0.49    5.97  4   3.92  4   3.18  3   2.92  2   2.78  2
141003          3.67  0.08  0.06  0.06  0.03    0.03  0.05    4.85 14   3.61  3   3.77  2   3.94  2   4.07  3
161797          3.42  0.39  0.75  0.53  0.38    0.36  0.37    5.40  3   3.74  3   3.22  3   3.07  2   3.01  2
203504          4.09  1.05  1.11  0.79  0.54    0.56  0.49    7.16  7   4.61  4   3.72  4   3.46  3   3.33  3
     6          6.29  1.02  1.10 99.99 99.99   99.99 99.99    9.31  7   6.80  4   5.89  4  99.99 99  99.99 99
   142          5.70 99.99  0.52 99.99 99.99   99.99 99.99   99.99 99   5.89  3   5.57  3  99.99 99  99.99 99
   203          6.18 99.99  0.38 99.99 99.99   99.99 99.99   99.99 99   6.29  3   6.12  2  99.99 99  99.99 99
   256          6.19 99.99  0.14 99.99 99.99   99.99 99.99   99.99 99   6.17  3   6.24  2  99.99 99  99.99 99
   315          6.43 -0.48 -0.14 99.99 99.99   99.99 99.99    6.74  7   6.26  2   6.61  2  99.99 99  99.99 99
   319          5.94 99.99  0.14 99.99 99.99   99.99 99.99   99.99 99   5.92  3   5.99  2  99.99 99  99.99 99
   360          5.99  0.82  1.04 99.99 99.99   99.99 99.99    8.71  4   6.47  4   5.62  4  99.99 99  99.99 99
   417          6.23  0.72  0.97 99.99 99.99   99.99 99.99    8.77  4   6.67  4   5.89  4  99.99 99  99.99 99
   469          6.33 99.99  0.74 99.99 99.99   99.99 99.99   99.99 99   6.64  3   6.10  3  99.99 99  99.99 99
   587          5.84  0.74  0.98 99.99 99.99   99.99 99.99    8.42  4   6.28  4   5.50  4  99.99 99  99.99 99
   636          5.28  0.92  1.05 99.99 99.99   99.99 99.99    8.14  6   5.76  4   4.91  4  99.99 99  99.99 99
   645          5.86 99.99  1.01 99.99 99.99   99.99 99.99   99.99 99   6.32  4   5.51  4  99.99 99  99.99 99
   661          6.64  0.06  0.37 99.99 99.99   99.99 99.99    7.96  5   6.75  3   6.58  2  99.99 99  99.99 99
   693          4.89 99.99  0.49 99.99 99.99   99.99 99.99   99.99 99   5.06  3   4.78  3  99.99 99  99.99 99
   739          5.25 99.99  0.44 99.99 99.99   99.99 99.99   99.99 99   5.40  3   5.16  3  99.99 99  99.99 99
   787          5.25  1.63  1.48 99.99 99.99   99.99 99.99    9.31 10   5.97  4   4.68  5  99.99 99  99.99 99
   942          5.94 99.99  1.55 99.99 99.99   99.99 99.99   99.99 99   6.70  5   5.33  5  99.99 99  99.99 99
  1032          5.77  2.10  1.72 99.99 99.99   99.99 99.99   10.60 14   6.62  5   5.09  5  99.99 99  99.99 99
  1064          5.75 99.99 -0.08 99.99 99.99   99.99 99.99   99.99 99   5.61  2   5.90  2  99.99 99  99.99 99
  1187          5.67 99.99  1.35 99.99 99.99   99.99 99.99   99.99 99   6.32  4   5.16  4  99.99 99  99.99 99
  1221          6.49  0.72  1.00 99.99 99.99   99.99 99.99    9.05  4   6.95  4   6.14  4  99.99 99  99.99 99
  1227          6.11  0.66  0.92 99.99 99.99   99.99 99.99    8.55  4   6.52  4   5.80  3  99.99 99  99.99 99
  1324          6.77  0.08  0.46 99.99 99.99   99.99 99.99    8.17  4   6.93  3   6.67  3  99.99 99  99.99 99
  1343          6.45 99.99  0.37 99.99 99.99   99.99 99.99   99.99 99   6.56  3   6.39  2  99.99 99  99.99 99
  1367          6.17  0.73  0.94 99.99 99.99   99.99 99.99    8.71  5   6.59  4   5.85  3  99.99 99  99.99 99
  1461          6.46  0.29  0.68 99.99 99.99   99.99 99.99    8.26  3   6.74  3   6.26  3  99.99 99  99.99 99
  1483          6.33 99.99  1.21 99.99 99.99   99.99 99.99   99.99 99   6.90  4   5.88  4  99.99 99  99.99 99
  1635          5.37  1.55  1.34 99.99 99.99   99.99 99.99    9.24 12   6.01  4   4.86  4  99.99 99  99.99 99
  1685          5.51 -0.14 -0.05 99.99 99.99   99.99 99.99    6.33 12   5.39  2   5.65  2  99.99 99  99.99 99
  1737          5.18 99.99  1.00 99.99 99.99   99.99 99.99   99.99 99   5.64  4   4.83  4  99.99 99  99.99 99
  1801          5.97  1.40  1.40 99.99 99.99   99.99 99.99    9.67  7   6.65  4   5.43  4  99.99 99  99.99 99
  2023          6.07  1.23  1.22 99.99 99.99   99.99 99.99    9.44  8   6.65  4   5.62  4  99.99 99  99.99 99
  2114          5.77  0.57  0.86 99.99 99.99   99.99 99.99    8.05  4   6.15  4   5.49  3  99.99 99  99.99 99
  2273          6.19  0.55  0.91 99.99 99.99   99.99 99.99    8.47  3   6.60  4   5.88  3  99.99 99  99.99 99
  2454          6.04 -0.06  0.44 99.99 99.99   99.99 99.99    7.24  3   6.19  3   5.95  3  99.99 99  99.99 99
```

```
2475     6.44 99.99  0.60 99.99 99.99    99.99 99.99    99.99 99    6.68  3    6.28  3    99.99 99    99.99 99
2490     5.43 99.99  1.56 99.99 99.99    99.99 99.99    99.99 99    6.19  5    4.82  5    99.99 99    99.99 99
2589     6.21  0.55  0.84 99.99 99.99    99.99 99.99     8.45  4    6.58  4    5.93  3    99.99 99    99.99 99
2630     6.14 99.99  0.37 99.99 99.99    99.99 99.99    99.99 99    6.25  3    6.08  2    99.99 99    99.99 99
2637     5.72  1.90  1.56 99.99 99.99    99.99 99.99    10.19 14    6.48  5    5.11  5    99.99 99    99.99 99
2696     5.19 99.99  0.12 99.99 99.99    99.99 99.99    99.99 99    5.16  3    5.25  2    99.99 99    99.99 99
2726     5.69 99.99  0.35 99.99 99.99    99.99 99.99    99.99 99    5.79  3    5.64  2    99.99 99    99.99 99
2774     5.60  1.12  1.15 99.99 99.99    99.99 99.99     8.79  8    6.14  4    5.18  4    99.99 99    99.99 99
2834     4.77  0.04  0.02 99.99 99.99    99.99 99.99     5.87 14    4.69  3    4.88  2    99.99 99    99.99 99
2884     4.37 -0.17 -0.07 99.99 99.99    99.99 99.99     5.14 12    4.24  2    4.52  2    99.99 99    99.99 99
2885     4.54  0.03  0.15 99.99 99.99    99.99 99.99     5.70 10    4.53  3    4.59  2    99.99 99    99.99 99
2913     5.67 -0.17  0.00 99.99 99.99    99.99 99.99     6.47 10    5.58  3    5.78  2    99.99 99    99.99 99
3059     5.55 99.99  1.27 99.99 99.99    99.99 99.99    99.99 99    6.15  4    5.07  4    99.99 99    99.99 99
3158     5.57 99.99  0.46 99.99 99.99    99.99 99.99    99.99 99    5.73  3    5.47  3    99.99 99    99.99 99
3229     5.93 -0.02  0.44 99.99 99.99    99.99 99.99     7.18  3    6.08  3    5.84  3    99.99 99    99.99 99
3303     6.06 99.99  1.01 99.99 99.99    99.99 99.99    99.99 99    6.52  4    5.71  4    99.99 99    99.99 99
3325     6.45  0.91  1.06 99.99 99.99    99.99 99.99     9.30  6    6.94  4    6.07  4    99.99 99    99.99 99
3443     5.57 99.99  0.71 99.99 99.99    99.99 99.99    99.99 99    5.87  3    5.35  3    99.99 99    99.99 99
3444     6.42  1.38  1.26 99.99 99.99    99.99 99.99    10.02 10    7.02  4    5.95  4    99.99 99    99.99 99
3457     6.39  1.52  1.34 99.99 99.99    99.99 99.99    10.22 11    7.03  4    5.88  4    99.99 99    99.99 99
3488     6.41 99.99  1.00 99.99 99.99    99.99 99.99    99.99 99    6.87  4    6.06  4    99.99 99    99.99 99
3719     6.85  0.12  0.11 99.99 99.99    99.99 99.99     8.11 13    6.82  3    6.91  2    99.99 99    99.99 99
3750     6.01 99.99  1.14 99.99 99.99    99.99 99.99    99.99 99    6.54  4    5.59  4    99.99 99    99.99 99
3794     6.49  0.53  0.92 99.99 99.99    99.99 99.99     8.75  3    6.90  4    6.18  3    99.99 99    99.99 99
3795     6.14 99.99  0.70 99.99 99.99    99.99 99.99    99.99 99    6.43  3    5.93  3    99.99 99    99.99 99
3807     5.91  0.89  1.10 99.99 99.99    99.99 99.99     8.76  4    6.42  4    5.51  4    99.99 99    99.99 99
3823     5.89 99.99  0.55 99.99 99.99    99.99 99.99    99.99 99    6.10  3    5.75  3    99.99 99    99.99 99
4065     6.06 99.99 -0.03 99.99 99.99    99.99 99.99    99.99 99    5.95  2    6.19  2    99.99 99    99.99 99
4088     5.98 99.99  1.32 99.99 99.99    99.99 99.99    99.99 99    6.61  4    5.48  4    99.99 99    99.99 99
4089     5.39  0.02  0.50 99.99 99.99    99.99 99.99     6.73  3    5.57  3    5.27  3    99.99 99    99.99 99
4142     5.68 -0.56 -0.13 99.99 99.99    99.99 99.99     5.88  5    5.51  2    5.86  2    99.99 99    99.99 99
4150     4.36 -0.02  0.00 99.99 99.99    99.99 99.99     5.37 13    4.27  3    4.47  2    99.99 99    99.99 99
4247     5.24 99.99  0.33 99.99 99.99    99.99 99.99    99.99 99    5.33  3    5.20  2    99.99 99    99.99 99
4293     5.94 99.99  0.28 99.99 99.99    99.99 99.99    99.99 99    6.00  3    5.92  2    99.99 99    99.99 99
4294     6.06 99.99  0.44 99.99 99.99    99.99 99.99    99.99 99    6.21  3    5.97  3    99.99 99    99.99 99
4301     6.15  2.00  1.62 99.99 99.99    99.99 99.99    10.79 15    6.95  5    5.51  5    99.99 99    99.99 99
4304     6.15 99.99  0.53 99.99 99.99    99.99 99.99    99.99 99    6.35  3    6.02  3    99.99 99    99.99 99
4338     6.47 99.99  0.31 99.99 99.99    99.99 99.99    99.99 99    6.55  3    6.44  2    99.99 99    99.99 99
4391     5.80 99.99  0.64 99.99 99.99    99.99 99.99    99.99 99    6.06  3    5.62  3    99.99 99    99.99 99
4398     5.50 99.99  0.98 99.99 99.99    99.99 99.99    99.99 99    5.94  4    5.16  4    99.99 99    99.99 99
4526     5.99  0.70  0.94 99.99 99.99    99.99 99.99     8.49  4    6.41  4    5.67  3    99.99 99    99.99 99
```

```
4622       5.57 99.99 -0.06 99.99 99.99    99.99 99.99    99.99 99    5.44  2    5.71  2    99.99 99    99.99 99
4627       5.93  0.99  1.10 99.99 99.99    99.99 99.99     8.91  6    6.44  4    5.53  4    99.99 99    99.99 99
4737       6.27 99.99  0.90 99.99 99.99    99.99 99.99    99.99 99    6.67  4    5.97  3    99.99 99    99.99 99
4772       6.28 99.99  0.13 99.99 99.99    99.99 99.99    99.99 99    6.26  3    6.33  2    99.99 99    99.99 99
4815       5.07  1.68  1.37 99.99 99.99    99.99 99.99     9.13 14    5.73  4    4.55  4    99.99 99    99.99 99
4928       6.37  0.88  1.07 99.99 99.99    99.99 99.99     9.19  5    6.86  4    5.99  4    99.99 99    99.99 99
5042       6.90 99.99  0.35 99.99 99.99    99.99 99.99    99.99 99    7.00  3    6.85  2    99.99 99    99.99 99
5190       6.22  0.08  0.56 99.99 99.99    99.99 99.99     7.67  3    6.43  3    6.07  3    99.99 99    99.99 99
5268       6.16  0.54  0.92 99.99 99.99    99.99 99.99     8.43  3    6.57  4    5.85  3    99.99 99    99.99 99
5384       5.85  1.93  1.52 99.99 99.99    99.99 99.99    10.34 16    6.59  5    5.26  5    99.99 99    99.99 99
5437       5.31 99.99  1.52 99.99 99.99    99.99 99.99    99.99 99    6.05  5    4.72  5    99.99 99    99.99 99
5457       5.45  1.00  1.09 99.99 99.99    99.99 99.99     8.44  7    5.95  4    5.06  4    99.99 99    99.99 99
5771       6.23 99.99  0.10 99.99 99.99    99.99 99.99    99.99 99    6.19  3    6.30  2    99.99 99    99.99 99
5914       6.49  0.07  0.10 99.99 99.99    99.99 99.99     7.67 12    6.45  3    6.56  2    99.99 99    99.99 99
6055       5.59 99.99  1.18 99.99 99.99    99.99 99.99    99.99 99    6.14  4    5.16  4    99.99 99    99.99 99
6192       6.11 99.99  0.94 99.99 99.99    99.99 99.99    99.99 99    6.53  4    5.79  3    99.99 99    99.99 99
6203       5.44  1.04  1.11 99.99 99.99    99.99 99.99     8.49  7    5.96  4    5.04  4    99.99 99    99.99 99
6245       5.36 99.99  0.90 99.99 99.99    99.99 99.99    99.99 99    5.76  4    5.06  3    99.99 99    99.99 99
6269       6.29 99.99  0.93 99.99 99.99    99.99 99.99    99.99 99    6.71  4    5.97  3    99.99 99    99.99 99
6288       6.04  0.13  0.27 99.99 99.99    99.99 99.99     7.40  9    6.09  3    6.03  2    99.99 99    99.99 99
6386       6.00  1.85  1.51 99.99 99.99    99.99 99.99    10.37 14    6.74  5    5.41  5    99.99 99    99.99 99
6479       6.34 -0.04  0.38 99.99 99.99    99.99 99.99     7.53  4    6.45  3    6.28  2    99.99 99    99.99 99
6480       7.24 -0.03  0.49 99.99 99.99    99.99 99.99     8.50  3    7.41  3    7.13  3    99.99 99    99.99 99
6497       6.43  1.27  1.18 99.99 99.99    99.99 99.99     9.83 10    6.98  4    6.00  4    99.99 99    99.99 99
6530       5.58 99.99  0.01 99.99 99.99    99.99 99.99    99.99 99    5.49  3    5.69  2    99.99 99    99.99 99
6668       6.37 99.99  0.23 99.99 99.99    99.99 99.99    99.99 99    6.40  3    6.38  2    99.99 99    99.99 99
6680       6.25 -0.01  0.40 99.99 99.99    99.99 99.99     7.49  4    6.38  3    6.18  3    99.99 99    99.99 99
6767       5.21  0.09  0.16 99.99 99.99    99.99 99.99     6.45 11    5.20  3    5.25  2    99.99 99    99.99 99
6976       6.40  0.87  1.04 99.99 99.99    99.99 99.99     9.19  5    6.88  4    6.03  4    99.99 99    99.99 99
7014       5.95  1.87  1.51 99.99 99.99    99.99 99.99    10.35 15    6.69  5    5.36  5    99.99 99    99.99 99
7147       5.95  1.66  1.40 99.99 99.99    99.99 99.99    10.00 13    6.63  4    5.41  4    99.99 99    99.99 99
7259       6.52 99.99  0.48 99.99 99.99    99.99 99.99    99.99 99    6.69  3    6.41  3    99.99 99    99.99 99
7344       4.86  0.09  0.32 99.99 99.99    99.99 99.99     6.19  7    4.94  3    4.83  2    99.99 99    99.99 99
7439       5.13 -0.05  0.46 99.99 99.99    99.99 99.99     6.35  3    5.29  3    5.03  3    99.99 99    99.99 99
7446       6.03  1.01  1.08 99.99 99.99    99.99 99.99     9.03  7    6.53  4    5.64  4    99.99 99    99.99 99
7476       5.70 -0.02  0.42 99.99 99.99    99.99 99.99     6.94  3    5.84  3    5.62  3    99.99 99    99.99 99
7788       4.86  0.03  0.47 99.99 99.99    99.99 99.99     6.19  3    5.02  3    4.76  3    99.99 99    99.99 99
7916       6.24 -0.02  0.05 99.99 99.99    99.99 99.99     7.28 12    6.17  3    6.33  2    99.99 99    99.99 99
8036       5.87  0.32  0.64 99.99 99.99    99.99 99.99     7.69  4    6.13  3    5.69  3    99.99 99    99.99 99
8120       6.23  0.84  1.02 99.99 99.99    99.99 99.99     8.96  5    6.70  4    5.87  4    99.99 99    99.99 99
8334       6.20  1.86  1.52 99.99 99.99    99.99 99.99    10.59 14    6.94  5    5.61  5    99.99 99    99.99 99
```

```
 8335      6.49  1.01  1.08 99.99 99.99     99.99 99.99      9.49  7    6.99  4    6.10  4    99.99 99    99.99 99
 8498      5.84 99.99  1.61 99.99 99.99     99.99 99.99     99.99 99    6.63  5    5.21  5    99.99 99    99.99 99
 8556      5.91 -0.05  0.41 99.99 99.99     99.99 99.99      7.10  3    6.04  3    5.83  3    99.99 99    99.99 99
 8599      6.15  0.71  0.96 99.99 99.99     99.99 99.99      8.68  4    6.58  4    5.82  4    99.99 99    99.99 99
 8651      5.42  0.85  1.04 99.99 99.99     99.99 99.99      8.18  5    5.90  4    5.05  4    99.99 99    99.99 99
 8681      6.26  1.08  1.14 99.99 99.99     99.99 99.99      9.39  7    6.79  4    5.84  4    99.99 99    99.99 99
 8705      4.90 99.99  1.23 99.99 99.99     99.99 99.99     99.99 99    5.48  4    4.44  4    99.99 99    99.99 99
 8779      6.41  1.27  1.24 99.99 99.99     99.99 99.99      9.85  9    7.00  4    5.95  4    99.99 99    99.99 99
 8803      6.58 -0.19 -0.04 99.99 99.99     99.99 99.99      7.34 10    6.46  2    6.71  2    99.99 99    99.99 99
 8810      5.93  1.91  1.56 99.99 99.99     99.99 99.99     10.41 14    6.69  5    5.32  5    99.99 99    99.99 99
 8949      6.20  1.07  1.12 99.99 99.99     99.99 99.99      9.30  7    6.72  4    5.79  4    99.99 99    99.99 99
 9132      5.13 99.99  0.02 99.99 99.99     99.99 99.99     99.99 99    5.05  3    5.24  2    99.99 99    99.99 99
 9228      5.93 99.99  1.34 99.99 99.99     99.99 99.99     99.99 99    6.57  4    5.42  4    99.99 99    99.99 99
 9377      5.82 99.99  1.08 99.99 99.99     99.99 99.99     99.99 99    6.32  4    5.43  4    99.99 99    99.99 99
 9414      6.17  0.09  0.06 99.99 99.99     99.99 99.99      7.36 14    6.11  3    6.26  2    99.99 99    99.99 99
 9484      6.59 -0.11 -0.04 99.99 99.99     99.99 99.99      7.45 12    6.47  2    6.72  2    99.99 99    99.99 99
 9525      5.51 99.99  1.02 99.99 99.99     99.99 99.99     99.99 99    5.98  4    5.15  4    99.99 99    99.99 99
 9544      6.28 99.99  0.46 99.99 99.99     99.99 99.99     99.99 99    6.44  3    6.18  3    99.99 99    99.99 99
 9562      5.76  0.21  0.64 99.99 99.99     99.99 99.99      7.43  3    6.02  3    5.58  3    99.99 99    99.99 99
 9896      6.01 99.99  0.38 99.99 99.99     99.99 99.99     99.99 99    6.12  3    5.95  2    99.99 99    99.99 99
 9906      5.69 99.99  0.33 99.99 99.99     99.99 99.99     99.99 99    5.78  3    5.65  2    99.99 99    99.99 99
10009      6.24  0.00  0.53 99.99 99.99     99.99 99.99      7.57  3    6.44  3    6.11  3    99.99 99    99.99 99
10042      6.11  0.70  0.97 99.99 99.99     99.99 99.99      8.63  4    6.55  4    5.77  4    99.99 99    99.99 99
10142      5.94 99.99  1.05 99.99 99.99     99.99 99.99     99.99 99    6.42  4    5.57  4    99.99 99    99.99 99
10161      6.69 99.99 -0.08 99.99 99.99     99.99 99.99     99.99 99    6.55  2    6.84  2    99.99 99    99.99 99
10241      6.84 99.99  0.44 99.99 99.99     99.99 99.99     99.99 99    6.99  3    6.75  3    99.99 99    99.99 99
10481      6.17 99.99  0.42 99.99 99.99     99.99 99.99     99.99 99    6.31  3    6.09  3    99.99 99    99.99 99
10538      5.72 99.99 -0.01 99.99 99.99     99.99 99.99     99.99 99    5.62  3    5.84  2    99.99 99    99.99 99
10550      4.99  1.58  1.38 99.99 99.99     99.99 99.99      8.92 12    5.65  4    4.46  4    99.99 99    99.99 99
10553      6.64  0.12  0.12 99.99 99.99     99.99 99.99      7.90 13    6.61  3    6.70  2    99.99 99    99.99 99
10615      5.71 99.99  1.27 99.99 99.99     99.99 99.99     99.99 99    6.31  4    5.23  4    99.99 99    99.99 99
10658      6.19  1.89  1.54 99.99 99.99     99.99 99.99     10.63 14    6.94  5    5.59  5    99.99 99    99.99 99
10800      5.87  0.10  0.61 99.99 99.99     99.99 99.99      7.38  3    6.11  3    5.70  3    99.99 99    99.99 99
10824      5.34  1.88  1.52 99.99 99.99     99.99 99.99      9.76 15    6.08  5    4.75  5    99.99 99    99.99 99
10830      5.31 99.99  0.39 99.99 99.99     99.99 99.99     99.99 99    5.43  3    5.24  2    99.99 99    99.99 99
10859      6.33  0.69  0.94 99.99 99.99     99.99 99.99      8.82  4    6.75  4    6.01  3    99.99 99    99.99 99
10863      6.39 99.99  0.36 99.99 99.99     99.99 99.99     99.99 99    6.49  3    6.34  2    99.99 99    99.99 99
10939      5.04 99.99  0.04 99.99 99.99     99.99 99.99     99.99 99    4.97  3    5.14  2    99.99 99    99.99 99
10975      5.94  0.71  0.97 99.99 99.99     99.99 99.99      8.47  4    6.38  4    5.60  4    99.99 99    99.99 99
11025      5.69  0.66  0.94 99.99 99.99     99.99 99.99      8.14  4    6.11  4    5.37  3    99.99 99    99.99 99
11037      5.91  0.74  0.97 99.99 99.99     99.99 99.99      8.48  5    6.35  4    5.57  4    99.99 99    99.99 99
```

```
11050      6.32 99.99  1.02 99.99 99.99     99.99 99.99    99.99 99   6.79  4   5.96  4   99.99 99   99.99 99
11154      5.86  0.50  0.74 99.99 99.99     99.99 99.99     7.98  5   6.17  3   5.63  3   99.99 99   99.99 99
11262      6.37 99.99  0.52 99.99 99.99     99.99 99.99    99.99 99   6.56  3   6.24  3   99.99 99   99.99 99
11332      6.14 99.99  1.01 99.99 99.99     99.99 99.99    99.99 99   6.60  4   5.79  4   99.99 99   99.99 99
11604      6.07  0.07  0.33 99.99 99.99     99.99 99.99     7.38  6   6.16  3   6.03  2   99.99 99   99.99 99
11749      5.67  0.91  1.06 99.99 99.99     99.99 99.99     8.52  6   6.16  4   5.29  4   99.99 99   99.99 99
11803      6.01  0.04  0.56 99.99 99.99     99.99 99.99     7.41  3   6.22  3   5.86  3   99.99 99   99.99 99
11937      3.70  0.46  0.85 99.99 99.99     99.99 99.99     5.83  3   4.07  4   3.42  3   99.99 99   99.99 99
11977      4.69  0.64  0.95 99.99 99.99     99.99 99.99     7.11  3   5.12  4   4.36  4   99.99 99   99.99 99
11995      6.06 99.99  0.37 99.99 99.99     99.99 99.99    99.99 99   6.17  3   6.00  2   99.99 99   99.99 99
12042      6.10 99.99  0.48 99.99 99.99     99.99 99.99    99.99 99   6.27  3   5.99  3   99.99 99   99.99 99
12135      6.35 99.99  1.02 99.99 99.99     99.99 99.99    99.99 99   6.82  4   5.99  4   99.99 99   99.99 99
12235      5.88  0.19  0.62 99.99 99.99     99.99 99.99     7.51  3   6.13  3   5.71  3   99.99 99   99.99 99
12270      6.37  0.62  0.90 99.99 99.99     99.99 99.99     8.74  4   6.77  4   6.07  3   99.99 99   99.99 99
12296      5.57 99.99  1.06 99.99 99.99     99.99 99.99    99.99 99   6.06  4   5.19  4   99.99 99   99.99 99
12301      5.58 -0.27  0.38 99.99 99.99     99.99 99.99     6.46  5   5.69  3   5.52  2   99.99 99   99.99 99
12311      2.87  0.14  0.28 99.99 99.99     99.99 99.99     4.25  9   2.93  3   2.85  2   99.99 99   99.99 99
12363      6.16  0.01  0.44 99.99 99.99     99.99 99.99     7.45  3   6.31  3   6.07  3   99.99 99   99.99 99
12438      5.35 99.99  0.88 99.99 99.99     99.99 99.99    99.99 99   5.74  4   5.06  3   99.99 99   99.99 99
12477      6.10  1.21  1.17 99.99 99.99     99.99 99.99     9.42  9   6.65  4   5.67  4   99.99 99   99.99 99
12563      6.42 99.99  0.14 99.99 99.99     99.99 99.99    99.99 99   6.40  3   6.47  2   99.99 99   99.99 99
12573      5.43  0.12  0.15 99.99 99.99     99.99 99.99     6.71 12   5.42  3   5.48  2   99.99 99   99.99 99
12641      5.93  0.50  0.88 99.99 99.99     99.99 99.99     8.13  3   6.32  4   5.64  3   99.99 99   99.99 99
12642      5.62  1.97  1.59 99.99 99.99     99.99 99.99    10.20 15   6.40  5   5.00  5   99.99 99   99.99 99
12923      6.28  0.63  0.90 99.99 99.99     99.99 99.99     8.66  4   6.68  4   5.98  3   99.99 99   99.99 99
13421      5.63  0.14  0.56 99.99 99.99     99.99 99.99     7.16  3   5.84  3   5.48  3   99.99 99   99.99 99
13423      6.32 99.99  0.90 99.99 99.99     99.99 99.99    99.99 99   6.72  4   6.02  3   99.99 99   99.99 99
13445      6.12 99.99  0.82 99.99 99.99     99.99 99.99    99.99 99   6.48  4   5.85  3   99.99 99   99.99 99
13456      6.01 99.99  0.39 99.99 99.99     99.99 99.99    99.99 99   6.13  3   5.94  2   99.99 99   99.99 99
13468      5.93  0.70  0.97 99.99 99.99     99.99 99.99     8.45  4   6.37  4   5.59  4   99.99 99   99.99 99
13530      5.32  0.62  0.93 99.99 99.99     99.99 99.99     7.71  3   5.74  4   5.00  3   99.99 99   99.99 99
13612      5.51  0.09  0.58 99.99 99.99     99.99 99.99     6.99  3   5.73  3   5.36  3   99.99 99   99.99 99
13709      5.28 99.99 -0.02 99.99 99.99     99.99 99.99    99.99 99   5.17  2   5.40  2   99.99 99   99.99 99
13936      6.55 -0.07 -0.02 99.99 99.99     99.99 99.99     7.48 12   6.44  2   6.67  2   99.99 99   99.99 99
13940      5.91 99.99  0.97 99.99 99.99     99.99 99.99    99.99 99   6.35  4   5.57  4   99.99 99   99.99 99
13974      4.87  0.02  0.61 99.99 99.99     99.99 99.99     6.27  4   5.11  3   4.70  3   99.99 99   99.99 99
14129      5.51  0.75  0.96 99.99 99.99     99.99 99.99     8.09  5   5.94  4   5.18  4   99.99 99   99.99 99
14141      5.55  1.84  1.55 99.99 99.99     99.99 99.99     9.93 13   6.31  5   4.94  5   99.99 99   99.99 99
14214      5.56  0.09  0.61 99.99 99.99     99.99 99.99     7.05  3   5.80  3   5.39  3   99.99 99   99.99 99
14228      3.56 -0.39 -0.12 99.99 99.99     99.99 99.99     4.00  8   3.40  2   3.73  2   99.99 99   99.99 99
14287      5.69  1.48  1.30 99.99 99.99     99.99 99.99     9.45 12   6.31  4   5.20  4   99.99 99   99.99 99
```

```
14412     6.34 99.99   0.73 99.99 99.99    99.99 99.99   99.99 99   6.65  3    6.12  3   99.99 99   99.99 99
14417     6.50  0.09   0.08 99.99 99.99    99.99 99.99    7.70 13   6.45  3    6.58  2   99.99 99   99.99 99
14509     6.37 99.99   1.16 99.99 99.99    99.99 99.99   99.99 99   6.91  4    5.95  4   99.99 99   99.99 99
14652     5.27  1.85   1.65 99.99 99.99    99.99 99.99    9.72 10   6.08  5    4.62  5   99.99 99   99.99 99
14690     5.42  0.10   0.31 99.99 99.99    99.99 99.99    6.76  8   5.50  3    5.39  2   99.99 99   99.99 99
14832     6.31 99.99   1.00 99.99 99.99    99.99 99.99   99.99 99   6.77  4    5.96  4   99.99 99   99.99 99
14890     6.53 99.99   1.61 99.99 99.99    99.99 99.99   99.99 99   7.32  5    5.90  5   99.99 99   99.99 99
14943     5.92 99.99   0.22 99.99 99.99    99.99 99.99   99.99 99   5.95  3    5.93  2   99.99 99   99.99 99
15008     4.09  0.05   0.03 99.99 99.99    99.99 99.99    5.21 14   4.01  3    4.19  2   99.99 99   99.99 99
15064     6.18 99.99   0.66 99.99 99.99    99.99 99.99   99.99 99   6.45  3    5.99  3   99.99 99   99.99 99
15220     5.88 99.99   1.26 99.99 99.99    99.99 99.99   99.99 99   6.48  4    5.41  4   99.99 99   99.99 99
15233     5.35 99.99   0.39 99.99 99.99    99.99 99.99   99.99 99   5.47  3    5.28  2   99.99 99   99.99 99
15248     5.01  1.04   1.09 99.99 99.99    99.99 99.99    8.05  8   5.51  4    4.62  4   99.99 99   99.99 99
15328     6.45  0.79   0.97 99.99 99.99    99.99 99.99    9.09  5   6.89  4    6.11  4   99.99 99   99.99 99
15427     5.14 99.99   0.10 99.99 99.99    99.99 99.99   99.99 99   5.10  3    5.21  2   99.99 99   99.99 99
15453     6.07  0.85   1.02 99.99 99.99    99.99 99.99    8.82  5   6.54  4    5.71  4   99.99 99   99.99 99
15588     6.77 99.99   0.19 99.99 99.99    99.99 99.99   99.99 99   6.78  3    6.80  2   99.99 99   99.99 99
15596     6.23  0.55   0.90 99.99 99.99    99.99 99.99    8.51  3   6.63  4    5.93  3   99.99 99   99.99 99
15633     6.00  0.17   0.17 99.99 99.99    99.99 99.99    7.36 13   6.00  3    6.04  2   99.99 99   99.99 99
15646     6.37 -0.05  -0.04 99.99 99.99    99.99 99.99    7.32 13   6.25  2    6.50  2   99.99 99   99.99 99
15694     5.25  1.41   1.27 99.99 99.99    99.99 99.99    8.89 11   5.85  4    4.77  4   99.99 99   99.99 99
15779     5.35  0.84   1.02 99.99 99.99    99.99 99.99    8.08  5   5.82  4    4.99  4   99.99 99   99.99 99
15889     6.30 99.99   1.02 99.99 99.99    99.99 99.99   99.99 99   6.77  4    5.94  4   99.99 99   99.99 99
15996     6.21 99.99   1.10 99.99 99.99    99.99 99.99   99.99 99   6.72  4    5.81  4   99.99 99   99.99 99
16060     6.18  0.88   1.06 99.99 99.99    99.99 99.99    8.99  5   6.67  4    5.80  4   99.99 99   99.99 99
16074     5.75  1.65   1.40 99.99 99.99    99.99 99.99    9.79 13   6.43  4    5.21  4   99.99 99   99.99 99
16212     5.53  1.93   1.59 99.99 99.99    99.99 99.99   10.05 14   6.31  5    4.91  5   99.99 99   99.99 99
16226     6.77 99.99  -0.06 99.99 99.99    99.99 99.99   99.99 99   6.64  2    6.91  2   99.99 99   99.99 99
16247     5.81  0.86   1.04 99.99 99.99    99.99 99.99    8.58  5   6.29  4    5.44  4   99.99 99   99.99 99
16307     5.75 99.99   1.02 99.99 99.99    99.99 99.99   99.99 99   6.22  4    5.39  4   99.99 99   99.99 99
16399     6.39  0.00   0.44 99.99 99.99    99.99 99.99    7.67  3   6.54  3    6.30  3   99.99 99   99.99 99
16400     5.65  0.85   1.02 99.99 99.99    99.99 99.99    8.40  5   6.12  4    5.29  4   99.99 99   99.99 99
16417     5.79 99.99   0.66 99.99 99.99    99.99 99.99   99.99 99   6.06  3    5.60  3   99.99 99   99.99 99
16467     6.21  0.76   1.00 99.99 99.99    99.99 99.99    8.82  4   6.67  4    5.86  4   99.99 99   99.99 99
16522     5.28  0.73   0.98 99.99 99.99    99.99 99.99    7.84  4   5.72  4    4.94  4   99.99 99   99.99 99
16538     5.83 99.99   0.48 99.99 99.99    99.99 99.99   99.99 99   6.00  3    5.72  3   99.99 99   99.99 99
16555     5.31 99.99   0.27 99.99 99.99    99.99 99.99   99.99 99   5.36  3    5.30  2   99.99 99   99.99 99
16589     6.49 99.99   0.52 99.99 99.99    99.99 99.99   99.99 99   6.68  3    6.36  3   99.99 99   99.99 99
16620     4.84 -0.01   0.45 99.99 99.99    99.99 99.99    6.11  3   4.99  3    4.75  3   99.99 99   99.99 99
16647     6.25 -0.03   0.40 99.99 99.99    99.99 99.99    7.46  3   6.38  3    6.18  3   99.99 99   99.99 99
16673     5.78  0.00   0.52 99.99 99.99    99.99 99.99    7.10  3   5.97  3    5.65  3   99.99 99   99.99 99
```

```
16733      6.52 99.99  1.04 99.99 99.99    99.99 99.99    99.99 99    7.00  4    6.15  4    99.99 99    99.99 99
16765      5.71 -0.02  0.52 99.99 99.99    99.99 99.99     7.00  3    5.90  3    5.58  3    99.99 99    99.99 99
16824      6.05  1.14  1.16 99.99 99.99    99.99 99.99     9.27  8    6.59  4    5.63  4    99.99 99    99.99 99
16891      6.55 -0.24 -0.08 99.99 99.99    99.99 99.99     7.22 10    6.41  2    6.70  2    99.99 99    99.99 99
16975      6.01 99.99  0.92 99.99 99.99    99.99 99.99    99.99 99    6.42  4    5.70  3    99.99 99    99.99 99
16978      4.11 -0.14 -0.06 99.99 99.99    99.99 99.99     4.92 12    3.98  2    4.25  2    99.99 99    99.99 99
17006      6.10 99.99  0.88 99.99 99.99    99.99 99.99    99.99 99    6.49  4    5.81  3    99.99 99    99.99 99
17051      5.41 99.99  0.56 99.99 99.99    99.99 99.99    99.99 99    5.62  3    5.26  3    99.99 99    99.99 99
17098      6.36 99.99 -0.02 99.99 99.99    99.99 99.99    99.99 99    6.25  2    6.48  2    99.99 99    99.99 99
17163      6.03  0.09  0.31 99.99 99.99    99.99 99.99     7.36  7    6.11  3    6.00  2    99.99 99    99.99 99
17168      6.22 99.99  0.04 99.99 99.99    99.99 99.99    99.99 99    6.15  3    6.32  2    99.99 99    99.99 99
17254      6.15 99.99  0.09 99.99 99.99    99.99 99.99    99.99 99    6.10  3    6.22  2    99.99 99    99.99 99
17325      6.85 99.99  1.36 99.99 99.99    99.99 99.99    99.99 99    7.50  4    6.33  4    99.99 99    99.99 99
17326      6.26  0.01  0.53 99.99 99.99    99.99 99.99     7.60  3    6.46  3    6.13  3    99.99 99    99.99 99
17390      6.49 99.99  0.38 99.99 99.99    99.99 99.99    99.99 99    6.60  3    6.43  2    99.99 99    99.99 99
17438      6.47 99.99  0.39 99.99 99.99    99.99 99.99    99.99 99    6.59  3    6.40  2    99.99 99    99.99 99
17566      4.84  0.09  0.06 99.99 99.99    99.99 99.99     6.03 14    4.78  3    4.93  2    99.99 99    99.99 99
17713      6.14 99.99  1.07 99.99 99.99    99.99 99.99    99.99 99    6.63  4    5.76  4    99.99 99    99.99 99
17729      5.39 99.99  0.02 99.99 99.99    99.99 99.99    99.99 99    5.31  3    5.50  2    99.99 99    99.99 99
17848      5.26 99.99  0.10 99.99 99.99    99.99 99.99    99.99 99    5.22  3    5.33  2    99.99 99    99.99 99
17864      6.36 99.99  0.05 99.99 99.99    99.99 99.99    99.99 99    6.29  3    6.45  2    99.99 99    99.99 99
17926      6.40 99.99  0.48 99.99 99.99    99.99 99.99    99.99 99    6.57  3    6.29  3    99.99 99    99.99 99
17943      6.32  0.12  0.19 99.99 99.99    99.99 99.99     7.62 11    6.33  3    6.35  2    99.99 99    99.99 99
18071      5.95 99.99  1.04 99.99 99.99    99.99 99.99    99.99 99    6.43  4    5.58  4    99.99 99    99.99 99
18149      5.92 99.99  0.44 99.99 99.99    99.99 99.99    99.99 99    6.07  3    5.83  3    99.99 99    99.99 99
18185      6.03 99.99  1.25 99.99 99.99    99.99 99.99    99.99 99    6.62  4    5.56  4    99.99 99    99.99 99
18262      5.97  0.06  0.48 99.99 99.99    99.99 99.99     7.35  3    6.14  3    5.86  3    99.99 99    99.99 99
18293      4.75  1.56  1.33 99.99 99.99    99.99 99.99     8.63 13    5.39  4    4.25  4    99.99 99    99.99 99
18423      6.56  1.59  1.39 99.99 99.99    99.99 99.99    10.51 12    7.23  4    6.03  4    99.99 99    99.99 99
18454      5.45 99.99  0.23 99.99 99.99    99.99 99.99    99.99 99    5.48  3    5.46  2    99.99 99    99.99 99
18535      5.84 99.99  1.33 99.99 99.99    99.99 99.99    99.99 99    6.48  4    5.34  4    99.99 99    99.99 99
18537      5.28 -0.45 -0.05 99.99 99.99    99.99 99.99     5.68  5    5.16  2    5.42  2    99.99 99    99.99 99
18538      6.74 -0.15  0.00 99.99 99.99    99.99 99.99     7.57 10    6.65  3    6.85  2    99.99 99    99.99 99
18543      5.23  0.04  0.00 99.99 99.99    99.99 99.99     6.32 14    5.14  3    5.34  2    99.99 99    99.99 99
18546      6.41 99.99 -0.03 99.99 99.99    99.99 99.99    99.99 99    6.30  2    6.54  2    99.99 99    99.99 99
18557      6.15 99.99  0.22 99.99 99.99    99.99 99.99    99.99 99    6.18  3    6.16  2    99.99 99    99.99 99
18633      5.56 -0.19 -0.08 99.99 99.99    99.99 99.99     6.29 11    5.42  2    5.71  2    99.99 99    99.99 99
18650      6.14 99.99  1.04 99.99 99.99    99.99 99.99    99.99 99    6.62  4    5.77  4    99.99 99    99.99 99
18692      5.71 99.99  0.40 99.99 99.99    99.99 99.99    99.99 99    5.84  3    5.64  3    99.99 99    99.99 99
18735      6.31 99.99  0.00 99.99 99.99    99.99 99.99    99.99 99    6.22  3    6.42  2    99.99 99    99.99 99
18784      5.75  0.87  1.05 99.99 99.99    99.99 99.99     8.54  5    6.23  4    5.38  4    99.99 99    99.99 99
```

```
18832      6.25   0.82   1.05 99.99 99.99      99.99 99.99      8.97  4      6.73  4      5.88  4     99.99 99     99.99 99
18866      4.99   0.15   0.13 99.99 99.99      99.99 99.99      6.30 13      4.97  3      5.04  2     99.99 99     99.99 99
18883      5.62  -0.41  -0.10 99.99 99.99      99.99 99.99      6.04  7      5.47  2      5.78  2     99.99 99     99.99 99
18885      5.83 99.99   1.11 99.99 99.99       99.99 99.99     99.99 99      6.35  4      5.43  4     99.99 99     99.99 99
18894      6.19   0.15   0.60 99.99 99.99      99.99 99.99      7.76  3      6.43  3      6.03  3     99.99 99     99.99 99
18907      5.89 99.99   0.79 99.99 99.99       99.99 99.99     99.99 99      6.23  3      5.64  3     99.99 99     99.99 99
18953      5.32   0.73   0.94 99.99 99.99      99.99 99.99      7.86  5      5.74  4      5.00  3     99.99 99     99.99 99
19107      5.26   0.09   0.20 99.99 99.99      99.99 99.99      6.53 10      5.28  3      5.28  2     99.99 99     99.99 99
19121      6.05   0.85   1.04 99.99 99.99      99.99 99.99      8.81  5      6.53  4      5.68  4     99.99 99     99.99 99
19319      5.11 99.99   0.34 99.99 99.99       99.99 99.99     99.99 99      5.20  3      5.07  2     99.99 99     99.99 99
19349      5.27   1.77   1.60 99.99 99.99      99.99 99.99      9.58 10      6.06  5      4.64  5     99.99 99     99.99 99
19400      5.53  -0.51  -0.14 99.99 99.99      99.99 99.99      5.80  6      5.36  2      5.71  2     99.99 99     99.99 99
19525      6.28   0.88   1.06 99.99 99.99      99.99 99.99      9.09  5      6.77  4      5.90  4     99.99 99     99.99 99
19545      6.19 99.99   0.16 99.99 99.99       99.99 99.99     99.99 99      6.18  3      6.23  2     99.99 99     99.99 99
19637      6.02   1.28   1.28 99.99 99.99      99.99 99.99      9.49  8      6.63  4      5.54  4     99.99 99     99.99 99
19826      6.38 99.99   0.93 99.99 99.99       99.99 99.99     99.99 99      6.80  4      6.06  3     99.99 99     99.99 99
19836      6.05   1.93   1.66 99.99 99.99      99.99 99.99     10.61 12      6.87  5      5.39  5     99.99 99     99.99 99
19887      6.26 99.99   1.20 99.99 99.99       99.99 99.99     99.99 99      6.83  4      5.82  4     99.99 99     99.99 99
19940      6.15   0.83   1.02 99.99 99.99      99.99 99.99      8.87  5      6.62  4      5.79  4     99.99 99     99.99 99
19948      6.12 99.99   1.12 99.99 99.99       99.99 99.99     99.99 99      6.64  4      5.71  4     99.99 99     99.99 99
19978      5.45   0.11   0.19 99.99 99.99      99.99 99.99      6.74 11      5.46  3      5.48  2     99.99 99     99.99 99
19994      5.06   0.12   0.57 99.99 99.99      99.99 99.99      6.57  3      5.28  3      4.91  3     99.99 99     99.99 99
20084      5.61   0.49   0.92 99.99 99.99      99.99 99.99      7.81  3      6.02  4      5.30  3     99.99 99     99.99 99
20121      5.93 99.99   0.44 99.99 99.99       99.99 99.99     99.99 99      6.08  3      5.84  3     99.99 99     99.99 99
20144      6.27 99.99  -0.08 99.99 99.99       99.99 99.99     99.99 99      6.13  2      6.42  2     99.99 99     99.99 99
20176      6.16 99.99   1.05 99.99 99.99       99.99 99.99     99.99 99      6.64  4      5.79  4     99.99 99     99.99 99
20293      6.25 99.99   0.04 99.99 99.99       99.99 99.99     99.99 99      6.18  3      6.35  2     99.99 99     99.99 99
20315      5.47  -0.34  -0.06 99.99 99.99      99.99 99.99      6.01  8      5.34  2      5.61  2     99.99 99     99.99 99
20319      6.17  -0.25  -0.02 99.99 99.99      99.99 99.99      6.86  9      6.06  2      6.29  2     99.99 99     99.99 99
20395      6.14  -0.04   0.40 99.99 99.99      99.99 99.99      7.34  3      6.27  3      6.07  3     99.99 99     99.99 99
20423      6.65 99.99  -0.07 99.99 99.99       99.99 99.99     99.99 99      6.52  2      6.80  2     99.99 99     99.99 99
20559      5.37   0.81   1.04 99.99 99.99      99.99 99.99      8.07  4      5.85  4      5.00  4     99.99 99     99.99 99
20606      5.91 99.99   0.33 99.99 99.99       99.99 99.99     99.99 99      6.00  3      5.87  2     99.99 99     99.99 99
20610      4.88 99.99   0.90 99.99 99.99       99.99 99.99     99.99 99      5.28  4      4.58  3     99.99 99     99.99 99
20631      5.71 99.99   0.37 99.99 99.99       99.99 99.99     99.99 99      5.82  3      5.65  2     99.99 99     99.99 99
20640      5.85 99.99   1.24 99.99 99.99       99.99 99.99     99.99 99      6.44  4      5.39  4     99.99 99     99.99 99
20791      5.69   0.76   0.97 99.99 99.99      99.99 99.99      8.29  5      6.13  4      5.35  4     99.99 99     99.99 99
20853      6.39 99.99   0.54 99.99 99.99       99.99 99.99     99.99 99      6.59  3      6.25  3     99.99 99     99.99 99
20888      6.05   0.10   0.13 99.99 99.99      99.99 99.99      7.29 12      6.03  3      6.10  2     99.99 99     99.99 99
20980      6.35 99.99   0.01 99.99 99.99       99.99 99.99     99.99 99      6.26  3      6.46  2     99.99 99     99.99 99
21011      6.39 99.99   1.00 99.99 99.99       99.99 99.99     99.99 99      6.85  4      6.04  4     99.99 99     99.99 99
```

```
21018      6.38  0.51  0.86 99.99 99.99     99.99 99.99      8.58  3   6.76  4   6.10  3   99.99 99   99.99 99
21019      6.20  0.16  0.70 99.99 99.99     99.99 99.99      7.84  4   6.49  3   5.99  3   99.99 99   99.99 99
21024      5.52  0.00  0.44 99.99 99.99     99.99 99.99      6.80  3   5.67  3   5.43  3   99.99 99   99.99 99
21149      6.51 99.99  1.37 99.99 99.99     99.99 99.99     99.99 99   7.17  4   5.99  4   99.99 99   99.99 99
21423      6.39 99.99  0.08 99.99 99.99     99.99 99.99     99.99 99   6.34  3   6.47  2   99.99 99   99.99 99
21430      5.93 99.99  0.94 99.99 99.99     99.99 99.99     99.99 99   6.35  4   5.61  3   99.99 99   99.99 99
21473      6.32 99.99  0.06 99.99 99.99     99.99 99.99     99.99 99   6.26  3   6.41  2   99.99 99   99.99 99
21563      6.15  0.25  0.48 99.99 99.99     99.99 99.99      7.79  6   6.32  3   6.04  3   99.99 99   99.99 99
21574      5.71 99.99  1.29 99.99 99.99     99.99 99.99     99.99 99   6.32  4   5.23  4   99.99 99   99.99 99
21635      6.50 99.99  0.13 99.99 99.99     99.99 99.99     99.99 99   6.48  3   6.55  2   99.99 99   99.99 99
21665      5.99  0.87  1.02 99.99 99.99     99.99 99.99      8.76  6   6.46  4   5.63  4   99.99 99   99.99 99
21688      5.59 99.99  0.18 99.99 99.99     99.99 99.99     99.99 99   5.59  3   5.62  2   99.99 99   99.99 99
21722      5.96  0.01  0.42 99.99 99.99     99.99 99.99      7.24  4   6.10  3   5.88  3   99.99 99   99.99 99
21755      5.94  0.63  0.96 99.99 99.99     99.99 99.99      8.36  3   6.37  4   5.61  4   99.99 99   99.99 99
21856      5.90 -0.86 -0.06 99.99 99.99     99.99 99.99      5.74  6   5.77  2   6.04  2   99.99 99   99.99 99
21882      5.78 99.99  0.22 99.99 99.99     99.99 99.99     99.99 99   5.81  3   5.79  2   99.99 99   99.99 99
21899      6.12 99.99  0.48 99.99 99.99     99.99 99.99     99.99 99   6.29  3   6.01  3   99.99 99   99.99 99
21933      5.77 -0.30 -0.08 99.99 99.99     99.99 99.99      6.35  9   5.63  2   5.92  2   99.99 99   99.99 99
21997      6.38 99.99  0.12 99.99 99.99     99.99 99.99     99.99 99   6.35  3   6.44  2   99.99 99   99.99 99
22211      6.49  0.16  0.63 99.99 99.99     99.99 99.99      8.09  3   6.74  3   6.31  3   99.99 99   99.99 99
22243      6.25 99.99  0.02 99.99 99.99     99.99 99.99     99.99 99   6.17  3   6.36  2   99.99 99   99.99 99
22252      5.83 -0.33 -0.06 99.99 99.99     99.99 99.99      6.38  8   5.70  2   5.97  2   99.99 99   99.99 99
22262      6.20 99.99  0.48 99.99 99.99     99.99 99.99     99.99 99   6.37  3   6.09  3   99.99 99   99.99 99
22634      6.75  0.15  0.17 99.99 99.99     99.99 99.99      8.08 12   6.75  3   6.79  2   99.99 99   99.99 99
22675      5.85  0.76  0.98 99.99 99.99     99.99 99.99      8.45  5   6.29  4   5.51  4   99.99 99   99.99 99
22676      5.70  0.65  0.93 99.99 99.99     99.99 99.99      8.13  4   6.12  4   5.38  3   99.99 99   99.99 99
22713      5.96  0.66  0.92 99.99 99.99     99.99 99.99      8.40  4   6.37  4   5.65  3   99.99 99   99.99 99
22789      6.01 99.99 -0.02 99.99 99.99     99.99 99.99     99.99 99   5.90  2   6.13  2   99.99 99   99.99 99
22796      5.57  0.67  0.94 99.99 99.99     99.99 99.99      8.03  4   5.99  4   5.25  3   99.99 99   99.99 99
22798      6.23  0.89  1.04 99.99 99.99     99.99 99.99      9.04  6   6.71  4   5.86  4   99.99 99   99.99 99
22819      6.12  0.77  1.00 99.99 99.99     99.99 99.99      8.75  4   6.58  4   5.77  4   99.99 99   99.99 99
22951      4.97 -0.84  0.00 99.99 99.99     99.99 99.99      4.87  7   4.88  3   5.08  2   99.99 99   99.99 99
23010      6.49 99.99  0.36 99.99 99.99     99.99 99.99     99.99 99   6.59  3   6.44  2   99.99 99   99.99 99
23055      6.59 99.99  0.09 99.99 99.99     99.99 99.99     99.99 99   6.54  3   6.66  2   99.99 99   99.99 99
23183      6.14  0.74  1.01 99.99 99.99     99.99 99.99      8.73  4   6.60  4   5.79  4   99.99 99   99.99 99
23281      5.59 99.99  0.23 99.99 99.99     99.99 99.99     99.99 99   5.62  3   5.60  2   99.99 99   99.99 99
23319      4.59  1.31  1.20 99.99 99.99     99.99 99.99      8.06 10   5.16  4   4.15  4   99.99 99   99.99 99
23363      5.25 -0.39 -0.10 99.99 99.99     99.99 99.99      5.70  8   5.10  2   5.41  2   99.99 99   99.99 99
23413      5.55  1.72  1.42 99.99 99.99     99.99 99.99      9.70 14   6.24  4   5.00  5   99.99 99   99.99 99
23474      6.29  1.03  1.15 99.99 99.99     99.99 99.99      9.35  6   6.83  4   5.87  4   99.99 99   99.99 99
23508      6.45 99.99  1.06 99.99 99.99     99.99 99.99     99.99 99   6.94  4   6.07  4   99.99 99   99.99 99
```

```
23526     5.91  0.75  1.00 99.99 99.99     99.99 99.99     8.51  4    6.37  4    5.56  4    99.99 99    99.99 99
23625     6.57 -0.61  0.08 99.99 99.99     99.99 99.99     6.82  4    6.52  3    6.65  2    99.99 99    99.99 99
23670     6.49 99.99  1.02 99.99 99.99     99.99 99.99    99.99 99    6.96  4    6.13  4    99.99 99    99.99 99
23697     6.30 99.99  1.04 99.99 99.99     99.99 99.99    99.99 99    6.78  4    5.93  4    99.99 99    99.99 99
23719     5.73 99.99  0.96 99.99 99.99     99.99 99.99    99.99 99    6.16  4    5.40  4    99.99 99    99.99 99
23738     5.90 99.99  0.12 99.99 99.99     99.99 99.99    99.99 99    5.87  3    5.96  2    99.99 99    99.99 99
23817     3.85  1.10  1.13 99.99 99.99     99.99 99.99     7.00  8    4.38  4    3.44  4    99.99 99    99.99 99
23856     6.55 99.99  0.49 99.99 99.99     99.99 99.99    99.99 99    6.72  3    6.44  3    99.99 99    99.99 99
23878     5.24 99.99  0.07 99.99 99.99     99.99 99.99    99.99 99    5.18  3    5.32  2    99.99 99    99.99 99
23887     5.91  1.31  1.24 99.99 99.99     99.99 99.99     9.40  9    6.50  4    5.45  4    99.99 99    99.99 99
23940     5.54 99.99  0.98 99.99 99.99     99.99 99.99    99.99 99    5.98  4    5.20  4    99.99 99    99.99 99
23958     6.21 99.99 -0.10 99.99 99.99     99.99 99.99    99.99 99    6.06  2    6.37  2    99.99 99    99.99 99
23978     5.81 99.99  1.64 99.99 99.99     99.99 99.99    99.99 99    6.62  5    5.16  5    99.99 99    99.99 99
24071     4.27 -0.04 -0.01 99.99 99.99     99.99 99.99     5.25 13    4.17  3    4.39  2    99.99 99    99.99 99
24131     5.77 -0.81  0.00 99.99 99.99     99.99 99.99     5.71  6    5.68  3    5.88  2    99.99 99    99.99 99
24263     5.67 -0.43  0.06 99.99 99.99     99.99 99.99     6.16  3    5.61  3    5.76  2    99.99 99    99.99 99
24305     6.86 99.99 -0.04 99.99 99.99     99.99 99.99    99.99 99    6.74  2    6.99  2    99.99 99    99.99 99
24388     5.48 -0.41 -0.10 99.99 99.99     99.99 99.99     5.90  7    5.33  2    5.64  2    99.99 99    99.99 99
24497     6.22 99.99  0.88 99.99 99.99     99.99 99.99    99.99 99    6.61  4    5.93  3    99.99 99    99.99 99
24512     3.25  1.98  1.62 99.99 99.99     99.99 99.99     7.86 14    4.05  5    2.61  5    99.99 99    99.99 99
24546     5.28  0.00  0.41 99.99 99.99     99.99 99.99     6.54  4    5.41  3    5.20  3    99.99 99    99.99 99
24626     5.11 99.99 -0.13 99.99 99.99     99.99 99.99    99.99 99    4.94  2    5.29  2    99.99 99    99.99 99
24640     5.49 -0.75 -0.03 99.99 99.99     99.99 99.99     5.49  4    5.38  2    5.62  2    99.99 99    99.99 99
24706     5.94  1.35  1.23 99.99 99.99     99.99 99.99     9.48 11    6.52  4    5.48  4    99.99 99    99.99 99
24744     5.71 99.99  0.60 99.99 99.99     99.99 99.99    99.99 99    5.95  3    5.55  3    99.99 99    99.99 99
24817     6.09  0.05  0.06 99.99 99.99     99.99 99.99     7.23 13    6.03  3    6.18  2    99.99 99    99.99 99
24863     6.46 99.99  0.16 99.99 99.99     99.99 99.99    99.99 99    6.45  3    6.50  2    99.99 99    99.99 99
25069     5.83  0.85  1.00 99.99 99.99     99.99 99.99     8.57  6    6.29  4    5.48  4    99.99 99    99.99 99
25330     5.67 -0.41  0.02 99.99 99.99     99.99 99.99     6.16  4    5.59  3    5.78  2    99.99 99    99.99 99
25340     5.28 -0.55 -0.15 99.99 99.99     99.99 99.99     5.49  6    5.10  2    5.46  2    99.99 99    99.99 99
25346     6.05 99.99  0.44 99.99 99.99     99.99 99.99    99.99 99    6.20  3    5.96  3    99.99 99    99.99 99
25371     5.93 99.99  0.04 99.99 99.99     99.99 99.99    99.99 99    5.86  3    6.03  2    99.99 99    99.99 99
25422     4.56  1.95  1.62 99.99 99.99     99.99 99.99     9.13 13    5.36  5    3.92  5    99.99 99    99.99 99
25457     5.38  0.00  0.50 99.99 99.99     99.99 99.99     6.69  3    5.56  3    5.26  3    99.99 99    99.99 99
25621     5.36  0.04  0.50 99.99 99.99     99.99 99.99     6.72  3    5.54  3    5.24  3    99.99 99    99.99 99
25631     6.46 99.99 -0.18 99.99 99.99     99.99 99.99    99.99 99    6.27  2    6.66  2    99.99 99    99.99 99
25680     5.90  0.12  0.62 99.99 99.99     99.99 99.99     7.44  3    6.15  3    5.73  3    99.99 99    99.99 99
25700     6.39 99.99  1.26 99.99 99.99     99.99 99.99    99.99 99    6.99  4    5.92  4    99.99 99    99.99 99
25728     4.97  1.70  1.42 99.99 99.99     99.99 99.99     9.09 13    5.66  4    4.42  5    99.99 99    99.99 99
25803     6.13 99.99  1.16 99.99 99.99     99.99 99.99    99.99 99    6.67  4    5.71  4    99.99 99    99.99 99
25887     6.41 -0.10 -0.01 99.99 99.99     99.99 99.99     7.30 12    6.31  3    6.53  2    99.99 99    99.99 99
```

```
25910      6.26  0.09   0.06 99.99 99.99     99.99 99.99     7.45 14    6.20  3    6.35  2    99.99 99    99.99 99
25944      6.34 99.99   0.92 99.99 99.99     99.99 99.99    99.99 99    6.75  4    6.03  3    99.99 99    99.99 99
25945      5.58 99.99   0.32 99.99 99.99     99.99 99.99    99.99 99    5.66  3    5.55  2    99.99 99    99.99 99
25975      6.09  0.75   0.95 99.99 99.99     99.99 99.99     8.66  5    6.52  4    5.76  4    99.99 99    99.99 99
26262      6.59  0.60   0.93 99.99 99.99     99.99 99.99     8.95  3    7.01  4    6.27  3    99.99 99    99.99 99
26409      5.44  0.67   0.94 99.99 99.99     99.99 99.99     7.90  4    5.86  4    5.12  3    99.99 99    99.99 99
26413      6.59 -0.01   0.37 99.99 99.99     99.99 99.99     7.81  4    6.70  3    6.53  2    99.99 99    99.99 99
26464      5.70  0.93   1.06 99.99 99.99     99.99 99.99     8.58  6    6.19  4    5.32  4    99.99 99    99.99 99
26491      6.38  0.11   0.64 99.99 99.99     99.99 99.99     7.92  4    6.64  3    6.20  3    99.99 99    99.99 99
26575      6.44 99.99   1.07 99.99 99.99     99.99 99.99    99.99 99    6.93  4    6.06  4    99.99 99    99.99 99
26659      5.46  0.46   0.87 99.99 99.99     99.99 99.99     7.60  3    5.84  4    5.17  3    99.99 99    99.99 99
26676      6.23 -0.34   0.05 99.99 99.99     99.99 99.99     6.83  5    6.16  3    6.32  2    99.99 99    99.99 99
26793      5.22 -0.34  -0.10 99.99 99.99     99.99 99.99     5.74  9    5.07  2    5.38  2    99.99 99    99.99 99
26820      6.71  1.80   1.47 99.99 99.99     99.99 99.99    10.99 14    7.42  4    6.14  5    99.99 99    99.99 99
26927      6.37 99.99   1.46 99.99 99.99     99.99 99.99    99.99 99    7.08  4    5.81  5    99.99 99    99.99 99
27179      5.94  0.94   1.08 99.99 99.99     99.99 99.99     8.84  6    6.44  4    5.55  4    99.99 99    99.99 99
27236      6.54  0.17   0.16 99.99 99.99     99.99 99.99     7.89 13    6.53  3    6.58  2    99.99 99    99.99 99
27256      3.35  0.62   0.91 99.99 99.99     99.99 99.99     5.73  4    3.76  4    3.04  3    99.99 99    99.99 99
27304      5.45 99.99   1.10 99.99 99.99     99.99 99.99    99.99 99    5.96  4    5.05  4    99.99 99    99.99 99
27386      6.31  1.60   1.43 99.99 99.99     99.99 99.99    10.30 11    7.00  4    5.76  5    99.99 99    99.99 99
27411      6.07 99.99   0.30 99.99 99.99     99.99 99.99    99.99 99    6.14  3    6.05  2    99.99 99    99.99 99
27490      6.37 99.99   0.13 99.99 99.99     99.99 99.99    99.99 99    6.35  3    6.42  2    99.99 99    99.99 99
27497      5.77  0.68   0.92 99.99 99.99     99.99 99.99     8.23  5    6.18  4    5.46  3    99.99 99    99.99 99
27505      6.53  0.11   0.15 99.99 99.99     99.99 99.99     7.80 12    6.52  3    6.58  2    99.99 99    99.99 99
27588      5.34  0.94   1.08 99.99 99.99     99.99 99.99     8.24  6    5.84  4    4.95  4    99.99 99    99.99 99
27604      6.09 99.99   0.49 99.99 99.99     99.99 99.99    99.99 99    6.26  3    5.98  3    99.99 99    99.99 99
27611      5.86  1.52   1.32 99.99 99.99     99.99 99.99     9.68 12    6.49  4    5.36  4    99.99 99    99.99 99
27710      5.96 99.99   0.35 99.99 99.99     99.99 99.99    99.99 99    6.06  3    5.91  2    99.99 99    99.99 99
27881      5.83 99.99   1.51 99.99 99.99     99.99 99.99    99.99 99    6.57  5    5.24  5    99.99 99    99.99 99
27941      6.39 99.99   1.24 99.99 99.99     99.99 99.99    99.99 99    6.98  4    5.93  4    99.99 99    99.99 99
28143      6.55 99.99   0.44 99.99 99.99     99.99 99.99    99.99 99    6.70  3    6.46  3    99.99 99    99.99 99
28191      6.23  1.03   1.09 99.99 99.99     99.99 99.99     9.26  7    6.73  4    5.84  4    99.99 99    99.99 99
28246      6.39 99.99   0.44 99.99 99.99     99.99 99.99    99.99 99    6.54  3    6.30  3    99.99 99    99.99 99
28255      6.28 99.99   0.66 99.99 99.99     99.99 99.99    99.99 99    6.55  3    6.09  3    99.99 99    99.99 99
28312      6.11 99.99   0.14 99.99 99.99     99.99 99.99    99.99 99    6.09  3    6.16  2    99.99 99    99.99 99
28322      6.15  0.81   1.02 99.99 99.99     99.99 99.99     8.84  5    6.62  4    5.79  4    99.99 99    99.99 99
28375      5.55 -0.55  -0.10 99.99 99.99     99.99 99.99     5.78  5    5.40  2    5.71  2    99.99 99    99.99 99
28413      5.95 99.99   1.54 99.99 99.99     99.99 99.99    99.99 99    6.70  5    5.35  5    99.99 99    99.99 99
28454      6.10 99.99   0.46 99.99 99.99     99.99 99.99    99.99 99    6.26  3    6.00  3    99.99 99    99.99 99
28479      5.96 99.99   1.22 99.99 99.99     99.99 99.99    99.99 99    6.54  4    5.51  4    99.99 99    99.99 99
28525      5.69  0.53   0.84 99.99 99.99     99.99 99.99     7.91  4    6.06  4    5.41  3    99.99 99    99.99 99
```

```
28625      6.24 99.99  1.00 99.99 99.99    99.99 99.99    99.99 99    6.70  4    5.89  4    99.99 99    99.99 99
28700      6.16 99.99  1.06 99.99 99.99    99.99 99.99    99.99 99    6.65  4    5.78  4    99.99 99    99.99 99
28763      6.21 99.99  0.12 99.99 99.99    99.99 99.99    99.99 99    6.18  3    6.27  2    99.99 99    99.99 99
28776      5.96 99.99  1.00 99.99 99.99    99.99 99.99    99.99 99    6.42  4    5.61  4    99.99 99    99.99 99
28873      5.07 99.99 -0.19 99.99 99.99    99.99 99.99    99.99 99    4.87  2    5.27  2    99.99 99    99.99 99
28930      6.01  0.81  1.06 99.99 99.99    99.99 99.99     8.73  4    6.50  4    5.63  4    99.99 99    99.99 99
29063      6.09  1.64  1.38 99.99 99.99    99.99 99.99    10.11 13    6.75  4    5.56  4    99.99 99    99.99 99
29065      5.27  1.70  1.46 99.99 99.99    99.99 99.99     9.41 12    5.98  4    4.71  5    99.99 99    99.99 99
29116      5.79  0.05  0.35 99.99 99.99    99.99 99.99     7.09  6    5.89  3    5.74  2    99.99 99    99.99 99
29173      6.37  0.10  0.11 99.99 99.99    99.99 99.99     7.60 13    6.34  3    6.43  2    99.99 99    99.99 99
29184      6.13 99.99  1.17 99.99 99.99    99.99 99.99    99.99 99    6.68  4    5.70  4    99.99 99    99.99 99
29227      6.33 -0.43 -0.10 99.99 99.99    99.99 99.99     6.73  7    6.18  2    6.49  2    99.99 99    99.99 99
29335      5.31 -0.44 -0.12 99.99 99.99    99.99 99.99     5.68  7    5.15  2    5.48  2    99.99 99    99.99 99
29391      5.23  0.04  0.28 99.99 99.99    99.99 99.99     6.47  7    5.29  3    5.21  2    99.99 99    99.99 99
29435      6.30 99.99 -0.10 99.99 99.99    99.99 99.99    99.99 99    6.15  2    6.46  2    99.99 99    99.99 99
29573      5.01 99.99  0.08 99.99 99.99    99.99 99.99    99.99 99    4.96  3    5.09  2    99.99 99    99.99 99
29598      6.76  0.10  0.20 99.99 99.99    99.99 99.99     8.04 10    6.78  3    6.78  2    99.99 99    99.99 99
29610      6.11  0.70  0.94 99.99 99.99    99.99 99.99     8.61  4    6.53  4    5.79  3    99.99 99    99.99 99
29613      5.45 99.99  1.06 99.99 99.99    99.99 99.99    99.99 99    5.94  4    5.07  4    99.99 99    99.99 99
29737      5.58 99.99  0.92 99.99 99.99    99.99 99.99    99.99 99    5.99  4    5.27  3    99.99 99    99.99 99
29992      5.05 99.99  0.37 99.99 99.99    99.99 99.99    99.99 99    5.16  3    4.99  2    99.99 99    99.99 99
30003      6.53 99.99  0.68 99.99 99.99    99.99 99.99    99.99 99    6.81  3    6.33  3    99.99 99    99.99 99
30080      5.68 99.99  1.41 99.99 99.99    99.99 99.99    99.99 99    6.36  4    5.14  4    99.99 99    99.99 99
30127      5.53 99.99  0.02 99.99 99.99    99.99 99.99    99.99 99    5.45  3    5.64  2    99.99 99    99.99 99
30185      5.31 99.99  0.98 99.99 99.99    99.99 99.99    99.99 99    5.75  4    4.97  4    99.99 99    99.99 99
30202      6.25 99.99  1.46 99.99 99.99    99.99 99.99    99.99 99    6.96  4    5.69  5    99.99 99    99.99 99
30238      5.72 99.99  1.47 99.99 99.99    99.99 99.99    99.99 99    6.43  4    5.15  5    99.99 99    99.99 99
30321      6.33  0.06  0.04 99.99 99.99    99.99 99.99     7.47 14    6.26  3    6.43  2    99.99 99    99.99 99
30397      6.86 99.99  0.00 99.99 99.99    99.99 99.99    99.99 99    6.77  3    6.97  2    99.99 99    99.99 99
30432      6.05 99.99  1.07 99.99 99.99    99.99 99.99    99.99 99    6.54  4    5.67  4    99.99 99    99.99 99
30479      6.05  0.95  1.10 99.99 99.99    99.99 99.99     8.98  5    6.56  4    5.65  4    99.99 99    99.99 99
30545      6.03  1.14  1.19 99.99 99.99    99.99 99.99     9.26  7    6.59  4    5.59  4    99.99 99    99.99 99
30562      5.78  0.20  0.62 99.99 99.99    99.99 99.99     7.43  3    6.03  3    5.61  3    99.99 99    99.99 99
30606      5.77 99.99  0.55 99.99 99.99    99.99 99.99    99.99 99    5.98  3    5.63  3    99.99 99    99.99 99
30608      6.37 99.99  1.07 99.99 99.99    99.99 99.99    99.99 99    6.86  4    5.99  4    99.99 99    99.99 99
30610      6.46 99.99  1.08 99.99 99.99    99.99 99.99    99.99 99    6.96  4    6.07  4    99.99 99    99.99 99
30612      5.54 -0.46 -0.12 99.99 99.99    99.99 99.99     5.89  7    5.38  2    5.71  2    99.99 99    99.99 99
30743      6.26 99.99  0.44 99.99 99.99    99.99 99.99    99.99 99    6.41  3    6.17  3    99.99 99    99.99 99
30788      6.72 99.99  0.95 99.99 99.99    99.99 99.99    99.99 99    7.15  4    6.39  4    99.99 99    99.99 99
30870      6.11 -0.44  0.08 99.99 99.99    99.99 99.99     6.59  3    6.06  3    6.19  2    99.99 99    99.99 99
30985      6.07 99.99  0.37 99.99 99.99    99.99 99.99    99.99 99    6.18  3    6.01  2    99.99 99    99.99 99
```

```
31093      5.86 99.99   0.08 99.99 99.99      99.99 99.99    99.99 99    5.81  3   5.94  2   99.99 99    99.99 99
31139      5.33  1.93   1.64 99.99 99.99      99.99 99.99     9.88 12    6.14  5   4.68  5   99.99 99    99.99 99
31203      5.23 99.99   0.33 99.99 99.99      99.99 99.99    99.99 99    5.32  3   5.19  2   99.99 99    99.99 99
31209      6.61  0.00   0.04 99.99 99.99      99.99 99.99     7.67 12    6.54  3   6.71  2   99.99 99    99.99 99
31296      5.33  1.18   1.22 99.99 99.99      99.99 99.99     8.63  7    5.91  4   4.88  4   99.99 99    99.99 99
31312      6.06  1.83   1.57 99.99 99.99      99.99 99.99    10.44 12    6.83  5   5.44  5   99.99 99    99.99 99
31327      6.06 -0.44   0.41 99.99 99.99      99.99 99.99     6.72  9    6.19  3   5.98  3   99.99 99    99.99 99
31331      5.99 -0.56  -0.12 99.99 99.99      99.99 99.99     6.20  5    5.83  2   6.16  2   99.99 99    99.99 99
31411      6.50 -0.01   0.02 99.99 99.99      99.99 99.99     7.53 13    6.42  3   6.61  2   99.99 99    99.99 99
31414      5.70 99.99   0.96 99.99 99.99      99.99 99.99    99.99 99    6.13  4   5.37  4   99.99 99    99.99 99
31512      5.51 -0.56  -0.13 99.99 99.99      99.99 99.99     5.71  5    5.34  2   5.69  2   99.99 99    99.99 99
31517      6.72 99.99   0.27 99.99 99.99      99.99 99.99    99.99 99    6.77  3   6.71  2   99.99 99    99.99 99
31529      6.10 99.99   1.42 99.99 99.99      99.99 99.99    99.99 99    6.79  4   5.55  5   99.99 99    99.99 99
31623      6.23  0.06   0.42 99.99 99.99      99.99 99.99     7.58  4    6.37  3   6.15  3   99.99 99    99.99 99
31726      6.15 99.99  -0.21 99.99 99.99      99.99 99.99    99.99 99    5.94  2   6.36  2   99.99 99    99.99 99
31739      6.35  0.11   0.10 99.99 99.99      99.99 99.99     7.59 13    6.31  3   6.42  2   99.99 99    99.99 99
31746      6.12 99.99   0.44 99.99 99.99      99.99 99.99    99.99 99    6.27  3   6.03  3   99.99 99    99.99 99
31754      6.41  1.96   1.63 99.99 99.99      99.99 99.99    11.00 13    7.21  5   5.77  5   99.99 99    99.99 99
31925      5.67 99.99   0.44 99.99 99.99      99.99 99.99    99.99 99    5.82  3   5.58  3   99.99 99    99.99 99
31975      6.28  0.01   0.52 99.99 99.99      99.99 99.99     7.61  3    6.47  3   6.15  3   99.99 99    99.99 99
32115      6.32  0.04   0.28 99.99 99.99      99.99 99.99     7.56  7    6.38  3   6.30  2   99.99 99    99.99 99
32263      5.92  1.39   1.27 99.99 99.99      99.99 99.99     9.54 10    6.52  4   5.44  4   99.99 99    99.99 99
32273      6.24 -0.40  -0.04 99.99 99.99      99.99 99.99     6.71  6    6.12  2   6.37  2   99.99 99    99.99 99
32393      5.85  1.36   1.21 99.99 99.99      99.99 99.99     9.39 11    6.42  4   5.40  4   99.99 99    99.99 99
32436      5.02  0.97   1.07 99.99 99.99      99.99 99.99     7.96  7    5.51  4   4.64  4   99.99 99    99.99 99
32440      5.47  1.82   1.52 99.99 99.99      99.99 99.99     9.81 13    6.21  5   4.88  5   99.99 99    99.99 99
32453      6.03 99.99   0.88 99.99 99.99      99.99 99.99    99.99 99    6.42  4   5.74  3   99.99 99    99.99 99
32503      5.75 99.99   1.20 99.99 99.99      99.99 99.99    99.99 99    6.32  4   5.31  4   99.99 99    99.99 99
32515      5.94 99.99   1.17 99.99 99.99      99.99 99.99    99.99 99    6.49  4   5.51  4   99.99 99    99.99 99
32612      6.41 99.99  -0.18 99.99 99.99      99.99 99.99    99.99 99    6.22  2   6.61  2   99.99 99    99.99 99
32667      5.61 99.99   0.10 99.99 99.99      99.99 99.99    99.99 99    5.57  3   5.68  2   99.99 99    99.99 99
32686      6.05 -0.53  -0.11 99.99 99.99      99.99 99.99     6.31  5    5.89  2   6.22  2   99.99 99    99.99 99
32743      5.39 99.99   0.42 99.99 99.99      99.99 99.99    99.99 99    5.53  3   5.31  3   99.99 99    99.99 99
32820      6.31 99.99   0.53 99.99 99.99      99.99 99.99    99.99 99    6.51  3   6.18  3   99.99 99    99.99 99
32890      5.73 99.99   1.17 99.99 99.99      99.99 99.99    99.99 99    6.28  4   5.30  4   99.99 99    99.99 99
33021      6.17  0.09   0.62 99.99 99.99      99.99 99.99     7.67  3    6.42  3   6.00  3   99.99 99    99.99 99
33054      5.34  0.11   0.32 99.99 99.99      99.99 99.99     6.70  8    5.42  3   5.31  2   99.99 99    99.99 99
33093      5.97 99.99   0.60 99.99 99.99      99.99 99.99    99.99 99    6.21  3   5.81  3   99.99 99    99.99 99
33224      5.78 -0.36  -0.06 99.99 99.99      99.99 99.99     6.29  7    5.65  2   5.92  2   99.99 99    99.99 99
33256      5.12 -0.05   0.44 99.99 99.99      99.99 99.99     6.33  3    5.27  3   5.03  3   99.99 99    99.99 99
33285      5.31  0.77   1.00 99.99 99.99      99.99 99.99     7.94  4    5.77  4   4.96  4   99.99 99    99.99 99
```

```
33377      6.52 99.99  1.08 99.99 99.99     99.99 99.99    99.99 99    7.02  4    6.13  4   99.99 99   99.99 99
33419      6.10  1.07  1.10 99.99 99.99     99.99 99.99     9.19  8    6.61  4    5.70  4   99.99 99   99.99 99
33519      6.29  1.87  1.51 99.99 99.99     99.99 99.99    10.69 15    7.03  5    5.70  5   99.99 99   99.99 99
33555      6.25  0.81  0.98 99.99 99.99     99.99 99.99     8.92  6    6.69  4    5.91  4   99.99 99   99.99 99
33608      5.90  0.04  0.46 99.99 99.99     99.99 99.99     7.24  3    6.06  3    5.80  3   99.99 99   99.99 99
33646      5.89  0.29  0.66 99.99 99.99     99.99 99.99     7.68  3    6.16  3    5.70  3   99.99 99   99.99 99
33667      6.41 99.99  1.25 99.99 99.99     99.99 99.99    99.99 99    7.00  4    5.94  4   99.99 99   99.99 99
33833      5.91  0.72  0.96 99.99 99.99     99.99 99.99     8.45  4    6.34  4    5.58  4   99.99 99   99.99 99
33872      6.57 99.99  1.62 99.99 99.99     99.99 99.99    99.99 99    7.37  5    5.93  5   99.99 99   99.99 99
33875      6.27  0.01 -0.01 99.99 99.99     99.99 99.99     7.31 14    6.17  3    6.39  2   99.99 99   99.99 99
33883      6.09  0.28  0.42 99.99 99.99     99.99 99.99     7.74  9    6.23  3    6.01  3   99.99 99   99.99 99
33946      6.32  1.34  1.46 99.99 99.99     99.99 99.99     9.97  5    7.03  4    5.76  5   99.99 99   99.99 99
33948      6.37 -0.54 -0.13 99.99 99.99     99.99 99.99     6.60  5    6.20  2    6.55  2   99.99 99   99.99 99
34043      5.50  1.55  1.37 99.99 99.99     99.99 99.99     9.39 11    6.16  4    4.98  4   99.99 99   99.99 99
34045      6.21 99.99  0.37 99.99 99.99     99.99 99.99    99.99 99    6.32  3    6.15  2   99.99 99   99.99 99
34172      5.85  0.64  0.93 99.99 99.99     99.99 99.99     8.26  4    6.27  4    5.53  3   99.99 99   99.99 99
34180      6.15 -0.03  0.39 99.99 99.99     99.99 99.99     7.36  3    6.27  3    6.08  2   99.99 99   99.99 99
34266      5.76 99.99  1.01 99.99 99.99     99.99 99.99    99.99 99    6.22  4    5.41  4   99.99 99   99.99 99
34310      5.07 99.99 -0.10 99.99 99.99     99.99 99.99    99.99 99    4.92  2    5.23  2   99.99 99   99.99 99
34347      6.05 99.99  1.39 99.99 99.99     99.99 99.99    99.99 99    6.72  4    5.52  4   99.99 99   99.99 99
34435      6.66 99.99  0.15 99.99 99.99     99.99 99.99    99.99 99    6.65  3    6.71  2   99.99 99   99.99 99
34538      5.51 99.99  0.93 99.99 99.99     99.99 99.99    99.99 99    5.93  4    5.19  3   99.99 99   99.99 99
34578      5.03  0.44  0.27 99.99 99.99     99.99 99.99     6.81 16    5.08  3    5.02  2   99.99 99   99.99 99
34587      6.49 99.99  1.21 99.99 99.99     99.99 99.99    99.99 99    7.06  4    6.04  4   99.99 99   99.99 99
34649      4.82 99.99  1.29 99.99 99.99     99.99 99.99    99.99 99    5.43  4    4.34  4   99.99 99   99.99 99
34658      5.34  0.08  0.41 99.99 99.99     99.99 99.99     6.71  5    5.47  3    5.26  3   99.99 99   99.99 99
34721      5.96 99.99  0.58 99.99 99.99     99.99 99.99    99.99 99    6.18  3    5.81  3   99.99 99   99.99 99
34868      5.99 99.99 -0.04 99.99 99.99     99.99 99.99    99.99 99    5.87  2    6.12  2   99.99 99   99.99 99
34880      6.39 -0.36 -0.03 99.99 99.99     99.99 99.99     6.92  7    6.28  2    6.52  2   99.99 99   99.99 99
35046      6.34 99.99  0.33 99.99 99.99     99.99 99.99    99.99 99    6.43  3    6.30  2   99.99 99   99.99 99
35072      5.45 99.99  0.51 99.99 99.99     99.99 99.99    99.99 99    5.64  3    5.33  3   99.99 99   99.99 99
35281      5.99 -0.36 -0.03 99.99 99.99     99.99 99.99     6.52  7    5.88  2    6.12  2   99.99 99   99.99 99
35317      6.11  0.02  0.51 99.99 99.99     99.99 99.99     7.45  3    6.30  3    5.99  3   99.99 99   99.99 99
35337      5.25 99.99 -0.21 99.99 99.99     99.99 99.99    99.99 99    5.04  2    5.46  2   99.99 99   99.99 99
35386      6.49 99.99  0.50 99.99 99.99     99.99 99.99    99.99 99    6.67  3    6.37  3   99.99 99   99.99 99
35410      5.09  0.68  0.96 99.99 99.99     99.99 99.99     7.57  4    5.52  4    4.76  4   99.99 99   99.99 99
35505      5.65 99.99  0.00 99.99 99.99     99.99 99.99    99.99 99    5.56  3    5.76  2   99.99 99   99.99 99
35528      6.82 99.99  1.04 99.99 99.99     99.99 99.99    99.99 99    7.30  4    6.45  4   99.99 99   99.99 99
35536      5.61 99.99  1.56 99.99 99.99     99.99 99.99    99.99 99    6.37  5    5.00  5   99.99 99   99.99 99
35548      6.57 -0.18 -0.05 99.99 99.99     99.99 99.99     7.33 11    6.45  2    6.71  2   99.99 99   99.99 99
35580      6.11 99.99 -0.10 99.99 99.99     99.99 99.99    99.99 99    5.96  2    6.27  2   99.99 99   99.99 99
```

```
35656      6.42 -0.03 -0.03 99.99 99.99    99.99 99.99    7.40 14    6.31  2    6.55  2    99.99 99    99.99 99
35736      5.65 99.99  0.44 99.99 99.99    99.99 99.99   99.99 99    5.80  3    5.56  3    99.99 99    99.99 99
35798      6.51  1.00  1.11 99.99 99.99    99.99 99.99    9.51  6    7.03  4    6.11  4    99.99 99    99.99 99
35991      6.07 99.99  1.04 99.99 99.99    99.99 99.99   99.99 99    6.55  4    5.70  4    99.99 99    99.99 99
36058      6.39 -0.06 -0.01 99.99 99.99    99.99 99.99    7.34 12    6.29  3    6.51  2    99.99 99    99.99 99
36060      5.87 99.99  0.23 99.99 99.99    99.99 99.99   99.99 99    5.90  3    5.88  2    99.99 99    99.99 99
36187      5.57 99.99  0.02 99.99 99.99    99.99 99.99   99.99 99    5.49  3    5.68  2    99.99 99    99.99 99
36189      5.14 99.99  1.00 99.99 99.99    99.99 99.99   99.99 99    5.60  4    4.79  4    99.99 99    99.99 99
36473      5.55 99.99  0.00 99.99 99.99    99.99 99.99   99.99 99    5.46  3    5.66  2    99.99 99    99.99 99
36553      5.46 99.99  0.62 99.99 99.99    99.99 99.99   99.99 99    5.71  3    5.29  3    99.99 99    99.99 99
36584      6.03  0.03  0.34 99.99 99.99    99.99 99.99    7.29  5    6.12  3    5.99  2    99.99 99    99.99 99
36689      6.59 99.99  1.53 99.99 99.99    99.99 99.99   99.99 99    7.34  5    5.99  5    99.99 99    99.99 99
36734      5.86 99.99  1.35 99.99 99.99    99.99 99.99   99.99 99    6.51  4    5.35  4    99.99 99    99.99 99
36780      5.93  1.87  1.55 99.99 99.99    99.99 99.99   10.35 13    6.69  5    5.32  5    99.99 99    99.99 99
36848      5.48 99.99  1.22 99.99 99.99    99.99 99.99   99.99 99    6.06  4    5.03  4    99.99 99    99.99 99
36876      6.19  0.16  0.22 99.99 99.99    99.99 99.99    7.56 11    6.22  3    6.20  2    99.99 99    99.99 99
37192      5.78 99.99  1.12 99.99 99.99    99.99 99.99   99.99 99    6.30  4    5.37  4    99.99 99    99.99 99
37226      6.43 99.99  0.55 99.99 99.99    99.99 99.99   99.99 99    6.64  3    6.29  3    99.99 99    99.99 99
37232      6.12 -0.82 -0.17 99.99 99.99    99.99 99.99    5.95  3    5.93  2    6.31  2    99.99 99    99.99 99
37297      5.34  0.85  1.04 99.99 99.99    99.99 99.99    8.10  5    5.82  4    4.97  4    99.99 99    99.99 99
37306      6.11 99.99  0.05 99.99 99.99    99.99 99.99   99.99 99    6.04  3    6.20  2    99.99 99    99.99 99
37320      5.88 -0.37 -0.07 99.99 99.99    99.99 99.99    6.38  7    5.75  2    6.03  2    99.99 99    99.99 99
37356      6.19 -0.72 -0.04 99.99 99.99    99.99 99.99    6.23  4    6.07  2    6.32  2    99.99 99    99.99 99
37481      5.96 -0.91 -0.23 99.99 99.99    99.99 99.99    5.64  3    5.74  2    6.18  2    99.99 99    99.99 99
37495      5.31  0.10  0.46 99.99 99.99    99.99 99.99    6.73  4    5.47  3    5.21  3    99.99 99    99.99 99
37501      6.32 99.99  0.85 99.99 99.99    99.99 99.99   99.99 99    6.69  4    6.04  3    99.99 99    99.99 99
37594      6.00  0.01  0.27 99.99 99.99    99.99 99.99    7.20  7    6.05  3    5.99  2    99.99 99    99.99 99
37635      6.49 -0.47 -0.10 99.99 99.99    99.99 99.99    6.83  6    6.34  2    6.65  2    99.99 99    99.99 99
37744      6.22 -0.87 -0.20 99.99 99.99    99.99 99.99    5.97  3    6.01  2    6.43  2    99.99 99    99.99 99
37756      4.95 -0.83 -0.21 99.99 99.99    99.99 99.99    4.74  3    4.74  2    5.16  2    99.99 99    99.99 99
37763      5.19  1.18  1.13 99.99 99.99    99.99 99.99    8.45  9    5.72  4    4.78  4    99.99 99    99.99 99
37788      5.93  0.04  0.30 99.99 99.99    99.99 99.99    7.18  7    6.00  3    5.91  2    99.99 99    99.99 99
37811      5.45 99.99  0.92 99.99 99.99    99.99 99.99   99.99 99    5.86  4    5.14  3    99.99 99    99.99 99
37904      6.42  0.07  0.30 99.99 99.99    99.99 99.99    7.71  7    6.49  3    6.40  2    99.99 99    99.99 99
37971      6.21 99.99 -0.13 99.99 99.99    99.99 99.99   99.99 99    6.04  2    6.39  2    99.99 99    99.99 99
38089      5.97 -0.02  0.44 99.99 99.99    99.99 99.99    7.22  3    6.12  3    5.88  3    99.99 99    99.99 99
38090      5.87 99.99  0.11 99.99 99.99    99.99 99.99   99.99 99    5.84  3    5.93  2    99.99 99    99.99 99
38138      6.19 99.99  0.01 99.99 99.99    99.99 99.99   99.99 99    6.10  3    6.30  2    99.99 99    99.99 99
38206      5.73 99.99 -0.01 99.99 99.99    99.99 99.99   99.99 99    5.63  3    5.85  2    99.99 99    99.99 99
38309      6.09  0.09  0.31 99.99 99.99    99.99 99.99    7.42  7    6.17  3    6.06  2    99.99 99    99.99 99
38382      6.34 99.99  0.58 99.99 99.99    99.99 99.99   99.99 99    6.56  3    6.19  3    99.99 99    99.99 99
```

```
38458      6.39 99.99   0.29 99.99 99.99     99.99 99.99    99.99 99    6.45  3    6.37  2    99.99 99    99.99 99
38495      6.25  0.81   1.03 99.99 99.99     99.99 99.99     8.95  4    6.72  4    5.89  4    99.99 99    99.99 99
38527      5.79  0.58   0.88 99.99 99.99     99.99 99.99     8.09  4    6.18  4    5.50  3    99.99 99    99.99 99
38529      5.95  0.41   0.78 99.99 99.99     99.99 99.99     7.97  3    6.28  3    5.70  3    99.99 99    99.99 99
38666      5.17 99.99  -0.28 99.99 99.99     99.99 99.99    99.99 99    4.92  2    5.42  2    99.99 99    99.99 99
38710      5.27  0.17   0.23 99.99 99.99     99.99 99.99     6.66 11    5.30  3    5.28  2    99.99 99    99.99 99
38858      5.97  0.10   0.64 99.99 99.99     99.99 99.99     7.49  4    6.23  3    5.79  3    99.99 99    99.99 99
38871      5.31 99.99   1.04 99.99 99.99     99.99 99.99    99.99 99    5.79  4    4.94  4    99.99 99    99.99 99
38885      6.32 99.99   1.18 99.99 99.99     99.99 99.99    99.99 99    6.87  4    5.89  4    99.99 99    99.99 99
39007      5.80  0.62   0.87 99.99 99.99     99.99 99.99     8.15  5    6.18  4    5.51  3    99.99 99    99.99 99
39014      4.35  0.12   0.21 99.99 99.99     99.99 99.99     5.66 11    4.37  3    4.37  2    99.99 99    99.99 99
39051      5.97  1.57   1.36 99.99 99.99     99.99 99.99     9.88 12    6.62  4    5.45  4    99.99 99    99.99 99
39060      3.85  0.10   0.17 99.99 99.99     99.99 99.99     5.11 11    3.85  3    3.89  2    99.99 99    99.99 99
39070      5.49 99.99   0.88 99.99 99.99     99.99 99.99    99.99 99    5.88  4    5.20  3    99.99 99    99.99 99
39091      5.65  0.11   0.60 99.99 99.99     99.99 99.99     7.16  3    5.89  3    5.49  3    99.99 99    99.99 99
39110      6.18 99.99   1.40 99.99 99.99     99.99 99.99    99.99 99    6.86  4    5.64  4    99.99 99    99.99 99
39118      5.98  0.30   0.91 99.99 99.99     99.99 99.99     7.92  6    6.39  4    5.67  3    99.99 99    99.99 99
39190      5.87 99.99   0.06 99.99 99.99     99.99 99.99    99.99 99    5.81  3    5.96  2    99.99 99    99.99 99
39291      5.35 -0.83  -0.20 99.99 99.99     99.99 99.99     5.15  3    5.14  2    5.56  2    99.99 99    99.99 99
39385      6.17 99.99   1.02 99.99 99.99     99.99 99.99    99.99 99    6.64  4    5.81  4    99.99 99    99.99 99
39400      4.78  1.46   1.38 99.99 99.99     99.99 99.99     8.55  9    5.44  4    4.25  4    99.99 99    99.99 99
39421      5.97  0.08   0.10 99.99 99.99     99.99 99.99     7.17 13    5.93  3    6.04  2    99.99 99    99.99 99
39523      4.51  0.98   1.10 99.99 99.99     99.99 99.99     7.48  6    5.02  4    4.11  4    99.99 99    99.99 99
39543      6.45 99.99   1.48 99.99 99.99     99.99 99.99    99.99 99    7.17  4    5.88  5    99.99 99    99.99 99
39547      6.35 99.99   0.76 99.99 99.99     99.99 99.99    99.99 99    6.67  3    6.11  3    99.99 99    99.99 99
39640      5.17  0.72   0.99 99.99 99.99     99.99 99.99     7.73  4    5.62  4    4.82  4    99.99 99    99.99 99
39685      6.31  1.47   1.29 99.99 99.99     99.99 99.99    10.05 12    6.92  4    5.83  4    99.99 99    99.99 99
39775      6.01  1.44   1.34 99.99 99.99     99.99 99.99     9.73 10    6.65  4    5.50  4    99.99 99    99.99 99
39777      6.57 -0.81  -0.18 99.99 99.99     99.99 99.99     6.41  3    6.38  2    6.77  2    99.99 99    99.99 99
39810      6.53  0.97   1.08 99.99 99.99     99.99 99.99     9.47  6    7.03  4    6.14  4    99.99 99    99.99 99
39844      5.11 -0.49  -0.14 99.99 99.99     99.99 99.99     5.40  7    4.94  2    5.29  2    99.99 99    99.99 99
39853      5.66  1.84   1.53 99.99 99.99     99.99 99.99    10.03 13    6.41  5    5.06  5    99.99 99    99.99 99
39881      6.60  0.13   0.65 99.99 99.99     99.99 99.99     8.17  3    6.86  3    6.41  3    99.99 99    99.99 99
39891      6.36 99.99   0.37 99.99 99.99     99.99 99.99    99.99 99    6.47  3    6.30  2    99.99 99    99.99 99
39910      5.87  1.21   1.18 99.99 99.99     99.99 99.99     9.19  9    6.42  4    5.44  4    99.99 99    99.99 99
39927      6.28  0.07   0.06 99.99 99.99     99.99 99.99     7.44 13    6.22  3    6.37  2    99.99 99    99.99 99
39937      5.94 99.99   0.66 99.99 99.99     99.99 99.99    99.99 99    6.21  3    5.75  3    99.99 99    99.99 99
39963      6.36  0.57   0.86 99.99 99.99     99.99 99.99     8.64  4    6.74  4    6.08  3    99.99 99    99.99 99
39970      6.02 -0.20   0.39 99.99 99.99     99.99 99.99     7.00  4    6.14  3    5.95  2    99.99 99    99.99 99
39985      5.99 -0.15  -0.07 99.99 99.99     99.99 99.99     6.78 12    5.86  2    6.14  2    99.99 99    99.99 99
40105      6.52 99.99   0.90 99.99 99.99     99.99 99.99    99.99 99    6.92  4    6.22  3    99.99 99    99.99 99
```

```
40176        4.97 99.99   1.11 99.99 99.99        99.99 99.99       99.99 99    5.49  4    4.57  4   99.99 99   99.99 99
40200        6.10 99.99  -0.13 99.99 99.99        99.99 99.99       99.99 99    5.93  2    6.28  2   99.99 99   99.99 99
40282        6.22  1.84   1.48 99.99 99.99        99.99 99.99       10.56 15    6.94  4    5.65  5   99.99 99   99.99 99
40292        5.29 99.99   0.31 99.99 99.99        99.99 99.99       99.99 99    5.37  3    5.26  2   99.99 99   99.99 99
40347        6.22  1.20   1.14 99.99 99.99        99.99 99.99        9.51 10    6.75  4    5.80  4   99.99 99   99.99 99
40359        6.44 99.99   1.07 99.99 99.99        99.99 99.99       99.99 99    6.93  4    6.06  4   99.99 99   99.99 99
40409        4.65  0.96   1.05 99.99 99.99        99.99 99.99        7.56  7    5.13  4    4.28  4   99.99 99   99.99 99
40446        5.22  0.01   0.01 99.99 99.99        99.99 99.99        6.27 13    5.13  3    5.33  2   99.99 99   99.99 99
40455        6.63  0.10   0.38 99.99 99.99        99.99 99.99        8.01  6    6.74  3    6.57  2   99.99 99   99.99 99
40536        5.04  0.16   0.19 99.99 99.99        99.99 99.99        6.40 12    5.05  3    5.07  2   99.99 99   99.99 99
40574        6.63 -0.39  -0.07 99.99 99.99        99.99 99.99        7.10  7    6.50  2    6.78  2   99.99 99   99.99 99
40589        6.05 -0.26   0.25 99.99 99.99        99.99 99.99        6.87  3    6.09  3    6.05  2   99.99 99   99.99 99
40665        6.45 99.99   1.48 99.99 99.99        99.99 99.99       99.99 99    7.17  4    5.88  5   99.99 99   99.99 99
40801        6.10  0.78   0.97 99.99 99.99        99.99 99.99        8.73  5    6.54  4    5.76  4   99.99 99   99.99 99
40953        5.47 -0.24  -0.08 99.99 99.99        99.99 99.99        6.14 10    5.33  2    5.62  2   99.99 99   99.99 99
40964        6.59 -0.27  -0.05 99.99 99.99        99.99 99.99        7.23  9    6.47  2    6.73  2   99.99 99   99.99 99
40972        6.05 99.99   0.02 99.99 99.99        99.99 99.99       99.99 99    5.97  3    6.16  2   99.99 99   99.99 99
41047        5.55 99.99   1.58 99.99 99.99        99.99 99.99       99.99 99    6.32  5    4.93  5   99.99 99   99.99 99
41125        6.20 99.99   0.96 99.99 99.99        99.99 99.99       99.99 99    6.63  4    5.87  4   99.99 99   99.99 99
41214        5.67 99.99   0.20 99.99 99.99        99.99 99.99       99.99 99    5.69  3    5.69  2   99.99 99   99.99 99
41312        5.05 99.99   1.34 99.99 99.99        99.99 99.99       99.99 99    5.69  4    4.54  4   99.99 99   99.99 99
41361        5.67  0.82   1.04 99.99 99.99        99.99 99.99        8.39  4    6.15  4    5.30  4   99.99 99   99.99 99
41380        5.63  0.74   1.04 99.99 99.99        99.99 99.99        8.24  3    6.11  4    5.26  4   99.99 99   99.99 99
41534        5.65 99.99  -0.19 99.99 99.99        99.99 99.99       99.99 99    5.45  2    5.85  2   99.99 99   99.99 99
41547        5.87 99.99   0.36 99.99 99.99        99.99 99.99       99.99 99    5.97  3    5.82  2   99.99 99   99.99 99
41636        6.36  0.82   1.05 99.99 99.99        99.99 99.99        9.08  4    6.84  4    5.99  4   99.99 99   99.99 99
41692        5.38 -0.54  -0.13 99.99 99.99        99.99 99.99        5.61  5    5.21  2    5.56  2   99.99 99   99.99 99
41700        6.35 99.99   0.52 99.99 99.99        99.99 99.99       99.99 99    6.54  3    6.22  3   99.99 99   99.99 99
41742        5.93 99.99   0.49 99.99 99.99        99.99 99.99       99.99 99    6.10  3    5.82  3   99.99 99   99.99 99
41759        5.80 99.99   0.02 99.99 99.99        99.99 99.99       99.99 99    5.72  3    5.91  2   99.99 99   99.99 99
41841        5.47 99.99   0.08 99.99 99.99        99.99 99.99       99.99 99    5.42  3    5.55  2   99.99 99   99.99 99
41843        5.81 99.99   0.05 99.99 99.99        99.99 99.99       99.99 99    5.74  3    5.90  2   99.99 99   99.99 99
42035        6.55 -0.17  -0.08 99.99 99.99        99.99 99.99        7.31 12    6.41  2    6.70  2   99.99 99   99.99 99
42111        5.73  0.05   0.07 99.99 99.99        99.99 99.99        6.87 13    5.67  3    5.81  2   99.99 99   99.99 99
42167        5.02 99.99  -0.11 99.99 99.99        99.99 99.99       99.99 99    4.86  2    5.19  2   99.99 99   99.99 99
42168        6.51 99.99   1.16 99.99 99.99        99.99 99.99       99.99 99    7.05  4    6.09  4   99.99 99   99.99 99
42278        6.17  0.01   0.34 99.99 99.99        99.99 99.99        7.41  5    6.26  3    6.13  2   99.99 99   99.99 99
42301        5.50 99.99  -0.01 99.99 99.99        99.99 99.99       99.99 99    5.40  3    5.62  2   99.99 99   99.99 99
42327        6.35 99.99  -0.03 99.99 99.99        99.99 99.99       99.99 99    6.24  2    6.48  2   99.99 99   99.99 99
42341        5.56 99.99   1.16 99.99 99.99        99.99 99.99       99.99 99    6.10  4    5.14  4   99.99 99   99.99 99
42443        5.71 99.99   0.44 99.99 99.99        99.99 99.99       99.99 99    5.86  3    5.62  3   99.99 99   99.99 99
```

```
42525       5.71 -0.07 -0.03 99.99 99.99     99.99 99.99     6.63 13    5.60  2    5.84  2    99.99 99    99.99 99
42683       6.49 99.99  0.52 99.99 99.99     99.99 99.99    99.99 99    6.68  3    6.36  3    99.99 99    99.99 99
42690       5.05 -0.78 -0.20 99.99 99.99     99.99 99.99     4.92  3    4.84  2    5.26  2    99.99 99    99.99 99
42729       6.09 99.99 -0.04 99.99 99.99     99.99 99.99    99.99 99    5.97  2    6.22  2    99.99 99    99.99 99
42824       6.62  0.08  0.05 99.99 99.99     99.99 99.99     7.79 14    6.55  3    6.71  2    99.99 99    99.99 99
42834       6.31 99.99 -0.03 99.99 99.99     99.99 99.99    99.99 99    6.20  2    6.44  2    99.99 99    99.99 99
43023       5.83  0.64  0.94 99.99 99.99     99.99 99.99     8.25  4    6.25  4    5.51  3    99.99 99    99.99 99
43042       5.20  0.00  0.44 99.99 99.99     99.99 99.99     6.48  3    5.35  3    5.11  3    99.99 99    99.99 99
43107       5.06 -0.21 -0.08 99.99 99.99     99.99 99.99     5.77 11    4.92  2    5.21  2    99.99 99    99.99 99
43112       5.90 -0.96 -0.24 99.99 99.99     99.99 99.99     5.50  4    5.67  2    6.13  2    99.99 99    99.99 99
43157       5.83 -0.65 -0.17 99.99 99.99     99.99 99.99     5.89  4    5.64  2    6.02  2    99.99 99    99.99 99
43285       6.07 -0.52 -0.13 99.99 99.99     99.99 99.99     6.33  6    5.90  2    6.25  2    99.99 99    99.99 99
43317       6.64 -0.65 -0.17 99.99 99.99     99.99 99.99     6.70  4    6.45  2    6.83  2    99.99 99    99.99 99
43318       5.65  0.00  0.50 99.99 99.99     99.99 99.99     6.96  3    5.83  3    5.53  3    99.99 99    99.99 99
43319       5.99  0.10  0.09 99.99 99.99     99.99 99.99     7.21 13    5.94  3    6.06  2    99.99 99    99.99 99
43358       6.37 -0.01  0.46 99.99 99.99     99.99 99.99     7.64  3    6.53  3    6.27  3    99.99 99    99.99 99
43362       6.10 -0.31 -0.08 99.99 99.99     99.99 99.99     6.67  9    5.96  2    6.25  2    99.99 99    99.99 99
43380       6.39  1.12  1.11 99.99 99.99     99.99 99.99     9.55  9    6.91  4    5.99  4    99.99 99    99.99 99
43386       5.04 -0.02  0.42 99.99 99.99     99.99 99.99     6.28  3    5.18  3    4.96  3    99.99 99    99.99 99
43455       5.01  1.85  1.61 99.99 99.99     99.99 99.99     9.44 11    5.80  5    4.38  5    99.99 99    99.99 99
43461       6.63 -0.44 -0.05 99.99 99.99     99.99 99.99     7.04  5    6.51  2    6.77  2    99.99 99    99.99 99
43525       5.39  0.09  0.10 99.99 99.99     99.99 99.99     6.60 13    5.35  3    5.46  2    99.99 99    99.99 99
43526       6.57 -0.51 -0.13 99.99 99.99     99.99 99.99     6.84  6    6.40  2    6.75  2    99.99 99    99.99 99
43544       5.94 99.99 -0.17 99.99 99.99     99.99 99.99    99.99 99    5.75  2    6.13  2    99.99 99    99.99 99
43587       5.71  0.10  0.61 99.99 99.99     99.99 99.99     7.22  3    5.95  3    5.54  3    99.99 99    99.99 99
43636       6.67 99.99  1.55 99.99 99.99     99.99 99.99    99.99 99    7.43  5    6.06  5    99.99 99    99.99 99
43821       6.24  0.52  0.87 99.99 99.99     99.99 99.99     8.46  3    6.62  4    5.95  3    99.99 99    99.99 99
43827       5.14 99.99  1.30 99.99 99.99     99.99 99.99    99.99 99    5.76  4    4.65  4    99.99 99    99.99 99
43847       6.00 99.99  0.16 99.99 99.99     99.99 99.99    99.99 99    5.99  3    6.04  2    99.99 99    99.99 99
43905       5.36  0.12  0.43 99.99 99.99     99.99 99.99     6.79  5    5.50  3    5.28  3    99.99 99    99.99 99
43993       5.36  1.32  1.24 99.99 99.99     99.99 99.99     8.87 10    5.95  4    4.90  4    99.99 99    99.99 99
44021       6.07 99.99  1.66 99.99 99.99     99.99 99.99    99.99 99    6.89  5    5.41  5    99.99 99    99.99 99
44037       6.22 -0.14 -0.04 99.99 99.99     99.99 99.99     7.04 11    6.10  2    6.35  2    99.99 99    99.99 99
44081       5.81 99.99 -0.15 99.99 99.99     99.99 99.99    99.99 99    5.63  2    5.99  2    99.99 99    99.99 99
44112       5.26 -0.75 -0.19 99.99 99.99     99.99 99.99     5.17  3    5.06  2    5.46  2    99.99 99    99.99 99
44120       6.43 99.99  0.59 99.99 99.99     99.99 99.99    99.99 99    6.66  3    6.27  3    99.99 99    99.99 99
44131       4.90  1.96  1.60 99.99 99.99     99.99 99.99     9.47 14    5.69  5    4.27  5    99.99 99    99.99 99
44267       6.41 99.99  1.46 99.99 99.99     99.99 99.99    99.99 99    7.12  4    5.85  5    99.99 99    99.99 99
44323       5.78 99.99 -0.08 99.99 99.99     99.99 99.99    99.99 99    5.64  2    5.93  2    99.99 99    99.99 99
44333       6.31  0.09  0.24 99.99 99.99     99.99 99.99     7.60  9    6.35  3    6.31  2    99.99 99    99.99 99
44362       7.04 99.99  0.83 99.99 99.99     99.99 99.99     7.40  4    6.77  3    99.99 99    99.99 99
```

```
44447       6.64  0.01  0.56 99.99 99.99    99.99 99.99     8.00  4    6.85  3    6.49  3    99.99 99    99.99 99
44506       5.52 99.99 -0.19 99.99 99.99    99.99 99.99    99.99 99    5.32  2    5.72  2    99.99 99    99.99 99
44594       6.60 99.99  0.66 99.99 99.99    99.99 99.99    99.99 99    6.87  3    6.41  3    99.99 99    99.99 99
44700       6.41 -0.63 -0.15 99.99 99.99    99.99 99.99     6.51  4    6.23  2    6.59  2    99.99 99    99.99 99
44756       6.67  0.02  0.06 99.99 99.99    99.99 99.99     7.77 12    6.61  3    6.76  2    99.99 99    99.99 99
44783       6.26 -0.31 -0.08 99.99 99.99    99.99 99.99     6.83  9    6.12  2    6.41  2    99.99 99    99.99 99
44951       5.22 99.99  1.24 99.99 99.99    99.99 99.99    99.99 99    5.81  4    4.76  4    99.99 99    99.99 99
44996       6.12 99.99 -0.08 99.99 99.99    99.99 99.99    99.99 99    5.98  2    6.27  2    99.99 99    99.99 99
45018       5.63 99.99  1.56 99.99 99.99    99.99 99.99    99.99 99    6.39  5    5.02  5    99.99 99    99.99 99
45050       6.66 -0.22 -0.04 99.99 99.99    99.99 99.99     7.37 10    6.54  2    6.79  2    99.99 99    99.99 99
45067       5.87  0.08  0.56 99.99 99.99    99.99 99.99     7.32  3    6.08  3    5.72  3    99.99 99    99.99 99
45137       6.51 -0.06 -0.03 99.99 99.99    99.99 99.99     7.45 13    6.40  2    6.64  2    99.99 99    99.99 99
45145       5.62 99.99  1.04 99.99 99.99    99.99 99.99    99.99 99    6.10  4    5.25  4    99.99 99    99.99 99
45168       6.35  0.77  1.02 99.99 99.99    99.99 99.99     8.99  4    6.82  4    5.99  4    99.99 99    99.99 99
45184       6.39 99.99  0.62 99.99 99.99    99.99 99.99    99.99 99    6.64  3    6.22  3    99.99 99    99.99 99
45229       5.61 99.99  0.23 99.99 99.99    99.99 99.99    99.99 99    5.64  3    5.62  2    99.99 99    99.99 99
45239       6.40  0.12  0.14 99.99 99.99    99.99 99.99     7.67 12    6.38  3    6.45  2    99.99 99    99.99 99
45291       5.98 99.99  1.04 99.99 99.99    99.99 99.99    99.99 99    6.46  4    5.61  4    99.99 99    99.99 99
45306       6.31 99.99 -0.05 99.99 99.99    99.99 99.99    99.99 99    6.19  2    6.45  2    99.99 99    99.99 99
45320       5.87  0.09  0.08 99.99 99.99    99.99 99.99     7.07 13    5.82  3    5.95  2    99.99 99    99.99 99
45357       6.71  0.00  0.04 99.99 99.99    99.99 99.99     7.77 12    6.64  3    6.81  2    99.99 99    99.99 99
45380       6.27 -0.10 -0.04 99.99 99.99    99.99 99.99     7.15 12    6.15  2    6.40  2    99.99 99    99.99 99
45383       6.25 99.99  1.36 99.99 99.99    99.99 99.99    99.99 99    6.90  4    5.73  4    99.99 99    99.99 99
45410       5.88  0.66  0.94 99.99 99.99    99.99 99.99     8.33  4    6.30  4    5.56  3    99.99 99    99.99 99
45415       5.55  0.86  1.04 99.99 99.99    99.99 99.99     8.32  5    6.03  4    5.18  4    99.99 99    99.99 99
45416       5.20  1.13  1.18 99.99 99.99    99.99 99.99     8.42  7    5.75  4    4.77  4    99.99 99    99.99 99
45433       5.55  1.61  1.38 99.99 99.99    99.99 99.99     9.53 12    6.21  4    5.02  4    99.99 99    99.99 99
45450       6.48 99.99  0.12 99.99 99.99    99.99 99.99    99.99 99    6.45  3    6.54  2    99.99 99    99.99 99
45509       6.51 99.99  1.70 99.99 99.99    99.99 99.99    99.99 99    7.35  5    5.83  5    99.99 99    99.99 99
45557       5.80 99.99  0.00 99.99 99.99    99.99 99.99    99.99 99    5.71  3    5.91  2    99.99 99    99.99 99
45563       6.48 -0.16 -0.07 99.99 99.99    99.99 99.99     7.26 12    6.35  2    6.63  2    99.99 99    99.99 99
45572       5.76 -0.17 -0.06 99.99 99.99    99.99 99.99     6.53 11    5.63  2    5.90  2    99.99 99    99.99 99
45588       6.07 99.99  0.54 99.99 99.99    99.99 99.99    99.99 99    6.27  3    5.93  3    99.99 99    99.99 99
45669       5.56  1.82  1.51 99.99 99.99    99.99 99.99     9.89 13    6.30  5    4.97  5    99.99 99    99.99 99
45701       6.46 99.99  0.66 99.99 99.99    99.99 99.99    99.99 99    6.73  3    6.27  3    99.99 99    99.99 99
45724       6.16  1.85  1.54 99.99 99.99    99.99 99.99    10.55 13    6.91  5    5.56  5    99.99 99    99.99 99
45765       5.77 99.99  1.12 99.99 99.99    99.99 99.99    99.99 99    6.29  4    5.36  4    99.99 99    99.99 99
45796       6.26 99.99 -0.13 99.99 99.99    99.99 99.99    99.99 99    6.09  2    6.44  2    99.99 99    99.99 99
45827       6.57 -0.06  0.14 99.99 99.99    99.99 99.99     7.60  8    6.55  3    6.62  2    99.99 99    99.99 99
45976       5.93 99.99  1.38 99.99 99.99    99.99 99.99    99.99 99    6.59  4    5.40  4    99.99 99    99.99 99
45983       6.32 99.99  0.40 99.99 99.99    99.99 99.99    99.99 99    6.45  3    6.25  3    99.99 99    99.99 99
```

```
45984      5.82 99.99  1.28 99.99 99.99     99.99 99.99    99.99 99    6.43  4    5.34  4    99.99 99    99.99 99
45995      6.15 -0.87 -0.08 99.99 99.99     99.99 99.99     5.96  5    6.01  2    6.30  2    99.99 99    99.99 99
46064      6.16 99.99 -0.15 99.99 99.99     99.99 99.99    99.99 99    5.98  2    6.34  2    99.99 99    99.99 99
46116      5.38  0.67  0.97 99.99 99.99     99.99 99.99     7.86  4    5.82  4    5.04  4    99.99 99    99.99 99
46189      5.94 99.99 -0.15 99.99 99.99     99.99 99.99    99.99 99    5.76  2    6.12  2    99.99 99    99.99 99
46229      5.43  1.52  1.38 99.99 99.99     99.99 99.99     9.28 10    6.09  4    4.90  4    99.99 99    99.99 99
46304      5.60  0.07  0.25 99.99 99.99     99.99 99.99     6.87  8    5.64  3    5.60  2    99.99 99    99.99 99
46355      5.22 99.99  1.09 99.99 99.99     99.99 99.99    99.99 99    5.72  4    4.83  4    99.99 99    99.99 99
46365      6.20 99.99  1.40 99.99 99.99     99.99 99.99    99.99 99    6.88  4    5.66  4    99.99 99    99.99 99
46487      5.10 -0.56 -0.13 99.99 99.99     99.99 99.99     5.30  5    4.93  2    5.28  2    99.99 99    99.99 99
46547      5.69 99.99 -0.17 99.99 99.99     99.99 99.99    99.99 99    5.50  2    5.88  2    99.99 99    99.99 99
46569      5.60 99.99  0.54 99.99 99.99     99.99 99.99    99.99 99    5.80  3    5.46  3    99.99 99    99.99 99
46642      6.45 -0.03 -0.02 99.99 99.99     99.99 99.99     7.43 13    6.34  2    6.57  2    99.99 99    99.99 99
46730      6.29  0.06  0.33 99.99 99.99     99.99 99.99     7.59  6    6.38  3    6.25  2    99.99 99    99.99 99
46769      5.80 -0.46  0.00 99.99 99.99     99.99 99.99     6.21  4    5.71  3    5.91  2    99.99 99    99.99 99
46815      5.42 99.99  1.44 99.99 99.99     99.99 99.99    99.99 99    6.12  4    4.87  5    99.99 99    99.99 99
46885      6.55 -0.28 -0.06 99.99 99.99     99.99 99.99     7.17  9    6.42  2    6.69  2    99.99 99    99.99 99
46936      5.62 99.99 -0.08 99.99 99.99     99.99 99.99    99.99 99    5.48  2    5.77  2    99.99 99    99.99 99
47001      6.19 99.99  1.09 99.99 99.99     99.99 99.99    99.99 99    6.69  4    5.80  4    99.99 99    99.99 99
47054      5.52 -0.39 -0.08 99.99 99.99     99.99 99.99     5.98  7    5.38  2    5.67  2    99.99 99    99.99 99
47182      5.97 99.99  1.56 99.99 99.99     99.99 99.99    99.99 99    6.73  5    5.36  5    99.99 99    99.99 99
47220      6.17  0.96  1.08 99.99 99.99     99.99 99.99     9.10  6    6.67  4    5.78  4    99.99 99    99.99 99
47230      6.35 99.99  0.48 99.99 99.99     99.99 99.99    99.99 99    6.52  3    6.24  3    99.99 99    99.99 99
47306      4.39 -0.15 -0.02 99.99 99.99     99.99 99.99     5.21 11    4.28  2    4.51  2    99.99 99    99.99 99
47366      6.12 99.99  1.00 99.99 99.99     99.99 99.99    99.99 99    6.58  4    5.77  4    99.99 99    99.99 99
47420      6.14  1.77  1.48 99.99 99.99     99.99 99.99    10.39 13    6.86  4    5.57  5    99.99 99    99.99 99
47431      6.57 -0.33 -0.07 99.99 99.99     99.99 99.99     7.12  8    6.44  2    6.72  2    99.99 99    99.99 99
47536      5.28 99.99  1.18 99.99 99.99     99.99 99.99    99.99 99    5.83  4    4.85  4    99.99 99    99.99 99
47756      6.51 -0.47 -0.14 99.99 99.99     99.99 99.99     6.83  7    6.34  2    6.69  2    99.99 99    99.99 99
47827      6.05 -0.06 -0.03 99.99 99.99     99.99 99.99     6.99 13    5.94  2    6.18  2    99.99 99    99.99 99
47914      5.02  1.83  1.48 99.99 99.99     99.99 99.99     9.35 14    5.74  4    4.45  5    99.99 99    99.99 99
47946      5.71 99.99  1.14 99.99 99.99     99.99 99.99    99.99 99    6.24  4    5.29  4    99.99 99    99.99 99
47964      5.79 -0.37 -0.10 99.99 99.99     99.99 99.99     6.27  8    5.64  2    5.95  2    99.99 99    99.99 99
47973      4.93  0.59  0.87 99.99 99.99     99.99 99.99     7.24  4    5.31  4    4.64  3    99.99 99    99.99 99
48217      5.19  1.89  1.54 99.99 99.99     99.99 99.99     9.63 14    5.94  5    4.59  5    99.99 99    99.99 99
48348      6.19  1.61  1.37 99.99 99.99     99.99 99.99    10.16 13    6.85  4    5.67  4    99.99 99    99.99 99
48383      6.12 99.99 -0.14 99.99 99.99     99.99 99.99    99.99 99    5.95  2    6.30  2    99.99 99    99.99 99
48434      5.91 -0.89 -0.02 99.99 99.99     99.99 99.99     5.73  7    5.80  2    6.03  2    99.99 99    99.99 99
48938      6.45 99.99  0.54 99.99 99.99     99.99 99.99    99.99 99    6.65  3    6.31  3    99.99 99    99.99 99
48977      5.93 -0.69 -0.17 99.99 99.99     99.99 99.99     5.94  4    5.74  2    6.12  2    99.99 99    99.99 99
49095      5.92 99.99  0.48 99.99 99.99     99.99 99.99    99.99 99    6.09  3    5.81  3    99.99 99    99.99 99
```

```
49147      5.66 -0.11 -0.05 99.99 99.99     99.99 99.99     6.52 12    5.54  2    5.80  2    99.99 99    99.99 99
49161      4.77  1.65  1.40 99.99 99.99     99.99 99.99     8.81 13    5.45  4    4.23  4    99.99 99    99.99 99
49229      5.29 99.99 -0.04 99.99 99.99     99.99 99.99    99.99 99    5.17  2    5.42  2    99.99 99    99.99 99
49268      6.51  1.06  1.11 99.99 99.99     99.99 99.99     9.59  7    7.03  4    6.11  4    99.99 99    99.99 99
49319      6.62 99.99 -0.12 99.99 99.99     99.99 99.99    99.99 99    6.46  2    6.79  2    99.99 99    99.99 99
49331      5.07  1.88  1.80 99.99 99.99     99.99 99.99     9.64  7    5.97  5    4.35  5    99.99 99    99.99 99
49434      5.75  0.05  0.28 99.99 99.99     99.99 99.99     7.01  7    5.81  3    5.73  2    99.99 99    99.99 99
49520      5.02  1.35  1.27 99.99 99.99     99.99 99.99     8.58 10    5.62  4    4.54  4    99.99 99    99.99 99
49591      5.26 99.99 -0.08 99.99 99.99     99.99 99.99    99.99 99    5.12  2    5.41  2    99.99 99    99.99 99
49662      5.39 99.99 -0.10 99.99 99.99     99.99 99.99    99.99 99    5.24  2    5.55  2    99.99 99    99.99 99
49689      5.40 99.99  1.34 99.99 99.99     99.99 99.99    99.99 99    6.04  4    4.89  4    99.99 99    99.99 99
49705      6.46 99.99  0.86 99.99 99.99     99.99 99.99    99.99 99    6.84  4    6.18  3    99.99 99    99.99 99
49933      5.77 -0.07  0.39 99.99 99.99     99.99 99.99     6.92  3    5.89  3    5.70  2    99.99 99    99.99 99
49947      6.37  0.66  0.96 99.99 99.99     99.99 99.99     8.83  4    6.80  4    6.04  4    99.99 99    99.99 99
49980      5.79 99.99  1.44 99.99 99.99     99.99 99.99    99.99 99    6.49  4    5.24  5    99.99 99    99.99 99
50002      6.11  1.50  1.33 99.99 99.99     99.99 99.99     9.91 11    6.75  4    5.61  4    99.99 99    99.99 99
50062      6.38  0.09  0.04 99.99 99.99     99.99 99.99     7.56 14    6.31  3    6.48  2    99.99 99    99.99 99
50196      6.55  1.82  1.51 99.99 99.99     99.99 99.99    10.88 13    7.29  5    5.96  5    99.99 99    99.99 99
50223      5.14 -0.03  0.44 99.99 99.99     99.99 99.99     6.38  3    5.29  3    5.05  3    99.99 99    99.99 99
50235      4.99 99.99  1.38 99.99 99.99     99.99 99.99    99.99 99    5.65  4    4.46  4    99.99 99    99.99 99
50241      3.27  0.13  0.21 99.99 99.99     99.99 99.99     4.60 11    3.29  3    3.29  2    99.99 99    99.99 99
50277      5.77  0.10  0.26 99.99 99.99     99.99 99.99     7.08  9    5.82  3    5.76  2    99.99 99    99.99 99
50282      6.30  0.76  0.99 99.99 99.99     99.99 99.99     8.91  4    6.75  4    5.95  4    99.99 99    99.99 99
50310      2.93  1.21  1.20 99.99 99.99     99.99 99.99     6.26  8    3.50  4    2.49  4    99.99 99    99.99 99
50506      5.64  0.15  0.20 99.99 99.99     99.99 99.99     6.99 11    5.66  3    5.66  2    99.99 99    99.99 99
50571      6.11 99.99  0.46 99.99 99.99     99.99 99.99    99.99 99    6.27  3    6.01  3    99.99 99    99.99 99
50700      6.41  0.16  0.17 99.99 99.99     99.99 99.99     7.76 12    6.41  3    6.45  2    99.99 99    99.99 99
50747      5.45  0.16  0.18 99.99 99.99     99.99 99.99     6.80 12    5.45  3    5.48  2    99.99 99    99.99 99
50785      6.52  0.11  0.42 99.99 99.99     99.99 99.99     7.94  5    6.66  3    6.44  3    99.99 99    99.99 99
50806      6.03 99.99  0.72 99.99 99.99     99.99 99.99    99.99 99    6.33  3    5.81  3    99.99 99    99.99 99
50820      6.21 -0.35  0.56 99.99 99.99     99.99 99.99     7.08 11    6.42  3    6.06  3    99.99 99    99.99 99
50860      6.46 -0.38 -0.10 99.99 99.99     99.99 99.99     6.93  8    6.31  2    6.62  2    99.99 99    99.99 99
50890      6.04  0.85  1.10 99.99 99.99     99.99 99.99     8.83  4    6.55  4    5.64  4    99.99 99    99.99 99
50931      6.29  0.06  0.04 99.99 99.99     99.99 99.99     7.43 14    6.22  3    6.39  2    99.99 99    99.99 99
51210      6.41 99.99  0.18 99.99 99.99     99.99 99.99    99.99 99    6.41  3    6.44  2    99.99 99    99.99 99
51250      5.00 99.99  1.18 99.99 99.99     99.99 99.99    99.99 99    5.55  4    4.57  4    99.99 99    99.99 99
51266      6.26 99.99  0.99 99.99 99.99     99.99 99.99    99.99 99    6.71  4    5.91  4    99.99 99    99.99 99
51424      6.34  0.26  0.64 99.99 99.99     99.99 99.99     8.08  3    6.60  3    6.16  3    99.99 99    99.99 99
51440      5.99  1.16  1.23 99.99 99.99     99.99 99.99     9.27  7    6.57  4    5.53  4    99.99 99    99.99 99
51557      5.40 -0.38 -0.11 99.99 99.99     99.99 99.99     5.86  8    5.24  2    5.57  2    99.99 99    99.99 99
51682      6.29 99.99  1.29 99.99 99.99     99.99 99.99    99.99 99    6.90  4    5.81  4    99.99 99    99.99 99
```

```
51693       6.27  0.10  0.11 99.99 99.99     99.99 99.99     7.50 13    6.24  3    6.33  2   99.99 99   99.99 99
51733       5.46  0.03  0.36 99.99 99.99     99.99 99.99     6.73  5    5.56  3    5.41  2   99.99 99   99.99 99
51814       5.97  0.88  1.06 99.99 99.99     99.99 99.99     8.78  5    6.46  4    5.59  4   99.99 99   99.99 99
51825       6.23 99.99  0.46 99.99 99.99     99.99 99.99    99.99 99    6.39  3    6.13  3   99.99 99   99.99 99
51892       6.35 -0.47 -0.11 99.99 99.99     99.99 99.99     6.69  6    6.19  2    6.52  2   99.99 99   99.99 99
52092       5.06 99.99 -0.16 99.99 99.99     99.99 99.99    99.99 99    4.88  2    5.25  2   99.99 99   99.99 99
52265       6.30  0.09  0.57 99.99 99.99     99.99 99.99     7.77  3    6.52  3    6.15  3   99.99 99   99.99 99
52312       5.96 -0.34 -0.08 99.99 99.99     99.99 99.99     6.49  8    5.82  2    6.11  2   99.99 99   99.99 99
52362       6.22 -0.06  0.00 99.99 99.99     99.99 99.99     7.17 12    6.13  3    6.33  2   99.99 99   99.99 99
52382       6.49 -0.75  0.22 99.99 99.99     99.99 99.99     6.63 10    6.52  3    6.50  2   99.99 99   99.99 99
52479       6.63  0.10  0.06 99.99 99.99     99.99 99.99     7.83 14    6.57  3    6.72  2   99.99 99   99.99 99
52559       6.59 -0.64 -0.02 99.99 99.99     99.99 99.99     6.75  3    6.48  2    6.71  2   99.99 99   99.99 99
52603       6.27 99.99  1.16 99.99 99.99     99.99 99.99    99.99 99    6.81  4    5.85  4   99.99 99   99.99 99
52611       6.17  1.36  1.28 99.99 99.99     99.99 99.99     9.75  9    6.78  4    5.69  4   99.99 99   99.99 99
52622       6.45 99.99  0.39 99.99 99.99     99.99 99.99    99.99 99    6.57  3    6.38  2   99.99 99   99.99 99
52666       5.21  2.04  1.68 99.99 99.99     99.99 99.99     9.93 14    6.04  5    4.54  5   99.99 99   99.99 99
52913       5.97  0.11  0.12 99.99 99.99     99.99 99.99     7.22 13    5.94  3    6.03  2   99.99 99   99.99 99
53205       6.57 -0.07  0.01 99.99 99.99     99.99 99.99     7.52 12    6.48  3    6.68  2   99.99 99   99.99 99
53208       5.62  1.24  1.29 99.99 99.99     99.99 99.99     9.04  7    6.23  4    5.14  4   99.99 99   99.99 99
53240       6.45 -0.31 -0.08 99.99 99.99     99.99 99.99     7.02  9    6.31  2    6.60  2   99.99 99   99.99 99
53253       6.43 -0.09 -0.04 99.99 99.99     99.99 99.99     7.32 13    6.31  2    6.56  2   99.99 99   99.99 99
53349       6.02 99.99  0.30 99.99 99.99     99.99 99.99    99.99 99    6.09  3    6.00  2   99.99 99   99.99 99
53501       5.17  1.65  1.40 99.99 99.99     99.99 99.99     9.21 13    5.85  4    4.63  4   99.99 99   99.99 99
53510       5.78  1.87  1.52 99.99 99.99     99.99 99.99    10.18 14    6.52  5    5.19  5   99.99 99   99.99 99
53704       5.20 99.99  0.20 99.99 99.99     99.99 99.99    99.99 99    5.22  3    5.22  2   99.99 99   99.99 99
53811       4.93 99.99  0.14 99.99 99.99     99.99 99.99    99.99 99    4.91  3    4.98  2   99.99 99   99.99 99
53929       6.11 -0.46 -0.13 99.99 99.99     99.99 99.99     6.45  7    5.94  2    6.29  2   99.99 99   99.99 99
53952       6.13 99.99  0.37 99.99 99.99     99.99 99.99    99.99 99    6.24  3    6.07  2   99.99 99   99.99 99
53975       6.47 -0.98 -0.10 99.99 99.99     99.99 99.99     6.12  7    6.32  2    6.63  2   99.99 99   99.99 99
54079       5.75  1.08  1.18 99.99 99.99     99.99 99.99     8.90  6    6.30  4    5.32  4   99.99 99   99.99 99
54153       6.11 99.99  0.70 99.99 99.99     99.99 99.99    99.99 99    6.40  3    5.90  3   99.99 99   99.99 99
54239       5.45 -0.07  0.05 99.99 99.99     99.99 99.99     6.42 11    5.38  3    5.54  2   99.99 99   99.99 99
54662       6.21 -0.88  0.03 99.99 99.99     99.99 99.99     6.07  8    6.13  3    6.31  2   99.99 99   99.99 99
55052       5.86 99.99  0.37 99.99 99.99     99.99 99.99    99.99 99    5.97  3    5.80  2   99.99 99   99.99 99
55111       6.09 -0.05 -0.02 99.99 99.99     99.99 99.99     7.05 13    5.98  2    6.21  2   99.99 99   99.99 99
55151       6.47  0.87  1.04 99.99 99.99     99.99 99.99     9.26  5    6.95  4    6.10  4   99.99 99   99.99 99
55184       6.16  0.97  1.14 99.99 99.99     99.99 99.99     9.14  5    6.69  4    5.74  4   99.99 99   99.99 99
55526       5.15 99.99  1.24 99.99 99.99     99.99 99.99    99.99 99    5.74  4    4.69  4   99.99 99   99.99 99
55568       6.10 99.99  0.27 99.99 99.99     99.99 99.99    99.99 99    6.15  3    6.09  2   99.99 99   99.99 99
55575       5.58  0.03  0.58 99.99 99.99     99.99 99.99     6.97  4    5.80  3    5.43  3   99.99 99   99.99 99
55621       5.82  1.83  1.56 99.99 99.99     99.99 99.99    10.19 12    6.58  5    5.21  5   99.99 99   99.99 99
```

```
55719       5.31   0.09   0.06 99.99 99.99      99.99 99.99      6.50 14    5.25  3    5.40  2    99.99 99    99.99 99
55751       5.35   1.18   1.19 99.99 99.99      99.99 99.99      8.64  8    5.91  4    4.91  4    99.99 99    99.99 99
55775       5.75   1.88   1.58 99.99 99.99      99.99 99.99     10.20 13    6.52  5    5.13  5    99.99 99    99.99 99
55832       5.91   1.70   1.52 99.99 99.99      99.99 99.99     10.08 11    6.65  5    5.32  5    99.99 99    99.99 99
55856       6.27  99.99  -0.13 99.99 99.99      99.99 99.99     99.99 99    6.10  2    6.45  2    99.99 99    99.99 99
55865       3.62   0.60   0.91 99.99 99.99      99.99 99.99      5.97  4    4.03  4    3.31  3    99.99 99    99.99 99
56003       6.41   0.57   0.90 99.99 99.99      99.99 99.99      8.71  3    6.81  4    6.11  3    99.99 99    99.99 99
56031       5.82   1.69   1.60 99.99 99.99      99.99 99.99     10.02  8    6.61  5    5.19  5    99.99 99    99.99 99
56239       6.02  99.99  -0.02 99.99 99.99      99.99 99.99     99.99 99    5.91  2    6.14  2    99.99 99    99.99 99
56342       5.35  99.99  -0.17 99.99 99.99      99.99 99.99     99.99 99    5.16  2    5.54  2    99.99 99    99.99 99
56446       6.65  -0.41  -0.11 99.99 99.99      99.99 99.99      7.07  7    6.49  2    6.82  2    99.99 99    99.99 99
56614       6.29   1.96   1.62 99.99 99.99      99.99 99.99     10.87 14    7.09  5    5.65  5    99.99 99    99.99 99
56705       5.97  99.99   1.10 99.99 99.99      99.99 99.99     99.99 99    6.48  4    5.57  4    99.99 99    99.99 99
56733       5.79  99.99  -0.12 99.99 99.99      99.99 99.99     99.99 99    5.63  2    5.96  2    99.99 99    99.99 99
56779       5.03  -0.69  -0.17 99.99 99.99      99.99 99.99      5.04  4    4.84  2    5.22  2    99.99 99    99.99 99
56813       5.66  99.99   1.44 99.99 99.99      99.99 99.99     99.99 99    6.36  4    5.11  5    99.99 99    99.99 99
56989       5.89   0.82   1.07 99.99 99.99      99.99 99.99      8.62  4    6.38  4    5.51  4    99.99 99    99.99 99
57006       5.91   0.07   0.53 99.99 99.99      99.99 99.99      7.33  3    6.11  3    5.78  3    99.99 99    99.99 99
57118       6.09   0.34   0.62 99.99 99.99      99.99 99.99      7.93  5    6.34  3    5.92  3    99.99 99    99.99 99
57146       5.28   0.65   0.96 99.99 99.99      99.99 99.99      7.72  3    5.71  4    4.95  4    99.99 99    99.99 99
57517       6.55   0.04   0.54 99.99 99.99      99.99 99.99      7.94  3    6.75  3    6.41  3    99.99 99    99.99 99
57573       6.61  99.99  -0.15 99.99 99.99      99.99 99.99     99.99 99    6.43  2    6.79  2    99.99 99    99.99 99
57608       5.99  -0.30  -0.07 99.99 99.99      99.99 99.99      6.58  9    5.86  2    6.14  2    99.99 99    99.99 99
57623       3.98   0.45   0.79 99.99 99.99      99.99 99.99      6.06  3    4.32  3    3.73  3    99.99 99    99.99 99
57682       6.44  -1.04  -0.19 99.99 99.99      99.99 99.99      5.96  6    6.24  2    6.64  2    99.99 99    99.99 99
57708       6.23   0.35   0.68 99.99 99.99      99.99 99.99      8.11  4    6.51  3    6.03  3    99.99 99    99.99 99
57749       5.82   0.19   0.35 99.99 99.99      99.99 99.99      7.30  8    5.92  3    5.77  2    99.99 99    99.99 99
58461       5.78  99.99   0.42 99.99 99.99      99.99 99.99     99.99 99    5.92  3    5.70  3    99.99 99    99.99 99
58526       5.97   0.62   0.92 99.99 99.99      99.99 99.99      8.35  4    6.38  4    5.66  3    99.99 99    99.99 99
58580       6.76  -0.11  -0.01 99.99 99.99      99.99 99.99      7.64 11    6.66  3    6.88  2    99.99 99    99.99 99
58585       6.05  99.99   0.23 99.99 99.99      99.99 99.99     99.99 99    6.08  3    6.06  2    99.99 99    99.99 99
58805       6.47  -0.07   0.42 99.99 99.99      99.99 99.99      7.64  3    6.61  3    6.39  3    99.99 99    99.99 99
58855       5.36  -0.06   0.45 99.99 99.99      99.99 99.99      6.56  3    5.51  3    5.27  3    99.99 99    99.99 99
58923       5.25   0.17   0.22 99.99 99.99      99.99 99.99      6.64 11    5.28  3    5.26  2    99.99 99    99.99 99
59067       5.79  99.99   0.58 99.99 99.99      99.99 99.99     99.99 99    6.01  3    5.64  3    99.99 99    99.99 99
59219       5.10  99.99   1.06 99.99 99.99      99.99 99.99     99.99 99    5.59  4    4.72  4    99.99 99    99.99 99
59311       5.59   1.76   1.50 99.99 99.99      99.99 99.99      9.83 12    6.32  5    5.01  5    99.99 99    99.99 99
59380       5.86  -0.02   0.48 99.99 99.99      99.99 99.99      7.13  3    6.03  3    5.75  3    99.99 99    99.99 99
59381       5.75  99.99   1.62 99.99 99.99      99.99 99.99     99.99 99    6.55  5    5.11  5    99.99 99    99.99 99
59438       6.04  99.99   0.48 99.99 99.99      99.99 99.99     99.99 99    6.21  3    5.93  3    99.99 99    99.99 99
59466       6.58  99.99   0.06 99.99 99.99      99.99 99.99     99.99 99    6.52  3    6.67  2    99.99 99    99.99 99
```

```
59550       5.77 99.99 -0.19 99.99 99.99     99.99 99.99    99.99 99   5.57  2   5.97  2   99.99 99   99.99 99
59635       5.43 99.99 -0.16 99.99 99.99     99.99 99.99    99.99 99   5.25  2   5.62  2   99.99 99   99.99 99
59669       6.24  1.06  1.18 99.99 99.99     99.99 99.99     9.36  6   6.79  4   5.81  4   99.99 99   99.99 99
59881       5.25  0.20  0.22 99.99 99.99     99.99 99.99     6.68 12   5.28  3   5.26  2   99.99 99   99.99 99
59967       6.65 99.99  0.63 99.99 99.99     99.99 99.99    99.99 99   6.90  3   6.47  3   99.99 99   99.99 99
59984       5.90 -0.08  0.54 99.99 99.99     99.99 99.99     7.12  5   6.10  3   5.76  3   99.99 99   99.99 99
60098       6.68 99.99 -0.12 99.99 99.99     99.99 99.99    99.99 99   6.52  2   6.85  2   99.99 99   99.99 99
60111       5.59  0.10  0.31 99.99 99.99     99.99 99.99     6.93  8   5.67  3   5.56  2   99.99 99   99.99 99
60150       6.39  1.84  1.56 99.99 99.99     99.99 99.99    10.78 13   7.15  5   5.78  5   99.99 99   99.99 99
60178       1.58  0.02  0.04 99.99 99.99     99.99 99.99     2.66 13   1.51  3   1.68  2   99.99 99   99.99 99
60345       5.85 99.99  0.17 99.99 99.99     99.99 99.99    99.99 99   5.85  3   5.89  2   99.99 99   99.99 99
60357       5.81 -0.09 -0.02 99.99 99.99     99.99 99.99     6.71 12   5.70  2   5.93  2   99.99 99   99.99 99
60489       6.56  0.15  0.22 99.99 99.99     99.99 99.99     7.92 11   6.59  3   6.57  2   99.99 99   99.99 99
60584       5.09  0.03  0.43 99.99 99.99     99.99 99.99     6.40  4   5.23  3   5.01  3   99.99 99   99.99 99
60629       6.65 99.99  0.00 99.99 99.99     99.99 99.99    99.99 99   6.56  3   6.76  2   99.99 99   99.99 99
60646       6.11 99.99  0.29 99.99 99.99     99.99 99.99    99.99 99   6.17  3   6.09  2   99.99 99   99.99 99
60666       5.77 99.99  1.06 99.99 99.99     99.99 99.99    99.99 99   6.26  4   5.39  4   99.99 99   99.99 99
60803       5.91  0.10  0.60 99.99 99.99     99.99 99.99     7.41  3   6.15  3   5.75  3   99.99 99   99.99 99
60816       5.53  1.42  1.28 99.99 99.99     99.99 99.99     9.19 11   6.14  4   5.05  4   99.99 99   99.99 99
60853       6.27  1.88  1.54 99.99 99.99     99.99 99.99    10.70 14   7.02  5   5.67  5   99.99 99   99.99 99
61064       5.12  0.12  0.44 99.99 99.99     99.99 99.99     6.56  5   5.27  3   5.03  3   99.99 99   99.99 99
61227       6.37 99.99  0.54 99.99 99.99     99.99 99.99    99.99 99   6.57  3   6.23  3   99.99 99   99.99 99
61248       4.94  1.63  1.40 99.99 99.99     99.99 99.99     8.95 12   5.62  4   4.40  4   99.99 99   99.99 99
61391       5.72 99.99 -0.06 99.99 99.99     99.99 99.99    99.99 99   5.59  2   5.86  2   99.99 99   99.99 99
61394       6.39 99.99  1.18 99.99 99.99     99.99 99.99    99.99 99   6.94  4   5.96  4   99.99 99   99.99 99
61409       6.60 99.99  1.14 99.99 99.99     99.99 99.99    99.99 99   7.13  4   6.18  4   99.99 99   99.99 99
61563       6.02 -0.18 -0.04 99.99 99.99     99.99 99.99     6.79 11   5.90  2   6.15  2   99.99 99   99.99 99
61623       6.59 -0.17 -0.05 99.99 99.99     99.99 99.99     7.37 11   6.47  2   6.73  2   99.99 99   99.99 99
61641       5.80 99.99 -0.16 99.99 99.99     99.99 99.99    99.99 99   5.62  2   5.99  2   99.99 99   99.99 99
61749       6.01  0.11  0.15 99.99 99.99     99.99 99.99     7.28 12   6.00  3   6.06  2   99.99 99   99.99 99
61887       5.95 -0.09 -0.04 99.99 99.99     99.99 99.99     6.84 13   5.83  2   6.08  2   99.99 99   99.99 99
62153       6.34 -0.24 -0.02 99.99 99.99     99.99 99.99     7.04  9   6.23  2   6.46  2   99.99 99   99.99 99
62264       6.19  0.74  1.02 99.99 99.99     99.99 99.99     8.79  4   6.66  4   5.83  4   99.99 99   99.99 99
62578       5.60 99.99 -0.13 99.99 99.99     99.99 99.99    99.99 99   5.43  2   5.78  2   99.99 99   99.99 99
62644       5.06 99.99  0.78 99.99 99.99     99.99 99.99    99.99 99   5.39  3   4.81  3   99.99 99   99.99 99
62689       6.18  2.03  1.73 99.99 99.99     99.99 99.99    10.92 12   7.04  5   5.49  5   99.99 99   99.99 99
62713       5.17 99.99  1.10 99.99 99.99     99.99 99.99    99.99 99   5.68  4   4.77  4   99.99 99   99.99 99
62781       5.81 99.99  0.31 99.99 99.99     99.99 99.99    99.99 99   5.89  3   5.78  2   99.99 99   99.99 99
62902       5.48  1.67  1.38 99.99 99.99     99.99 99.99     9.54 14   6.14  4   4.95  4   99.99 99   99.99 99
62952       5.04  0.09  0.33 99.99 99.99     99.99 99.99     6.38  7   5.13  3   5.00  2   99.99 99   99.99 99
63077       5.37 99.99  0.58 99.99 99.99     99.99 99.99    99.99 99   5.59  3   5.22  3   99.99 99   99.99 99
```

```
63112      6.39 -0.08 -0.02 99.99 99.99    99.99 99.99    7.31 12    6.28  2    6.51  2    99.99 99    99.99 99
63118      6.03 99.99 -0.07 99.99 99.99    99.99 99.99   99.99 99    5.90  2    6.18  2    99.99 99    99.99 99
63271      5.90 99.99 -0.19 99.99 99.99    99.99 99.99   99.99 99    5.70  2    6.10  2    99.99 99    99.99 99
63295      3.95  0.83  1.04 99.99 99.99    99.99 99.99    6.68  5    4.43  4    3.58  4    99.99 99    99.99 99
63382      6.12 99.99  0.39 99.99 99.99    99.99 99.99   99.99 99    6.24  3    6.05  2    99.99 99    99.99 99
63435      6.53  0.38  0.78 99.99 99.99    99.99 99.99    8.51  3    6.86  3    6.28  3    99.99 99    99.99 99
63513      6.38  0.65  0.95 99.99 99.99    99.99 99.99    8.82  4    6.81  4    6.05  4    99.99 99    99.99 99
63584      6.18 -0.09 -0.06 99.99 99.99    99.99 99.99    7.06 13    6.05  2    6.32  2    99.99 99    99.99 99
63660      5.33 99.99  0.76 99.99 99.99    99.99 99.99   99.99 99    5.65  3    5.09  3    99.99 99    99.99 99
63752      5.61  1.56  1.44 99.99 99.99    99.99 99.99    9.55 10    6.31  4    5.06  5    99.99 99    99.99 99
63754      6.56 99.99  0.58 99.99 99.99    99.99 99.99   99.99 99    6.78  3    6.41  3    99.99 99    99.99 99
63799      6.18  1.01  1.12 99.99 99.99    99.99 99.99    9.20  6    6.70  4    5.77  4    99.99 99    99.99 99
63975      5.14 -0.49 -0.11 99.99 99.99    99.99 99.99    5.45  6    4.98  2    5.31  2    99.99 99    99.99 99
64042      6.45 99.99 -0.01 99.99 99.99    99.99 99.99   99.99 99    6.35  3    6.57  2    99.99 99    99.99 99
64152      5.63 99.99  0.96 99.99 99.99    99.99 99.99   99.99 99    6.06  4    5.30  4    99.99 99    99.99 99
64185      5.78 99.99  0.42 99.99 99.99    99.99 99.99   99.99 99    5.92  3    5.70  3    99.99 99    99.99 99
64235      5.76  0.00  0.41 99.99 99.99    99.99 99.99    7.02  4    5.89  3    5.68  3    99.99 99    99.99 99
64287      6.32 -0.76 -0.17 99.99 99.99    99.99 99.99    6.23  3    6.13  2    6.51  2    99.99 99    99.99 99
64320      6.72 99.99  1.24 99.99 99.99    99.99 99.99   99.99 99    7.31  4    6.26  4    99.99 99    99.99 99
64379      5.01 -0.06  0.44 99.99 99.99    99.99 99.99    6.21  3    5.16  3    4.92  3    99.99 99    99.99 99
64484      5.79 -0.16 -0.04 99.99 99.99    99.99 99.99    6.59 11    5.67  2    5.92  2    99.99 99    99.99 99
64572      5.43 99.99  1.16 99.99 99.99    99.99 99.99   99.99 99    5.97  4    5.01  4    99.99 99    99.99 99
64685      5.86  0.02  0.35 99.99 99.99    99.99 99.99    7.11  5    5.96  3    5.81  2    99.99 99    99.99 99
64802      5.50 99.99 -0.19 99.99 99.99    99.99 99.99   99.99 99    5.30  2    5.70  2    99.99 99    99.99 99
64938      6.17  0.73  0.98 99.99 99.99    99.99 99.99    8.73  4    6.61  4    5.83  4    99.99 99    99.99 99
65066      6.05  0.86  1.00 99.99 99.99    99.99 99.99    8.80  6    6.51  4    5.70  4    99.99 99    99.99 99
65123      6.35  0.00  0.50 99.99 99.99    99.99 99.99    7.66  3    6.53  3    6.23  3    99.99 99    99.99 99
65189      6.38 99.99 -0.01 99.99 99.99    99.99 99.99   99.99 99    6.28  3    6.50  2    99.99 99    99.99 99
65211      6.03 99.99 -0.11 99.99 99.99    99.99 99.99   99.99 99    5.87  2    6.20  2    99.99 99    99.99 99
65273      5.63 99.99  1.30 99.99 99.99    99.99 99.99   99.99 99    6.25  4    5.14  4    99.99 99    99.99 99
65315      6.79 99.99 -0.18 99.99 99.99    99.99 99.99   99.99 99    6.60  2    6.99  2    99.99 99    99.99 99
65345      5.28  0.72  0.91 99.99 99.99    99.99 99.99    7.79  5    5.69  4    4.97  3    99.99 99    99.99 99
65551      5.09 99.99 -0.17 99.99 99.99    99.99 99.99   99.99 99    4.90  2    5.28  2    99.99 99    99.99 99
65598      6.22 99.99 -0.10 99.99 99.99    99.99 99.99   99.99 99    6.07  2    6.38  2    99.99 99    99.99 99
65685      5.17 99.99  1.27 99.99 99.99    99.99 99.99   99.99 99    5.77  4    4.69  4    99.99 99    99.99 99
65695      4.93  1.21  1.21 99.99 99.99    99.99 99.99    8.27  8    5.50  4    4.48  4    99.99 99    99.99 99
65699      5.11 99.99  1.12 99.99 99.99    99.99 99.99   99.99 99    5.63  4    4.70  4    99.99 99    99.99 99
65900      5.65  0.01  0.00 99.99 99.99    99.99 99.99    6.70 14    5.56  3    5.76  2    99.99 99    99.99 99
65904      5.99 99.99 -0.14 99.99 99.99    99.99 99.99   99.99 99    5.82  2    6.17  2    99.99 99    99.99 99
65908      6.14 99.99 -0.10 99.99 99.99    99.99 99.99   99.99 99    5.99  2    6.30  2    99.99 99    99.99 99
65925      5.24 99.99  0.39 99.99 99.99    99.99 99.99   99.99 99    5.36  3    5.17  2    99.99 99    99.99 99
```

```
66011         6.22   0.13   0.57 99.99 99.99      99.99 99.99      7.74  3    6.44  3    6.07  3    99.99 99    99.99 99
66210         6.03 99.99   0.04 99.99 99.99      99.99 99.99     99.99 99    5.96  3    6.13  2    99.99 99    99.99 99
66242         6.33   0.13   0.62 99.99 99.99      99.99 99.99      7.88  3    6.58  3    6.16  3    99.99 99    99.99 99
66358         5.96 99.99   0.14 99.99 99.99      99.99 99.99     99.99 99    5.94  3    6.01  2    99.99 99    99.99 99
66591         4.82  -0.62  -0.17 99.99 99.99      99.99 99.99      4.92  5    4.63  2    5.01  2    99.99 99    99.99 99
66834         6.13 99.99  -0.16 99.99 99.99      99.99 99.99     99.99 99    5.95  2    6.32  2    99.99 99    99.99 99
66920         6.34   0.13   0.14 99.99 99.99      99.99 99.99      7.63 13    6.32  3    6.39  2    99.99 99    99.99 99
66950         6.41   1.01   1.05 99.99 99.99      99.99 99.99      9.39  8    6.89  4    6.04  4    99.99 99    99.99 99
67159         6.01  -0.13  -0.04 99.99 99.99      99.99 99.99      6.85 12    5.89  2    6.14  2    99.99 99    99.99 99
67228         5.30   0.21   0.63 99.99 99.99      99.99 99.99      6.97  3    5.55  3    5.12  3    99.99 99    99.99 99
67364         5.53 99.99   1.34 99.99 99.99      99.99 99.99     99.99 99    6.17  4    5.02  4    99.99 99    99.99 99
67456         5.38 99.99   0.10 99.99 99.99      99.99 99.99     99.99 99    5.34  3    5.45  2    99.99 99    99.99 99
67725         6.32 99.99   0.00 99.99 99.99      99.99 99.99     99.99 99    6.23  3    6.43  2    99.99 99    99.99 99
67751         6.36 99.99   0.16 99.99 99.99      99.99 99.99     99.99 99    6.35  3    6.40  2    99.99 99    99.99 99
67880         5.68 99.99  -0.17 99.99 99.99      99.99 99.99     99.99 99    5.49  2    5.87  2    99.99 99    99.99 99
67921         6.65 99.99   1.40 99.99 99.99      99.99 99.99     99.99 99    7.33  4    6.11  4    99.99 99    99.99 99
68146         5.54 99.99   0.49 99.99 99.99      99.99 99.99     99.99 99    5.71  3    5.43  3    99.99 99    99.99 99
68217         5.21 99.99  -0.19 99.99 99.99      99.99 99.99     99.99 99    5.01  2    5.41  2    99.99 99    99.99 99
68312         5.36   0.60   0.89 99.99 99.99      99.99 99.99      7.70  4    5.75  4    5.06  3    99.99 99    99.99 99
68423         6.28 99.99  -0.06 99.99 99.99      99.99 99.99     99.99 99    6.15  2    6.42  2    99.99 99    99.99 99
68456         4.76  -0.03   0.43 99.99 99.99      99.99 99.99      5.99  3    4.90  3    4.68  3    99.99 99    99.99 99
68520         4.35  -0.46  -0.11 99.99 99.99      99.99 99.99      4.70  6    4.19  2    4.52  2    99.99 99    99.99 99
68752         4.99 99.99   1.07 99.99 99.99      99.99 99.99     99.99 99    5.48  4    4.61  4    99.99 99    99.99 99
68758         6.52 99.99   0.06 99.99 99.99      99.99 99.99     99.99 99    6.46  3    6.61  2    99.99 99    99.99 99
68895         6.03 99.99  -0.11 99.99 99.99      99.99 99.99     99.99 99    5.87  2    6.20  2    99.99 99    99.99 99
69080         6.06 99.99  -0.16 99.99 99.99      99.99 99.99     99.99 99    5.88  2    6.25  2    99.99 99    99.99 99
69082         6.11  -0.71  -0.18 99.99 99.99      99.99 99.99      6.08  4    5.92  2    6.31  2    99.99 99    99.99 99
69123         5.78 99.99   1.02 99.99 99.99      99.99 99.99     99.99 99    6.25  4    5.42  4    99.99 99    99.99 99
69478         6.29   0.75   0.98 99.99 99.99      99.99 99.99      8.88  4    6.73  4    5.95  4    99.99 99    99.99 99
69589         6.60 99.99   0.02 99.99 99.99      99.99 99.99     99.99 99    6.52  3    6.71  2    99.99 99    99.99 99
69830         6.00   0.32   0.76 99.99 99.99      99.99 99.99      7.89  3    6.32  3    5.76  3    99.99 99    99.99 99
69863         5.16 99.99   0.08 99.99 99.99      99.99 99.99     99.99 99    5.11  3    5.24  2    99.99 99    99.99 99
69879         6.45 99.99   1.04 99.99 99.99      99.99 99.99     99.99 99    6.93  4    6.08  4    99.99 99    99.99 99
69897         5.14  -0.06   0.47 99.99 99.99      99.99 99.99      6.35  3    5.30  3    5.04  3    99.99 99    99.99 99
69994         5.83 99.99   1.13 99.99 99.99      99.99 99.99     99.99 99    6.36  4    5.42  4    99.99 99    99.99 99
70003         6.70 99.99   0.24 99.99 99.99      99.99 99.99     99.99 99    6.74  3    6.70  2    99.99 99    99.99 99
70013         6.05   0.63   0.97 99.99 99.99      99.99 99.99      8.47  3    6.49  4    5.71  4    99.99 99    99.99 99
70110         6.18   0.15   0.60 99.99 99.99      99.99 99.99      7.75  3    6.42  3    6.02  3    99.99 99    99.99 99
70148         6.13   1.52   1.33 99.99 99.99      99.99 99.99      9.96 12    6.77  4    5.63  4    99.99 99    99.99 99
70235         6.43 99.99  -0.08 99.99 99.99      99.99 99.99     99.99 99    6.29  2    6.58  2    99.99 99    99.99 99
70267         6.42 99.99   0.39 99.99 99.99      99.99 99.99     99.99 99    6.54  3    6.35  2    99.99 99    99.99 99
```

```
70302      6.13 99.99  1.04 99.99 99.99    99.99 99.99   99.99 99   6.61  4   5.76  4   99.99 99   99.99 99
70340      6.50 -0.02  0.02 99.99 99.99    99.99 99.99    7.52 12   6.42  3   6.61  2   99.99 99   99.99 99
70514      5.07  1.19  1.15 99.99 99.99    99.99 99.99    8.35  9   5.61  4   4.65  4   99.99 99   99.99 99
70523      5.77  0.99  1.05 99.99 99.99    99.99 99.99    8.72  7   6.25  4   5.40  4   99.99 99   99.99 99
70556      5.20 99.99 -0.19 99.99 99.99    99.99 99.99   99.99 99   5.00  2   5.40  2   99.99 99   99.99 99
70574      6.15  0.13  0.22 99.99 99.99    99.99 99.99    7.48 11   6.18  3   6.16  2   99.99 99   99.99 99
70612      6.16 99.99  0.17 99.99 99.99    99.99 99.99   99.99 99   6.16  3   6.20  2   99.99 99   99.99 99
70652      5.96  1.85  1.67 99.99 99.99    99.99 99.99   10.42 10   6.78  5   5.30  5   99.99 99   99.99 99
70673      6.13 99.99  1.00 99.99 99.99    99.99 99.99   99.99 99   6.59  4   5.78  4   99.99 99   99.99 99
70761      5.90 99.99  0.37 99.99 99.99    99.99 99.99   99.99 99   6.01  3   5.84  2   99.99 99   99.99 99
70937      6.01  0.04  0.46 99.99 99.99    99.99 99.99    7.35  3   6.17  3   5.91  3   99.99 99   99.99 99
70958      5.62 -0.06  0.46 99.99 99.99    99.99 99.99    6.83  3   5.78  3   5.52  3   99.99 99   99.99 99
70982      6.12 99.99  0.93 99.99 99.99    99.99 99.99   99.99 99   6.54  4   5.80  3   99.99 99   99.99 99
71046      5.37 -0.31 -0.06 99.99 99.99    99.99 99.99    5.95  8   5.24  2   5.51  2   99.99 99   99.99 99
71066      5.65 -0.31 -0.10 99.99 99.99    99.99 99.99    6.21  9   5.50  2   5.81  2   99.99 99   99.99 99
71095      5.73  1.86  1.53 99.99 99.99    99.99 99.99   10.13 14   6.48  5   5.13  5   99.99 99   99.99 99
71129      1.86  0.19  1.27 99.99 99.99    99.99 99.99    3.85 18   2.46  4   1.38  4   99.99 99   99.99 99
71141      5.68 99.99  0.06 99.99 99.99    99.99 99.99   99.99 99   5.62  3   5.77  2   99.99 99   99.99 99
71243      4.07 -0.02  0.39 99.99 99.99    99.99 99.99    5.29  4   4.19  3   4.00  2   99.99 99   99.99 99
71267      5.98 99.99  0.17 99.99 99.99    99.99 99.99   99.99 99   5.98  3   6.02  2   99.99 99   99.99 99
71377      5.54  1.14  1.19 99.99 99.99    99.99 99.99    8.77  7   6.10  4   5.10  4   99.99 99   99.99 99
71433      6.59  0.06  0.51 99.99 99.99    99.99 99.99    7.99  3   6.78  3   6.47  3   99.99 99   99.99 99
71459      5.47 99.99 -0.14 99.99 99.99    99.99 99.99   99.99 99   5.30  2   5.65  2   99.99 99   99.99 99
71576      5.29 -0.01  0.01 99.99 99.99    99.99 99.99    6.32 13   5.20  3   5.40  2   99.99 99   99.99 99
71665      6.43  1.23  1.20 99.99 99.99    99.99 99.99    9.79  9   7.00  4   5.99  4   99.99 99   99.99 99
71701      4.35  1.19  1.16 99.99 99.99    99.99 99.99    7.64  9   4.89  4   3.93  4   99.99 99   99.99 99
71766      6.00  0.20  0.42 99.99 99.99    99.99 99.99    7.54  7   6.14  3   5.92  3   99.99 99   99.99 99
71815      6.51  0.08  0.05 99.99 99.99    99.99 99.99    7.68 14   6.44  3   6.60  2   99.99 99   99.99 99
71833      6.67 -0.41 -0.06 99.99 99.99    99.99 99.99    7.12  6   6.54  2   6.81  2   99.99 99   99.99 99
71863      5.97  0.73  0.96 99.99 99.99    99.99 99.99    8.52  5   6.40  4   5.64  4   99.99 99   99.99 99
71878      3.77  1.13  1.13 99.99 99.99    99.99 99.99    6.96  8   4.30  4   3.36  4   99.99 99   99.99 99
72108      5.33 99.99 -0.14 99.99 99.99    99.99 99.99   99.99 99   5.16  2   5.51  2   99.99 99   99.99 99
72184      5.90  1.17  1.11 99.99 99.99    99.99 99.99    9.13 10   6.42  4   5.50  4   99.99 99   99.99 99
72227      5.65 99.99  1.49 99.99 99.99    99.99 99.99   99.99 99   6.37  4   5.07  5   99.99 99   99.99 99
72310      5.42 99.99 -0.06 99.99 99.99    99.99 99.99   99.99 99   5.29  2   5.56  2   99.99 99   99.99 99
72322      6.36 99.99  0.80 99.99 99.99    99.99 99.99   99.99 99   6.71  3   6.10  3   99.99 99   99.99 99
72324      6.36  0.88  1.02 99.99 99.99    99.99 99.99    9.15  6   6.83  4   6.00  4   99.99 99   99.99 99
72337      5.53 -0.04 -0.03 99.99 99.99    99.99 99.99    6.49 13   5.42  2   5.66  2   99.99 99   99.99 99
72436      6.28 99.99 -0.14 99.99 99.99    99.99 99.99   99.99 99   6.11  2   6.46  2   99.99 99   99.99 99
72462      6.38 99.99  0.27 99.99 99.99    99.99 99.99   99.99 99   6.43  3   6.37  2   99.99 99   99.99 99
72485      6.39 99.99 -0.15 99.99 99.99    99.99 99.99   99.99 99   6.21  2   6.57  2   99.99 99   99.99 99
```

```
72561      5.87  0.87   1.07 99.99 99.99      99.99 99.99       8.67  5    6.36  4    5.49  4    99.99 99    99.99 99
72617      6.03  0.03   0.33 99.99 99.99      99.99 99.99       7.29  6    6.12  3    5.99  2    99.99 99    99.99 99
72626      6.19 99.99   0.27 99.99 99.99      99.99 99.99      99.99 99    6.24  3    6.18  2    99.99 99    99.99 99
72650      6.34 99.99   1.31 99.99 99.99      99.99 99.99      99.99 99    6.97  4    5.85  4    99.99 99    99.99 99
72660      5.81  0.00   0.00 99.99 99.99      99.99 99.99       6.84 13    5.72  3    5.92  2    99.99 99    99.99 99
72673      6.38 99.99   0.79 99.99 99.99      99.99 99.99      99.99 99    6.72  3    6.13  3    99.99 99    99.99 99
72738      5.69 99.99   0.58 99.99 99.99      99.99 99.99      99.99 99    5.91  3    5.54  3    99.99 99    99.99 99
72787      6.49 99.99  -0.18 99.99 99.99      99.99 99.99      99.99 99    6.30  2    6.69  2    99.99 99    99.99 99
72832      5.96 99.99  -0.14 99.99 99.99      99.99 99.99      99.99 99    5.79  2    6.14  2    99.99 99    99.99 99
72908      6.33  0.87   1.02 99.99 99.99      99.99 99.99       9.10  6    6.80  4    5.97  4    99.99 99    99.99 99
72922      5.69  0.74   1.02 99.99 99.99      99.99 99.99       8.29  4    6.16  4    5.33  4    99.99 99    99.99 99
72945      5.63  0.09   0.56 99.99 99.99      99.99 99.99       7.09  3    5.84  3    5.48  3    99.99 99    99.99 99
72954      6.43 99.99   0.75 99.99 99.99      99.99 99.99      99.99 99    6.75  3    6.20  3    99.99 99    99.99 99
73072      5.96 99.99   0.37 99.99 99.99      99.99 99.99      99.99 99    6.07  3    5.90  2    99.99 99    99.99 99
73121      6.47 99.99   0.58 99.99 99.99      99.99 99.99      99.99 99    6.69  3    6.32  3    99.99 99    99.99 99
73281      6.19  0.89   1.05 99.99 99.99      99.99 99.99       9.01  5    6.67  4    5.82  4    99.99 99    99.99 99
73316      6.53 -0.04  -0.02 99.99 99.99      99.99 99.99       7.50 13    6.42  2    6.65  2    99.99 99    99.99 99
73389      4.86  0.81   1.00 99.99 99.99      99.99 99.99       7.54  5    5.32  4    4.51  4    99.99 99    99.99 99
73390      5.26 99.99  -0.14 99.99 99.99      99.99 99.99      99.99 99    5.09  2    5.44  2    99.99 99    99.99 99
73451      6.51  0.17   0.44 99.99 99.99      99.99 99.99       8.02  6    6.66  3    6.42  3    99.99 99    99.99 99
73468      6.12  0.65   0.95 99.99 99.99      99.99 99.99       8.56  4    6.55  4    5.79  4    99.99 99    99.99 99
73476      6.48 99.99   0.33 99.99 99.99      99.99 99.99      99.99 99    6.57  3    6.44  2    99.99 99    99.99 99
73495      5.27 99.99  -0.04 99.99 99.99      99.99 99.99      99.99 99    5.15  2    5.40  2    99.99 99    99.99 99
73524      6.55 99.99   0.60 99.99 99.99      99.99 99.99      99.99 99    6.79  3    6.39  3    99.99 99    99.99 99
73599      6.45  0.99   1.08 99.99 99.99      99.99 99.99       9.42  7    6.95  4    6.06  4    99.99 99    99.99 99
73603      6.33 99.99   1.59 99.99 99.99      99.99 99.99      99.99 99    7.11  5    5.71  5    99.99 99    99.99 99
73752      5.05 99.99   0.73 99.99 99.99      99.99 99.99      99.99 99    5.36  3    4.83  3    99.99 99    99.99 99
73840      4.98 99.99   1.42 99.99 99.99      99.99 99.99      99.99 99    5.67  4    4.43  5    99.99 99    99.99 99
73887      5.47 99.99   1.02 99.99 99.99      99.99 99.99      99.99 99    5.94  4    5.11  4    99.99 99    99.99 99
73898      4.89 99.99   0.90 99.99 99.99      99.99 99.99      99.99 99    5.29  4    4.59  3    99.99 99    99.99 99
73900      6.13 99.99   0.42 99.99 99.99      99.99 99.99      99.99 99    6.27  3    6.05  3    99.99 99    99.99 99
73997      6.63  0.02  -0.02 99.99 99.99      99.99 99.99       7.68 14    6.52  2    6.75  2    99.99 99    99.99 99
74148      6.36 99.99   0.00 99.99 99.99      99.99 99.99      99.99 99    6.27  3    6.47  2    99.99 99    99.99 99
74167      5.71 99.99   1.65 99.99 99.99      99.99 99.99      99.99 99    6.52  5    5.06  5    99.99 99    99.99 99
74190      6.45  0.13   0.16 99.99 99.99      99.99 99.99       7.75 12    6.44  3    6.49  2    99.99 99    99.99 99
74273      5.90 99.99  -0.21 99.99 99.99      99.99 99.99      99.99 99    5.69  2    6.11  2    99.99 99    99.99 99
74341      6.34  0.14   0.20 99.99 99.99      99.99 99.99       7.67 11    6.36  3    6.36  2    99.99 99    99.99 99
74393      6.37 -0.13  -0.06 99.99 99.99      99.99 99.99       7.20 12    6.24  2    6.51  2    99.99 99    99.99 99
74405      5.20 -0.03   0.01 99.99 99.99      99.99 99.99       6.20 13    5.11  3    5.31  2    99.99 99    99.99 99
74475      6.42 99.99   0.02 99.99 99.99      99.99 99.99      99.99 99    6.34  3    6.53  2    99.99 99    99.99 99
74591      6.13  0.10   0.20 99.99 99.99      99.99 99.99       7.41 10    6.15  3    6.15  2    99.99 99    99.99 99
```

```
74622      6.29 99.99  1.18 99.99 99.99     99.99 99.99    99.99 99   6.84  4   5.86  4    99.99 99    99.99 99
74706      6.11 99.99  0.22 99.99 99.99     99.99 99.99    99.99 99   6.14  3   6.12  2    99.99 99    99.99 99
74753      5.16 -1.02 -0.20 99.99 99.99     99.99 99.99     4.70  5   4.95  2   5.37  2    99.99 99    99.99 99
74794      5.70  1.03  1.10 99.99 99.99     99.99 99.99     8.74  7   6.21  4   5.30  4    99.99 99    99.99 99
74824      5.76 99.99 -0.15 99.99 99.99     99.99 99.99    99.99 99   5.58  2   5.94  2    99.99 99    99.99 99
74860      6.25 99.99  1.61 99.99 99.99     99.99 99.99    99.99 99   7.04  5   5.62  5    99.99 99    99.99 99
74879      6.10 99.99  0.08 99.99 99.99     99.99 99.99    99.99 99   6.05  3   6.18  2    99.99 99    99.99 99
74988      5.29  0.08  0.04 99.99 99.99     99.99 99.99     6.46 14   5.22  3   5.39  2    99.99 99    99.99 99
75081      6.21 99.99 -0.05 99.99 99.99     99.99 99.99    99.99 99   6.09  2   6.35  2    99.99 99    99.99 99
75086      6.22 99.99 -0.10 99.99 99.99     99.99 99.99    99.99 99   6.07  2   6.38  2    99.99 99    99.99 99
75116      6.32  1.78  1.50 99.99 99.99     99.99 99.99    10.59 13   7.05  5   5.74  5    99.99 99    99.99 99
75140      6.09  1.42  1.28 99.99 99.99     99.99 99.99     9.75 11   6.70  4   5.61  4    99.99 99    99.99 99
75171      6.05  0.08  0.20 99.99 99.99     99.99 99.99     7.30 10   6.07  3   6.07  2    99.99 99    99.99 99
75276      5.75  0.37  0.56 99.99 99.99     99.99 99.99     7.59  7   5.96  3   5.60  3    99.99 99    99.99 99
75289      6.37 99.99  0.58 99.99 99.99     99.99 99.99    99.99 99   6.59  3   6.22  3    99.99 99    99.99 99
75416      5.47 -0.35 -0.10 99.99 99.99     99.99 99.99     5.98  8   5.32  2   5.63  2    99.99 99    99.99 99
75495      6.47 99.99  0.23 99.99 99.99     99.99 99.99    99.99 99   6.50  3   6.48  2    99.99 99    99.99 99
75528      6.38 99.99  0.64 99.99 99.99     99.99 99.99    99.99 99   6.64  3   6.20  3    99.99 99    99.99 99
75605      5.21 99.99  0.87 99.99 99.99     99.99 99.99    99.99 99   5.59  4   4.92  3    99.99 99    99.99 99
75630      5.49 99.99  0.06 99.99 99.99     99.99 99.99    99.99 99   5.43  3   5.58  2    99.99 99    99.99 99
75649      6.17 -0.19 -0.08 99.99 99.99     99.99 99.99     6.90 11   6.03  2   6.32  2    99.99 99    99.99 99
75737      5.54  0.14  0.15 99.99 99.99     99.99 99.99     6.85 13   5.53  3   5.59  2    99.99 99    99.99 99
75811      6.33  0.14  0.12 99.99 99.99     99.99 99.99     7.62 13   6.30  3   6.39  2    99.99 99    99.99 99
75926      6.55 99.99  0.04 99.99 99.99     99.99 99.99    99.99 99   6.48  3   6.65  2    99.99 99    99.99 99
76001      6.50 99.99  1.46 99.99 99.99     99.99 99.99    99.99 99   7.21  4   5.94  5    99.99 99    99.99 99
76073      6.42 99.99  0.56 99.99 99.99     99.99 99.99    99.99 99   6.63  3   6.27  3    99.99 99    99.99 99
76113      5.60 99.99 -0.11 99.99 99.99     99.99 99.99    99.99 99   5.44  2   5.77  2    99.99 99    99.99 99
76143      5.35  0.07  0.42 99.99 99.99     99.99 99.99     6.71  4   5.49  3   5.27  3    99.99 99    99.99 99
76151      6.00  0.22  0.67 99.99 99.99     99.99 99.99     7.70  3   6.27  3   5.80  3    99.99 99    99.99 99
76161      5.91 99.99 -0.15 99.99 99.99     99.99 99.99    99.99 99   5.73  2   6.09  2    99.99 99    99.99 99
76230      6.39 99.99  0.00 99.99 99.99     99.99 99.99    99.99 99   6.30  3   6.50  2    99.99 99    99.99 99
76236      5.79  1.96  1.60 99.99 99.99     99.99 99.99    10.36 14   6.58  5   5.16  5    99.99 99    99.99 99
76270      6.11  0.17  0.20 99.99 99.99     99.99 99.99     7.49 12   6.13  3   6.13  2    99.99 99    99.99 99
76291      5.74  1.06  1.09 99.99 99.99     99.99 99.99     8.81  8   6.24  4   5.35  4    99.99 99    99.99 99
76304      6.47 99.99  0.96 99.99 99.99     99.99 99.99    99.99 99   6.90  4   6.14  4    99.99 99    99.99 99
76346      6.03 99.99 -0.02 99.99 99.99     99.99 99.99    99.99 99   5.92  2   6.15  2    99.99 99    99.99 99
76360      5.33 99.99  0.26 99.99 99.99     99.99 99.99    99.99 99   5.38  3   5.32  2    99.99 99    99.99 99
76376      5.75 99.99  1.32 99.99 99.99     99.99 99.99    99.99 99   6.38  4   5.25  4    99.99 99    99.99 99
76494      6.14  0.75  1.00 99.99 99.99     99.99 99.99     8.74  4   6.60  4   5.79  4    99.99 99    99.99 99
76508      6.17 99.99  1.00 99.99 99.99     99.99 99.99    99.99 99   6.63  4   5.82  4    99.99 99    99.99 99
76512      6.39  0.12  0.16 99.99 99.99     99.99 99.99     7.68 12   6.38  3   6.43  2    99.99 99    99.99 99
```

```
76538      5.78 99.99 -0.08 99.99 99.99    99.99 99.99    99.99 99    5.64  2    5.93  2    99.99 99    99.99 99
76579      5.96 99.99  1.54 99.99 99.99    99.99 99.99    99.99 99    6.71  5    5.36  5    99.99 99    99.99 99
76629      6.19  0.76  0.98 99.99 99.99    99.99 99.99     8.79  5    6.63  4    5.85  4    99.99 99    99.99 99
76653      5.71  0.00  0.48 99.99 99.99    99.99 99.99     7.01  3    5.88  3    5.60  3    99.99 99    99.99 99
76728      3.84 -0.45 -0.10 99.99 99.99    99.99 99.99     4.21  6    3.69  2    4.00  2    99.99 99    99.99 99
76757      6.59  0.09  0.06 99.99 99.99    99.99 99.99     7.78 14    6.53  3    6.68  2    99.99 99    99.99 99
76805      4.69 -0.47 -0.12 99.99 99.99    99.99 99.99     5.02  6    4.53  2    4.86  2    99.99 99    99.99 99
76932      5.86  0.08  0.53 99.99 99.99    99.99 99.99     7.29  3    6.06  3    5.73  3    99.99 99    99.99 99
77087      6.25 99.99  1.00 99.99 99.99    99.99 99.99    99.99 99    6.71  4    5.90  4    99.99 99    99.99 99
77250      6.07  1.02  1.12 99.99 99.99    99.99 99.99     9.10  6    6.59  4    5.66  4    99.99 99    99.99 99
77353      5.68  1.04  1.14 99.99 99.99    99.99 99.99     8.75  6    6.21  4    5.26  4    99.99 99    99.99 99
77370      5.16 99.99  0.42 99.99 99.99    99.99 99.99    99.99 99    5.30  3    5.08  3    99.99 99    99.99 99
77445      5.85  0.98  1.11 99.99 99.99    99.99 99.99     8.82  6    6.37  4    5.45  4    99.99 99    99.99 99
77580      6.27 99.99  1.00 99.99 99.99    99.99 99.99    99.99 99    6.73  4    5.92  4    99.99 99    99.99 99
77615      5.79 99.99  1.21 99.99 99.99    99.99 99.99    99.99 99    6.36  4    5.34  4    99.99 99    99.99 99
77665      6.74 99.99 -0.03 99.99 99.99    99.99 99.99    99.99 99    6.63  2    6.87  2    99.99 99    99.99 99
77887      5.88  1.96  1.63 99.99 99.99    99.99 99.99    10.47 13    6.68  5    5.24  5    99.99 99    99.99 99
77907      6.41 99.99 -0.10 99.99 99.99    99.99 99.99    99.99 99    6.26  2    6.57  2    99.99 99    99.99 99
77996      4.97  1.21  1.22 99.99 99.99    99.99 99.99     8.32  8    5.55  4    4.52  4    99.99 99    99.99 99
78045      4.01  0.13  0.14 99.99 99.99    99.99 99.99     5.30 13    3.99  3    4.06  2    99.99 99    99.99 99
78293      6.44  0.19  0.25 99.99 99.99    99.99 99.99     7.87 11    6.48  3    6.44  2    99.99 99    99.99 99
78548      6.11 99.99 -0.15 99.99 99.99    99.99 99.99    99.99 99    5.93  2    6.29  2    99.99 99    99.99 99
78556      5.60 -0.19 -0.06 99.99 99.99    99.99 99.99     6.34 11    5.47  2    5.74  2    99.99 99    99.99 99
78599      6.73 99.99  1.62 99.99 99.99    99.99 99.99    99.99 99    7.53  5    6.09  5    99.99 99    99.99 99
78632      6.37  1.58  1.35 99.99 99.99    99.99 99.99    10.29 12    7.02  4    5.86  4    99.99 99    99.99 99
78676      6.15 99.99  0.17 99.99 99.99    99.99 99.99    99.99 99    6.15  3    6.19  2    99.99 99    99.99 99
78702      5.73  0.01  0.00 99.99 99.99    99.99 99.99     6.78 14    5.64  3    5.84  2    99.99 99    99.99 99
78732      5.46  0.78  1.01 99.99 99.99    99.99 99.99     8.11  4    5.92  4    5.11  4    99.99 99    99.99 99
78791      4.48  0.22  0.61 99.99 99.99    99.99 99.99     6.15  3    4.72  3    4.31  3    99.99 99    99.99 99
78922      5.59 99.99  0.19 99.99 99.99    99.99 99.99    99.99 99    5.60  3    5.62  2    99.99 99    99.99 99
78955      6.53 99.99  0.00 99.99 99.99    99.99 99.99    99.99 99    6.44  3    6.64  2    99.99 99    99.99 99
79025      6.48 99.99  0.17 99.99 99.99    99.99 99.99    99.99 99    6.48  3    6.52  2    99.99 99    99.99 99
79066      6.35  0.01  0.32 99.99 99.99    99.99 99.99     7.57  6    6.43  3    6.32  2    99.99 99    99.99 99
79096      6.51 99.99  0.73 99.99 99.99    99.99 99.99    99.99 99    6.82  3    6.29  3    99.99 99    99.99 99
79108      6.14  0.01 -0.01 99.99 99.99    99.99 99.99     7.18 14    6.04  3    6.26  2    99.99 99    99.99 99
79181      5.73 99.99  0.98 99.99 99.99    99.99 99.99    99.99 99    6.17  4    5.39  4    99.99 99    99.99 99
79241      6.00 99.99 -0.11 99.99 99.99    99.99 99.99    99.99 99    5.84  2    6.17  2    99.99 99    99.99 99
79275      5.79 -0.83 -0.21 99.99 99.99    99.99 99.99     5.58  3    5.58  2    6.00  2    99.99 99    99.99 99
79447      3.97 -0.66 -0.18 99.99 99.99    99.99 99.99     4.01  4    3.78  2    4.17  2    99.99 99    99.99 99
79452      5.97  0.37  0.86 99.99 99.99    99.99 99.99     7.98  4    6.35  4    5.69  3    99.99 99    99.99 99
79523      6.31  0.01  0.00 99.99 99.99    99.99 99.99     7.36 14    6.22  3    6.42  2    99.99 99    99.99 99
```

```
79524      6.29 99.99  1.25 99.99 99.99    99.99 99.99    99.99 99    6.88  4    5.82  4    99.99 99    99.99 99
79621      5.92 -0.10 -0.05 99.99 99.99    99.99 99.99     6.79 13    5.80  2    6.06  2    99.99 99    99.99 99
79694      5.85 -0.48 -0.13 99.99 99.99    99.99 99.99     6.16  7    5.68  2    6.03  2    99.99 99    99.99 99
79698      5.54 99.99  0.85 99.99 99.99    99.99 99.99    99.99 99    5.91  4    5.26  3    99.99 99    99.99 99
79735      5.25 -0.56 -0.14 99.99 99.99    99.99 99.99     5.45  5    5.08  2    5.43  2    99.99 99    99.99 99
79752      6.35 99.99  0.02 99.99 99.99    99.99 99.99    99.99 99    6.27  3    6.46  2    99.99 99    99.99 99
79807      5.86 99.99  0.83 99.99 99.99    99.99 99.99    99.99 99    6.22  4    5.59  3    99.99 99    99.99 99
79837      5.42  0.07  0.31 99.99 99.99    99.99 99.99     6.72  7    5.50  3    5.39  2    99.99 99    99.99 99
79846      5.27 99.99  0.99 99.99 99.99    99.99 99.99    99.99 99    5.72  4    4.92  4    99.99 99    99.99 99
79900      6.25 -0.27 -0.08 99.99 99.99    99.99 99.99     6.88 10    6.11  2    6.40  2    99.99 99    99.99 99
79910      5.24  1.22  1.17 99.99 99.99    99.99 99.99     8.57  9    5.79  4    4.81  4    99.99 99    99.99 99
79931      5.47 -0.30 -0.09 99.99 99.99    99.99 99.99     6.05  9    5.33  2    5.63  2    99.99 99    99.99 99
80050      5.85 99.99  1.06 99.99 99.99    99.99 99.99    99.99 99    6.34  4    5.47  4    99.99 99    99.99 99
80057      6.04 -0.16  0.28 99.99 99.99    99.99 99.99     7.01  4    6.10  3    6.02  2    99.99 99    99.99 99
80094      6.02 99.99 -0.10 99.99 99.99    99.99 99.99    99.99 99    5.87  2    6.18  2    99.99 99    99.99 99
80108      5.12  1.96  1.66 99.99 99.99    99.99 99.99     9.72 13    5.94  5    4.46  5    99.99 99    99.99 99
80126      6.32 99.99  1.04 99.99 99.99    99.99 99.99    99.99 99    6.80  4    5.95  4    99.99 99    99.99 99
80170      5.33 99.99  1.17 99.99 99.99    99.99 99.99    99.99 99    5.88  4    4.90  4    99.99 99    99.99 99
80194      6.14  0.99  1.09 99.99 99.99    99.99 99.99     9.12  7    6.64  4    5.75  4    99.99 99    99.99 99
80230      4.34  1.98  1.63 99.99 99.99    99.99 99.99     8.95 14    5.14  5    3.70  5    99.99 99    99.99 99
80404      2.25  0.16  0.18 99.99 99.99    99.99 99.99     3.60 12    2.25  3    2.28  2    99.99 99    99.99 99
80435      6.33 99.99  1.41 99.99 99.99    99.99 99.99    99.99 99    7.01  4    5.79  4    99.99 99    99.99 99
80447      6.62 99.99  0.08 99.99 99.99    99.99 99.99    99.99 99    6.57  3    6.70  2    99.99 99    99.99 99
80456      5.26 99.99 -0.08 99.99 99.99    99.99 99.99    99.99 99    5.12  2    5.41  2    99.99 99    99.99 99
80479      5.78 99.99  1.29 99.99 99.99    99.99 99.99    99.99 99    6.39  4    5.30  4    99.99 99    99.99 99
80590      6.39 99.99 -0.10 99.99 99.99    99.99 99.99    99.99 99    6.24  2    6.55  2    99.99 99    99.99 99
80671      5.39 -0.03  0.42 99.99 99.99    99.99 99.99     6.62  3    5.53  3    5.31  3    99.99 99    99.99 99
80719      6.33  0.02  0.46 99.99 99.99    99.99 99.99     7.65  3    6.49  3    6.23  3    99.99 99    99.99 99
80773      6.82 99.99  0.00 99.99 99.99    99.99 99.99    99.99 99    6.73  3    6.93  2    99.99 99    99.99 99
80774      6.05 99.99  1.39 99.99 99.99    99.99 99.99    99.99 99    6.72  4    5.52  4    99.99 99    99.99 99
80781      6.28 99.99 -0.10 99.99 99.99    99.99 99.99    99.99 99    6.13  2    6.44  2    99.99 99    99.99 99
80950      5.87 -0.04 -0.02 99.99 99.99    99.99 99.99     6.84 13    5.76  2    5.99  2    99.99 99    99.99 99
80951      5.29  0.04  0.02 99.99 99.99    99.99 99.99     6.39 14    5.21  3    5.40  2    99.99 99    99.99 99
81034      5.58 99.99  1.63 99.99 99.99    99.99 99.99    99.99 99    6.38  5    4.94  5    99.99 99    99.99 99
81101      4.81  0.62  0.94 99.99 99.99    99.99 99.99     7.20  3    5.23  4    4.49  3    99.99 99    99.99 99
81134      6.54 99.99  1.13 99.99 99.99    99.99 99.99    99.99 99    7.07  4    6.13  4    99.99 99    99.99 99
81136      5.74  0.65  0.91 99.99 99.99    99.99 99.99     8.16  4    6.15  4    5.43  3    99.99 99    99.99 99
81157      5.64 99.99  0.19 99.99 99.99    99.99 99.99    99.99 99    5.65  3    5.67  2    99.99 99    99.99 99
81309      6.48 99.99  0.18 99.99 99.99    99.99 99.99    99.99 99    6.48  3    6.51  2    99.99 99    99.99 99
81361      6.29 99.99  0.97 99.99 99.99    99.99 99.99    99.99 99    6.73  4    5.95  4    99.99 99    99.99 99
81411      6.06 99.99  0.20 99.99 99.99    99.99 99.99    99.99 99    6.08  3    6.08  2    99.99 99    99.99 99
```

```
81420      5.59   1.82   1.52 99.99 99.99     99.99 99.99      9.93 13    6.33   5     5.00   5   99.99 99     99.99 99
81471      6.07  99.99   0.57 99.99 99.99     99.99 99.99     99.99 99    6.29   3     5.92   3   99.99 99     99.99 99
81502      6.30  99.99   1.48 99.99 99.99     99.99 99.99     99.99 99    7.02   4     5.73   5   99.99 99     99.99 99
81567      6.01   1.44   1.32 99.99 99.99     99.99 99.99      9.72 10    6.64   4     5.51   4   99.99 99     99.99 99
81613      5.99  99.99   1.06 99.99 99.99     99.99 99.99     99.99 99    6.48   4     5.61   4   99.99 99     99.99 99
81728      6.54   0.02   0.05 99.99 99.99     99.99 99.99      7.63 13    6.47   3     6.63   2   99.99 99     99.99 99
81753      6.10  -0.55  -0.10 99.99 99.99     99.99 99.99      6.33  5    5.95   2     6.26   2   99.99 99     99.99 99
81780      6.20  99.99   0.25 99.99 99.99     99.99 99.99     99.99 99    6.24   3     6.20   2   99.99 99     99.99 99
81809      5.38   0.13   0.64 99.99 99.99     99.99 99.99      6.94  3    5.64   3     5.20   3   99.99 99     99.99 99
81830      5.76  99.99   0.16 99.99 99.99     99.99 99.99     99.99 99    5.75   3     5.80   2   99.99 99     99.99 99
81848      5.11  99.99  -0.11 99.99 99.99     99.99 99.99     99.99 99    4.95   2     5.28   2   99.99 99     99.99 99
81858      5.41   0.14   0.60 99.99 99.99     99.99 99.99      6.96  3    5.65   3     5.25   3   99.99 99     99.99 99
81873      5.71   0.90   1.04 99.99 99.99     99.99 99.99      8.54  6    6.19   4     5.34   4   99.99 99     99.99 99
81919      6.65  99.99   0.20 99.99 99.99     99.99 99.99     99.99 99    6.67   3     6.67   2   99.99 99     99.99 99
81980      6.27   0.11   0.27 99.99 99.99     99.99 99.99      7.60  9    6.32   3     6.26   2   99.99 99     99.99 99
82043      6.14   0.16   0.22 99.99 99.99     99.99 99.99      7.51 11    6.17   3     6.15   2   99.99 99     99.99 99
82068      6.05   0.10   0.15 99.99 99.99     99.99 99.99      7.30 12    6.04   3     6.10   2   99.99 99     99.99 99
82074      6.26   0.40   0.84 99.99 99.99     99.99 99.99      8.30  3    6.63   4     5.98   3   99.99 99     99.99 99
82165      6.19  99.99   0.22 99.99 99.99     99.99 99.99     99.99 99    6.22   3     6.20   2   99.99 99     99.99 99
82180      6.24  99.99   1.57 99.99 99.99     99.99 99.99     99.99 99    7.01   5     5.62   5   99.99 99     99.99 99
82205      5.48  99.99   1.36 99.99 99.99     99.99 99.99     99.99 99    6.13   4     4.96   4   99.99 99     99.99 99
82232      5.85  99.99   1.20 99.99 99.99     99.99 99.99     99.99 99    6.42   4     5.41   4   99.99 99     99.99 99
82347      5.92  99.99   1.10 99.99 99.99     99.99 99.99     99.99 99    6.43   4     5.52   4   99.99 99     99.99 99
82350      5.47   0.98   1.08 99.99 99.99     99.99 99.99      8.43  7    5.97   4     5.08   4   99.99 99     99.99 99
82406      5.91  -0.02   0.01 99.99 99.99     99.99 99.99      6.92 13    5.82   3     6.02   2   99.99 99     99.99 99
82419      5.45  99.99  -0.10 99.99 99.99     99.99 99.99     99.99 99    5.30   2     5.61   2   99.99 99     99.99 99
82428      6.14   0.10   0.24 99.99 99.99     99.99 99.99      7.44  9    6.18   3     6.14   2   99.99 99     99.99 99
82477      6.13   1.17   1.19 99.99 99.99     99.99 99.99      9.40  8    6.69   4     5.69   4   99.99 99     99.99 99
82513      5.93  99.99   0.26 99.99 99.99     99.99 99.99     99.99 99    5.98   3     5.92   2   99.99 99     99.99 99
82514      5.87  99.99   1.29 99.99 99.99     99.99 99.99     99.99 99    6.48   4     5.39   4   99.99 99     99.99 99
82543      6.11   0.26   0.62 99.99 99.99     99.99 99.99      7.84  4    6.36   3     5.94   3   99.99 99     99.99 99
82554      5.37   0.04   0.44 99.99 99.99     99.99 99.99      6.70  4    5.52   3     5.28   3   99.99 99     99.99 99
82638      6.12   0.71   0.98 99.99 99.99     99.99 99.99      8.66  4    6.56   4     5.78   4   99.99 99     99.99 99
82660      5.94  99.99   1.50 99.99 99.99     99.99 99.99     99.99 99    6.67   5     5.36   5   99.99 99     99.99 99
82674      6.25   1.25   1.17 99.99 99.99     99.99 99.99      9.62 10    6.80   4     5.82   4   99.99 99     99.99 99
82694      5.35  99.99   0.90 99.99 99.99     99.99 99.99     99.99 99    5.75   4     5.05   3   99.99 99     99.99 99
82734      5.01  99.99   1.02 99.99 99.99     99.99 99.99     99.99 99    5.48   4     4.65   4   99.99 99     99.99 99
82747      5.91  99.99   0.02 99.99 99.99     99.99 99.99     99.99 99    5.83   3     6.02   2   99.99 99     99.99 99
82785      6.43  99.99   0.33 99.99 99.99     99.99 99.99     99.99 99    6.52   3     6.39   2   99.99 99     99.99 99
82858      6.27   1.38   1.35 99.99 99.99     99.99 99.99      9.92  8    6.92   4     5.76   4   99.99 99     99.99 99
82870      5.56   1.13   1.16 99.99 99.99     99.99 99.99      8.76  8    6.10   4     5.14   4   99.99 99     99.99 99
```

```
82984       5.11 99.99 -0.12 99.99 99.99      99.99 99.99     99.99 99    4.95  2    5.28  2    99.99 99    99.99 99
83058       5.01 99.99 -0.19 99.99 99.99      99.99 99.99     99.99 99    4.81  2    5.21  2    99.99 99    99.99 99
83095       5.47  1.74  1.56 99.99 99.99      99.99 99.99      9.72 10    6.23  5    4.86  5    99.99 99    99.99 99
83108       6.49 99.99  0.42 99.99 99.99      99.99 99.99     99.99 99    6.63  3    6.41  3    99.99 99    99.99 99
83183       4.08 -0.56  0.01 99.99 99.99      99.99 99.99      4.36  3    3.99  3    4.19  2    99.99 99    99.99 99
83240       5.00  0.87  1.05 99.99 99.99      99.99 99.99      7.79  5    5.48  4    4.63  4    99.99 99    99.99 99
83261       6.53 99.99  0.38 99.99 99.99      99.99 99.99     99.99 99    6.64  3    6.47  2    99.99 99    99.99 99
83332       5.70 99.99  1.12 99.99 99.99      99.99 99.99     99.99 99    6.22  4    5.29  4    99.99 99    99.99 99
83373       6.40 -0.09 -0.04 99.99 99.99      99.99 99.99      7.29 13    6.28  2    6.53  2    99.99 99    99.99 99
83380       5.63 99.99  1.02 99.99 99.99      99.99 99.99     99.99 99    6.10  4    5.27  4    99.99 99    99.99 99
83441       5.98 99.99  1.12 99.99 99.99      99.99 99.99     99.99 99    6.50  4    5.57  4    99.99 99    99.99 99
83465       6.19 99.99  1.05 99.99 99.99      99.99 99.99     99.99 99    6.67  4    5.82  4    99.99 99    99.99 99
83520       5.45 99.99  0.15 99.99 99.99      99.99 99.99     99.99 99    5.44  3    5.50  2    99.99 99    99.99 99
83523       6.56  0.11  0.10 99.99 99.99      99.99 99.99      7.80 13    6.52  3    6.63  2    99.99 99    99.99 99
83548       5.50  0.67  1.00 99.99 99.99      99.99 99.99      7.99  3    5.96  4    5.15  4    99.99 99    99.99 99
83610       6.70 99.99  0.49 99.99 99.99      99.99 99.99     99.99 99    6.87  3    6.59  3    99.99 99    99.99 99
83731       6.37 99.99  0.08 99.99 99.99      99.99 99.99     99.99 99    6.32  3    6.45  2    99.99 99    99.99 99
83944       4.52 -0.20 -0.07 99.99 99.99      99.99 99.99      5.25 11    4.39  2    4.67  2    99.99 99    99.99 99
84117       4.94 99.99  0.53 99.99 99.99      99.99 99.99     99.99 99    5.14  3    4.81  3    99.99 99    99.99 99
84121       5.32 99.99  0.20 99.99 99.99      99.99 99.99     99.99 99    5.34  3    5.34  2    99.99 99    99.99 99
84152       5.80 99.99  1.09 99.99 99.99      99.99 99.99     99.99 99    6.30  4    5.41  4    99.99 99    99.99 99
84224       6.41 -0.37 -0.06 99.99 99.99      99.99 99.99      6.91  7    6.28  2    6.55  2    99.99 99    99.99 99
84228       6.00 99.99 -0.13 99.99 99.99      99.99 99.99     99.99 99    5.83  2    6.18  2    99.99 99    99.99 99
84447       6.82 99.99  0.30 99.99 99.99      99.99 99.99     99.99 99    6.89  3    6.80  2    99.99 99    99.99 99
84461       5.56 99.99 -0.04 99.99 99.99      99.99 99.99     99.99 99    5.44  2    5.69  2    99.99 99    99.99 99
84542       5.79  1.95  1.64 99.99 99.99      99.99 99.99     10.37 13    6.60  5    5.14  5    99.99 99    99.99 99
84567       6.45 -0.94 -0.13 99.99 99.99      99.99 99.99      6.14  6    6.28  2    6.63  2    99.99 99    99.99 99
84607       5.65  0.12  0.34 99.99 99.99      99.99 99.99      7.03  7    5.74  3    5.61  2    99.99 99    99.99 99
84722       6.45 99.99  0.43 99.99 99.99      99.99 99.99     99.99 99    6.59  3    6.37  3    99.99 99    99.99 99
84816       5.55 99.99 -0.18 99.99 99.99      99.99 99.99     99.99 99    5.36  2    5.75  2    99.99 99    99.99 99
84850       6.22 99.99  0.46 99.99 99.99      99.99 99.99     99.99 99    6.38  3    6.12  3    99.99 99    99.99 99
85123       2.97  0.13  0.27 99.99 99.99      99.99 99.99      4.33  9    3.02  3    2.96  2    99.99 99    99.99 99
85206       5.97 99.99  1.24 99.99 99.99      99.99 99.99     99.99 99    6.56  4    5.51  4    99.99 99    99.99 99
85217       6.24 -0.02  0.48 99.99 99.99      99.99 99.99      7.51  3    6.41  3    6.13  3    99.99 99    99.99 99
85296       6.37 99.99  1.01 99.99 99.99      99.99 99.99     99.99 99    6.83  4    6.02  4    99.99 99    99.99 99
85355       5.08 99.99 -0.10 99.99 99.99      99.99 99.99     99.99 99    4.93  2    5.24  2    99.99 99    99.99 99
85364       6.01  0.11  0.17 99.99 99.99      99.99 99.99      7.29 11    6.01  3    6.05  2    99.99 99    99.99 99
85380       6.42  0.11  0.58 99.99 99.99      99.99 99.99      7.92  3    6.64  3    6.27  3    99.99 99    99.99 99
85396       5.45  0.57  0.89 99.99 99.99      99.99 99.99      7.75  3    5.84  4    5.15  3    99.99 99    99.99 99
85483       5.73 99.99  1.08 99.99 99.99      99.99 99.99     99.99 99    6.23  4    5.34  4    99.99 99    99.99 99
85504       6.02 -0.09 -0.04 99.99 99.99      99.99 99.99      6.91 13    5.90  2    6.15  2    99.99 99    99.99 99
```

```
85505      6.35  0.68   0.94 99.99 99.99      99.99 99.99     8.82  4    6.77  4    6.03  3    99.99 99    99.99 99
85558      5.05  0.06   0.04 99.99 99.99      99.99 99.99     6.19 14    4.98  3    5.15  2    99.99 99    99.99 99
85563      5.62 99.99   1.17 99.99 99.99      99.99 99.99    99.99 99    6.17  4    5.19  4    99.99 99    99.99 99
85655      5.79 99.99   1.36 99.99 99.99      99.99 99.99    99.99 99    6.44  4    5.27  4    99.99 99    99.99 99
85656      5.56 99.99   1.32 99.99 99.99      99.99 99.99    99.99 99    6.19  4    5.06  4    99.99 99    99.99 99
85709      5.95  1.93   1.66 99.99 99.99      99.99 99.99    10.51 12    6.77  5    5.29  5    99.99 99    99.99 99
85725      6.30 99.99   0.62 99.99 99.99      99.99 99.99    99.99 99    6.55  3    6.13  3    99.99 99    99.99 99
85859      4.88  1.30   1.23 99.99 99.99      99.99 99.99     8.35  9    5.46  4    4.42  4    99.99 99    99.99 99
85905      6.24 99.99   0.04 99.99 99.99      99.99 99.99    99.99 99    6.17  3    6.34  2    99.99 99    99.99 99
85980      5.71 99.99  -0.11 99.99 99.99      99.99 99.99    99.99 99    5.55  2    5.88  2    99.99 99    99.99 99
86080      5.85  1.00   1.13 99.99 99.99      99.99 99.99     8.86  6    6.38  4    5.44  4    99.99 99    99.99 99
86087      5.72 -0.01  -0.01 99.99 99.99      99.99 99.99     6.74 14    5.62  3    5.84  2    99.99 99    99.99 99
86146      5.14  0.00   0.46 99.99 99.99      99.99 99.99     6.43  3    5.30  3    5.04  3    99.99 99    99.99 99
86211      6.41 99.99   1.61 99.99 99.99      99.99 99.99    99.99 99    7.20  5    5.78  5    99.99 99    99.99 99
86266      6.28  0.11   0.22 99.99 99.99      99.99 99.99     7.59 10    6.31  3    6.29  2    99.99 99    99.99 99
86267      5.84 99.99   1.20 99.99 99.99      99.99 99.99    99.99 99    6.41  4    5.40  4    99.99 99    99.99 99
86352      6.37 99.99  -0.17 99.99 99.99      99.99 99.99    99.99 99    6.18  2    6.56  2    99.99 99    99.99 99
86369      6.04  1.55   1.36 99.99 99.99      99.99 99.99     9.92 11    6.69  4    5.52  4    99.99 99    99.99 99
86440      3.54 -0.63  -0.08 99.99 99.99      99.99 99.99     3.68  3    3.40  2    3.69  2    99.99 99    99.99 99
86523      6.05 99.99  -0.14 99.99 99.99      99.99 99.99    99.99 99    5.88  2    6.23  2    99.99 99    99.99 99
86606      6.35 -0.92  -0.08 99.99 99.99      99.99 99.99     6.09  6    6.21  2    6.50  2    99.99 99    99.99 99
86611      6.70  0.09   0.27 99.99 99.99      99.99 99.99     8.01  8    6.75  3    6.69  2    99.99 99    99.99 99
86629      5.23 99.99   0.30 99.99 99.99      99.99 99.99    99.99 99    5.30  3    5.21  2    99.99 99    99.99 99
86634      6.58  1.07   1.13 99.99 99.99      99.99 99.99     9.69  7    7.11  4    6.17  4    99.99 99    99.99 99
86728      5.35  0.28   0.66 99.99 99.99      99.99 99.99     7.13  3    5.62  3    5.16  3    99.99 99    99.99 99
87030      6.52 99.99   0.98 99.99 99.99      99.99 99.99    99.99 99    6.96  4    6.18  4    99.99 99    99.99 99
87152      6.20 -0.66  -0.13 99.99 99.99      99.99 99.99     6.27  3    6.03  2    6.38  2    99.99 99    99.99 99
87238      6.20 99.99   1.12 99.99 99.99      99.99 99.99    99.99 99    6.72  4    5.79  4    99.99 99    99.99 99
87262      6.12  2.01   1.66 99.99 99.99      99.99 99.99    10.79 14    6.94  5    5.46  5    99.99 99    99.99 99
87301      6.45 -0.02   0.40 99.99 99.99      99.99 99.99     7.68  3    6.58  3    6.38  3    99.99 99    99.99 99
87344      5.86 99.99  -0.06 99.99 99.99      99.99 99.99    99.99 99    5.73  2    6.00  2    99.99 99    99.99 99
87363      6.12 99.99   0.02 99.99 99.99      99.99 99.99    99.99 99    6.04  3    6.23  2    99.99 99    99.99 99
87477      6.43 99.99   1.30 99.99 99.99      99.99 99.99    99.99 99    7.05  4    5.94  4    99.99 99    99.99 99
87682      6.21  0.72   0.94 99.99 99.99      99.99 99.99     8.74  5    6.63  4    5.89  3    99.99 99    99.99 99
87783      5.08 99.99   0.88 99.99 99.99      99.99 99.99    99.99 99    5.47  4    4.79  3    99.99 99    99.99 99
87971      5.53  0.05   0.03 99.99 99.99      99.99 99.99     6.65 14    5.45  3    5.63  2    99.99 99    99.99 99
88013      6.36 99.99   0.98 99.99 99.99      99.99 99.99    99.99 99    6.80  4    6.02  4    99.99 99    99.99 99
88024      6.53 99.99   0.02 99.99 99.99      99.99 99.99    99.99 99    6.45  3    6.64  2    99.99 99    99.99 99
88182      6.24  0.15   0.18 99.99 99.99      99.99 99.99     7.58 12    6.24  3    6.27  2    99.99 99    99.99 99
88195      5.91 -0.05   0.02 99.99 99.99      99.99 99.99     6.89 12    5.83  3    6.02  2    99.99 99    99.99 99
88206      4.86 99.99  -0.12 99.99 99.99      99.99 99.99    99.99 99    4.70  2    5.03  2    99.99 99    99.99 99
```

```
88215      5.31  0.00   0.36 99.99 99.99     99.99 99.99     6.54  4   5.41  3   5.26  2   99.99 99   99.99 99
88218      6.13  0.16   0.60 99.99 99.99     99.99 99.99     7.71  3   6.37  3   5.97  3   99.99 99   99.99 99
88323      5.28  0.75   0.98 99.99 99.99     99.99 99.99     7.87  4   5.72  4   4.94  4   99.99 99   99.99 99
88333      5.65  1.42   1.30 99.99 99.99     99.99 99.99     9.32 10   6.27  4   5.16  4   99.99 99   99.99 99
88351      6.60  0.53   0.92 99.99 99.99     99.99 99.99     8.86  3   7.01  4   6.29  3   99.99 99   99.99 99
88372      6.25  0.01   0.01 99.99 99.99     99.99 99.99     7.30 13   6.16  3   6.36  2   99.99 99   99.99 99
88399      5.98 99.99   1.24 99.99 99.99     99.99 99.99    99.99 99   6.57  4   5.52  4   99.99 99   99.99 99
88473      5.81 -0.02   0.02 99.99 99.99     99.99 99.99     6.83 12   5.73  3   5.92  2   99.99 99   99.99 99
88522      6.28 99.99   0.01 99.99 99.99     99.99 99.99    99.99 99   6.19  3   6.39  2   99.99 99   99.99 99
88547      5.77  1.11   1.18 99.99 99.99     99.99 99.99     8.96  7   6.32  4   5.34  4   99.99 99   99.99 99
88693      6.16 99.99   1.17 99.99 99.99     99.99 99.99    99.99 99   6.71  4   5.73  4   99.99 99   99.99 99
88699      6.25 99.99   0.31 99.99 99.99     99.99 99.99    99.99 99   6.33  3   6.22  2   99.99 99   99.99 99
88742      6.38 99.99   0.60 99.99 99.99     99.99 99.99    99.99 99   6.62  3   6.22  3   99.99 99   99.99 99
88809      5.91 99.99   1.21 99.99 99.99     99.99 99.99    99.99 99   6.48  4   5.46  4   99.99 99   99.99 99
88825      6.10 99.99  -0.08 99.99 99.99     99.99 99.99    99.99 99   5.96  2   6.25  2   99.99 99   99.99 99
88836      6.35 99.99   0.94 99.99 99.99     99.99 99.99    99.99 99   6.77  4   6.03  3   99.99 99   99.99 99
88842      5.78 99.99   0.14 99.99 99.99     99.99 99.99    99.99 99   5.76  3   5.83  2   99.99 99   99.99 99
88907      6.39 99.99  -0.10 99.99 99.99     99.99 99.99    99.99 99   6.24  2   6.55  2   99.99 99   99.99 99
88981      5.16  0.18   0.22 99.99 99.99     99.99 99.99     6.56 12   5.19  3   5.17  2   99.99 99   99.99 99
89015      6.19 99.99   1.06 99.99 99.99     99.99 99.99    99.99 99   6.68  4   5.81  4   99.99 99   99.99 99
89062      5.60 99.99   1.52 99.99 99.99     99.99 99.99    99.99 99   6.34  5   5.01  5   99.99 99   99.99 99
89080      3.32 -0.33  -0.08 99.99 99.99     99.99 99.99     3.86  8   3.18  2   3.47  2   99.99 99   99.99 99
89104      6.17 99.99  -0.17 99.99 99.99     99.99 99.99    99.99 99   5.98  2   6.36  2   99.99 99   99.99 99
89169      6.57 99.99   0.48 99.99 99.99     99.99 99.99    99.99 99   6.74  3   6.46  3   99.99 99   99.99 99
89254      5.24  0.14   0.31 99.99 99.99     99.99 99.99     6.64  8   5.32  3   5.21  2   99.99 99   99.99 99
89263      6.22 99.99   0.20 99.99 99.99     99.99 99.99    99.99 99   6.24  3   6.24  2   99.99 99   99.99 99
89461      5.96 99.99  -0.06 99.99 99.99     99.99 99.99    99.99 99   5.83  2   6.10  2   99.99 99   99.99 99
89490      6.37  0.60   0.90 99.99 99.99     99.99 99.99     8.71  4   6.77  4   6.07  3   99.99 99   99.99 99
89565      6.32  0.03   0.33 99.99 99.99     99.99 99.99     7.58  6   6.41  3   6.28  2   99.99 99   99.99 99
89569      5.81 99.99   0.48 99.99 99.99     99.99 99.99    99.99 99   5.98  3   5.70  3   99.99 99   99.99 99
89715      5.67  0.05   0.05 99.99 99.99     99.99 99.99     6.80 13   5.60  3   5.76  2   99.99 99   99.99 99
89736      5.65 99.99   1.67 99.99 99.99     99.99 99.99    99.99 99   6.47  5   4.99  5   99.99 99   99.99 99
89747      6.51 99.99   0.40 99.99 99.99     99.99 99.99    99.99 99   6.64  3   6.44  3   99.99 99   99.99 99
89816      6.50 99.99   0.20 99.99 99.99     99.99 99.99    99.99 99   6.52  3   6.52  2   99.99 99   99.99 99
89962      6.07  1.13   1.12 99.99 99.99     99.99 99.99     9.25  9   6.59  4   5.66  4   99.99 99   99.99 99
89995      6.54 -0.04   0.46 99.99 99.99     99.99 99.99     7.77  3   6.70  3   6.44  3   99.99 99   99.99 99
90071      6.27  0.05   0.31 99.99 99.99     99.99 99.99     7.54  7   6.35  3   6.24  2   99.99 99   99.99 99
90089      5.26 -0.05   0.37 99.99 99.99     99.99 99.99     6.43  4   5.37  3   5.20  2   99.99 99   99.99 99
90125      6.32  0.74   1.00 99.99 99.99     99.99 99.99     8.91  4   6.78  4   5.97  4   99.99 99   99.99 99
90132      5.33 99.99   0.25 99.99 99.99     99.99 99.99    99.99 99   5.37  3   5.33  2   99.99 99   99.99 99
90170      6.27 99.99   0.88 99.99 99.99     99.99 99.99    99.99 99   6.66  4   5.98  3   99.99 99   99.99 99
```

```
90264      4.99 -0.51 -0.13 99.99 99.99     99.99 99.99      5.26  6    4.82  2    5.17  2    99.99 99    99.99 99
90289      6.35 99.99  1.52 99.99 99.99     99.99 99.99     99.99 99    7.09  5    5.76  5    99.99 99    99.99 99
90454      5.95 99.99  0.32 99.99 99.99     99.99 99.99     99.99 99    6.03  3    5.92  2    99.99 99    99.99 99
90508      6.45  0.05  0.60 99.99 99.99     99.99 99.99      7.88  4    6.69  3    6.29  3    99.99 99    99.99 99
90518      6.13 99.99  1.13 99.99 99.99     99.99 99.99     99.99 99    6.66  4    5.72  4    99.99 99    99.99 99
90589      4.00 -0.01  0.35 99.99 99.99     99.99 99.99      5.21  4    4.10  3    3.95  2    99.99 99    99.99 99
90630      6.19  0.11  0.07 99.99 99.99     99.99 99.99      7.41 14    6.13  3    6.27  2    99.99 99    99.99 99
90677      5.58 99.99  1.56 99.99 99.99     99.99 99.99     99.99 99    6.34  5    4.97  5    99.99 99    99.99 99
90763      6.05  0.04  0.05 99.99 99.99     99.99 99.99      7.17 13    5.98  3    6.14  2    99.99 99    99.99 99
90798      6.11 99.99  1.51 99.99 99.99     99.99 99.99     99.99 99    6.85  5    5.52  5    99.99 99    99.99 99
90853      3.82  0.24  0.31 99.99 99.99     99.99 99.99      5.35 11    3.90  3    3.79  2    99.99 99    99.99 99
90874      6.01  0.11  0.09 99.99 99.99     99.99 99.99      7.24 14    5.96  3    6.08  2    99.99 99    99.99 99
90882      5.21 -0.12 -0.06 99.99 99.99     99.99 99.99      6.05 12    5.08  2    5.35  2    99.99 99    99.99 99
90972      5.56 99.99 -0.04 99.99 99.99     99.99 99.99     99.99 99    5.44  2    5.69  2    99.99 99    99.99 99
91056      5.29 99.99  1.86 99.99 99.99     99.99 99.99     99.99 99    6.22  5    4.54  6    99.99 99    99.99 99
91106      6.20  1.18  1.38 99.99 99.99     99.99 99.99      9.59  4    6.86  4    5.67  4    99.99 99    99.99 99
91135      6.51 99.99  0.54 99.99 99.99     99.99 99.99     99.99 99    6.71  3    6.37  3    99.99 99    99.99 99
91280      6.05 99.99  0.51 99.99 99.99     99.99 99.99     99.99 99    6.24  3    5.93  3    99.99 99    99.99 99
91324      4.89 99.99  0.50 99.99 99.99     99.99 99.99     99.99 99    5.07  3    4.77  3    99.99 99    99.99 99
91355      5.16 99.99 -0.15 99.99 99.99     99.99 99.99     99.99 99    4.98  2    5.34  2    99.99 99    99.99 99
91375      4.74  0.07  0.04 99.99 99.99     99.99 99.99      5.89 14    4.67  3    4.84  2    99.99 99    99.99 99
91437      5.91 99.99  0.92 99.99 99.99     99.99 99.99     99.99 99    6.32  4    5.60  3    99.99 99    99.99 99
91496      4.93  1.82  1.68 99.99 99.99     99.99 99.99      9.36  9    5.76  5    4.26  5    99.99 99    99.99 99
91504      5.02 99.99  1.04 99.99 99.99     99.99 99.99     99.99 99    5.50  4    4.65  4    99.99 99    99.99 99
91533      6.00 99.99  0.31 99.99 99.99     99.99 99.99     99.99 99    6.08  3    5.97  2    99.99 99    99.99 99
91550      5.08 99.99  1.60 99.99 99.99     99.99 99.99     99.99 99    5.87  5    4.45  5    99.99 99    99.99 99
91612      5.08  0.64  0.93 99.99 99.99     99.99 99.99      7.49  4    5.50  4    4.76  3    99.99 99    99.99 99
91706      6.10 99.99  0.50 99.99 99.99     99.99 99.99     99.99 99    6.28  3    5.98  3    99.99 99    99.99 99
91767      6.23 99.99  1.40 99.99 99.99     99.99 99.99     99.99 99    6.91  4    5.69  4    99.99 99    99.99 99
91790      6.49 99.99  0.20 99.99 99.99     99.99 99.99     99.99 99    6.51  3    6.51  2    99.99 99    99.99 99
91805      6.08 99.99  0.94 99.99 99.99     99.99 99.99     99.99 99    6.50  4    5.76  3    99.99 99    99.99 99
91858      6.57 99.99  0.29 99.99 99.99     99.99 99.99     99.99 99    6.63  3    6.55  2    99.99 99    99.99 99
91880      6.04 99.99  1.64 99.99 99.99     99.99 99.99     99.99 99    6.85  5    5.39  5    99.99 99    99.99 99
91881      6.29 99.99  0.48 99.99 99.99     99.99 99.99     99.99 99    6.46  3    6.18  3    99.99 99    99.99 99
91889      5.72  0.00  0.52 99.99 99.99     99.99 99.99      7.04  3    5.91  3    5.59  3    99.99 99    99.99 99
91942      4.45  1.79  1.62 99.99 99.99     99.99 99.99      8.80 10    5.25  5    3.81  5    99.99 99    99.99 99
91992      6.52 99.99  0.29 99.99 99.99     99.99 99.99     99.99 99    6.58  3    6.50  2    99.99 99    99.99 99
92029      7.07 -0.51 -0.08 99.99 99.99     99.99 99.99      7.37  5    6.93  2    7.22  2    99.99 99    99.99 99
92036      4.89 99.99  1.62 99.99 99.99     99.99 99.99     99.99 99    5.69  5    4.25  5    99.99 99    99.99 99
92063      5.08 99.99  1.18 99.99 99.99     99.99 99.99     99.99 99    5.63  4    4.65  4    99.99 99    99.99 99
92095      5.52  1.34  1.27 99.99 99.99     99.99 99.99      9.07  9    6.12  4    5.04  4    99.99 99    99.99 99
```

```
92209      6.30  1.27  1.20 99.99 99.99     99.99 99.99      9.72 10    6.87  4    5.86  4    99.99 99    99.99 99
92214      4.91 99.99  0.92 99.99 99.99     99.99 99.99     99.99 99    5.32  4    4.60  3    99.99 99    99.99 99
92245      6.04 99.99  0.00 99.99 99.99     99.99 99.99     99.99 99    5.95  3    6.15  2    99.99 99    99.99 99
92305      4.11  1.94  1.58 99.99 99.99     99.99 99.99      8.64 14    4.88  5    3.49  5    99.99 99    99.99 99
92328      6.12 99.99  0.66 99.99 99.99     99.99 99.99     99.99 99    6.39  3    5.93  3    99.99 99    99.99 99
92397      4.66  0.97  1.48 99.99 99.99     99.99 99.99      7.82  6    5.38  4    4.09  5    99.99 99    99.99 99
92436      5.87 99.99  1.44 99.99 99.99     99.99 99.99     99.99 99    6.57  4    5.32  5    99.99 99    99.99 99
92449      4.28  0.75  1.04 99.99 99.99     99.99 99.99      6.90  3    4.76  4    3.91  4    99.99 99    99.99 99
92588      6.26  0.59  0.88 99.99 99.99     99.99 99.99      8.58  4    6.65  4    5.97  3    99.99 99    99.99 99
92589      6.37 99.99  0.92 99.99 99.99     99.99 99.99     99.99 99    6.78  4    6.06  3    99.99 99    99.99 99
92682      6.07  1.91  1.71 99.99 99.99     99.99 99.99     10.63 10    6.92  5    5.39  5    99.99 99    99.99 99
92787      5.18  0.01  0.33 99.99 99.99     99.99 99.99      6.41  5    5.27  3    5.14  2    99.99 99    99.99 99
92841      5.79  1.09  1.17 99.99 99.99     99.99 99.99      8.95  7    6.34  4    5.36  4    99.99 99    99.99 99
92845      5.64 99.99  0.00 99.99 99.99     99.99 99.99     99.99 99    5.55  3    5.75  2    99.99 99    99.99 99
93102      6.28  1.28  1.21 99.99 99.99     99.99 99.99      9.71 10    6.85  4    5.83  4    99.99 99    99.99 99
93132      6.34  1.90  1.56 99.99 99.99     99.99 99.99     10.81 14    7.10  5    5.73  5    99.99 99    99.99 99
93244      6.37  1.11  1.12 99.99 99.99     99.99 99.99      9.53  8    6.89  4    5.96  4    99.99 99    99.99 99
93344      6.26  0.17  0.20 99.99 99.99     99.99 99.99      7.64 12    6.28  3    6.28  2    99.99 99    99.99 99
93359      6.46  0.14  0.23 99.99 99.99     99.99 99.99      7.81 10    6.49  3    6.47  2    99.99 99    99.99 99
93372      6.27  0.01  0.49 99.99 99.99     99.99 99.99      7.59  3    6.44  3    6.16  3    99.99 99    99.99 99
93502      6.25 99.99  0.04 99.99 99.99     99.99 99.99     99.99 99    6.18  3    6.35  2    99.99 99    99.99 99
93526      6.67 99.99 -0.01 99.99 99.99     99.99 99.99     99.99 99    6.57  3    6.79  2    99.99 99    99.99 99
93563      5.26 99.99 -0.08 99.99 99.99     99.99 99.99     99.99 99    5.12  2    5.41  2    99.99 99    99.99 99
93655      5.93  1.93  1.60 99.99 99.99     99.99 99.99     10.46 13    6.72  5    5.30  5    99.99 99    99.99 99
93657      5.88  0.04  0.03 99.99 99.99     99.99 99.99      6.99 14    5.80  3    5.98  2    99.99 99    99.99 99
93662      6.35 99.99  1.63 99.99 99.99     99.99 99.99     99.99 99    7.15  5    5.71  5    99.99 99    99.99 99
93742      6.61  0.08  0.22 99.99 99.99     99.99 99.99      7.87  9    6.64  3    6.62  2    99.99 99    99.99 99
93779      5.47  0.74  0.95 99.99 99.99     99.99 99.99      8.03  5    5.90  4    5.14  4    99.99 99    99.99 99
93833      5.86  0.91  1.07 99.99 99.99     99.99 99.99      8.72  5    6.35  4    5.48  4    99.99 99    99.99 99
93845      4.45 -0.70 -0.19 99.99 99.99     99.99 99.99      4.43  4    4.25  2    4.65  2    99.99 99    99.99 99
93903      5.79  0.13  0.16 99.99 99.99     99.99 99.99      7.09 12    5.78  3    5.83  2    99.99 99    99.99 99
93905      5.61  0.05  0.04 99.99 99.99     99.99 99.99      6.73 14    5.54  3    5.71  2    99.99 99    99.99 99
93943      5.85 99.99  0.01 99.99 99.99     99.99 99.99     99.99 99    5.76  3    5.96  2    99.99 99    99.99 99
94014      5.95  1.83  1.48 99.99 99.99     99.99 99.99     10.28 14    6.67  4    5.38  5    99.99 99    99.99 99
94084      6.44  1.01  1.11 99.99 99.99     99.99 99.99      9.45  6    6.96  4    6.04  4    99.99 99    99.99 99
94180      6.38  0.13  0.08 99.99 99.99     99.99 99.99      7.64 14    6.33  3    6.46  2    99.99 99    99.99 99
94237      6.31  1.82  1.50 99.99 99.99     99.99 99.99     10.64 14    7.04  5    5.73  5    99.99 99    99.99 99
94386      6.38 99.99  1.16 99.99 99.99     99.99 99.99     99.99 99    6.92  4    5.96  4    99.99 99    99.99 99
94388      5.24  0.06  0.46 99.99 99.99     99.99 99.99      6.61  4    5.40  3    5.14  3    99.99 99    99.99 99
94402      5.45  0.77  0.96 99.99 99.99     99.99 99.99      8.06  5    5.88  4    5.12  4    99.99 99    99.99 99
94481      5.66 99.99  0.83 99.99 99.99     99.99 99.99     99.99 99    6.02  4    5.39  3    99.99 99    99.99 99
```

```
94510      3.78  0.65  0.95 99.99 99.99    99.99 99.99     6.22  4   4.21  4   3.45  4   99.99 99   99.99 99
94650      5.99 -0.46 -0.02 99.99 99.99    99.99 99.99     6.39  4   5.88  2   6.11  2   99.99 99   99.99 99
94669      6.03  1.08  1.13 99.99 99.99    99.99 99.99     9.15  7   6.56  4   5.62  4   99.99 99   99.99 99
94672      5.91 -0.02  0.42 99.99 99.99    99.99 99.99     7.15  3   6.05  3   5.83  3   99.99 99   99.99 99
94683      5.93 99.99  1.75 99.99 99.99    99.99 99.99    99.99 99   6.80  5   5.23  5   99.99 99   99.99 99
94717      6.33  1.57  1.46 99.99 99.99    99.99 99.99    10.30  9   7.04  4   5.77  5   99.99 99   99.99 99
95128      5.05  0.13  0.61 99.99 99.99    99.99 99.99     6.60  3   5.29  3   4.88  3   99.99 99   99.99 99
95208      6.13  1.63  1.53 99.99 99.99    99.99 99.99    10.22  9   6.88  5   5.53  5   99.99 99   99.99 99
95221      5.71 99.99  0.37 99.99 99.99    99.99 99.99    99.99 99   5.82  3   5.65  2   99.99 99   99.99 99
95234      5.89 99.99  1.63 99.99 99.99    99.99 99.99    99.99 99   6.69  5   5.25  5   99.99 99   99.99 99
95310      5.08  0.17  0.24 99.99 99.99    99.99 99.99     6.48 11   5.12  3   5.08  2   99.99 99   99.99 99
95314      5.88 99.99  1.50 99.99 99.99    99.99 99.99    99.99 99   6.61  5   5.30  5   99.99 99   99.99 99
95324      6.15 99.99 -0.06 99.99 99.99    99.99 99.99    99.99 99   6.02  2   6.29  2   99.99 99   99.99 99
95345      4.84  1.12  1.16 99.99 99.99    99.99 99.99     8.03  7   5.38  4   4.42  4   99.99 99   99.99 99
95347      5.81 99.99 -0.08 99.99 99.99    99.99 99.99    99.99 99   5.67  2   5.96  2   99.99 99   99.99 99
95382      4.99  0.11  0.16 99.99 99.99    99.99 99.99     6.26 12   4.98  3   5.03  2   99.99 99   99.99 99
95429      6.15 99.99  0.18 99.99 99.99    99.99 99.99    99.99 99   6.15  3   6.18  2   99.99 99   99.99 99
95456      6.07 99.99  0.52 99.99 99.99    99.99 99.99    99.99 99   6.26  3   5.94  3   99.99 99   99.99 99
95698      6.23 99.99  0.31 99.99 99.99    99.99 99.99    99.99 99   6.31  3   6.20  2   99.99 99   99.99 99
95771      6.14  0.06  0.26 99.99 99.99    99.99 99.99     7.40  8   6.19  3   6.13  2   99.99 99   99.99 99
95788      6.71  0.04  0.55 99.99 99.99    99.99 99.99     8.10  3   6.92  3   6.57  3   99.99 99   99.99 99
95849      5.95  1.32  1.22 99.99 99.99    99.99 99.99     9.44 10   6.53  4   5.50  4   99.99 99   99.99 99
95870      6.35 99.99  0.88 99.99 99.99    99.99 99.99    99.99 99   6.74  4   6.06  3   99.99 99   99.99 99
96113      5.67 99.99  0.24 99.99 99.99    99.99 99.99    99.99 99   5.71  3   5.67  2   99.99 99   99.99 99
96124      6.19  0.12  0.11 99.99 99.99    99.99 99.99     7.45 13   6.16  3   6.25  2   99.99 99   99.99 99
96146      5.43 99.99  0.03 99.99 99.99    99.99 99.99    99.99 99   5.35  3   5.53  2   99.99 99   99.99 99
96202      4.94  0.04  0.36 99.99 99.99    99.99 99.99     6.23  5   5.04  3   4.89  2   99.99 99   99.99 99
96220      6.09 99.99  0.30 99.99 99.99    99.99 99.99    99.99 99   6.16  3   6.07  2   99.99 99   99.99 99
96224      6.13 99.99 -0.02 99.99 99.99    99.99 99.99    99.99 99   6.02  2   6.25  2   99.99 99   99.99 99
96407      6.30 99.99  0.95 99.99 99.99    99.99 99.99    99.99 99   6.73  4   5.97  4   99.99 99   99.99 99
96436      5.52  0.66  0.96 99.99 99.99    99.99 99.99     7.98  4   5.95  4   5.19  4   99.99 99   99.99 99
96441      6.77 99.99  0.04 99.99 99.99    99.99 99.99    99.99 99   6.70  3   6.87  2   99.99 99   99.99 99
96484      6.32 99.99  1.16 99.99 99.99    99.99 99.99    99.99 99   6.86  4   5.90  4   99.99 99   99.99 99
96557      6.59 99.99  0.34 99.99 99.99    99.99 99.99    99.99 99   6.68  3   6.55  2   99.99 99   99.99 99
96566      4.61  0.81  1.03 99.99 99.99    99.99 99.99     7.31  4   5.08  4   4.25  4   99.99 99   99.99 99
96568      6.41  0.13  0.12 99.99 99.99    99.99 99.99     7.69 13   6.38  3   6.47  2   99.99 99   99.99 99
96700      6.54 99.99  0.60 99.99 99.99    99.99 99.99    99.99 99   6.78  3   6.38  3   99.99 99   99.99 99
96706      5.57 -0.63 -0.05 99.99 99.99    99.99 99.99     5.72  3   5.45  2   5.71  2   99.99 99   99.99 99
96723      6.49  0.05  0.03 99.99 99.99    99.99 99.99     7.61 14   6.41  3   6.59  2   99.99 99   99.99 99
96819      5.44  0.06  0.07 99.99 99.99    99.99 99.99     6.59 13   5.38  3   5.52  2   99.99 99   99.99 99
96834      5.89  1.90  1.57 99.99 99.99    99.99 99.99    10.36 14   6.66  5   5.27  5   99.99 99   99.99 99
```

```
 97023        5.81 99.99  0.04 99.99 99.99       99.99 99.99      99.99 99    5.74  3    5.91  2    99.99 99    99.99 99
 97271        6.88 99.99 -0.08 99.99 99.99       99.99 99.99      99.99 99    6.74  2    7.03  2    99.99 99    99.99 99
 97411        6.13 99.99 -0.01 99.99 99.99       99.99 99.99      99.99 99    6.03  3    6.25  2    99.99 99    99.99 99
 97472        6.35  1.54  1.37 99.99 99.99       99.99 99.99      10.23 11    7.01  4    5.83  4    99.99 99    99.99 99
 97495        5.36 99.99  0.18 99.99 99.99       99.99 99.99      99.99 99    5.36  3    5.39  2    99.99 99    99.99 99
 97501        6.33  1.12  1.15 99.99 99.99       99.99 99.99       9.52  8    6.87  4    5.91  4    99.99 99    99.99 99
 97550        6.11 99.99  1.06 99.99 99.99       99.99 99.99      99.99 99    6.60  4    5.73  4    99.99 99    99.99 99
 97576        5.80 99.99  1.66 99.99 99.99       99.99 99.99      99.99 99    6.62  5    5.14  5    99.99 99    99.99 99
 97585        5.42 -0.05 -0.03 99.99 99.99       99.99 99.99       6.37 13    5.31  2    5.55  2    99.99 99    99.99 99
 97605        5.79  1.13  1.12 99.99 99.99       99.99 99.99       8.97  9    6.31  4    5.38  4    99.99 99    99.99 99
 97651        5.76 99.99  1.32 99.99 99.99       99.99 99.99      99.99 99    6.39  4    5.26  4    99.99 99    99.99 99
 97670        5.74 99.99 -0.10 99.99 99.99       99.99 99.99      99.99 99    5.59  2    5.90  2    99.99 99    99.99 99
 98096        6.31 99.99  0.40 99.99 99.99       99.99 99.99      99.99 99    6.44  3    6.24  3    99.99 99    99.99 99
 98161        6.27 99.99  0.10 99.99 99.99       99.99 99.99      99.99 99    6.23  3    6.34  2    99.99 99    99.99 99
 98221        6.45 99.99  0.40 99.99 99.99       99.99 99.99      99.99 99    6.58  3    6.38  3    99.99 99    99.99 99
 98366        5.91  0.87  1.04 99.99 99.99       99.99 99.99       8.70  5    6.39  4    5.54  4    99.99 99    99.99 99
 98560        5.99  0.00  0.46 99.99 99.99       99.99 99.99       7.28  3    6.15  3    5.89  3    99.99 99    99.99 99
 98617        6.35  0.08  0.26 99.99 99.99       99.99 99.99       7.64  8    6.40  3    6.34  2    99.99 99    99.99 99
 98672        6.27 -0.12 -0.03 99.99 99.99       99.99 99.99       7.13 12    6.16  2    6.40  2    99.99 99    99.99 99
 98695        6.41 -0.48  0.05 99.99 99.99       99.99 99.99       6.82  3    6.34  3    6.50  2    99.99 99    99.99 99
 98718        3.89 -0.59 -0.15 99.99 99.99       99.99 99.99       4.04  5    3.71  2    4.07  2    99.99 99    99.99 99
 98960        6.05  1.78  1.46 99.99 99.99       99.99 99.99      10.30 14    6.76  4    5.49  5    99.99 99    99.99 99
 98991        5.09 -0.06  0.42 99.99 99.99       99.99 99.99       6.27  3    5.23  3    5.01  3    99.99 99    99.99 99
 99015        6.43  0.11  0.20 99.99 99.99       99.99 99.99       7.72 11    6.45  3    6.45  2    99.99 99    99.99 99
 99022        5.79 99.99  0.01 99.99 99.99       99.99 99.99      99.99 99    5.70  3    5.90  2    99.99 99    99.99 99
 99055        5.39  0.66  0.94 99.99 99.99       99.99 99.99       7.84  4    5.81  4    5.07  3    99.99 99    99.99 99
 99104        5.11 -0.42 -0.08 99.99 99.99       99.99 99.99       5.53  7    4.97  2    5.26  2    99.99 99    99.99 99
 99167        4.83 99.99  1.56 99.99 99.99       99.99 99.99      99.99 99    5.59  5    4.22  5    99.99 99    99.99 99
 99171        6.12 -0.81 -0.18 99.99 99.99       99.99 99.99       5.96  3    5.93  2    6.32  2    99.99 99    99.99 99
 99196        5.79  1.57  1.38 99.99 99.99       99.99 99.99       9.71 11    6.45  4    5.26  4    99.99 99    99.99 99
 99264        5.59 -0.57  0.06 99.99 99.99       99.99 99.99       5.89  3    5.53  3    5.68  2    99.99 99    99.99 99
 99453        5.17 99.99  0.50 99.99 99.99       99.99 99.99      99.99 99    5.35  3    5.05  3    99.99 99    99.99 99
 99491        6.16  0.62  0.85 99.99 99.99       99.99 99.99       8.50  5    6.53  4    5.88  3    99.99 99    99.99 99
 99564        5.94 99.99  0.49 99.99 99.99       99.99 99.99      99.99 99    6.11  3    5.83  3    99.99 99    99.99 99
 99575        5.81 99.99  0.52 99.99 99.99       99.99 99.99      99.99 99    6.00  3    5.68  3    99.99 99    99.99 99
 99651        6.25  0.73  1.04 99.99 99.99       99.99 99.99       8.85  3    6.73  4    5.88  4    99.99 99    99.99 99
 99803        5.08 99.99 -0.03 99.99 99.99       99.99 99.99      99.99 99    4.97  2    5.21  2    99.99 99    99.99 99
 99872        6.09 -0.42  0.16 99.99 99.99       99.99 99.99       6.64  3    6.08  3    6.13  2    99.99 99    99.99 99
 99922        5.76 99.99  0.07 99.99 99.99       99.99 99.99      99.99 99    5.70  3    5.84  2    99.99 99    99.99 99
 99998        4.77  1.82  1.54 99.99 99.99       99.99 99.99       9.12 13    5.52  5    4.17  5    99.99 99    99.99 99
100203        5.48 -0.01  0.50 99.99 99.99       99.99 99.99       6.78  3    5.66  3    5.36  3    99.99 99    99.99 99
```

```
100219      6.24 99.99  0.54 99.99 99.99   99.99 99.99   99.99 99    6.44  3    6.10  3   99.99 99   99.99 99
100343      5.95  1.64  1.38 99.99 99.99   99.99 99.99    9.97 13    6.61  4    5.42  4   99.99 99   99.99 99
100378      5.64 99.99  1.58 99.99 99.99   99.99 99.99   99.99 99    6.41  5    5.02  5   99.99 99   99.99 99
100382      5.90  1.11  1.13 99.99 99.99   99.99 99.99    9.06  8    6.43  4    5.49  4   99.99 99   99.99 99
100393      5.04 99.99  1.58 99.99 99.99   99.99 99.99   99.99 99    5.81  5    4.42  5   99.99 99   99.99 99
100470      6.40  0.93  1.05 99.99 99.99   99.99 99.99    9.27  6    6.88  4    6.03  4   99.99 99   99.99 99
100493      5.39 99.99  0.12 99.99 99.99   99.99 99.99   99.99 99    5.36  3    5.45  2   99.99 99   99.99 99
100563      5.77  0.01  0.46 99.99 99.99   99.99 99.99    7.07  3    5.93  3    5.67  3   99.99 99   99.99 99
100623      5.98  0.39  0.81 99.99 99.99   99.99 99.99    7.99  3    6.33  4    5.72  3   99.99 99   99.99 99
100673      4.62 -0.21 -0.08 99.99 99.99   99.99 99.99    5.33 11    4.48  2    4.77  2   99.99 99   99.99 99
100708      5.50 99.99  1.04 99.99 99.99   99.99 99.99   99.99 99    5.98  4    5.13  4   99.99 99   99.99 99
100825      5.25  0.12  0.25 99.99 99.99   99.99 99.99    6.59 10    5.29  3    5.25  2   99.99 99   99.99 99
100841      3.13 -0.19 -0.04 99.99 99.99   99.99 99.99    3.89 10    3.01  2    3.26  2   99.99 99   99.99 99
100911      6.31 99.99  0.06 99.99 99.99   99.99 99.99   99.99 99    6.25  3    6.40  2   99.99 99   99.99 99
100929      5.82 99.99 -0.08 99.99 99.99   99.99 99.99   99.99 99    5.68  2    5.97  2   99.99 99   99.99 99
100953      6.29 99.99  0.46 99.99 99.99   99.99 99.99   99.99 99    6.45  3    6.19  3   99.99 99   99.99 99
101067      5.44 99.99  1.24 99.99 99.99   99.99 99.99   99.99 99    6.03  4    4.98  4   99.99 99   99.99 99
101112      6.17  0.95  1.08 99.99 99.99   99.99 99.99    9.09  6    6.67  4    5.78  4   99.99 99   99.99 99
101132      5.65 -0.01  0.35 99.99 99.99   99.99 99.99    6.86  4    5.75  3    5.60  2   99.99 99   99.99 99
101154      6.22  1.07  1.12 99.99 99.99   99.99 99.99    9.32  7    6.74  4    5.81  4   99.99 99   99.99 99
101162      5.96  0.83  1.02 99.99 99.99   99.99 99.99    8.68  5    6.43  4    5.60  4   99.99 99   99.99 99
101259      6.42 99.99  0.82 99.99 99.99   99.99 99.99   99.99 99    6.78  4    6.15  3   99.99 99   99.99 99
101369      6.21 99.99  0.00 99.99 99.99   99.99 99.99   99.99 99    6.12  3    6.32  2   99.99 99   99.99 99
101563      6.44 99.99  0.66 99.99 99.99   99.99 99.99   99.99 99    6.71  3    6.25  3   99.99 99   99.99 99
101570      4.94  0.81  1.15 99.99 99.99   99.99 99.99    7.71  3    5.48  4    4.52  4   99.99 99   99.99 99
101666      5.22 99.99  1.48 99.99 99.99   99.99 99.99   99.99 99    5.94  4    4.65  5   99.99 99   99.99 99
101782      6.33  0.88  1.08 99.99 99.99   99.99 99.99    9.15  5    6.83  4    5.94  4   99.99 99   99.99 99
101883      5.98 99.99  1.46 99.99 99.99   99.99 99.99   99.99 99    6.69  4    5.42  5   99.99 99   99.99 99
101917      6.39  0.59  0.90 99.99 99.99   99.99 99.99    8.72  4    6.79  4    6.09  3   99.99 99   99.99 99
101933      6.07  0.71  0.96 99.99 99.99   99.99 99.99    8.60  4    6.50  4    5.74  4   99.99 99   99.99 99
102124      4.85  0.10  0.18 99.99 99.99   99.99 99.99    6.12 11    4.85  3    4.88  2   99.99 99   99.99 99
102150      6.26 99.99  1.17 99.99 99.99   99.99 99.99   99.99 99    6.81  4    5.83  4   99.99 99   99.99 99
102232      5.29 -0.55 -0.12 99.99 99.99   99.99 99.99    5.51  5    5.13  2    5.46  2   99.99 99   99.99 99
102249      3.64  0.15  0.16 99.99 99.99   99.99 99.99    4.97 13    3.63  3    3.68  2   99.99 99   99.99 99
102350      4.11  0.58  0.90 99.99 99.99   99.99 99.99    6.43  3    4.51  4    3.81  3   99.99 99   99.99 99
102365      4.91 99.99  0.66 99.99 99.99   99.99 99.99   99.99 99    5.18  3    4.72  3   99.99 99   99.99 99
102397      6.17 99.99  0.96 99.99 99.99   99.99 99.99   99.99 99    6.60  4    5.84  4   99.99 99   99.99 99
102438      6.48  0.24  0.68 99.99 99.99   99.99 99.99    8.21  3    6.76  3    6.28  3   99.99 99   99.99 99
102510      5.32  0.04  0.02 99.99 99.99   99.99 99.99    6.42 14    5.24  3    5.43  2   99.99 99   99.99 99
102574      6.26  0.13  0.58 99.99 99.99   99.99 99.99    7.79  3    6.48  3    6.11  3   99.99 99   99.99 99
102634      6.15  0.08  0.52 99.99 99.99   99.99 99.99    7.58  3    6.34  3    6.02  3   99.99 99   99.99 99
```

```
102776      4.32 -0.62 -0.15 99.99 99.99    99.99 99.99     4.43  4    4.14  2    4.50  2    99.99 99    99.99 99
102839      4.97  1.22  1.40 99.99 99.99    99.99 99.99     8.43  4    5.65  4    4.43  4    99.99 99    99.99 99
102878      5.70 99.99  0.26 99.99 99.99    99.99 99.99    99.99 99    5.75  3    5.69  2    99.99 99    99.99 99
102928      5.64  0.88  1.06 99.99 99.99    99.99 99.99     8.45  5    6.13  4    5.26  4    99.99 99    99.99 99
102990      6.35 99.99  0.40 99.99 99.99    99.99 99.99    99.99 99    6.48  3    6.28  3    99.99 99    99.99 99
103026      5.85 99.99  0.56 99.99 99.99    99.99 99.99    99.99 99    6.06  3    5.70  3    99.99 99    99.99 99
103079      4.90 -0.53 -0.11 99.99 99.99    99.99 99.99     5.16  5    4.74  2    5.07  2    99.99 99    99.99 99
103101      5.57 99.99  0.07 99.99 99.99    99.99 99.99    99.99 99    5.51  3    5.65  2    99.99 99    99.99 99
103266      6.17 99.99  0.08 99.99 99.99    99.99 99.99    99.99 99    6.12  3    6.25  2    99.99 99    99.99 99
103400      6.06 99.99  0.05 99.99 99.99    99.99 99.99    99.99 99    5.99  3    6.15  2    99.99 99    99.99 99
103437      6.46 99.99  0.52 99.99 99.99    99.99 99.99    99.99 99    6.65  3    6.33  3    99.99 99    99.99 99
103484      5.58  0.67  0.94 99.99 99.99    99.99 99.99     8.04  4    6.00  4    5.26  3    99.99 99    99.99 99
103596      5.93 99.99  1.50 99.99 99.99    99.99 99.99    99.99 99    6.66  5    5.35  5    99.99 99    99.99 99
103632      5.18 99.99 -0.03 99.99 99.99    99.99 99.99    99.99 99    5.07  2    5.31  2    99.99 99    99.99 99
103637      6.13 99.99  1.02 99.99 99.99    99.99 99.99    99.99 99    6.60  4    5.77  4    99.99 99    99.99 99
103746      6.26  0.04  0.40 99.99 99.99    99.99 99.99     7.57  4    6.39  3    6.19  3    99.99 99    99.99 99
103884      5.57 99.99 -0.15 99.99 99.99    99.99 99.99    99.99 99    5.39  2    5.75  2    99.99 99    99.99 99
103961      5.44 99.99 -0.08 99.99 99.99    99.99 99.99    99.99 99    5.30  2    5.59  2    99.99 99    99.99 99
103974      6.79 99.99  0.97 99.99 99.99    99.99 99.99    99.99 99    7.23  4    6.45  4    99.99 99    99.99 99
104035      5.61 -0.12  0.19 99.99 99.99    99.99 99.99     6.59  6    5.62  3    5.64  2    99.99 99    99.99 99
104039      6.43 99.99  0.03 99.99 99.99    99.99 99.99    99.99 99    6.35  3    6.53  2    99.99 99    99.99 99
104055      6.17  1.42  1.26 99.99 99.99    99.99 99.99     9.82 11    6.77  4    5.70  4    99.99 99    99.99 99
104081      6.05 99.99  1.28 99.99 99.99    99.99 99.99    99.99 99    6.66  4    5.57  4    99.99 99    99.99 99
104174      4.91 -0.16 -0.06 99.99 99.99    99.99 99.99     5.70 12    4.78  2    5.05  2    99.99 99    99.99 99
104181      5.37  0.00  0.00 99.99 99.99    99.99 99.99     6.40 13    5.28  3    5.48  2    99.99 99    99.99 99
104304      5.54  0.43  0.76 99.99 99.99    99.99 99.99     7.58  4    5.86  3    5.30  3    99.99 99    99.99 99
104356      6.31  1.06  1.21 99.99 99.99    99.99 99.99     9.45  5    6.88  4    5.86  4    99.99 99    99.99 99
104430      6.16 99.99  0.00 99.99 99.99    99.99 99.99    99.99 99    6.07  3    6.27  2    99.99 99    99.99 99
104555      6.05  1.54  1.29 99.99 99.99    99.99 99.99     9.88 13    6.66  4    5.57  4    99.99 99    99.99 99
104570      6.42  1.14  1.17 99.99 99.99    99.99 99.99     9.64  8    6.97  4    5.99  4    99.99 99    99.99 99
104600      5.89 -0.28 -0.08 99.99 99.99    99.99 99.99     6.50  9    5.75  2    6.04  2    99.99 99    99.99 99
104625      6.22  1.83  1.49 99.99 99.99    99.99 99.99    10.55 14    6.94  4    5.64  5    99.99 99    99.99 99
104671      4.33  0.04  0.27 99.99 99.99    99.99 99.99     5.57  7    4.38  3    4.32  2    99.99 99    99.99 99
104731      5.15 -0.04  0.42 99.99 99.99    99.99 99.99     6.36  3    5.29  3    5.07  3    99.99 99    99.99 99
104752      6.44  0.97  1.23 99.99 99.99    99.99 99.99     9.47  3    7.02  4    5.98  4    99.99 99    99.99 99
104878      5.35 -0.17 -0.01 99.99 99.99    99.99 99.99     6.15 10    5.25  3    5.47  2    99.99 99    99.99 99
104902      5.04  1.78  1.49 99.99 99.99    99.99 99.99     9.31 13    5.76  4    4.46  5    99.99 99    99.99 99
104933      5.96 99.99  1.67 99.99 99.99    99.99 99.99    99.99 99    6.78  5    5.30  5    99.99 99    99.99 99
104985      5.80  0.85  1.01 99.99 99.99    99.99 99.99     8.54  6    6.26  4    5.45  4    99.99 99    99.99 99
105043      6.13  1.15  0.17 99.99 99.99    99.99 99.99     8.82 35    6.13  3    6.17  2    99.99 99    99.99 99
105071      6.33 -0.48  0.22 99.99 99.99    99.99 99.99     6.84  5    6.36  3    6.34  2    99.99 99    99.99 99
```

```
105078      6.23 99.99 -0.08 99.99 99.99     99.99 99.99    99.99 99    6.09  2    6.38  2    99.99 99    99.99 99
105089      6.37  0.79  1.00 99.99 99.99     99.99 99.99     9.03  5    6.83  4    6.02  4    99.99 99    99.99 99
105138      6.23  1.04  1.24 99.99 99.99     99.99 99.99     9.36  4    6.82  4    5.77  4    99.99 99    99.99 99
105151      5.93  0.32  0.60 99.99 99.99     99.99 99.99     7.73  5    6.17  3    5.77  3    99.99 99    99.99 99
105211      4.15  0.03  0.34 99.99 99.99     99.99 99.99     5.41  5    4.24  3    4.11  2    99.99 99    99.99 99
105340      5.18  1.37  1.30 99.99 99.99     99.99 99.99     8.79  9    5.80  4    4.69  4    99.99 99    99.99 99
105383      6.37 99.99 -0.05 99.99 99.99     99.99 99.99    99.99 99    6.25  2    6.51  2    99.99 99    99.99 99
105416      5.34 -0.04  0.00 99.99 99.99     99.99 99.99     6.32 13    5.25  3    5.45  2    99.99 99    99.99 99
105437      6.22 99.99  1.74 99.99 99.99     99.99 99.99    99.99 99    7.08  5    5.53  5    99.99 99    99.99 99
105639      5.95  1.13  1.12 99.99 99.99     99.99 99.99     9.13  9    6.47  4    5.54  4    99.99 99    99.99 99
105686      6.17 99.99  0.03 99.99 99.99     99.99 99.99    99.99 99    6.09  3    6.27  2    99.99 99    99.99 99
105702      5.73  0.17  0.35 99.99 99.99     99.99 99.99     7.19  8    5.83  3    5.68  2    99.99 99    99.99 99
105776      6.06 99.99  0.20 99.99 99.99     99.99 99.99    99.99 99    6.08  3    6.08  2    99.99 99    99.99 99
105841      6.08 99.99  0.39 99.99 99.99     99.99 99.99    99.99 99    6.20  3    6.01  2    99.99 99    99.99 99
105850      5.46 99.99  0.06 99.99 99.99     99.99 99.99    99.99 99    5.40  3    5.55  2    99.99 99    99.99 99
105852      6.62  0.77  1.08 99.99 99.99     99.99 99.99     9.29  3    7.12  4    6.23  4    99.99 99    99.99 99
105920      6.23 99.99  0.82 99.99 99.99     99.99 99.99    99.99 99    6.59  4    5.96  3    99.99 99    99.99 99
106068      5.92 99.99  0.29 99.99 99.99     99.99 99.99    99.99 99    5.98  3    5.90  2    99.99 99    99.99 99
106231      5.76 99.99 -0.13 99.99 99.99     99.99 99.99    99.99 99    5.59  2    5.94  2    99.99 99    99.99 99
106248      6.35  1.41  1.24 99.99 99.99     99.99 99.99     9.98 12    6.94  4    5.89  4    99.99 99    99.99 99
106321      5.31  1.59  1.43 99.99 99.99     99.99 99.99     9.29 11    6.00  4    4.76  5    99.99 99    99.99 99
106485      5.83 99.99  1.05 99.99 99.99     99.99 99.99    99.99 99    6.31  4    5.46  4    99.99 99    99.99 99
106516      6.12 -0.14  0.45 99.99 99.99     99.99 99.99     7.21  4    6.27  3    6.03  3    99.99 99    99.99 99
106572      6.26 99.99  1.01 99.99 99.99     99.99 99.99    99.99 99    6.72  4    5.91  4    99.99 99    99.99 99
106676      6.22 -0.06 -0.01 99.99 99.99     99.99 99.99     7.17 12    6.12  3    6.34  2    99.99 99    99.99 99
106797      6.06 -0.05  0.03 99.99 99.99     99.99 99.99     7.04 12    5.98  3    6.16  2    99.99 99    99.99 99
106819      6.07 99.99  0.10 99.99 99.99     99.99 99.99    99.99 99    6.03  3    6.14  2    99.99 99    99.99 99
106911      4.26 -0.51 -0.12 99.99 99.99     99.99 99.99     4.54  6    4.10  2    4.43  2    99.99 99    99.99 99
106922      6.15 99.99  0.01 99.99 99.99     99.99 99.99    99.99 99    6.06  3    6.26  2    99.99 99    99.99 99
106975      5.99  0.03  0.35 99.99 99.99     99.99 99.99     7.26  5    6.09  3    5.94  2    99.99 99    99.99 99
106983      4.04 -0.68 -0.17 99.99 99.99     99.99 99.99     4.06  4    3.85  2    4.23  2    99.99 99    99.99 99
107070      5.90  0.13  0.17 99.99 99.99     99.99 99.99     7.20 12    5.90  3    5.94  2    99.99 99    99.99 99
107079      5.00  1.95  1.59 99.99 99.99     99.99 99.99     9.55 14    5.78  5    4.38  5    99.99 99    99.99 99
107113      6.33 -0.08  0.43 99.99 99.99     99.99 99.99     7.49  3    6.47  3    6.25  3    99.99 99    99.99 99
107295      5.97 99.99  0.82 99.99 99.99     99.99 99.99    99.99 99    6.33  4    5.70  3    99.99 99    99.99 99
107301      6.21 -0.15 -0.04 99.99 99.99     99.99 99.99     7.02 11    6.09  2    6.34  2    99.99 99    99.99 99
107348      5.21 99.99 -0.10 99.99 99.99     99.99 99.99    99.99 99    5.06  2    5.37  2    99.99 99    99.99 99
107446      3.59  1.63  1.42 99.99 99.99     99.99 99.99     7.61 12    4.28  4    3.04  5    99.99 99    99.99 99
107566      5.15  0.17  0.19 99.99 99.99     99.99 99.99     6.52 12    5.16  3    5.18  2    99.99 99    99.99 99
107567      5.74  0.82  1.04 99.99 99.99     99.99 99.99     8.46  4    6.22  4    5.37  4    99.99 99    99.99 99
107696      5.39 99.99 -0.10 99.99 99.99     99.99 99.99    99.99 99    5.24  2    5.55  2    99.99 99    99.99 99
```

```
107739      6.33  0.89  1.08 99.99 99.99    99.99 99.99     9.16  5    6.83  4    5.94  4    99.99 99    99.99 99
107773      6.36  0.53  0.89 99.99 99.99    99.99 99.99     8.60  3    6.75  4    6.06  3    99.99 99    99.99 99
107815      5.68 99.99  1.16 99.99 99.99    99.99 99.99    99.99 99    6.22  4    5.26  4    99.99 99    99.99 99
107832      5.32 99.99 -0.08 99.99 99.99    99.99 99.99    99.99 99    5.18  2    5.47  2    99.99 99    99.99 99
107833      6.40 99.99  0.29 99.99 99.99    99.99 99.99    99.99 99    6.46  3    6.38  2    99.99 99    99.99 99
107860      5.79 99.99 -0.08 99.99 99.99    99.99 99.99    99.99 99    5.65  2    5.94  2    99.99 99    99.99 99
107998      6.25 99.99  1.18 99.99 99.99    99.99 99.99    99.99 99    6.80  4    5.82  4    99.99 99    99.99 99
108054      6.30  0.74  0.96 99.99 99.99    99.99 99.99     8.87  5    6.73  4    5.97  4    99.99 99    99.99 99
108063      6.11 99.99  0.65 99.99 99.99    99.99 99.99    99.99 99    6.37  3    5.92  3    99.99 99    99.99 99
108107      5.95 99.99  0.03 99.99 99.99    99.99 99.99    99.99 99    5.87  3    6.05  2    99.99 99    99.99 99
108250      4.86 -0.59 -0.12 99.99 99.99    99.99 99.99     5.03  4    4.70  2    5.03  2    99.99 99    99.99 99
108309      6.26 99.99  0.68 99.99 99.99    99.99 99.99    99.99 99    6.54  3    6.06  3    99.99 99    99.99 99
108355      6.00 99.99  0.07 99.99 99.99    99.99 99.99    99.99 99    5.94  3    6.08  2    99.99 99    99.99 99
108471      6.37  0.68  0.93 99.99 99.99    99.99 99.99     8.84  4    6.79  4    6.05  3    99.99 99    99.99 99
108501      6.04 -0.03  0.02 99.99 99.99    99.99 99.99     7.05 12    5.96  3    6.15  2    99.99 99    99.99 99
108530      6.22 99.99  1.26 99.99 99.99    99.99 99.99    99.99 99    6.82  4    5.75  4    99.99 99    99.99 99
108570      6.15 99.99  0.92 99.99 99.99    99.99 99.99    99.99 99    6.56  4    5.84  3    99.99 99    99.99 99
108732      5.80 99.99  1.56 99.99 99.99    99.99 99.99    99.99 99    6.56  5    5.19  5    99.99 99    99.99 99
108903      1.63  1.77  1.59 99.99 99.99    99.99 99.99     5.94 10    2.41  5    1.01  5    99.99 99    99.99 99
108925      6.42 99.99  0.16 99.99 99.99    99.99 99.99    99.99 99    6.41  3    6.46  2    99.99 99    99.99 99
108970      5.88  1.05  1.11 99.99 99.99    99.99 99.99     8.95  7    6.40  4    5.48  4    99.99 99    99.99 99
108985      6.05  1.88  1.52 99.99 99.99    99.99 99.99    10.47 15    6.79  5    5.46  5    99.99 99    99.99 99
109000      5.95 99.99  0.27 99.99 99.99    99.99 99.99    99.99 99    6.00  3    5.94  2    99.99 99    99.99 99
109014      6.19  0.86  1.04 99.99 99.99    99.99 99.99     8.96  5    6.67  4    5.82  4    99.99 99    99.99 99
109074      6.46 99.99  0.20 99.99 99.99    99.99 99.99    99.99 99    6.48  3    6.48  2    99.99 99    99.99 99
109238      6.26 99.99  0.29 99.99 99.99    99.99 99.99    99.99 99    6.32  3    6.24  2    99.99 99    99.99 99
109309      5.48 -0.10 -0.03 99.99 99.99    99.99 99.99     6.36 12    5.37  2    5.61  2    99.99 99    99.99 99
109312      6.38 99.99  0.46 99.99 99.99    99.99 99.99    99.99 99    6.54  3    6.28  3    99.99 99    99.99 99
109409      5.77 99.99  0.70 99.99 99.99    99.99 99.99    99.99 99    6.06  3    5.56  3    99.99 99    99.99 99
109492      6.22 99.99  0.73 99.99 99.99    99.99 99.99    99.99 99    6.53  3    6.00  3    99.99 99    99.99 99
109536      5.13 99.99  0.22 99.99 99.99    99.99 99.99    99.99 99    5.16  3    5.14  2    99.99 99    99.99 99
109573      5.80 99.99  0.01 99.99 99.99    99.99 99.99    99.99 99    5.71  3    5.91  2    99.99 99    99.99 99
109857      6.49 -0.24  0.08 99.99 99.99    99.99 99.99     7.24  6    6.44  3    6.57  2    99.99 99    99.99 99
109860      6.33  0.01  0.01 99.99 99.99    99.99 99.99     7.38 13    6.24  3    6.44  2    99.99 99    99.99 99
109931      6.00 99.99  0.29 99.99 99.99    99.99 99.99    99.99 99    6.06  3    5.98  2    99.99 99    99.99 99
109960      5.89 99.99  1.21 99.99 99.99    99.99 99.99    99.99 99    6.46  4    5.44  4    99.99 99    99.99 99
110287      5.84 99.99  1.52 99.99 99.99    99.99 99.99    99.99 99    6.58  5    5.25  5    99.99 99    99.99 99
110385      6.04 99.99  0.39 99.99 99.99    99.99 99.99    99.99 99    6.16  3    5.97  2    99.99 99    99.99 99
110423      5.59 -0.02  0.00 99.99 99.99    99.99 99.99     6.60 13    5.50  3    5.70  2    99.99 99    99.99 99
110506      5.99 99.99 -0.08 99.99 99.99    99.99 99.99    99.99 99    5.85  2    6.14  2    99.99 99    99.99 99
110532      6.39 99.99  1.09 99.99 99.99    99.99 99.99    99.99 99    6.89  4    6.00  4    99.99 99    99.99 99
```

```
110575       6.44 99.99   0.25 99.99 99.99     99.99 99.99   99.99 99    6.48  3    6.44  2   99.99 99   99.99 99
110653       6.39 99.99  -0.06 99.99 99.99     99.99 99.99   99.99 99    6.26  2    6.53  2   99.99 99   99.99 99
110716       6.16  0.38   0.69 99.99 99.99     99.99 99.99    8.09  4    6.44  3    5.95  3   99.99 99   99.99 99
110829       4.69  0.93   1.05 99.99 99.99     99.99 99.99    7.56  6    5.17  4    4.32  4   99.99 99   99.99 99
110879       3.05 -0.74  -0.18 99.99 99.99     99.99 99.99    2.98  3    2.86  2    3.25  2   99.99 99   99.99 99
111028       5.67  0.74   0.99 99.99 99.99     99.99 99.99    8.25  4    6.12  4    5.32  4   99.99 99   99.99 99
111032       5.86 99.99   1.34 99.99 99.99     99.99 99.99   99.99 99    6.50  4    5.35  4   99.99 99   99.99 99
111199       6.26  0.07   0.55 99.99 99.99     99.99 99.99    7.69  3    6.47  3    6.12  3   99.99 99   99.99 99
111226       6.44 99.99  -0.06 99.99 99.99     99.99 99.99   99.99 99    6.31  2    6.58  2   99.99 99   99.99 99
111239       6.41  1.82   1.60 99.99 99.99     99.99 99.99   10.79 11    7.20  5    5.78  5   99.99 99   99.99 99
111295       5.66 99.99   0.95 99.99 99.99     99.99 99.99   99.99 99    6.09  4    5.33  4   99.99 99   99.99 99
111315       5.55  0.95   1.17 99.99 99.99     99.99 99.99    8.52  4    6.10  4    5.12  4   99.99 99   99.99 99
111482       5.46  0.79   1.02 99.99 99.99     99.99 99.99    8.13  4    5.93  4    5.10  4   99.99 99   99.99 99
111519       6.24 99.99   0.05 99.99 99.99     99.99 99.99   99.99 99    6.17  3    6.33  2   99.99 99   99.99 99
111588       5.73 99.99   0.13 99.99 99.99     99.99 99.99   99.99 99    5.71  3    5.78  2   99.99 99   99.99 99
111597       4.91 99.99  -0.04 99.99 99.99     99.99 99.99   99.99 99    4.79  2    5.04  2   99.99 99   99.99 99
111720       6.41  0.79   1.02 99.99 99.99     99.99 99.99    9.08  4    6.88  4    6.05  4   99.99 99   99.99 99
111765       6.02  1.44   1.29 99.99 99.99     99.99 99.99    9.72 11    6.63  4    5.54  4   99.99 99   99.99 99
111774       5.98 99.99  -0.10 99.99 99.99     99.99 99.99   99.99 99    5.83  2    6.14  2   99.99 99   99.99 99
111775       6.33 99.99   0.03 99.99 99.99     99.99 99.99   99.99 99    6.25  3    6.43  2   99.99 99   99.99 99
111884       5.93 99.99   1.31 99.99 99.99     99.99 99.99   99.99 99    6.56  4    5.44  4   99.99 99   99.99 99
111998       6.11  0.04   0.50 99.99 99.99     99.99 99.99    7.47  3    6.29  3    5.99  3   99.99 99   99.99 99
112131       6.00 99.99   0.08 99.99 99.99     99.99 99.99   99.99 99    5.95  3    6.08  2   99.99 99   99.99 99
112164       5.89 99.99   0.64 99.99 99.99     99.99 99.99   99.99 99    6.15  3    5.71  3   99.99 99   99.99 99
112213       5.47 99.99   1.68 99.99 99.99     99.99 99.99   99.99 99    6.30  5    4.80  5   99.99 99   99.99 99
112219       5.93  0.89   1.13 99.99 99.99     99.99 99.99    8.79  4    6.46  4    5.52  4   99.99 99   99.99 99
112519       6.31 99.99   1.07 99.99 99.99     99.99 99.99   99.99 99    6.80  4    5.93  4   99.99 99   99.99 99
112846       5.79  0.08   0.18 99.99 99.99     99.99 99.99    7.03 10    5.79  3    5.82  2   99.99 99   99.99 99
112935       6.02 99.99   0.38 99.99 99.99     99.99 99.99   99.99 99    6.13  3    5.96  2   99.99 99   99.99 99
112985       3.62  1.26   1.18 99.99 99.99     99.99 99.99    7.01 10    4.17  4    3.19  4   99.99 99   99.99 99
112992       5.99  1.09   1.12 99.99 99.99     99.99 99.99    9.12  8    6.51  4    5.58  4   99.99 99   99.99 99
113092       5.32  1.29   1.29 99.99 99.99     99.99 99.99    8.81  8    5.93  4    4.84  4   99.99 99   99.99 99
113314       4.85 99.99   0.02 99.99 99.99     99.99 99.99   99.99 99    4.77  3    4.96  2   99.99 99   99.99 99
113415       5.58 99.99   0.56 99.99 99.99     99.99 99.99   99.99 99    5.79  3    5.43  3   99.99 99   99.99 99
113459       6.59  0.11   0.29 99.99 99.99     99.99 99.99    7.93  8    6.65  3    6.57  2   99.99 99   99.99 99
113823       5.99 99.99   0.48 99.99 99.99     99.99 99.99   99.99 99    6.16  3    5.88  3   99.99 99   99.99 99
113852       5.65 99.99   0.03 99.99 99.99     99.99 99.99   99.99 99    5.57  3    5.75  2   99.99 99   99.99 99
113902       5.71 99.99  -0.07 99.99 99.99     99.99 99.99   99.99 99    5.58  2    5.86  2   99.99 99   99.99 99
114113       5.55  1.18   1.18 99.99 99.99     99.99 99.99    8.83  8    6.10  4    5.12  4   99.99 99   99.99 99
114203       6.32  0.79   1.02 99.99 99.99     99.99 99.99    8.99  4    6.79  4    5.96  4   99.99 99   99.99 99
114287       5.94  1.69   1.49 99.99 99.99     99.99 99.99   10.08 11    6.66  4    5.36  5   99.99 99   99.99 99
```

```
114371       5.91 -0.03    0.42 99.99 99.99    99.99 99.99      7.14  3    6.05  3    5.83  3    99.99 99    99.99 99
114378       4.32 -0.06    0.45 99.99 99.99    99.99 99.99      5.52  3    4.47  3    4.23  3    99.99 99    99.99 99
114435       5.79 99.99    0.52 99.99 99.99    99.99 99.99     99.99 99    5.98  3    5.66  3    99.99 99    99.99 99
114461       6.33 99.99    0.44 99.99 99.99    99.99 99.99     99.99 99    6.48  3    6.24  3    99.99 99    99.99 99
114533       5.85  0.72    1.07 99.99 99.99    99.99 99.99      8.45  3    6.34  4    5.47  4    99.99 99    99.99 99
114570       5.90  0.04    0.06 99.99 99.99    99.99 99.99      7.02 13    5.84  3    5.99  2    99.99 99    99.99 99
114576       6.51 99.99    0.19 99.99 99.99    99.99 99.99     99.99 99    6.52  3    6.54  2    99.99 99    99.99 99
114630       6.16 99.99    0.60 99.99 99.99    99.99 99.99     99.99 99    6.40  3    6.00  3    99.99 99    99.99 99
114642       5.04  0.02    0.46 99.99 99.99    99.99 99.99      6.36  3    5.20  3    4.94  3    99.99 99    99.99 99
114707       6.22 99.99    1.06 99.99 99.99    99.99 99.99     99.99 99    6.71  4    5.84  4    99.99 99    99.99 99
114837       4.92 99.99    0.48 99.99 99.99    99.99 99.99     99.99 99    5.09  3    4.81  3    99.99 99    99.99 99
114873       6.16 99.99    1.38 99.99 99.99    99.99 99.99     99.99 99    6.82  4    5.63  4    99.99 99    99.99 99
114912       6.37  1.33    1.21 99.99 99.99    99.99 99.99      9.87 11    6.94  4    5.92  4    99.99 99    99.99 99
114971       5.89 99.99    1.06 99.99 99.99    99.99 99.99     99.99 99    6.38  4    5.51  4    99.99 99    99.99 99
115050       6.19 99.99    0.97 99.99 99.99    99.99 99.99     99.99 99    6.63  4    5.85  4    99.99 99    99.99 99
115149       6.07 -0.02    0.44 99.99 99.99    99.99 99.99      7.32  3    6.22  3    5.98  3    99.99 99    99.99 99
115202       5.22 99.99    1.04 99.99 99.99    99.99 99.99     99.99 99    5.70  4    4.85  4    99.99 99    99.99 99
115211       4.87  1.58    1.50 99.99 99.99    99.99 99.99      8.87  8    5.60  5    4.29  5    99.99 99    99.99 99
115331       5.85 99.99    0.20 99.99 99.99    99.99 99.99     99.99 99    5.87  3    5.87  2    99.99 99    99.99 99
115383       5.22  0.09    0.58 99.99 99.99    99.99 99.99      6.70  3    5.44  3    5.07  3    99.99 99    99.99 99
115439       6.04  1.46    1.35 99.99 99.99    99.99 99.99      9.80 10    6.69  4    5.53  4    99.99 99    99.99 99
115488       6.37  0.04    0.26 99.99 99.99    99.99 99.99      7.60  8    6.42  3    6.36  2    99.99 99    99.99 99
115521       4.81  1.95    1.67 99.99 99.99    99.99 99.99      9.41 12    5.63  5    4.15  5    99.99 99    99.99 99
115529       6.19 99.99    0.01 99.99 99.99    99.99 99.99     99.99 99    6.10  3    6.30  2    99.99 99    99.99 99
115709       6.62  0.03    0.06 99.99 99.99    99.99 99.99      7.73 13    6.56  3    6.71  2    99.99 99    99.99 99
115842       6.02 99.99    0.29 99.99 99.99    99.99 99.99     99.99 99    6.08  3    6.00  2    99.99 99    99.99 99
115912       5.77 99.99    1.12 99.99 99.99    99.99 99.99     99.99 99    6.29  4    5.36  4    99.99 99    99.99 99
115967       6.05 -0.34    0.09 99.99 99.99    99.99 99.99      6.67  4    6.00  3    6.12  2    99.99 99    99.99 99
115995       6.26  0.10    0.10 99.99 99.99    99.99 99.99      7.49 13    6.22  3    6.33  2    99.99 99    99.99 99
116061       6.21 99.99    0.10 99.99 99.99    99.99 99.99     99.99 99    6.17  3    6.28  2    99.99 99    99.99 99
116087       4.53 -0.59   -0.13 99.99 99.99    99.99 99.99      4.69  4    4.36  2    4.71  2    99.99 99    99.99 99
116160       5.69  0.03    0.06 99.99 99.99    99.99 99.99      6.80 13    5.63  3    5.78  2    99.99 99    99.99 99
116235       5.87  0.10    0.12 99.99 99.99    99.99 99.99      7.11 12    5.84  3    5.93  2    99.99 99    99.99 99
116243       4.53  0.47    0.85 99.99 99.99    99.99 99.99      6.67  3    4.90  4    4.25  3    99.99 99    99.99 99
116244       5.05  1.01    1.11 99.99 99.99    99.99 99.99      8.06  6    5.57  4    4.65  4    99.99 99    99.99 99
116338       6.48 99.99    0.96 99.99 99.99    99.99 99.99     99.99 99    6.91  4    6.15  4    99.99 99    99.99 99
116365       5.89  1.66    1.43 99.99 99.99    99.99 99.99      9.96 12    6.58  4    5.34  5    99.99 99    99.99 99
116457       5.31 -0.01    0.40 99.99 99.99    99.99 99.99      6.55  4    5.44  3    5.24  3    99.99 99    99.99 99
116458       5.67 -0.18   -0.03 99.99 99.99    99.99 99.99      6.44 10    5.56  2    5.80  2    99.99 99    99.99 99
116579       6.63 -0.24   -0.06 99.99 99.99    99.99 99.99      7.31 10    6.50  2    6.77  2    99.99 99    99.99 99
116831       5.97  0.13    0.20 99.99 99.99    99.99 99.99      7.29 11    5.99  3    5.99  2    99.99 99    99.99 99
```

```
116862     6.28 99.99 -0.12 99.99 99.99    99.99 99.99    99.99 99    6.12  2    6.45  2    99.99 99    99.99 99
116870     5.26  1.76  1.53 99.99 99.99    99.99 99.99     9.52 12    6.01  5    4.66  5    99.99 99    99.99 99
117025     6.11  0.05  0.11 99.99 99.99    99.99 99.99     7.27 12    6.08  3    6.17  2    99.99 99    99.99 99
117150     5.06 99.99  0.07 99.99 99.99    99.99 99.99    99.99 99    5.00  3    5.14  2    99.99 99    99.99 99
117267     6.43  1.11  1.11 99.99 99.99    99.99 99.99     9.58  9    6.95  4    6.03  4    99.99 99    99.99 99
117374     5.58  0.16  0.18 99.99 99.99    99.99 99.99     6.93 12    5.58  3    5.61  2    99.99 99    99.99 99
117404     6.17  1.80  1.47 99.99 99.99    99.99 99.99    10.45 14    6.88  4    5.60  5    99.99 99    99.99 99
117405     6.51  0.59  0.96 99.99 99.99    99.99 99.99     8.87  3    6.94  4    6.18  4    99.99 99    99.99 99
117558     6.47 99.99  0.10 99.99 99.99    99.99 99.99    99.99 99    6.43  3    6.54  2    99.99 99    99.99 99
117597     6.16 99.99  1.04 99.99 99.99    99.99 99.99    99.99 99    6.64  4    5.79  4    99.99 99    99.99 99
117651     6.37 -0.08 -0.02 99.99 99.99    99.99 99.99     7.29 12    6.26  2    6.49  2    99.99 99    99.99 99
117718     6.45 99.99  0.44 99.99 99.99    99.99 99.99    99.99 99    6.60  3    6.36  3    99.99 99    99.99 99
117818     5.21  0.60  0.96 99.99 99.99    99.99 99.99     7.59  3    5.64  4    4.88  4    99.99 99    99.99 99
117876     6.11  0.70  0.96 99.99 99.99    99.99 99.99     8.62  4    6.54  4    5.78  4    99.99 99    99.99 99
117919     6.33 99.99 -0.05 99.99 99.99    99.99 99.99    99.99 99    6.21  2    6.47  2    99.99 99    99.99 99
118219     5.73  0.65  0.95 99.99 99.99    99.99 99.99     8.17  4    6.16  4    5.40  4    99.99 99    99.99 99
118261     5.63 99.99  0.50 99.99 99.99    99.99 99.99    99.99 99    5.81  3    5.51  3    99.99 99    99.99 99
118338     5.98 99.99  0.94 99.99 99.99    99.99 99.99    99.99 99    6.40  4    5.66  3    99.99 99    99.99 99
118344     6.10  1.73  1.42 99.99 99.99    99.99 99.99    10.26 14    6.79  4    5.55  5    99.99 99    99.99 99
118384     6.42 99.99  1.12 99.99 99.99    99.99 99.99    99.99 99    6.94  4    6.01  4    99.99 99    99.99 99
118522     6.59  1.09  1.30 99.99 99.99    99.99 99.99     9.82  4    7.21  4    6.10  4    99.99 99    99.99 99
118646     5.83 99.99  0.40 99.99 99.99    99.99 99.99    99.99 99    5.96  3    5.76  3    99.99 99    99.99 99
118666     5.79  0.02  0.39 99.99 99.99    99.99 99.99     7.07  4    5.91  3    5.72  2    99.99 99    99.99 99
118978     5.38 99.99 -0.03 99.99 99.99    99.99 99.99    99.99 99    5.27  2    5.51  2    99.99 99    99.99 99
119086     6.59 99.99  0.07 99.99 99.99    99.99 99.99    99.99 99    6.53  3    6.67  2    99.99 99    99.99 99
119288     6.16 -0.04  0.42 99.99 99.99    99.99 99.99     7.37  3    6.30  3    6.08  3    99.99 99    99.99 99
119425     5.37  1.05  1.10 99.99 99.99    99.99 99.99     8.43  7    5.88  4    4.97  4    99.99 99    99.99 99
119537     6.51  0.01  0.05 99.99 99.99    99.99 99.99     7.59 12    6.44  3    6.60  2    99.99 99    99.99 99
119752     5.81 99.99  0.02 99.99 99.99    99.99 99.99    99.99 99    5.73  3    5.92  2    99.99 99    99.99 99
119834     4.65  0.72  0.96 99.99 99.99    99.99 99.99     7.19  4    5.08  4    4.32  4    99.99 99    99.99 99
119921     5.15 99.99 -0.02 99.99 99.99    99.99 99.99    99.99 99    5.04  2    5.27  2    99.99 99    99.99 99
119938     5.91 99.99  0.29 99.99 99.99    99.99 99.99    99.99 99    5.97  3    5.89  2    99.99 99    99.99 99
119971     5.45 99.99  1.36 99.99 99.99    99.99 99.99    99.99 99    6.10  4    4.93  4    99.99 99    99.99 99
120033     6.05  1.66  1.42 99.99 99.99    99.99 99.99    10.12 12    6.74  4    5.50  5    99.99 99    99.99 99
120052     5.44  1.93  1.61 99.99 99.99    99.99 99.99     9.98 13    6.23  5    4.81  5    99.99 99    99.99 99
120066     6.33  0.16  0.63 99.99 99.99    99.99 99.99     7.93  3    6.58  3    6.15  3    99.99 99    99.99 99
120213     5.95  1.57  1.46 99.99 99.99    99.99 99.99     9.92  9    6.66  4    5.39  5    99.99 99    99.99 99
120237     6.52 99.99  0.57 99.99 99.99    99.99 99.99    99.99 99    6.74  3    6.37  3    99.99 99    99.99 99
120404     5.75  2.02  1.73 99.99 99.99    99.99 99.99    10.47 12    6.61  5    5.06  5    99.99 99    99.99 99
120602     6.01  0.59  0.90 99.99 99.99    99.99 99.99     8.34  4    6.41  4    5.71  3    99.99 99    99.99 99
120672     6.35 99.99  0.48 99.99 99.99    99.99 99.99    99.99 99    6.52  3    6.24  3    99.99 99    99.99 99
```

```
120690       6.45  0.27  0.69 99.99 99.99    99.99 99.99    8.23  3    6.73  3    6.24  3    99.99 99    99.99 99
120759       6.12 99.99  0.48 99.99 99.99    99.99 99.99   99.99 99    6.29  3    6.01  3    99.99 99    99.99 99
120913       5.71  1.67  1.49 99.99 99.99    99.99 99.99    9.83 11    6.43  4    5.13  5    99.99 99    99.99 99
121056       6.19 99.99  1.02 99.99 99.99    99.99 99.99   99.99 99    6.66  4    5.83  4    99.99 99    99.99 99
121299       5.15  1.08  1.08 99.99 99.99    99.99 99.99    8.24  9    5.65  4    4.76  4    99.99 99    99.99 99
121336       6.13 99.99  0.08 99.99 99.99    99.99 99.99   99.99 99    6.08  3    6.21  2    99.99 99    99.99 99
121416       5.83 99.99  1.14 99.99 99.99    99.99 99.99   99.99 99    6.36  4    5.41  4    99.99 99    99.99 99
121439       6.09 -0.17  0.03 99.99 99.99    99.99 99.99    6.91  9    6.01  3    6.19  2    99.99 99    99.99 99
121474       4.71  1.04  1.11 99.99 99.99    99.99 99.99    7.76  7    5.23  4    4.31  4    99.99 99    99.99 99
121557       6.20  0.81  1.05 99.99 99.99    99.99 99.99    8.91  4    6.68  4    5.83  4    99.99 99    99.99 99
121607       5.91  0.14  0.20 99.99 99.99    99.99 99.99    7.24 11    5.93  3    5.93  2    99.99 99    99.99 99
121901       6.49 99.99  0.33 99.99 99.99    99.99 99.99   99.99 99    6.58  3    6.45  2    99.99 99    99.99 99
121932       5.97  0.07  0.35 99.99 99.99    99.99 99.99    7.29  6    6.07  3    5.92  2    99.99 99    99.99 99
122106       6.40  0.06  0.49 99.99 99.99    99.99 99.99    7.79  3    6.57  3    6.29  3    99.99 99    99.99 99
122365       5.99  0.10  0.09 99.99 99.99    99.99 99.99    7.21 13    5.94  3    6.06  2    99.99 99    99.99 99
122438       5.92 99.99  1.22 99.99 99.99    99.99 99.99   99.99 99    6.50  4    5.47  4    99.99 99    99.99 99
122510       6.18 99.99  0.48 99.99 99.99    99.99 99.99   99.99 99    6.35  3    6.07  3    99.99 99    99.99 99
122563       6.20  0.38  0.90 99.99 99.99    99.99 99.99    8.24  4    6.60  4    5.90  3    99.99 99    99.99 99
122744       6.26  0.70  0.94 99.99 99.99    99.99 99.99    8.76  4    6.68  4    5.94  3    99.99 99    99.99 99
122797       6.24 -0.02  0.39 99.99 99.99    99.99 99.99    7.46  4    6.36  3    6.17  2    99.99 99    99.99 99
122862       6.02  0.06  0.58 99.99 99.99    99.99 99.99    7.46  3    6.24  3    5.87  3    99.99 99    99.99 99
122879       6.42 99.99  0.12 99.99 99.99    99.99 99.99   99.99 99    6.39  3    6.48  2    99.99 99    99.99 99
122910       6.28  0.87  1.02 99.99 99.99    99.99 99.99    9.05  6    6.75  4    5.92  4    99.99 99    99.99 99
123151       6.40 99.99  1.02 99.99 99.99    99.99 99.99   99.99 99    6.87  4    6.04  4    99.99 99    99.99 99
123255       5.46 99.99  0.34 99.99 99.99    99.99 99.99   99.99 99    5.55  3    5.42  2    99.99 99    99.99 99
123377       6.05  1.84  1.74 99.99 99.99    99.99 99.99   10.54  8    6.91  5    5.36  5    99.99 99    99.99 99
123492       6.06  0.15  0.18 99.99 99.99    99.99 99.99    7.40 12    6.06  3    6.09  2    99.99 99    99.99 99
123569       4.75  0.72  0.94 99.99 99.99    99.99 99.99    7.28  5    5.17  4    4.43  3    99.99 99    99.99 99
123998       4.91  0.11  0.25 99.99 99.99    99.99 99.99    6.23  9    4.95  3    4.91  2    99.99 99    99.99 99
124099       6.47  1.42  1.42 99.99 99.99    99.99 99.99   10.21  7    7.16  4    5.92  5    99.99 99    99.99 99
124115       6.43  0.04  0.48 99.99 99.99    99.99 99.99    7.78  3    6.60  3    6.32  3    99.99 99    99.99 99
124425       5.93  0.02  0.47 99.99 99.99    99.99 99.99    7.25  3    6.09  3    5.83  3    99.99 99    99.99 99
124433       5.61 99.99  0.93 99.99 99.99    99.99 99.99   99.99 99    6.03  4    5.29  3    99.99 99    99.99 99
124471       5.75 -0.86 -0.06 99.99 99.99    99.99 99.99    5.59  6    5.62  2    5.89  2    99.99 99    99.99 99
124553       6.36  0.16  0.60 99.99 99.99    99.99 99.99    7.94  3    6.60  3    6.20  3    99.99 99    99.99 99
124580       6.31 99.99  0.60 99.99 99.99    99.99 99.99   99.99 99    6.55  3    6.15  3    99.99 99    99.99 99
124639       6.42 -0.33  0.02 99.99 99.99    99.99 99.99    7.02  6    6.34  3    6.53  2    99.99 99    99.99 99
124882       4.32  1.45  1.31 99.99 99.99    99.99 99.99    8.04 11    4.95  4    3.83  4    99.99 99    99.99 99
125150       6.54 99.99  0.27 99.99 99.99    99.99 99.99   99.99 99    6.59  3    6.53  2    99.99 99    99.99 99
125158       5.23 99.99  0.29 99.99 99.99    99.99 99.99   99.99 99    5.29  3    5.21  2    99.99 99    99.99 99
125184       6.48 99.99  0.72 99.99 99.99    99.99 99.99   99.99 99    6.78  3    6.26  3    99.99 99    99.99 99
```

```
125276        5.87 99.99   0.50 99.99 99.99     99.99 99.99    99.99 99    6.05   3    5.75   3    99.99 99    99.99 99
125283        5.94 99.99   0.08 99.99 99.99     99.99 99.99    99.99 99    5.89   3    6.02   2    99.99 99    99.99 99
125288        4.33 99.99   0.12 99.99 99.99     99.99 99.99    99.99 99    4.30   3    4.39   2    99.99 99    99.99 99
125442        4.77 99.99   0.31 99.99 99.99     99.99 99.99    99.99 99    4.85   3    4.74   2    99.99 99    99.99 99
125489        6.19  0.09   0.20 99.99 99.99     99.99 99.99     7.46 10    6.21   3    6.21   2    99.99 99    99.99 99
125835        5.61  0.01   0.49 99.99 99.99     99.99 99.99     6.93  3    5.78   3    5.50   3    99.99 99    99.99 99
125869        6.00 99.99   1.10 99.99 99.99     99.99 99.99    99.99 99    6.51   4    5.60   4    99.99 99    99.99 99
125990        6.36  0.06   0.13 99.99 99.99     99.99 99.99     7.55 11    6.34   3    6.41   2    99.99 99    99.99 99
126128        4.87 -0.03   0.05 99.99 99.99     99.99 99.99     5.89 11    4.80   3    4.96   2    99.99 99    99.99 99
126200        5.95  0.08   0.06 99.99 99.99     99.99 99.99     7.13 14    5.89   3    6.04   2    99.99 99    99.99 99
126209        6.07  1.11   1.18 99.99 99.99     99.99 99.99     9.26  7    6.62   4    5.64   4    99.99 99    99.99 99
126241        5.85  1.76   1.50 99.99 99.99     99.99 99.99    10.09 12    6.58   5    5.27   5    99.99 99    99.99 99
126251        6.49 99.99   0.42 99.99 99.99     99.99 99.99    99.99 99    6.63   3    6.41   3    99.99 99    99.99 99
126271        6.19  1.29   1.20 99.99 99.99     99.99 99.99     9.63 10    6.76   4    5.75   4    99.99 99    99.99 99
126386        6.32 99.99   1.19 99.99 99.99     99.99 99.99    99.99 99    6.88   4    5.88   4    99.99 99    99.99 99
126475        6.35 99.99  -0.08 99.99 99.99     99.99 99.99    99.99 99    6.21   2    6.50   2    99.99 99    99.99 99
126504        5.83  0.16   0.30 99.99 99.99     99.99 99.99     7.25  9    5.90   3    5.81   2    99.99 99    99.99 99
126610        6.45 99.99   0.14 99.99 99.99     99.99 99.99    99.99 99    6.43   3    6.50   2    99.99 99    99.99 99
126769        4.97 -0.41  -0.07 99.99 99.99     99.99 99.99     5.41  6    4.84   2    5.12   2    99.99 99    99.99 99
126862        5.83  0.78   1.00 99.99 99.99     99.99 99.99     8.47  5    6.29   4    5.48   4    99.99 99    99.99 99
126981        5.50 -0.27  -0.08 99.99 99.99     99.99 99.99     6.13 10    5.36   2    5.65   2    99.99 99    99.99 99
126983        5.37 99.99   0.05 99.99 99.99     99.99 99.99    99.99 99    5.30   3    5.46   2    99.99 99    99.99 99
127167        5.94  0.11   0.16 99.99 99.99     99.99 99.99     7.21 12    5.93   3    5.98   2    99.99 99    99.99 99
127486        5.87 99.99   0.48 99.99 99.99     99.99 99.99    99.99 99    6.04   3    5.76   3    99.99 99    99.99 99
127501        5.87 99.99   1.09 99.99 99.99     99.99 99.99    99.99 99    6.37   4    5.48   4    99.99 99    99.99 99
127724        6.40 99.99   1.26 99.99 99.99     99.99 99.99    99.99 99    7.00   4    5.93   4    99.99 99    99.99 99
128020        6.04 -0.02   0.50 99.99 99.99     99.99 99.99     7.32  3    6.22   3    5.92   3    99.99 99    99.99 99
128068        5.55 99.99   1.48 99.99 99.99     99.99 99.99    99.99 99    6.27   4    4.98   5    99.99 99    99.99 99
128266        5.41  0.71   1.00 99.99 99.99     99.99 99.99     7.96  4    5.87   4    5.06   4    99.99 99    99.99 99
128429        6.20 -0.04   0.47 99.99 99.99     99.99 99.99     7.44  3    6.36   3    6.10   3    99.99 99    99.99 99
128582        6.07  0.04   0.51 99.99 99.99     99.99 99.99     7.44  3    6.26   3    5.95   3    99.99 99    99.99 99
128617        6.39 99.99   0.44 99.99 99.99     99.99 99.99    99.99 99    6.54   3    6.30   3    99.99 99    99.99 99
128713        6.30 99.99   1.18 99.99 99.99     99.99 99.99    99.99 99    6.85   4    5.87   4    99.99 99    99.99 99
128902        5.70  1.69   1.48 99.99 99.99     99.99 99.99     9.84 11    6.42   4    5.13   5    99.99 99    99.99 99
128917        6.22 99.99   0.45 99.99 99.99     99.99 99.99    99.99 99    6.37   3    6.13   3    99.99 99    99.99 99
128974        5.67 99.99  -0.08 99.99 99.99     99.99 99.99    99.99 99    5.53   2    5.82   2    99.99 99    99.99 99
129078        3.83  1.68   1.43 99.99 99.99     99.99 99.99     7.93 12    4.52   4    3.28   5    99.99 99    99.99 99
129245        6.26  1.46   1.30 99.99 99.99     99.99 99.99     9.99 11    6.88   4    5.77   4    99.99 99    99.99 99
129246        3.78  0.05   0.05 99.99 99.99     99.99 99.99     4.91 13    3.71   3    3.87   2    99.99 99    99.99 99
129336        5.56  0.67   0.94 99.99 99.99     99.99 99.99     8.02  4    5.98   4    5.24   3    99.99 99    99.99 99
129422        5.36 99.99   0.29 99.99 99.99     99.99 99.99    99.99 99    5.42   3    5.34   2    99.99 99    99.99 99
```

```
129462      6.11 99.99  1.02 99.99 99.99     99.99 99.99     99.99 99    6.58  4    5.75  4    99.99 99    99.99 99
129858      5.74  0.09  0.07 99.99 99.99     99.99 99.99      6.94 14    5.68  3    5.82  2    99.99 99    99.99 99
130073      6.30  0.88  1.08 99.99 99.99     99.99 99.99      9.12  5    6.80  4    5.91  4    99.99 99    99.99 99
130227      6.23 99.99  1.13 99.99 99.99     99.99 99.99     99.99 99    6.76  4    5.82  4    99.99 99    99.99 99
130458      5.60  0.42  0.82 99.99 99.99     99.99 99.99      7.66  3    5.96  4    5.33  3    99.99 99    99.99 99
130650      5.65  0.74  0.95 99.99 99.99     99.99 99.99      8.21  5    6.08  4    5.32  4    99.99 99    99.99 99
130952      4.95  0.70  0.98 99.99 99.99     99.99 99.99      7.47  4    5.39  4    4.61  4    99.99 99    99.99 99
131058      6.09 -0.60 -0.06 99.99 99.99     99.99 99.99      6.28  3    5.96  2    6.23  2    99.99 99    99.99 99
131111      5.48  0.84  1.02 99.99 99.99     99.99 99.99      8.21  5    5.95  4    5.12  4    99.99 99    99.99 99
131117      6.29 99.99  0.60 99.99 99.99     99.99 99.99     99.99 99    6.53  3    6.13  3    99.99 99    99.99 99
131246      5.65  1.15  1.30 99.99 99.99     99.99 99.99      8.96  5    6.27  4    5.16  4    99.99 99    99.99 99
131342      5.20 99.99  1.16 99.99 99.99     99.99 99.99     99.99 99    5.74  4    4.78  4    99.99 99    99.99 99
131425      5.93  0.81  1.05 99.99 99.99     99.99 99.99      8.64  4    6.41  4    5.56  4    99.99 99    99.99 99
131507      5.46  1.60  1.36 99.99 99.99     99.99 99.99      9.41 13    6.11  4    4.94  4    99.99 99    99.99 99
131551      6.20 -0.20 -0.04 99.99 99.99     99.99 99.99      6.94 10    6.08  2    6.33  2    99.99 99    99.99 99
131596      5.91 -0.13 -0.06 99.99 99.99     99.99 99.99      6.74 12    5.78  2    6.05  2    99.99 99    99.99 99
131657      5.64 -0.27 -0.04 99.99 99.99     99.99 99.99      6.29  9    5.52  2    5.77  2    99.99 99    99.99 99
131923      6.35 99.99  0.71 99.99 99.99     99.99 99.99     99.99 99    6.65  3    6.13  3    99.99 99    99.99 99
132242      6.10 99.99  0.60 99.99 99.99     99.99 99.99     99.99 99    6.34  3    5.94  3    99.99 99    99.99 99
132375      6.09  0.05  0.50 99.99 99.99     99.99 99.99      7.47  3    6.27  3    5.97  3    99.99 99    99.99 99
132851      5.85 99.99  0.16 99.99 99.99     99.99 99.99     99.99 99    5.84  3    5.89  2    99.99 99    99.99 99
133049      6.52  1.86  1.59 99.99 99.99     99.99 99.99     10.95 12    7.30  5    5.90  5    99.99 99    99.99 99
133456      6.17  1.70  1.47 99.99 99.99     99.99 99.99     10.32 12    6.88  4    5.60  5    99.99 99    99.99 99
133529      6.67 99.99 -0.01 99.99 99.99     99.99 99.99     99.99 99    6.57  3    6.79  2    99.99 99    99.99 99
133631      5.77 99.99  0.92 99.99 99.99     99.99 99.99     99.99 99    6.18  4    5.46  3    99.99 99    99.99 99
133683      5.76  0.39  0.69 99.99 99.99     99.99 99.99      7.70  4    6.04  3    5.55  3    99.99 99    99.99 99
133981      6.01 -0.24  0.00 99.99 99.99     99.99 99.99      6.72  8    5.92  3    6.12  2    99.99 99    99.99 99
134047      6.16  0.71  0.94 99.99 99.99     99.99 99.99      8.67  5    6.58  4    5.84  3    99.99 99    99.99 99
134060      6.30 99.99  0.63 99.99 99.99     99.99 99.99     99.99 99    6.55  3    6.12  3    99.99 99    99.99 99
134482      5.69  0.09  0.14 99.99 99.99     99.99 99.99      6.92 12    5.67  3    5.74  2    99.99 99    99.99 99
134505      3.41  0.66  0.92 99.99 99.99     99.99 99.99      5.85  4    3.82  4    3.10  3    99.99 99    99.99 99
134597      6.33 99.99  1.12 99.99 99.99     99.99 99.99     99.99 99    6.85  4    5.92  4    99.99 99    99.99 99
134837      6.10 99.99 -0.08 99.99 99.99     99.99 99.99     99.99 99    5.96  2    6.25  2    99.99 99    99.99 99
134946      6.47 99.99 -0.04 99.99 99.99     99.99 99.99     99.99 99    6.35  2    6.60  2    99.99 99    99.99 99
134967      6.08 99.99  0.12 99.99 99.99     99.99 99.99     99.99 99    6.05  3    6.14  2    99.99 99    99.99 99
134987      6.45 99.99  0.70 99.99 99.99     99.99 99.99     99.99 99    6.74  3    6.24  3    99.99 99    99.99 99
135160      5.73 -0.87 -0.08 99.99 99.99     99.99 99.99      5.54  5    5.59  2    5.88  2    99.99 99    99.99 99
135291      4.86  1.32  1.25 99.99 99.99     99.99 99.99      8.37  9    5.45  4    4.39  4    99.99 99    99.99 99
135379      4.07  0.09  0.09 99.99 99.99     99.99 99.99      5.28 13    4.02  3    4.14  2    99.99 99    99.99 99
135382      2.89 -0.02  0.00 99.99 99.99     99.99 99.99      3.90 13    2.80  3    3.00  2    99.99 99    99.99 99
135559      5.63  0.09  0.18 99.99 99.99     99.99 99.99      6.89 11    5.63  3    5.66  2    99.99 99    99.99 99
```

```
135730      6.28 99.99   0.18 99.99 99.99      99.99 99.99     99.99 99    6.28  3    6.31  2    99.99 99    99.99 99
135737      6.28 -0.60  -0.09 99.99 99.99      99.99 99.99      6.45  4    6.14  2    6.44  2    99.99 99    99.99 99
136014      6.20 99.99   0.96 99.99 99.99      99.99 99.99     99.99 99    6.63  4    5.87  4    99.99 99    99.99 99
136064      5.13  0.08   0.53 99.99 99.99      99.99 99.99      6.56  3    5.33  3    5.00  3    99.99 99    99.99 99
136351      5.00 99.99   0.50 99.99 99.99      99.99 99.99     99.99 99    5.18  3    4.88  3    99.99 99    99.99 99
136352      5.65 99.99   0.65 99.99 99.99      99.99 99.99     99.99 99    5.91  3    5.46  3    99.99 99    99.99 99
136359      5.67 99.99   0.48 99.99 99.99      99.99 99.99     99.99 99    5.84  3    5.56  3    99.99 99    99.99 99
136366      6.17 99.99   1.02 99.99 99.99      99.99 99.99     99.99 99    6.64  4    5.81  4    99.99 99    99.99 99
136512      5.51  0.77   1.02 99.99 99.99      99.99 99.99      8.15  4    5.98  4    5.15  4    99.99 99    99.99 99
136514      5.35  1.22   1.19 99.99 99.99      99.99 99.99      8.69  9    5.91  4    4.91  4    99.99 99    99.99 99
136672      5.89  0.77   1.02 99.99 99.99      99.99 99.99      8.53  4    6.36  4    5.53  4    99.99 99    99.99 99
137052      4.94  0.02   0.44 99.99 99.99      99.99 99.99      6.24  3    5.09  3    4.85  3    99.99 99    99.99 99
137058      4.60 -0.06   0.00 99.99 99.99      99.99 99.99      5.55 12    4.51  3    4.71  2    99.99 99    99.99 99
137333      5.57  0.08   0.11 99.99 99.99      99.99 99.99      6.77 12    5.54  3    5.63  2    99.99 99    99.99 99
137392      6.50  0.13   0.59 99.99 99.99      99.99 99.99      8.04  3    6.73  3    6.34  3    99.99 99    99.99 99
137471      5.17  1.95   1.66 99.99 99.99      99.99 99.99      9.76 12    5.99  5    4.51  5    99.99 99    99.99 99
137704      5.46  1.64   1.40 99.99 99.99      99.99 99.99      9.49 12    6.14  4    4.92  4    99.99 99    99.99 99
137709      5.24 99.99   1.74 99.99 99.99      99.99 99.99     99.99 99    6.10  5    4.55  5    99.99 99    99.99 99
138204      6.25 99.99   0.20 99.99 99.99      99.99 99.99     99.99 99    6.27  3    6.27  2    99.99 99    99.99 99
138268      6.22 99.99   0.22 99.99 99.99      99.99 99.99     99.99 99    6.25  3    6.23  2    99.99 99    99.99 99
138289      6.18  1.36   1.21 99.99 99.99      99.99 99.99      9.72 11    6.75  4    5.73  4    99.99 99    99.99 99
138498      6.51  0.03   0.35 99.99 99.99      99.99 99.99      7.78  5    6.61  3    6.46  2    99.99 99    99.99 99
138538      4.11  1.16   1.17 99.99 99.99      99.99 99.99      7.36  8    4.66  4    3.68  4    99.99 99    99.99 99
138763      6.51  0.04   0.58 99.99 99.99      99.99 99.99      7.92  4    6.73  3    6.36  3    99.99 99    99.99 99
138800      5.65 -0.38  -0.04 99.99 99.99      99.99 99.99      6.15  6    5.53  2    5.78  2    99.99 99    99.99 99
138816      5.43 99.99   1.50 99.99 99.99      99.99 99.99     99.99 99    6.16  5    4.85  5    99.99 99    99.99 99
138867      5.95 -0.16  -0.04 99.99 99.99      99.99 99.99      6.75 11    5.83  2    6.08  2    99.99 99    99.99 99
138965      6.44  0.08   0.08 99.99 99.99      99.99 99.99      7.63 13    6.39  3    6.52  2    99.99 99    99.99 99
139129      5.44 99.99   0.00 99.99 99.99      99.99 99.99     99.99 99    5.35  3    5.55  2    99.99 99    99.99 99
139211      5.95 99.99   0.48 99.99 99.99      99.99 99.99     99.99 99    6.12  3    5.84  3    99.99 99    99.99 99
139446      5.38  0.54   0.86 99.99 99.99      99.99 99.99      7.62  4    5.76  4    5.10  3    99.99 99    99.99 99
139641      5.24  0.53   0.88 99.99 99.99      99.99 99.99      7.48  3    5.63  4    4.95  3    99.99 99    99.99 99
139777      6.58  0.13   0.67 99.99 99.99      99.99 99.99      8.16  4    6.85  3    6.38  3    99.99 99    99.99 99
139915      6.49 99.99   1.05 99.99 99.99      99.99 99.99     99.99 99    6.97  4    6.12  4    99.99 99    99.99 99
140285      5.94 -0.20   0.00 99.99 99.99      99.99 99.99      6.70  9    5.85  3    6.05  2    99.99 99    99.99 99
140417      5.41 99.99   0.23 99.99 99.99      99.99 99.99     99.99 99    5.44  3    5.42  2    99.99 99    99.99 99
140483      5.53  0.10   0.23 99.99 99.99      99.99 99.99      6.83 10    5.56  3    5.54  2    99.99 99    99.99 99
140722      6.51 99.99   0.32 99.99 99.99      99.99 99.99     99.99 99    6.59  3    6.48  2    99.99 99    99.99 99
140784      5.61 -0.42  -0.11 99.99 99.99      99.99 99.99      6.02  7    5.45  2    5.78  2    99.99 99    99.99 99
140815      6.33  1.20   1.19 99.99 99.99      99.99 99.99      9.65  8    6.89  4    5.89  4    99.99 99    99.99 99
140901      6.01 99.99   0.72 99.99 99.99      99.99 99.99     99.99 99    6.31  3    5.79  3    99.99 99    99.99 99
```

```
140979       6.07 99.99  1.38 99.99 99.99    99.99 99.99   99.99 99   6.73  4   5.54  4   99.99 99   99.99 99
141168       5.77 -0.23 -0.08 99.99 99.99    99.99 99.99    6.45 11   5.63  2   5.92  2   99.99 99   99.99 99
141194       5.84  0.06  0.07 99.99 99.99    99.99 99.99    6.99 13   5.78  3   5.92  2   99.99 99   99.99 99
141296       6.12 99.99  0.29 99.99 99.99    99.99 99.99   99.99 99   6.18  3   6.10  2   99.99 99   99.99 99
141378       5.53  0.11  0.12 99.99 99.99    99.99 99.99    6.78 13   5.50  3   5.59  2   99.99 99   99.99 99
141413       6.54  0.12  0.19 99.99 99.99    99.99 99.99    7.84 11   6.55  3   6.57  2   99.99 99   99.99 99
141544       6.01 99.99  1.16 99.99 99.99    99.99 99.99   99.99 99   6.55  4   5.59  4   99.99 99   99.99 99
141585       6.19 99.99  1.47 99.99 99.99    99.99 99.99   99.99 99   6.90  4   5.62  5   99.99 99   99.99 99
141767       5.09  0.93  1.13 99.99 99.99    99.99 99.99    8.01  4   5.62  4   4.68  4   99.99 99   99.99 99
141832       6.40 99.99  0.98 99.99 99.99    99.99 99.99   99.99 99   6.84  4   6.06  4   99.99 99   99.99 99
141851       5.11  0.07  0.12 99.99 99.99    99.99 99.99    6.31 12   5.08  3   5.17  2   99.99 99   99.99 99
141913       6.15 99.99  0.10 99.99 99.99    99.99 99.99   99.99 99   6.11  3   6.22  2   99.99 99   99.99 99
142049       5.77 99.99  0.34 99.99 99.99    99.99 99.99   99.99 99   5.86  3   5.73  2   99.99 99   99.99 99
142139       5.76 99.99  0.06 99.99 99.99    99.99 99.99   99.99 99   5.70  3   5.85  2   99.99 99   99.99 99
142267       6.10  0.00  0.60 99.99 99.99    99.99 99.99    7.47  4   6.34  3   5.94  3   99.99 99   99.99 99
142514       5.75 -0.40 -0.06 99.99 99.99    99.99 99.99    6.21  6   5.62  2   5.89  2   99.99 99   99.99 99
142529       6.30 99.99  0.37 99.99 99.99    99.99 99.99   99.99 99   6.41  3   6.24  2   99.99 99   99.99 99
142629       4.59  0.08  0.09 99.99 99.99    99.99 99.99    5.78 13   4.54  3   4.66  2   99.99 99   99.99 99
142908       5.45  0.03  0.33 99.99 99.99    99.99 99.99    6.71  6   5.54  3   5.41  2   99.99 99   99.99 99
143009       4.99 99.99  1.00 99.99 99.99    99.99 99.99   99.99 99   5.45  4   4.64  4   99.99 99   99.99 99
143101       6.13 99.99  0.25 99.99 99.99    99.99 99.99   99.99 99   6.17  3   6.13  2   99.99 99   99.99 99
143238       6.25 99.99 -0.04 99.99 99.99    99.99 99.99   99.99 99   6.13  2   6.38  2   99.99 99   99.99 99
143248       6.21 -0.01  0.01 99.99 99.99    99.99 99.99    7.24 13   6.12  3   6.32  2   99.99 99   99.99 99
143333       5.48  0.02  0.52 99.99 99.99    99.99 99.99    6.83  3   5.67  3   5.35  3   99.99 99   99.99 99
143346       5.70  1.26  1.17 99.99 99.99    99.99 99.99    9.09 10   6.25  4   5.27  4   99.99 99   99.99 99
143459       5.55 -0.05  0.05 99.99 99.99    99.99 99.99    6.54 11   5.48  3   5.64  2   99.99 99   99.99 99
143474       4.63  0.09  0.24 99.99 99.99    99.99 99.99    5.92  9   4.67  3   4.63  2   99.99 99   99.99 99
143666       5.12  0.74  0.98 99.99 99.99    99.99 99.99    7.70  4   5.56  4   4.78  4   99.99 99   99.99 99
143761       5.40  0.09  0.60 99.99 99.99    99.99 99.99    6.89  3   5.64  3   5.24  3   99.99 99   99.99 99
143902       6.10 99.99  0.34 99.99 99.99    99.99 99.99   99.99 99   6.19  3   6.06  2   99.99 99   99.99 99
144069       4.17  0.01  0.47 99.99 99.99    99.99 99.99    5.48  3   4.33  3   4.07  3   99.99 99   99.99 99
144218       4.92 -0.70 -0.02 99.99 99.99    99.99 99.99    5.00  4   4.81  2   5.04  2   99.99 99   99.99 99
144415       5.73 99.99  0.29 99.99 99.99    99.99 99.99   99.99 99   5.79  3   5.71  2   99.99 99   99.99 99
144426       6.29  0.11  0.08 99.99 99.99    99.99 99.99    7.52 14   6.24  3   6.37  2   99.99 99   99.99 99
144585       6.32 99.99  0.66 99.99 99.99    99.99 99.99   99.99 99   6.59  3   6.13  3   99.99 99   99.99 99
144889       6.14  1.51  1.37 99.99 99.99    99.99 99.99    9.97 10   6.80  4   5.62  4   99.99 99   99.99 99
145100       6.47 99.99  0.44 99.99 99.99    99.99 99.99   99.99 99   6.62  3   6.38  3   99.99 99   99.99 99
145148       5.98  0.76  1.02 99.99 99.99    99.99 99.99    8.61  4   6.45  4   5.62  4   99.99 99   99.99 99
145191       5.86 99.99  0.27 99.99 99.99    99.99 99.99   99.99 99   5.91  3   5.85  2   99.99 99   99.99 99
145206       5.37  1.46  1.45 99.99 99.99    99.99 99.99    9.18  7   6.07  4   4.81  5   99.99 99   99.99 99
145361       5.81 99.99  0.34 99.99 99.99    99.99 99.99   99.99 99   5.90  3   5.77  2   99.99 99   99.99 99
```

```
145388       5.27   1.62   1.41 99.99 99.99    99.99 99.99     9.28 12   5.95   4    4.73   4   99.99 99   99.99 99
145483       5.69 99.99   0.02 99.99 99.99    99.99 99.99    99.99 99   5.61   3    5.80   2   99.99 99   99.99 99
145544       3.85   0.86   1.11 99.99 99.99    99.99 99.99     6.66  4   4.37   4    3.45   4   99.99 99   99.99 99
145589       6.53   0.10   0.24 99.99 99.99    99.99 99.99     7.83  9   6.57   3    6.53   2   99.99 99   99.99 99
145689       5.75   0.09   0.15 99.99 99.99    99.99 99.99     6.99 11   5.74   3    5.80   2   99.99 99   99.99 99
145782       5.63 99.99   0.14 99.99 99.99    99.99 99.99    99.99 99   5.61   3    5.68   2   99.99 99   99.99 99
146143       4.99   0.51   0.80 99.99 99.99    99.99 99.99     7.16  4   5.34   3    4.73   3   99.99 99   99.99 99
146145       6.33 99.99   0.28 99.99 99.99    99.99 99.99    99.99 99   6.39   3    6.31   2   99.99 99   99.99 99
146233       5.49   0.17   0.65 99.99 99.99    99.99 99.99     7.11  3   5.75   3    5.30   3   99.99 99   99.99 99
146667       5.45 99.99   0.10 99.99 99.99    99.99 99.99    99.99 99   5.41   3    5.52   2   99.99 99   99.99 99
147349       6.15 99.99   0.03 99.99 99.99    99.99 99.99    99.99 99   6.07   3    6.25   2   99.99 99   99.99 99
147513       5.39 99.99   0.62 99.99 99.99    99.99 99.99    99.99 99   5.64   3    5.22   3   99.99 99   99.99 99
147675       3.89   0.62   0.91 99.99 99.99    99.99 99.99     6.27  4   4.30   4    3.58   3   99.99 99   99.99 99
147722       5.41 99.99   0.60 99.99 99.99    99.99 99.99    99.99 99   5.65   3    5.25   3   99.99 99   99.99 99
147869       5.85   0.01  -0.01 99.99 99.99    99.99 99.99     6.89 14   5.75   3    5.97   2   99.99 99   99.99 99
147977       5.69 99.99   0.00 99.99 99.99    99.99 99.99    99.99 99   5.60   3    5.80   2   99.99 99   99.99 99
148451       6.57   0.56   0.91 99.99 99.99    99.99 99.99     8.86  3   6.98   4    6.26   3   99.99 99   99.99 99
148488       5.50   1.24   1.22 99.99 99.99    99.99 99.99     8.89  8   6.08   4    5.05   4   99.99 99   99.99 99
148513       5.39   1.80   1.46 99.99 99.99    99.99 99.99     9.67 14   6.10   4    4.83   5   99.99 99   99.99 99
148890       5.52   0.73   0.93 99.99 99.99    99.99 99.99     8.06  5   5.94   4    5.20   3   99.99 99   99.99 99
149324       4.24   0.95   1.06 99.99 99.99    99.99 99.99     7.15  6   4.73   4    3.86   4   99.99 99   99.99 99
149671       5.91  -0.42  -0.08 99.99 99.99    99.99 99.99     6.33  7   5.77   2    6.06   2   99.99 99   99.99 99
149837       6.18 99.99   0.48 99.99 99.99    99.99 99.99    99.99 99   6.35   3    6.07   3   99.99 99   99.99 99
150026       6.03   0.02   0.02 99.99 99.99    99.99 99.99     7.10 13   5.95   3    6.14   2   99.99 99   99.99 99
150168       5.66 99.99  -0.03 99.99 99.99    99.99 99.99    99.99 99   5.55   2    5.79   2   99.99 99   99.99 99
150275       6.34   0.71   1.00 99.99 99.99    99.99 99.99     8.89  4   6.80   4    5.99   4   99.99 99   99.99 99
150331       5.87 99.99   0.65 99.99 99.99    99.99 99.99    99.99 99   6.13   3    5.68   3   99.99 99   99.99 99
150366       6.09 99.99   0.20 99.99 99.99    99.99 99.99    99.99 99   6.11   3    6.11   2   99.99 99   99.99 99
150451       6.24   0.07   0.30 99.99 99.99    99.99 99.99     7.53  7   6.31   3    6.22   2   99.99 99   99.99 99
150798       1.92   1.56   1.44 99.99 99.99    99.99 99.99     5.86 10   2.62   4    1.37   5   99.99 99   99.99 99
151199       6.16   0.11   0.07 99.99 99.99    99.99 99.99     7.38 14   6.10   3    6.24   2   99.99 99   99.99 99
151217       5.15   1.94   1.53 99.99 99.99    99.99 99.99     9.66 16   5.90   5    4.55   5   99.99 99   99.99 99
151249       3.76   1.93   1.57 99.99 99.99    99.99 99.99     8.27 14   4.53   5    3.14   5   99.99 99   99.99 99
151404       6.32   1.51   1.28 99.99 99.99    99.99 99.99    10.11 13   6.93   4    5.84   4   99.99 99   99.99 99
151441       6.13  -0.28  -0.02 99.99 99.99    99.99 99.99     6.77  8   6.02   2    6.25   2   99.99 99   99.99 99
151566       6.47 99.99   0.31 99.99 99.99    99.99 99.99    99.99 99   6.55   3    6.44   2   99.99 99   99.99 99
151900       6.32  -0.05   0.42 99.99 99.99    99.99 99.99     7.52  3   6.46   3    6.24   3   99.99 99   99.99 99
151956       5.49   0.11   0.10 99.99 99.99    99.99 99.99     6.73 13   5.45   3    5.56   2   99.99 99   99.99 99
151967       5.94 99.99   1.59 99.99 99.99    99.99 99.99    99.99 99   6.72   5    5.32   5   99.99 99   99.99 99
152127       5.51   0.03   0.05 99.99 99.99    99.99 99.99     6.61 13   5.44   3    5.60   2   99.99 99   99.99 99
152636       6.37 99.99   1.71 99.99 99.99    99.99 99.99    99.99 99   7.22   5    5.69   5   99.99 99   99.99 99
```

```
152786       3.13  1.96  1.60 99.99 99.99    99.99 99.99    7.70 14    3.92  5    2.50  5    99.99 99    99.99 99
152820       5.48 99.99  1.59 99.99 99.99    99.99 99.99   99.99 99    6.26  5    4.86  5    99.99 99    99.99 99
152879       5.35  1.66  1.41 99.99 99.99    99.99 99.99    9.41 13    6.03  4    4.81  4    99.99 99    99.99 99
152980       4.06  1.71  1.45 99.99 99.99    99.99 99.99    8.21 13    4.76  4    3.50  5    99.99 99    99.99 99
153021       6.19 99.99  1.00 99.99 99.99    99.99 99.99   99.99 99    6.65  4    5.84  4    99.99 99    99.99 99
153053       5.65 99.99  0.19 99.99 99.99    99.99 99.99   99.99 99    5.66  3    5.68  2    99.99 99    99.99 99
153072       6.09 99.99  0.19 99.99 99.99    99.99 99.99   99.99 99    6.10  3    6.12  2    99.99 99    99.99 99
153221       6.00 99.99  0.88 99.99 99.99    99.99 99.99   99.99 99    6.39  4    5.71  3    99.99 99    99.99 99
153258       6.65 99.99  1.80 99.99 99.99    99.99 99.99   99.99 99    7.55  5    5.93  5    99.99 99    99.99 99
153370       6.45 99.99  0.27 99.99 99.99    99.99 99.99   99.99 99    6.50  3    6.44  2    99.99 99    99.99 99
153472       6.34  1.41  1.28 99.99 99.99    99.99 99.99    9.99 11    6.95  4    5.86  4    99.99 99    99.99 99
153687       4.83  1.80  1.48 99.99 99.99    99.99 99.99    9.12 14    5.55  4    4.26  5    99.99 99    99.99 99
153914       6.33  0.06  0.09 99.99 99.99    99.99 99.99    7.50 12    6.28  3    6.40  2    99.99 99    99.99 99
154278       6.09  0.85  1.02 99.99 99.99    99.99 99.99    8.84  5    6.56  4    5.73  4    99.99 99    99.99 99
154556       6.22  1.04  1.06 99.99 99.99    99.99 99.99    9.25  8    6.71  4    5.84  4    99.99 99    99.99 99
154610       6.37  1.73  1.45 99.99 99.99    99.99 99.99   10.55 13    7.07  4    5.81  5    99.99 99    99.99 99
154733       5.55  1.52  1.30 99.99 99.99    99.99 99.99    9.36 12    6.17  4    5.06  4    99.99 99    99.99 99
154783       5.97 99.99  0.25 99.99 99.99    99.99 99.99   99.99 99    6.01  3    5.97  2    99.99 99    99.99 99
154903       5.89  0.90  1.06 99.99 99.99    99.99 99.99    8.73  5    6.38  4    5.51  4    99.99 99    99.99 99
154948       5.08  0.58  0.86 99.99 99.99    99.99 99.99    7.37  4    5.46  4    4.80  3    99.99 99    99.99 99
154972       6.25  0.00 -0.01 99.99 99.99    99.99 99.99    7.28 14    6.15  3    6.37  2    99.99 99    99.99 99
155259       5.67 99.99  0.04 99.99 99.99    99.99 99.99   99.99 99    5.60  3    5.77  2    99.99 99    99.99 99
155450       6.01 99.99  0.07 99.99 99.99    99.99 99.99   99.99 99    5.95  3    6.09  2    99.99 99    99.99 99
155500       6.33  0.86  1.04 99.99 99.99    99.99 99.99    9.10  5    6.81  4    5.96  4    99.99 99    99.99 99
155826       5.96 99.99  0.58 99.99 99.99    99.99 99.99   99.99 99    6.18  3    5.81  3    99.99 99    99.99 99
155875       6.53  0.15  0.60 99.99 99.99    99.99 99.99    8.10  3    6.77  3    6.37  3    99.99 99    99.99 99
155974       6.12 99.99  0.48 99.99 99.99    99.99 99.99   99.99 99    6.29  3    6.01  3    99.99 99    99.99 99
156190       5.41 -0.23 -0.04 99.99 99.99    99.99 99.99    6.11  9    5.29  2    5.54  2    99.99 99    99.99 99
156277       4.78  1.27  1.21 99.99 99.99    99.99 99.99    8.20  9    5.35  4    4.33  4    99.99 99    99.99 99
156293       5.77 -0.14 -0.05 99.99 99.99    99.99 99.99    6.59 12    5.65  2    5.91  2    99.99 99    99.99 99
156398       6.65  0.00  0.20 99.99 99.99    99.99 99.99    7.79  8    6.67  3    6.67  2    99.99 99    99.99 99
156681       5.03  1.90  1.55 99.99 99.99    99.99 99.99    9.49 14    5.79  5    4.42  5    99.99 99    99.99 99
156768       5.88 99.99  1.07 99.99 99.99    99.99 99.99   99.99 99    6.37  4    5.50  4    99.99 99    99.99 99
156826       6.32  0.47  0.85 99.99 99.99    99.99 99.99    8.46  3    6.69  4    6.04  3    99.99 99    99.99 99
156846       6.52 99.99  0.58 99.99 99.99    99.99 99.99   99.99 99    6.74  3    6.37  3    99.99 99    99.99 99
156854       5.80 99.99  1.00 99.99 99.99    99.99 99.99   99.99 99    6.26  4    5.45  4    99.99 99    99.99 99
156942       5.77 99.99 -0.08 99.99 99.99    99.99 99.99   99.99 99    5.63  2    5.92  2    99.99 99    99.99 99
156971       6.45 99.99  0.33 99.99 99.99    99.99 99.99   99.99 99    6.54  3    6.41  2    99.99 99    99.99 99
157060       6.47 99.99  0.54 99.99 99.99    99.99 99.99   99.99 99    6.67  3    6.33  3    99.99 99    99.99 99
157097       5.93 99.99  1.08 99.99 99.99    99.99 99.99   99.99 99    6.43  4    5.54  4    99.99 99    99.99 99
157214       5.39  0.07  0.62 99.99 99.99    99.99 99.99    6.86  4    5.64  3    5.22  3    99.99 99    99.99 99
```

```
157243     5.12 -0.40 -0.06 99.99 99.99    99.99 99.99     5.58  6   4.99  2   5.26  2   99.99 99   99.99 99
157244     2.85  1.56  1.46 99.99 99.99    99.99 99.99     6.80  9   3.56  4   2.29  5   99.99 99   99.99 99
157246     3.34 -0.95 -0.13 99.99 99.99    99.99 99.99     3.02  6   3.17  2   3.52  2   99.99 99   99.99 99
157347     6.29  0.24  0.68 99.99 99.99    99.99 99.99     8.02  3   6.57  3   6.09  3   99.99 99   99.99 99
157457     5.23 99.99  1.06 99.99 99.99    99.99 99.99    99.99 99   5.72  4   4.85  4   99.99 99   99.99 99
157617     5.77  1.27  1.25 99.99 99.99    99.99 99.99     9.21  8   6.36  4   5.30  4   99.99 99   99.99 99
157661     5.29 -0.34 -0.07 99.99 99.99    99.99 99.99     5.83  8   5.16  2   5.44  2   99.99 99   99.99 99
157778     4.17 -0.06  0.00 99.99 99.99    99.99 99.99     5.12 12   4.08  3   4.28  2   99.99 99   99.99 99
157856     6.44 -0.02  0.46 99.99 99.99    99.99 99.99     7.70  3   6.60  3   6.34  3   99.99 99   99.99 99
158094     3.62 -0.31 -0.10 99.99 99.99    99.99 99.99     4.18  9   3.47  2   3.78  2   99.99 99   99.99 99
158156     6.39 99.99  0.09 99.99 99.99    99.99 99.99    99.99 99   6.34  3   6.46  2   99.99 99   99.99 99
158170     6.37  0.16  0.58 99.99 99.99    99.99 99.99     7.94  3   6.59  3   6.22  3   99.99 99   99.99 99
158463     6.37  0.67  0.93 99.99 99.99    99.99 99.99     8.82  4   6.79  4   6.05  3   99.99 99   99.99 99
158614     5.31  0.31  0.72 99.99 99.99    99.99 99.99     7.16  3   5.61  3   5.09  3   99.99 99   99.99 99
158633     6.43  0.29  0.76 99.99 99.99    99.99 99.99     8.28  3   6.75  3   6.19  3   99.99 99   99.99 99
158895     6.28 99.99 -0.08 99.99 99.99    99.99 99.99    99.99 99   6.14  2   6.43  2   99.99 99   99.99 99
159463     5.93 99.99  1.10 99.99 99.99    99.99 99.99    99.99 99   6.44  4   5.53  4   99.99 99   99.99 99
159480     5.49  0.01  0.02 99.99 99.99    99.99 99.99     6.55 13   5.41  3   5.60  2   99.99 99   99.99 99
159492     5.25 99.99  0.20 99.99 99.99    99.99 99.99    99.99 99   5.27  3   5.27  2   99.99 99   99.99 99
159517     6.45  0.02  0.44 99.99 99.99    99.99 99.99     7.75  3   6.60  3   6.36  3   99.99 99   99.99 99
159707     6.10 -0.26 -0.06 99.99 99.99    99.99 99.99     6.75  9   5.97  2   6.24  2   99.99 99   99.99 99
159964     6.49 -0.04  0.48 99.99 99.99    99.99 99.99     7.73  3   6.66  3   6.38  3   99.99 99   99.99 99
159966     5.05  0.92  1.08 99.99 99.99    99.99 99.99     7.93  5   5.55  4   4.66  4   99.99 99   99.99 99
160263     5.79 -0.06 -0.01 99.99 99.99    99.99 99.99     6.74 12   5.69  3   5.91  2   99.99 99   99.99 99
160269     5.23  0.10  0.61 99.99 99.99    99.99 99.99     6.74  3   5.47  3   5.06  3   99.99 99   99.99 99
160315     6.25  0.83  1.02 99.99 99.99    99.99 99.99     8.97  5   6.72  4   5.89  4   99.99 99   99.99 99
160635     3.62  1.17  1.19 99.99 99.99    99.99 99.99     6.89  8   4.18  4   3.18  4   99.99 99   99.99 99
160691     5.15 99.99  0.70 99.99 99.99    99.99 99.99    99.99 99   5.44  3   4.94  3   99.99 99   99.99 99
160781     5.95  1.18  1.26 99.99 99.99    99.99 99.99     9.28  6   6.55  4   5.48  4   99.99 99   99.99 99
161023     6.38 99.99  0.37 99.99 99.99    99.99 99.99    99.99 99   6.49  3   6.32  2   99.99 99   99.99 99
161074     5.51  1.68  1.46 99.99 99.99    99.99 99.99     9.62 12   6.22  4   4.95  5   99.99 99   99.99 99
161390     6.43 99.99 -0.02 99.99 99.99    99.99 99.99    99.99 99   6.32  2   6.55  2   99.99 99   99.99 99
161471     3.03  0.26  0.51 99.99 99.99    99.99 99.99     4.70  6   3.22  3   2.91  3   99.99 99   99.99 99
161840     4.83 -0.29 -0.04 99.99 99.99    99.99 99.99     5.45  8   4.71  2   4.96  2   99.99 99   99.99 99
161892     3.21  1.19  1.17 99.99 99.99    99.99 99.99     6.50  9   3.76  4   2.78  4   99.99 99   99.99 99
161912     4.81  0.00  0.26 99.99 99.99    99.99 99.99     5.99  7   4.86  3   4.80  2   99.99 99   99.99 99
161941     6.22 -0.03  0.16 99.99 99.99    99.99 99.99     7.30  9   6.21  3   6.26  2   99.99 99   99.99 99
161955     6.49  1.00  1.08 99.99 99.99    99.99 99.99     9.47  7   6.99  4   6.10  4   99.99 99   99.99 99
161988     6.07  1.28  1.20 99.99 99.99    99.99 99.99     9.50 10   6.64  4   5.63  4   99.99 99   99.99 99
162337     6.35  1.75  1.50 99.99 99.99    99.99 99.99    10.58 12   7.08  5   5.77  5   99.99 99   99.99 99
162396     6.20 99.99  0.54 99.99 99.99    99.99 99.99    99.99 99   6.40  3   6.06  3   99.99 99   99.99 99
```

```
162917      5.77 -0.01  0.42 99.99 99.99     99.99 99.99      7.02  3   5.91  3   5.69  3   99.99 99   99.99 99
162978      6.20 -0.89  0.04 99.99 99.99     99.99 99.99      6.05  9   6.13  3   6.30  2   99.99 99   99.99 99
163145      4.86  1.22  1.21 99.99 99.99     99.99 99.99      8.21  8   5.43  4   4.41  4   99.99 99   99.99 99
163376      4.88  1.96  1.65 99.99 99.99     99.99 99.99      9.48 13   5.69  5   4.23  5   99.99 99   99.99 99
163641      6.29 -0.25  0.00 99.99 99.99     99.99 99.99      6.99  8   6.20  3   6.40  2   99.99 99   99.99 99
163755      4.99 99.99  1.62 99.99 99.99     99.99 99.99     99.99 99   5.79  5   4.35  5   99.99 99   99.99 99
164432      6.33 -0.75 -0.08 99.99 99.99     99.99 99.99      6.30  3   6.19  2   6.48  2   99.99 99   99.99 99
164461      5.28  1.60  1.28 99.99 99.99     99.99 99.99      9.19 15   5.89  4   4.80  4   99.99 99   99.99 99
164637      6.74 99.99 -0.05 99.99 99.99     99.99 99.99     99.99 99   6.62  2   6.88  2   99.99 99   99.99 99
164712      5.86  1.43  1.24 99.99 99.99     99.99 99.99      9.51 12   6.45  4   5.40  4   99.99 99   99.99 99
164764      4.79  0.05  0.38 99.99 99.99     99.99 99.99      6.10  5   4.90  3   4.73  2   99.99 99   99.99 99
164871      6.41  1.41  1.26 99.99 99.99     99.99 99.99     10.05 11   7.01  4   5.94  4   99.99 99   99.99 99
165024      3.66 -0.84 -0.08 99.99 99.99     99.99 99.99      3.51  5   3.52  2   3.81  2   99.99 99   99.99 99
165040      4.35  0.18  0.22 99.99 99.99     99.99 99.99      5.75 12   4.38  3   4.36  2   99.99 99   99.99 99
165185      5.95 99.99  0.62 99.99 99.99     99.99 99.99     99.99 99   6.20  3   5.78  3   99.99 99   99.99 99
165190      4.95 99.99  0.22 99.99 99.99     99.99 99.99     99.99 99   4.98  3   4.96  2   99.99 99   99.99 99
165259      5.85  0.05  0.46 99.99 99.99     99.99 99.99      7.21  3   6.01  3   5.75  3   99.99 99   99.99 99
165493      6.15 99.99 -0.08 99.99 99.99     99.99 99.99     99.99 99   6.01  2   6.30  2   99.99 99   99.99 99
165499      5.49 99.99  0.58 99.99 99.99     99.99 99.99     99.99 99   5.71  3   5.34  3   99.99 99   99.99 99
165516      6.29 -0.78  0.12 99.99 99.99     99.99 99.99      6.33  8   6.26  3   6.35  2   99.99 99   99.99 99
165687      5.52  1.08  1.11 99.99 99.99     99.99 99.99      8.63  8   6.04  4   5.12  4   99.99 99   99.99 99
165861      6.73 -0.36 -0.03 99.99 99.99     99.99 99.99      7.26  7   6.62  2   6.86  2   99.99 99   99.99 99
165908      5.04 -0.10  0.52 99.99 99.99     99.99 99.99      6.23  4   5.23  3   4.91  3   99.99 99   99.99 99
166006      6.07 99.99  1.20 99.99 99.99     99.99 99.99     99.99 99   6.64  4   5.63  4   99.99 99   99.99 99
166023      5.53 99.99  0.97 99.99 99.99     99.99 99.99     99.99 99   5.97  4   5.19  4   99.99 99   99.99 99
166045      5.86  0.08  0.12 99.99 99.99     99.99 99.99      7.07 12   5.83  3   5.92  2   99.99 99   99.99 99
166046      5.90  0.08  0.14 99.99 99.99     99.99 99.99      7.12 12   5.88  3   5.95  2   99.99 99   99.99 99
166114      5.86 99.99  0.29 99.99 99.99     99.99 99.99     99.99 99   5.92  3   5.84  2   99.99 99   99.99 99
166197      6.16 99.99 -0.15 99.99 99.99     99.99 99.99     99.99 99   5.98  2   6.34  2   99.99 99   99.99 99
166229      5.47  1.21  1.17 99.99 99.99     99.99 99.99      8.79  9   6.02  4   5.04  4   99.99 99   99.99 99
166841      6.33 -0.23 -0.04 99.99 99.99     99.99 99.99      7.03  9   6.21  2   6.46  2   99.99 99   99.99 99
166865      6.04 -0.01  0.51 99.99 99.99     99.99 99.99      7.34  3   6.23  3   5.92  3   99.99 99   99.99 99
166866      5.68 -0.01  0.50 99.99 99.99     99.99 99.99      6.98  3   5.86  3   5.56  3   99.99 99   99.99 99
166926      5.79  0.07  0.25 99.99 99.99     99.99 99.99      7.06  8   5.83  3   5.79  2   99.99 99   99.99 99
166960      6.59  0.14  0.27 99.99 99.99     99.99 99.99      7.96  9   6.64  3   6.58  2   99.99 99   99.99 99
167042      5.94  0.72  0.94 99.99 99.99     99.99 99.99      8.47  5   6.36  4   5.62  3   99.99 99   99.99 99
167193      6.12  1.72  1.47 99.99 99.99     99.99 99.99     10.29 12   6.83  4   5.55  5   99.99 99   99.99 99
167263      5.98 -0.88  0.04 99.99 99.99     99.99 99.99      5.84  8   5.91  3   6.08  2   99.99 99   99.99 99
167264      5.38 -0.87  0.07 99.99 99.99     99.99 99.99      5.27  9   5.32  3   5.46  2   99.99 99   99.99 99
167425      6.18 99.99  0.58 99.99 99.99     99.99 99.99     99.99 99   6.40  3   6.03  3   99.99 99   99.99 99
167468      5.47  0.04  0.02 99.99 99.99     99.99 99.99      6.57 14   5.39  3   5.58  2   99.99 99   99.99 99
```

```
167665      6.40 99.99  0.54 99.99 99.99    99.99 99.99    99.99 99    6.60  3    6.26  3    99.99 99    99.99 99
167714      5.95  1.27  1.16 99.99 99.99    99.99 99.99     9.34 11    6.49  4    5.53  4    99.99 99    99.99 99
167768      6.00  0.55  0.89 99.99 99.99    99.99 99.99     8.27  3    6.39  4    5.70  3    99.99 99    99.99 99
167771      6.54 -0.84  0.12 99.99 99.99    99.99 99.99     6.50 10    6.51  3    6.60  2    99.99 99    99.99 99
168322      6.10  0.70  1.00 99.99 99.99    99.99 99.99     8.63  3    6.56  4    5.75  4    99.99 99    99.99 99
168339      4.36  1.55  1.48 99.99 99.99    99.99 99.99     8.31  8    5.08  4    3.79  5    99.99 99    99.99 99
168387      5.39  1.06  1.07 99.99 99.99    99.99 99.99     8.45  8    5.88  4    5.01  4    99.99 99    99.99 99
168905      5.25 99.99 -0.19 99.99 99.99    99.99 99.99    99.99 99    5.05  2    5.45  2    99.99 99    99.99 99
169233      5.60 99.99  1.14 99.99 99.99    99.99 99.99    99.99 99    6.13  4    5.18  4    99.99 99    99.99 99
169370      6.31  1.09  1.16 99.99 99.99    99.99 99.99     9.46  7    6.85  4    5.89  4    99.99 99    99.99 99
169570      5.89  0.83  0.98 99.99 99.99    99.99 99.99     8.59  6    6.33  4    5.55  4    99.99 99    99.99 99
169767      4.13  0.82  1.02 99.99 99.99    99.99 99.99     6.84  5    4.60  4    3.77  4    99.99 99    99.99 99
169904      6.27 -0.37 -0.13 99.99 99.99    99.99 99.99     6.73  9    6.10  2    6.45  2    99.99 99    99.99 99
170040      6.63 99.99 -0.02 99.99 99.99    99.99 99.99    99.99 99    6.52  2    6.75  2    99.99 99    99.99 99
170137      6.07  1.83  1.62 99.99 99.99    99.99 99.99    10.48 11    6.87  5    5.43  5    99.99 99    99.99 99
170465      4.96 99.99 -0.11 99.99 99.99    99.99 99.99    99.99 99    4.80  2    5.13  2    99.99 99    99.99 99
170525      6.44 99.99  0.68 99.99 99.99    99.99 99.99    99.99 99    6.72  3    6.24  3    99.99 99    99.99 99
170920      5.94  0.16  0.16 99.99 99.99    99.99 99.99     7.28 13    5.93  3    5.98  2    99.99 99    99.99 99
171161      7.16  1.30  1.26 99.99 99.99    99.99 99.99    10.65  9    7.76  4    6.69  4    99.99 99    99.99 99
171391      5.14 99.99  0.92 99.99 99.99    99.99 99.99    99.99 99    5.55  4    4.83  3    99.99 99    99.99 99
171759      4.01  1.02  1.14 99.99 99.99    99.99 99.99     7.05  6    4.54  4    3.59  4    99.99 99    99.99 99
171802      5.39 -0.02  0.37 99.99 99.99    99.99 99.99     6.60  4    5.50  3    5.33  2    99.99 99    99.99 99
171834      5.45 -0.04  0.37 99.99 99.99    99.99 99.99     6.63  4    5.56  3    5.39  2    99.99 99    99.99 99
171990      6.39  0.13  0.60 99.99 99.99    99.99 99.99     7.93  3    6.63  3    6.23  3    99.99 99    99.99 99
172021      6.37  0.14  0.15 99.99 99.99    99.99 99.99     7.68 13    6.36  3    6.42  2    99.99 99    99.99 99
172211      5.78  0.77  0.96 99.99 99.99    99.99 99.99     8.39  5    6.21  4    5.45  4    99.99 99    99.99 99
172223      6.49 99.99  1.22 99.99 99.99    99.99 99.99    99.99 99    7.07  4    6.04  4    99.99 99    99.99 99
172424      6.28  0.69  0.96 99.99 99.99    99.99 99.99     8.78  4    6.71  4    5.95  4    99.99 99    99.99 99
172555      4.79  0.08  0.20 99.99 99.99    99.99 99.99     6.04 10    4.81  3    4.81  2    99.99 99    99.99 99
172881      6.06 -0.09  0.00 99.99 99.99    99.99 99.99     6.97 11    5.97  3    6.17  2    99.99 99    99.99 99
173009      4.89 99.99  1.12 99.99 99.99    99.99 99.99    99.99 99    5.41  4    4.48  4    99.99 99    99.99 99
173093      6.31  0.02  0.48 99.99 99.99    99.99 99.99     7.64  3    6.48  3    6.20  3    99.99 99    99.99 99
173168      5.73  0.11  0.24 99.99 99.99    99.99 99.99     7.05 10    5.77  3    5.73  2    99.99 99    99.99 99
173495      5.83  0.03  0.04 99.99 99.99    99.99 99.99     6.93 13    5.76  3    5.93  2    99.99 99    99.99 99
173540      5.24 99.99  0.78 99.99 99.99    99.99 99.99    99.99 99    5.57  3    4.99  3    99.99 99    99.99 99
173649      5.74  0.06  0.28 99.99 99.99    99.99 99.99     7.01  7    5.80  3    5.72  2    99.99 99    99.99 99
173715      5.49 99.99  0.13 99.99 99.99    99.99 99.99    99.99 99    5.47  3    5.54  2    99.99 99    99.99 99
173791      5.81 99.99  0.90 99.99 99.99    99.99 99.99    99.99 99    6.21  4    5.51  3    99.99 99    99.99 99
174115      6.75 99.99  0.20 99.99 99.99    99.99 99.99    99.99 99    6.77  3    6.77  2    99.99 99    99.99 99
174866      6.34  0.11  0.20 99.99 99.99    99.99 99.99     7.63 11    6.36  3    6.36  2    99.99 99    99.99 99
175219      5.36 99.99  1.00 99.99 99.99    99.99 99.99    99.99 99    5.82  4    5.01  4    99.99 99    99.99 99
```

```
175329      5.14 99.99  1.37 99.99 99.99    99.99 99.99   99.99 99    5.80  4    4.62  4   99.99 99   99.99 99
175401      6.01  0.80  0.97 99.99 99.99    99.99 99.99    8.66  6    6.45  4    5.67  4   99.99 99   99.99 99
175515      5.57  0.87  1.04 99.99 99.99    99.99 99.99    8.36  5    6.05  4    5.20  4   99.99 99   99.99 99
175679      6.15  0.71  0.97 99.99 99.99    99.99 99.99    8.68  4    6.59  4    5.81  4   99.99 99   99.99 99
175751      4.80  1.02  1.08 99.99 99.99    99.99 99.99    7.81  7    5.30  4    4.41  4   99.99 99   99.99 99
175986      5.88  0.13  0.56 99.99 99.99    99.99 99.99    7.40  3    6.09  3    5.73  3   99.99 99   99.99 99
176095      6.21  0.01  0.46 99.99 99.99    99.99 99.99    7.51  3    6.37  3    6.11  3   99.99 99   99.99 99
176304      6.76 -0.44  0.25 99.99 99.99    99.99 99.99    7.34  5    6.80  3    6.76  2   99.99 99   99.99 99
176425      6.23 99.99  0.00 99.99 99.99    99.99 99.99   99.99 99    6.14  3    6.34  2   99.99 99   99.99 99
176593      6.32  0.78  1.00 99.99 99.99    99.99 99.99    8.96  5    6.78  4    5.97  4   99.99 99   99.99 99
176638      4.75 -0.07 -0.02 99.99 99.99    99.99 99.99    5.68 12    4.64  2    4.87  2   99.99 99   99.99 99
176664      5.93 99.99  1.24 99.99 99.99    99.99 99.99   99.99 99    6.52  4    5.47  4   99.99 99   99.99 99
176704      5.65 99.99  1.23 99.99 99.99    99.99 99.99   99.99 99    6.23  4    5.19  4   99.99 99   99.99 99
176981      6.30  1.82  1.62 99.99 99.99    99.99 99.99   10.69 11    7.10  5    5.66  5   99.99 99   99.99 99
176984      5.42 -0.07  0.00 99.99 99.99    99.99 99.99    6.36 12    5.33  3    5.53  2   99.99 99   99.99 99
177171      5.16 99.99  0.53 99.99 99.99    99.99 99.99   99.99 99    5.36  3    5.03  3   99.99 99   99.99 99
177332      6.73  0.14  0.13 99.99 99.99    99.99 99.99    8.03 13    6.71  3    6.78  2   99.99 99   99.99 99
177389      5.33  0.61  0.91 99.99 99.99    99.99 99.99    7.69  4    5.74  4    5.02  3   99.99 99   99.99 99
177474      4.21  0.02  0.52 99.99 99.99    99.99 99.99    5.56  3    4.40  3    4.08  3   99.99 99   99.99 99
177565      6.16 99.99  0.72 99.99 99.99    99.99 99.99   99.99 99    6.46  3    5.94  3   99.99 99   99.99 99
177693      6.49 99.99  1.10 99.99 99.99    99.99 99.99   99.99 99    7.00  4    6.09  4   99.99 99   99.99 99
177808      5.54  1.90  1.54 99.99 99.99    99.99 99.99   10.00 14    6.29  5    4.94  5   99.99 99   99.99 99
177873      4.59  1.03  1.10 99.99 99.99    99.99 99.99    7.63  7    5.10  4    4.19  4   99.99 99   99.99 99
178322      5.88 99.99 -0.08 99.99 99.99    99.99 99.99   99.99 99    5.74  2    6.03  2   99.99 99   99.99 99
178845      6.13 99.99  0.95 99.99 99.99    99.99 99.99   99.99 99    6.56  4    5.80  4   99.99 99   99.99 99
178937      6.57 99.99  1.02 99.99 99.99    99.99 99.99   99.99 99    7.04  4    6.21  4   99.99 99   99.99 99
179009      6.27  0.14  0.17 99.99 99.99    99.99 99.99    7.59 12    6.27  3    6.31  2   99.99 99   99.99 99
179366      5.53  0.13  0.18 99.99 99.99    99.99 99.99    6.84 12    5.53  3    5.56  2   99.99 99   99.99 99
179957      6.75  0.17  0.64 99.99 99.99    99.99 99.99    8.37  3    7.01  3    6.57  3   99.99 99   99.99 99
179958      6.57  0.21  0.65 99.99 99.99    99.99 99.99    8.25  3    6.83  3    6.38  3   99.99 99   99.99 99
180134      6.38 99.99  0.49 99.99 99.99    99.99 99.99   99.99 99    6.55  3    6.27  3   99.99 99   99.99 99
181019      6.34  1.25  1.23 99.99 99.99    99.99 99.99    9.75  8    6.92  4    5.88  4   99.99 99   99.99 99
181109      6.58 99.99  1.67 99.99 99.99    99.99 99.99   99.99 99    7.40  5    5.92  5   99.99 99   99.99 99
181122      6.32  0.88  1.05 99.99 99.99    99.99 99.99    9.12  5    6.80  4    5.95  4   99.99 99   99.99 99
181296      5.05 99.99  0.02 99.99 99.99    99.99 99.99   99.99 99    4.97  3    5.16  2   99.99 99   99.99 99
181321      6.48 99.99  0.63 99.99 99.99    99.99 99.99   99.99 99    6.73  3    6.30  3   99.99 99   99.99 99
181401      6.34 99.99  1.14 99.99 99.99    99.99 99.99   99.99 99    6.87  4    5.92  4   99.99 99   99.99 99
181440      5.49 -0.23 -0.04 99.99 99.99    99.99 99.99    6.19  9    5.37  2    5.62  2   99.99 99   99.99 99
181454      3.93 -0.32 -0.08 99.99 99.99    99.99 99.99    4.49  9    3.79  2    4.08  2   99.99 99   99.99 99
181623      4.29  0.07  0.34 99.99 99.99    99.99 99.99    5.61  6    4.38  3    4.25  2   99.99 99   99.99 99
181858      6.67 99.99 -0.03 99.99 99.99    99.99 99.99    6.56  2    6.80  2   99.99 99   99.99 99
```

```
181869      3.97 -0.33 -0.10 99.99 99.99    99.99 99.99     4.50  9    3.82  2    4.13  2    99.99 99    99.99 99
181907      5.83  0.97  1.09 99.99 99.99    99.99 99.99     8.78  6    6.33  4    5.44  4    99.99 99    99.99 99
182101      6.35 -0.03  0.44 99.99 99.99    99.99 99.99     7.59  3    6.50  3    6.26  3    99.99 99    99.99 99
182369      5.03 99.99  0.23 99.99 99.99    99.99 99.99    99.99 99    5.06  3    5.04  2    99.99 99    99.99 99
182572      5.16  0.42  0.78 99.99 99.99    99.99 99.99     7.19  3    5.49  3    4.91  3    99.99 99    99.99 99
182629      5.59 99.99  1.22 99.99 99.99    99.99 99.99    99.99 99    6.17  4    5.14  4    99.99 99    99.99 99
182709      5.96  1.92  1.64 99.99 99.99    99.99 99.99    10.50 12    6.77  5    5.31  5    99.99 99    99.99 99
182807      6.19 -0.03  0.52 99.99 99.99    99.99 99.99     7.47  3    6.38  3    6.06  3    99.99 99    99.99 99
183007      5.72 99.99  0.22 99.99 99.99    99.99 99.99    99.99 99    5.75  3    5.73  2    99.99 99    99.99 99
183030      6.38  1.79  1.57 99.99 99.99    99.99 99.99    10.70 11    7.15  5    5.76  5    99.99 99    99.99 99
183275      5.46 99.99  1.12 99.99 99.99    99.99 99.99    99.99 99    5.98  4    5.05  4    99.99 99    99.99 99
183312      6.60 99.99  0.39 99.99 99.99    99.99 99.99    99.99 99    6.72  3    6.53  2    99.99 99    99.99 99
183387      6.25  1.38  1.32 99.99 99.99    99.99 99.99     9.88  9    6.88  4    5.75  4    99.99 99    99.99 99
183914      5.11 -0.32 -0.10 99.99 99.99    99.99 99.99     5.66  9    4.96  2    5.27  2    99.99 99    99.99 99
184127      4.90 99.99  1.09 99.99 99.99    99.99 99.99    99.99 99    5.40  4    4.51  4    99.99 99    99.99 99
184586      6.39  0.04  0.02 99.99 99.99    99.99 99.99     7.49 14    6.31  3    6.50  2    99.99 99    99.99 99
184996      6.09  1.95  1.55 99.99 99.99    99.99 99.99    10.62 15    6.85  5    5.48  5    99.99 99    99.99 99
185075      6.26 99.99  1.00 99.99 99.99    99.99 99.99    99.99 99    6.72  4    5.91  4    99.99 99    99.99 99
185859      6.49 -0.62  0.39 99.99 99.99    99.99 99.99     6.90 12    6.61  3    6.42  2    99.99 99    99.99 99
186005      5.06 99.99  0.33 99.99 99.99    99.99 99.99    99.99 99    5.15  3    5.02  2    99.99 99    99.99 99
186154      6.39  1.68  1.40 99.99 99.99    99.99 99.99    10.47 13    7.07  4    5.85  4    99.99 99    99.99 99
186219      5.41  0.12  0.22 99.99 99.99    99.99 99.99     6.73 10    5.44  3    5.42  2    99.99 99    99.99 99
186543      5.35 99.99  0.20 99.99 99.99    99.99 99.99    99.99 99    5.37  3    5.37  2    99.99 99    99.99 99
186584      6.45  1.72  1.48 99.99 99.99    99.99 99.99    10.63 12    7.17  4    5.88  5    99.99 99    99.99 99
186689      5.91  0.09  0.18 99.99 99.99    99.99 99.99     7.17 11    5.91  3    5.94  2    99.99 99    99.99 99
187013      4.99  0.00  0.46 99.99 99.99    99.99 99.99     6.28  3    5.15  3    4.89  3    99.99 99    99.99 99
187098      6.05 99.99  0.40 99.99 99.99    99.99 99.99    99.99 99    6.18  3    5.98  3    99.99 99    99.99 99
187753      6.26  0.09  0.10 99.99 99.99    99.99 99.99     7.47 13    6.22  3    6.33  2    99.99 99    99.99 99
187923      5.78  0.21  0.53 99.99 99.99    99.99 99.99     7.39  5    5.98  3    5.65  3    99.99 99    99.99 99
188097      5.75  0.17  0.22 99.99 99.99    99.99 99.99     7.14 11    5.78  3    5.76  2    99.99 99    99.99 99
188107      6.53 -0.11  0.02 99.99 99.99    99.99 99.99     7.43 11    6.45  3    6.64  2    99.99 99    99.99 99
188114      4.13  0.89  1.08 99.99 99.99    99.99 99.99     6.96  5    4.63  4    3.74  4    99.99 99    99.99 99
188164      6.39  0.09  0.16 99.99 99.99    99.99 99.99     7.63 11    6.38  3    6.43  2    99.99 99    99.99 99
188228      3.96 -0.05 -0.03 99.99 99.99    99.99 99.99     4.91 13    3.85  2    4.09  2    99.99 99    99.99 99
188293      5.70 -0.49 -0.08 99.99 99.99    99.99 99.99     6.03  5    5.56  2    5.85  2    99.99 99    99.99 99
188294      6.48 -0.27 -0.04 99.99 99.99    99.99 99.99     7.13  9    6.36  2    6.61  2    99.99 99    99.99 99
188385      6.15  0.04  0.03 99.99 99.99    99.99 99.99     7.26 14    6.07  3    6.25  2    99.99 99    99.99 99
188584      5.76  0.80  1.04 99.99 99.99    99.99 99.99     8.45  4    6.24  4    5.39  4    99.99 99    99.99 99
188642      6.55 99.99  0.39 99.99 99.99    99.99 99.99    99.99 99    6.67  3    6.48  2    99.99 99    99.99 99
188887      5.31  1.29  1.22 99.99 99.99    99.99 99.99     8.76 10    5.89  4    4.86  4    99.99 99    99.99 99
189103      4.36 -0.67 -0.15 99.99 99.99    99.99 99.99     4.40  4    4.18  2    4.54  2    99.99 99    99.99 99
```

```
189140      6.14 99.99   1.64 99.99 99.99    99.99 99.99   99.99 99   6.95  5   5.49  5   99.99 99   99.99 99
189198      5.81  0.11   0.27 99.99 99.99    99.99 99.99    7.14  9   5.86  3   5.80  2   99.99 99   99.99 99
189245      5.66 99.99   0.49 99.99 99.99    99.99 99.99   99.99 99   5.83  3   5.55  3   99.99 99   99.99 99
189322      6.17  0.89   1.13 99.99 99.99    99.99 99.99    9.03  4   6.70  4   5.76  4   99.99 99   99.99 99
189340      5.87  0.05   0.58 99.99 99.99    99.99 99.99    7.29  3   6.09  3   5.72  3   99.99 99   99.99 99
189388      6.29 99.99   0.11 99.99 99.99    99.99 99.99   99.99 99   6.26  3   6.35  2   99.99 99   99.99 99
189567      6.07  0.08   0.64 99.99 99.99    99.99 99.99    7.57  4   6.33  3   5.89  3   99.99 99   99.99 99
189695      5.91  1.89   1.52 99.99 99.99    99.99 99.99   10.34 15   6.65  5   5.32  5   99.99 99   99.99 99
189831      4.77  1.68   1.41 99.99 99.99    99.99 99.99    8.86 13   5.45  4   4.23  4   99.99 99   99.99 99
190009      6.45 99.99   0.50 99.99 99.99    99.99 99.99   99.99 99   6.63  3   6.33  3   99.99 99   99.99 99
190222      6.45  1.98   1.58 99.99 99.99    99.99 99.99   11.04 15   7.22  5   5.83  5   99.99 99   99.99 99
190299      5.68  1.35   1.30 99.99 99.99    99.99 99.99    9.26  9   6.30  4   5.19  4   99.99 99   99.99 99
190327      5.52  0.86   1.06 99.99 99.99    99.99 99.99    8.30  5   6.01  4   5.14  4   99.99 99   99.99 99
190360      5.70  0.38   0.72 99.99 99.99    99.99 99.99    7.65  4   6.00  3   5.48  3   99.99 99   99.99 99
190406      5.80  0.09   0.61 99.99 99.99    99.99 99.99    7.29  3   6.04  3   5.63  3   99.99 99   99.99 99
190421      4.94  1.84   1.62 99.99 99.99    99.99 99.99    9.36 11   5.74  5   4.30  5   99.99 99   99.99 99
190422      6.26 99.99   0.53 99.99 99.99    99.99 99.99   99.99 99   6.46  3   6.13  3   99.99 99   99.99 99
190664      6.47  1.00   1.16 99.99 99.99    99.99 99.99    9.50  5   7.01  4   6.05  4   99.99 99   99.99 99
190960      6.20  1.89   1.61 99.99 99.99    99.99 99.99   10.68 12   6.99  5   5.57  5   99.99 99   99.99 99
191067      5.99  0.90   1.02 99.99 99.99    99.99 99.99    8.81  6   6.46  4   5.63  4   99.99 99   99.99 99
191095      6.37 99.99   0.06 99.99 99.99    99.99 99.99   99.99 99   6.31  3   6.46  2   99.99 99   99.99 99
191104      6.43 -0.02   0.46 99.99 99.99    99.99 99.99    7.69  3   6.59  3   6.33  3   99.99 99   99.99 99
191220      6.17  0.13   0.20 99.99 99.99    99.99 99.99    7.49 11   6.19  3   6.19  2   99.99 99   99.99 99
191570      6.48 -0.04   0.38 99.99 99.99    99.99 99.99    7.67  4   6.59  3   6.42  2   99.99 99   99.99 99
191584      6.22 99.99   1.23 99.99 99.99    99.99 99.99   99.99 99   6.80  4   5.76  4   99.99 99   99.99 99
191603      6.09 99.99   0.31 99.99 99.99    99.99 99.99   99.99 99   6.17  3   6.06  2   99.99 99   99.99 99
191984      6.27 -0.01   0.02 99.99 99.99    99.99 99.99    7.30 13   6.19  3   6.38  2   99.99 99   99.99 99
192486      6.53 99.99   0.37 99.99 99.99    99.99 99.99   99.99 99   6.64  3   6.47  2   99.99 99   99.99 99
192879      5.87 99.99   1.00 99.99 99.99    99.99 99.99   99.99 99   6.33  4   5.52  4   99.99 99   99.99 99
192886      6.13  0.01   0.46 99.99 99.99    99.99 99.99    7.43  3   6.29  3   6.03  3   99.99 99   99.99 99
193307      6.27 99.99   0.55 99.99 99.99    99.99 99.99   99.99 99   6.48  3   6.13  3   99.99 99   99.99 99
193429      6.63  1.83   1.55 99.99 99.99    99.99 99.99   11.00 13   7.39  5   6.02  5   99.99 99   99.99 99
193571      5.59  0.00   0.00 99.99 99.99    99.99 99.99    6.62 13   5.50  3   5.70  2   99.99 99   99.99 99
193664      5.94  0.06   0.58 99.99 99.99    99.99 99.99    7.38  3   6.16  3   5.79  3   99.99 99   99.99 99
193721      5.77  0.85   1.14 99.99 99.99    99.99 99.99    8.58  3   6.30  4   5.35  4   99.99 99   99.99 99
193807      5.64  0.10   0.20 99.99 99.99    99.99 99.99    6.92 10   5.66  3   5.66  2   99.99 99   99.99 99
194013      5.31  0.77   0.96 99.99 99.99    99.99 99.99    7.92  5   5.74  4   4.98  4   99.99 99   99.99 99
194215      5.85 99.99   1.10 99.99 99.99    99.99 99.99   99.99 99   6.36  4   5.45  4   99.99 99   99.99 99
194433      6.25 99.99   0.97 99.99 99.99    99.99 99.99   99.99 99   6.69  4   5.91  4   99.99 99   99.99 99
194526      6.33  1.91   1.56 99.99 99.99    99.99 99.99   10.81 14   7.09  5   5.72  5   99.99 99   99.99 99
194612      5.91  2.02   1.71 99.99 99.99    99.99 99.99   10.62 13   6.76  5   5.23  5   99.99 99   99.99 99
```

```
194937       6.25   0.92   1.08 99.99 99.99     99.99 99.99       9.13  5    6.75  4    5.86  4    99.99 99    99.99 99
195066       6.36   0.00   0.00 99.99 99.99     99.99 99.99       7.39 13    6.27  3    6.47  2    99.99 99    99.99 99
195093       6.74   0.04   0.22 99.99 99.99     99.99 99.99       7.95  9    6.77  3    6.75  2    99.99 99    99.99 99
195094       5.94   0.30   0.08 99.99 99.99     99.99 99.99       7.43 18    5.89  3    6.02  2    99.99 99    99.99 99
195402       6.11   1.41   1.29 99.99 99.99     99.99 99.99       9.77 10    6.72  4    5.63  4    99.99 99    99.99 99
195506       6.40   1.07   1.12 99.99 99.99     99.99 99.99       9.50  7    6.92  4    5.99  4    99.99 99    99.99 99
195564       5.65  99.99   0.69 99.99 99.99     99.99 99.99      99.99 99    5.93  3    5.44  3    99.99 99    99.99 99
195627       4.76   0.05   0.28 99.99 99.99     99.99 99.99       6.02  7    4.82  3    4.74  2    99.99 99    99.99 99
195922       6.56   0.05   0.08 99.99 99.99     99.99 99.99       7.71 12    6.51  3    6.64  2    99.99 99    99.99 99
196051       6.00   0.11   0.44 99.99 99.99     99.99 99.99       7.43  5    6.15  3    5.91  3    99.99 99    99.99 99
196067       6.03   0.19   0.62 99.99 99.99     99.99 99.99       7.66  3    6.28  3    5.86  3    99.99 99    99.99 99
196078       6.19  99.99   0.20 99.99 99.99     99.99 99.99      99.99 99    6.21  3    6.21  2    99.99 99    99.99 99
196171       3.11   0.79   1.00 99.99 99.99     99.99 99.99       5.77  5    3.57  4    2.76  4    99.99 99    99.99 99
196378       5.12  99.99   0.53 99.99 99.99     99.99 99.99      99.99 99    5.32  3    4.99  3    99.99 99    99.99 99
196519       5.15  -0.28  -0.06 99.99 99.99     99.99 99.99       5.77  9    5.02  2    5.29  2    99.99 99    99.99 99
196755       5.02   0.19   0.74 99.99 99.99     99.99 99.99       6.72  4    5.33  3    4.79  3    99.99 99    99.99 99
196761       6.37  99.99   0.72 99.99 99.99     99.99 99.99      99.99 99    6.67  3    6.15  3    99.99 99    99.99 99
196917       5.76  99.99   1.53 99.99 99.99     99.99 99.99      99.99 99    6.51  5    5.16  5    99.99 99    99.99 99
197051       3.43   0.12   0.16 99.99 99.99     99.99 99.99       4.72 12    3.42  3    3.47  2    99.99 99    99.99 99
197093       6.29  99.99   1.08 99.99 99.99     99.99 99.99      99.99 99    6.79  4    5.90  4    99.99 99    99.99 99
197157       4.51   0.09   0.27 99.99 99.99     99.99 99.99       5.82  8    4.56  3    4.50  2    99.99 99    99.99 99
197635       5.41   1.09   1.12 99.99 99.99     99.99 99.99       8.54  8    5.93  4    5.00  4    99.99 99    99.99 99
197649       6.49  99.99   0.39 99.99 99.99     99.99 99.99      99.99 99    6.61  3    6.42  2    99.99 99    99.99 99
197937       5.10  99.99   0.35 99.99 99.99     99.99 99.99      99.99 99    5.20  3    5.05  2    99.99 99    99.99 99
198048       4.89   1.87   1.52 99.99 99.99     99.99 99.99       9.29 14    5.63  5    4.30  5    99.99 99    99.99 99
198069       5.58  -0.09  -0.02 99.99 99.99     99.99 99.99       6.48 12    5.47  2    5.70  2    99.99 99    99.99 99
198160       5.65  99.99   0.18 99.99 99.99     99.99 99.99      99.99 99    5.65  3    5.68  2    99.99 99    99.99 99
198232       4.89   0.73   1.00 99.99 99.99     99.99 99.99       7.46  4    5.35  4    4.54  4    99.99 99    99.99 99
198391       6.33  -0.02   0.02 99.99 99.99     99.99 99.99       7.35 12    6.25  3    6.44  2    99.99 99    99.99 99
198571       5.99   0.03   0.46 99.99 99.99     99.99 99.99       7.32  3    6.15  3    5.89  3    99.99 99    99.99 99
198667       5.55  -0.27  -0.08 99.99 99.99     99.99 99.99       6.18 10    5.41  2    5.70  2    99.99 99    99.99 99
198700       3.65   1.23   1.25 99.99 99.99     99.99 99.99       7.04  7    4.24  4    3.18  4    99.99 99    99.99 99
198716       5.35  99.99   1.32 99.99 99.99     99.99 99.99      99.99 99    5.98  4    4.85  4    99.99 99    99.99 99
198781       6.45  -0.77   0.07 99.99 99.99     99.99 99.99       6.48  7    6.39  3    6.53  2    99.99 99    99.99 99
199223       6.05   0.49   0.82 99.99 99.99     99.99 99.99       8.20  4    6.41  4    5.78  3    99.99 99    99.99 99
199260       5.69  99.99   0.50 99.99 99.99     99.99 99.99      99.99 99    5.87  3    5.57  3    99.99 99    99.99 99
199280       6.57  -0.29  -0.10 99.99 99.99     99.99 99.99       7.16 10    6.42  2    6.73  2    99.99 99    99.99 99
199442       6.05   1.36   1.22 99.99 99.99     99.99 99.99       9.60 11    6.63  4    5.60  4    99.99 99    99.99 99
199443       5.87  99.99   0.18 99.99 99.99     99.99 99.99      99.99 99    5.87  3    5.90  2    99.99 99    99.99 99
199475       6.37   0.11   0.10 99.99 99.99     99.99 99.99       7.61 13    6.33  3    6.44  2    99.99 99    99.99 99
199579       5.96  -0.85   0.05 99.99 99.99     99.99 99.99       5.87  8    5.89  3    6.05  2    99.99 99    99.99 99
```

```
199684     6.11 99.99  0.39 99.99 99.99    99.99 99.99   99.99 99    6.23  3    6.04  2   99.99 99   99.99 99
199942     5.99  0.09  0.26 99.99 99.99    99.99 99.99    7.29  9    6.04  3    5.98  2   99.99 99   99.99 99
199951     4.67  0.54  0.89 99.99 99.99    99.99 99.99    6.93  3    5.06  4    4.37  3   99.99 99   99.99 99
200011     6.64 99.99  0.68 99.99 99.99    99.99 99.99   99.99 99    6.92  3    6.44  3   99.99 99   99.99 99
200073     5.94 99.99  1.11 99.99 99.99    99.99 99.99   99.99 99    6.46  4    5.54  4   99.99 99   99.99 99
200266     6.58  1.18  1.23 99.99 99.99    99.99 99.99    9.89  7    7.16  4    6.12  4   99.99 99   99.99 99
200340     6.50 -0.49 -0.10 99.99 99.99    99.99 99.99    6.82  6    6.35  2    6.66  2   99.99 99   99.99 99
200375     6.21  0.02  0.48 99.99 99.99    99.99 99.99    7.54  3    6.38  3    6.10  3   99.99 99   99.99 99
200525     5.68  0.10  0.59 99.99 99.99    99.99 99.99    7.18  3    5.91  3    5.52  3   99.99 99   99.99 99
200644     5.60  2.01  1.65 99.99 99.99    99.99 99.99   10.27 14    6.41  5    4.95  5   99.99 99   99.99 99
200663     6.33  0.61  0.97 99.99 99.99    99.99 99.99    8.73  3    6.77  4    5.99  4   99.99 99   99.99 99
200751     5.76  1.15  1.18 99.99 99.99    99.99 99.99    9.00  8    6.31  4    5.33  4   99.99 99   99.99 99
200790     5.94  0.06  0.54 99.99 99.99    99.99 99.99    7.35  3    6.14  3    5.80  3   99.99 99   99.99 99
200924     6.20  1.03  1.08 99.99 99.99    99.99 99.99    9.22  8    6.70  4    5.81  4   99.99 99   99.99 99
201298     6.15  1.97  1.66 99.99 99.99    99.99 99.99   10.77 13    6.97  5    5.49  5   99.99 99   99.99 99
201371     5.02  1.55  1.58 99.99 99.99    99.99 99.99    9.02  6    5.79  5    4.40  5   99.99 99   99.99 99
201507     6.45  0.06  0.37 99.99 99.99    99.99 99.99    7.77  5    6.56  3    6.39  2   99.99 99   99.99 99
201567     6.27  1.21  1.16 99.99 99.99    99.99 99.99    9.58  9    6.81  4    5.85  4   99.99 99   99.99 99
201616     6.07  0.04  0.02 99.99 99.99    99.99 99.99    7.17 14    5.99  3    6.18  2   99.99 99   99.99 99
201647     5.83 99.99  0.45 99.99 99.99    99.99 99.99   99.99 99    5.98  3    5.74  3   99.99 99   99.99 99
201901     5.42 99.99  1.42 99.99 99.99    99.99 99.99   99.99 99    6.11  4    4.87  5   99.99 99   99.99 99
201906     6.63  0.00  0.03 99.99 99.99    99.99 99.99    7.68 13    6.55  3    6.73  2   99.99 99   99.99 99
202103     5.75 99.99  0.19 99.99 99.99    99.99 99.99   99.99 99    5.76  3    5.78  2   99.99 99   99.99 99
202135     6.21 99.99  1.14 99.99 99.99    99.99 99.99   99.99 99    6.74  4    5.79  4   99.99 99   99.99 99
202214     5.63 -0.77  0.11 99.99 99.99    99.99 99.99    5.68  8    5.60  3    5.69  2   99.99 99   99.99 99
202259     6.38  1.92  1.61 99.99 99.99    99.99 99.99   10.90 13    7.17  5    5.75  5   99.99 99   99.99 99
202299     6.31 -0.22 -0.07 99.99 99.99    99.99 99.99    7.01 10    6.18  2    6.46  2   99.99 99   99.99 99
202418     6.45  1.66  1.40 99.99 99.99    99.99 99.99   10.50 13    7.13  4    5.91  4   99.99 99   99.99 99
202448     3.90  0.30  0.52 99.99 99.99    99.99 99.99    5.63  6    4.09  3    3.77  3   99.99 99   99.99 99
202627     4.71  0.02  0.06 99.99 99.99    99.99 99.99    5.81 12    4.65  3    4.80  2   99.99 99   99.99 99
202730     4.39  0.12  0.19 99.99 99.99    99.99 99.99    5.69 11    4.40  3    4.42  2   99.99 99   99.99 99
202753     5.82 -0.51 -0.13 99.99 99.99    99.99 99.99    6.09  6    5.65  2    6.00  2   99.99 99   99.99 99
202773     6.40 99.99  0.97 99.99 99.99    99.99 99.99   99.99 99    6.84  4    6.06  4   99.99 99   99.99 99
202940     6.56 99.99  0.73 99.99 99.99    99.99 99.99   99.99 99    6.87  3    6.34  3   99.99 99   99.99 99
203010     6.38 99.99  1.32 99.99 99.99    99.99 99.99   99.99 99    7.01  4    5.88  4   99.99 99   99.99 99
203212     6.09  1.41  1.26 99.99 99.99    99.99 99.99    9.73 11    6.69  4    5.62  4   99.99 99   99.99 99
203222     5.87  0.64  0.92 99.99 99.99    99.99 99.99    8.28  4    6.28  4    5.56  3   99.99 99   99.99 99
203291     5.82  1.97  1.66 99.99 99.99    99.99 99.99   10.44 13    6.64  5    5.16  5   99.99 99   99.99 99
203344     5.57  0.91  1.05 99.99 99.99    99.99 99.99    8.42  6    6.05  4    5.20  4   99.99 99   99.99 99
203525     5.99  1.88  1.54 99.99 99.99    99.99 99.99   10.42 14    6.74  5    5.39  5   99.99 99   99.99 99
203532     6.38 -0.36  0.13 99.99 99.99    99.99 99.99    7.00  3    6.36  3    6.43  2   99.99 99   99.99 99
```

```
203548      6.31 99.99   0.20 99.99 99.99     99.99 99.99    99.99 99   6.33  3   6.33  2   99.99 99   99.99 99
203562      5.17  0.07   0.05 99.99 99.99     99.99 99.99     6.33 14   5.10  3   5.26  2   99.99 99   99.99 99
203585      5.76 99.99  -0.04 99.99 99.99     99.99 99.99    99.99 99   5.64  2   5.89  2   99.99 99   99.99 99
203705      5.49 99.99   0.29 99.99 99.99     99.99 99.99    99.99 99   5.55  3   5.47  2   99.99 99   99.99 99
203842      6.34  0.13   0.47 99.99 99.99     99.99 99.99     7.81  4   6.50  3   6.24  3   99.99 99   99.99 99
203843      6.36  0.13   0.33 99.99 99.99     99.99 99.99     7.75  8   6.45  3   6.32  2   99.99 99   99.99 99
203875      5.70 99.99   0.20 99.99 99.99     99.99 99.99    99.99 99   5.72  3   5.72  2   99.99 99   99.99 99
203949      5.63 99.99   1.19 99.99 99.99     99.99 99.99    99.99 99   6.19  4   5.19  4   99.99 99   99.99 99
203955      6.47  0.05   0.04 99.99 99.99     99.99 99.99     7.59 14   6.40  3   6.57  2   99.99 99   99.99 99
204041      6.46  0.05   0.16 99.99 99.99     99.99 99.99     7.65 10   6.45  3   6.50  2   99.99 99   99.99 99
204121      6.13  0.01   0.44 99.99 99.99     99.99 99.99     7.42  3   6.28  3   6.04  3   99.99 99   99.99 99
204445      6.40  1.90   1.64 99.99 99.99     99.99 99.99    10.91 12   7.21  5   5.75  5   99.99 99   99.99 99
204783      5.29 99.99   1.10 99.99 99.99     99.99 99.99    99.99 99   5.80  4   4.89  4   99.99 99   99.99 99
204904      6.18  0.02   0.46 99.99 99.99     99.99 99.99     7.50  3   6.34  3   6.08  3   99.99 99   99.99 99
204943      6.43 99.99   0.20 99.99 99.99     99.99 99.99    99.99 99   6.45  3   6.45  2   99.99 99   99.99 99
204960      5.57 99.99   1.04 99.99 99.99     99.99 99.99    99.99 99   6.05  4   5.20  4   99.99 99   99.99 99
205417      6.20  0.05   0.03 99.99 99.99     99.99 99.99     7.32 14   6.12  3   6.30  2   99.99 99   99.99 99
205423      5.77  1.04   1.11 99.99 99.99     99.99 99.99     8.82  7   6.29  4   5.37  4   99.99 99   99.99 99
205471      5.73 99.99   0.22 99.99 99.99     99.99 99.99    99.99 99   5.76  3   5.74  2   99.99 99   99.99 99
205478      3.76  0.89   1.00 99.99 99.99     99.99 99.99     6.55  7   4.22  4   3.41  4   99.99 99   99.99 99
205529      6.11 99.99   0.22 99.99 99.99     99.99 99.99    99.99 99   6.14  3   6.12  2   99.99 99   99.99 99
205811      6.17  0.02   0.03 99.99 99.99     99.99 99.99     7.25 13   6.09  3   6.27  2   99.99 99   99.99 99
205924      5.67  0.08   0.25 99.99 99.99     99.99 99.99     6.95  9   5.71  3   5.67  2   99.99 99   99.99 99
205935      6.33 99.99   1.06 99.99 99.99     99.99 99.99    99.99 99   6.82  4   5.95  4   99.99 99   99.99 99
206240      5.29  0.47   0.75 99.99 99.99     99.99 99.99     7.37  4   5.61  3   5.06  3   99.99 99   99.99 99
206267      5.62 -0.74   0.21 99.99 99.99     99.99 99.99     5.77 10   5.64  3   5.64  2   99.99 99   99.99 99
206399      6.01 -0.35  -0.10 99.99 99.99     99.99 99.99     6.52  8   5.86  2   6.17  2   99.99 99   99.99 99
206445      5.67  1.67   1.44 99.99 99.99     99.99 99.99     9.76 12   6.37  4   5.12  5   99.99 99   99.99 99
206487      5.29  1.94   1.64 99.99 99.99     99.99 99.99     9.86 13   6.10  5   4.64  5   99.99 99   99.99 99
206546      6.23 99.99   0.27 99.99 99.99     99.99 99.99    99.99 99   6.28  3   6.22  2   99.99 99   99.99 99
206642      6.30 99.99   1.12 99.99 99.99     99.99 99.99    99.99 99   6.82  4   5.89  4   99.99 99   99.99 99
206690      6.45 99.99   1.15 99.99 99.99     99.99 99.99    99.99 99   6.99  4   6.03  4   99.99 99   99.99 99
206834      5.09  0.94   1.11 99.99 99.99     99.99 99.99     8.01  5   5.61  4   4.69  4   99.99 99   99.99 99
207129      5.59 99.99   0.60 99.99 99.99     99.99 99.99    99.99 99   5.83  3   5.43  3   99.99 99   99.99 99
207134      6.28  1.25   1.21 99.99 99.99     99.99 99.99     9.67  9   6.85  4   5.83  4   99.99 99   99.99 99
207155      5.01 99.99   0.04 99.99 99.99     99.99 99.99    99.99 99   4.94  3   5.11  2   99.99 99   99.99 99
207203      5.64 -0.02   0.00 99.99 99.99     99.99 99.99     6.65 13   5.55  3   5.75  2   99.99 99   99.99 99
207229      5.62  0.83   1.02 99.99 99.99     99.99 99.99     8.34  5   6.09  4   5.26  4   99.99 99   99.99 99
207235      6.17  0.05   0.22 99.99 99.99     99.99 99.99     7.39  9   6.20  3   6.18  2   99.99 99   99.99 99
207241      5.53  1.63   1.37 99.99 99.99     99.99 99.99     9.53 13   6.19  4   5.01  4   99.99 99   99.99 99
207958      5.08 99.99   0.37 99.99 99.99     99.99 99.99    99.99 99   5.19  3   5.02  2   99.99 99   99.99 99
```

```
207964      5.90 99.99  0.39 99.99 99.99    99.99 99.99   99.99 99   6.02  3   5.83  2   99.99 99   99.99 99
208110      6.15  0.26  0.80 99.99 99.99    99.99 99.99    7.98  4   6.50  3   5.89  3   99.99 99   99.99 99
208111      5.71  1.27  1.18 99.99 99.99    99.99 99.99    9.11 10   6.26  4   5.28  4   99.99 99   99.99 99
208177      6.20  0.07  0.48 99.99 99.99    99.99 99.99    7.59  3   6.37  3   6.09  3   99.99 99   99.99 99
208285      6.41 99.99  0.93 99.99 99.99    99.99 99.99   99.99 99   6.83  4   6.09  3   99.99 99   99.99 99
208321      5.46 99.99  0.08 99.99 99.99    99.99 99.99   99.99 99   5.41  3   5.54  2   99.99 99   99.99 99
208450      4.40  0.10  0.28 99.99 99.99    99.99 99.99    5.72  8   4.46  3   4.38  2   99.99 99   99.99 99
208500      6.41  0.08  0.22 99.99 99.99    99.99 99.99    7.67  9   6.44  3   6.42  2   99.99 99   99.99 99
208703      6.33 -0.03  0.37 99.99 99.99    99.99 99.99    7.53  4   6.44  3   6.27  2   99.99 99   99.99 99
208737      5.50 99.99  1.00 99.99 99.99    99.99 99.99   99.99 99   5.96  4   5.15  4   99.99 99   99.99 99
208741      5.95  0.11  0.39 99.99 99.99    99.99 99.99    7.35  6   6.07  3   5.88  2   99.99 99   99.99 99
208796      6.01 99.99 -0.10 99.99 99.99    99.99 99.99   99.99 99   5.86  2   6.17  2   99.99 99   99.99 99
208801      6.22  0.87  1.00 99.99 99.99    99.99 99.99    8.98  6   6.68  4   5.87  4   99.99 99   99.99 99
209008      6.00 -0.57 -0.12 99.99 99.99    99.99 99.99    6.20  5   5.84  2   6.17  2   99.99 99   99.99 99
209014      5.42 99.99 -0.10 99.99 99.99    99.99 99.99   99.99 99   5.27  2   5.58  2   99.99 99   99.99 99
209128      5.58  1.40  1.28 99.99 99.99    99.99 99.99    9.22 10   6.19  4   5.10  4   99.99 99   99.99 99
209396      5.54  0.73  0.96 99.99 99.99    99.99 99.99    8.09  5   5.97  4   5.21  4   99.99 99   99.99 99
209476      6.47 99.99  1.63 99.99 99.99    99.99 99.99   99.99 99   7.27  5   5.83  5   99.99 99   99.99 99
209855      6.55  1.30  1.18 99.99 99.99    99.99 99.99   10.00 11   7.10  4   6.12  4   99.99 99   99.99 99
209942      6.98 -0.02  0.52 99.99 99.99    99.99 99.99    8.27  3   7.17  3   6.85  3   99.99 99   99.99 99
210056      6.15  0.82  1.00 99.99 99.99    99.99 99.99    8.85  5   6.61  4   5.80  4   99.99 99   99.99 99
210129      5.78 -0.41 -0.10 99.99 99.99    99.99 99.99    6.20  7   5.63  2   5.94  2   99.99 99   99.99 99
210204      6.43 99.99  1.39 99.99 99.99    99.99 99.99   99.99 99   7.10  4   5.90  4   99.99 99   99.99 99
210302      4.92 99.99  0.48 99.99 99.99    99.99 99.99   99.99 99   5.09  3   4.81  3   99.99 99   99.99 99
210419      6.27 -0.07 -0.01 99.99 99.99    99.99 99.99    7.20 12   6.17  3   6.39  2   99.99 99   99.99 99
210434      6.01  0.83  0.98 99.99 99.99    99.99 99.99    8.71  6   6.45  4   5.67  4   99.99 99   99.99 99
210464      6.09 99.99  0.50 99.99 99.99    99.99 99.99   99.99 99   6.27  3   5.97  3   99.99 99   99.99 99
210739      6.17 99.99  0.17 99.99 99.99    99.99 99.99   99.99 99   6.17  3   6.21  2   99.99 99   99.99 99
210763      6.39  0.06  0.50 99.99 99.99    99.99 99.99    7.78  3   6.57  3   6.27  3   99.99 99   99.99 99
210853      5.51  0.12  0.31 99.99 99.99    99.99 99.99    6.88  8   5.59  3   5.48  2   99.99 99   99.99 99
210918      6.23 99.99  0.64 99.99 99.99    99.99 99.99   99.99 99   6.49  3   6.05  3   99.99 99   99.99 99
211053      6.09  0.81  1.02 99.99 99.99    99.99 99.99    8.78  5   6.56  4   5.73  4   99.99 99   99.99 99
211088      4.79  0.47  0.80 99.99 99.99    99.99 99.99    6.90  4   5.14  3   4.53  3   99.99 99   99.99 99
211202      5.10 99.99  0.92 99.99 99.99    99.99 99.99   99.99 99   5.51  4   4.79  3   99.99 99   99.99 99
211287      6.21 -0.04  0.02 99.99 99.99    99.99 99.99    7.20 12   6.13  3   6.32  2   99.99 99   99.99 99
211291      6.15 99.99  1.11 99.99 99.99    99.99 99.99   99.99 99   6.67  4   5.75  4   99.99 99   99.99 99
211356      6.15  0.09  0.19 99.99 99.99    99.99 99.99    7.41 10   6.16  3   6.18  2   99.99 99   99.99 99
211392      5.79  1.14  1.16 99.99 99.99    99.99 99.99    9.01  8   6.33  4   5.37  4   99.99 99   99.99 99
211415      5.37 99.99  0.60 99.99 99.99    99.99 99.99   99.99 99   5.61  3   5.21  3   99.99 99   99.99 99
211416      2.86  1.54  1.39 99.99 99.99    99.99 99.99    6.75 10   3.53  4   2.33  4   99.99 99   99.99 99
211434      5.75  0.51  0.88 99.99 99.99    99.99 99.99    7.96  3   6.14  4   5.46  3   99.99 99   99.99 99
```

```
211539       5.77  0.88  1.02 99.99 99.99     99.99 99.99     8.56  6    6.24  4    5.41  4    99.99 99    99.99 99
211575       6.39  0.00  0.44 99.99 99.99     99.99 99.99     7.67  3    6.54  3    6.30  3    99.99 99    99.99 99
211797       6.17  0.11  0.28 99.99 99.99     99.99 99.99     7.51  9    6.23  3    6.15  2    99.99 99    99.99 99
211838       5.37 -0.37 -0.06 99.99 99.99     99.99 99.99     5.87  7    5.24  2    5.51  2    99.99 99    99.99 99
211924       5.37 -0.47 -0.02 99.99 99.99     99.99 99.99     5.76  4    5.26  2    5.49  2    99.99 99    99.99 99
211976       6.17 -0.03  0.44 99.99 99.99     99.99 99.99     7.41  3    6.32  3    6.08  3    99.99 99    99.99 99
211998       5.29 -0.06  0.65 99.99 99.99     99.99 99.99     6.60  7    5.55  3    5.10  3    99.99 99    99.99 99
212132       5.62  0.01  0.36 99.99 99.99     99.99 99.99     6.87  5    5.72  3    5.57  2    99.99 99    99.99 99
212168       6.04  0.14  0.64 99.99 99.99     99.99 99.99     7.62  3    6.30  3    5.86  3    99.99 99    99.99 99
212211       5.78  0.03  0.38 99.99 99.99     99.99 99.99     7.06  5    5.89  3    5.72  2    99.99 99    99.99 99
212320       5.93  0.69  1.00 99.99 99.99     99.99 99.99     8.45  3    6.39  4    5.58  4    99.99 99    99.99 99
212330       5.32 99.99  0.67 99.99 99.99     99.99 99.99     99.99 99   5.59  3    5.12  3    99.99 99    99.99 99
212404       5.78 -0.11 -0.04 99.99 99.99     99.99 99.99     6.64 12    5.66  2    5.91  2    99.99 99    99.99 99
212581       4.47 -0.07 -0.03 99.99 99.99     99.99 99.99     5.39 13    4.36  2    4.60  2    99.99 99    99.99 99
212710       5.27 -0.11 -0.03 99.99 99.99     99.99 99.99     6.14 12    5.16  2    5.40  2    99.99 99    99.99 99
212728       5.55  0.08  0.20 99.99 99.99     99.99 99.99     6.80 10    5.57  3    5.57  2    99.99 99    99.99 99
212754       5.75  0.04  0.52 99.99 99.99     99.99 99.99     7.13  3    5.94  3    5.62  3    99.99 99    99.99 99
212883       6.45 -0.75 -0.13 99.99 99.99     99.99 99.99     6.40  3    6.28  2    6.63  2    99.99 99    99.99 99
212953       5.47 99.99  0.95 99.99 99.99     99.99 99.99     99.99 99   5.90  4    5.14  4    99.99 99    99.99 99
212978       6.14 -0.77 -0.14 99.99 99.99     99.99 99.99     6.05  3    5.97  2    6.32  2    99.99 99    99.99 99
213087       5.46 -0.59  0.37 99.99 99.99     99.99 99.99     5.90 11    5.57  3    5.40  2    99.99 99    99.99 99
213119       5.58  1.90  1.55 99.99 99.99     99.99 99.99    10.04 14    6.34  5    4.97  5    99.99 99    99.99 99
213135       5.95 99.99  0.33 99.99 99.99     99.99 99.99     99.99 99   6.04  3    5.91  2    99.99 99    99.99 99
213235       5.48  0.09  0.38 99.99 99.99     99.99 99.99     6.85  6    5.59  3    5.42  2    99.99 99    99.99 99
213402       6.15  1.35  1.38 99.99 99.99     99.99 99.99     9.77  7    6.81  4    5.62  4    99.99 99    99.99 99
213428       6.16  1.00  1.08 99.99 99.99     99.99 99.99     9.14  7    6.66  4    5.77  4    99.99 99    99.99 99
213429       6.14  0.06  0.56 99.99 99.99     99.99 99.99     7.56  3    6.35  3    5.99  3    99.99 99    99.99 99
213789       5.89  0.73  1.00 99.99 99.99     99.99 99.99     8.46  4    6.35  4    5.54  4    99.99 99    99.99 99
213845       5.21 99.99  0.44 99.99 99.99     99.99 99.99     99.99 99   5.36  3    5.12  3    99.99 99    99.99 99
213884       6.23 99.99  0.20 99.99 99.99     99.99 99.99     99.99 99   6.25  3    6.25  2    99.99 99    99.99 99
214085       6.28 99.99  0.12 99.99 99.99     99.99 99.99     99.99 99   6.25  3    6.34  2    99.99 99    99.99 99
214150       5.86 99.99  0.06 99.99 99.99     99.99 99.99     99.99 99   5.80  3    5.95  2    99.99 99    99.99 99
214167       5.73 -0.90 -0.15 99.99 99.99     99.99 99.99     5.46  4    5.55  2    5.91  2    99.99 99    99.99 99
214240       6.29 -0.54 -0.05 99.99 99.99     99.99 99.99     6.57  4    6.17  2    6.43  2    99.99 99    99.99 99
214376       5.03  1.16  1.14 99.99 99.99     99.99 99.99     8.26  9    5.56  4    4.61  4    99.99 99    99.99 99
214448       6.23  0.49  0.78 99.99 99.99     99.99 99.99     8.36  4    6.56  3    5.98  3    99.99 99    99.99 99
214462       6.47 99.99  1.04 99.99 99.99     99.99 99.99     99.99 99   6.95  4    6.10  4    99.99 99    99.99 99
214599       6.31 99.99  1.01 99.99 99.99     99.99 99.99     99.99 99   6.77  4    5.96  4    99.99 99    99.99 99
214632       5.97 99.99  1.46 99.99 99.99     99.99 99.99     99.99 99   6.68  4    5.41  5    99.99 99    99.99 99
214690       5.87 99.99  1.30 99.99 99.99     99.99 99.99     99.99 99   6.49  4    5.38  4    99.99 99    99.99 99
214810       6.31  0.00  0.52 99.99 99.99     99.99 99.99     7.63  3    6.50  3    6.18  3    99.99 99    99.99 99
```

```
214846      4.15   0.11   0.20 99.99 99.99    99.99 99.99    5.44 11    4.17  3    4.17  2    99.99 99    99.99 99
214953      5.98   0.07   0.58 99.99 99.99    99.99 99.99    7.43  3    6.20  3    5.83  3    99.99 99    99.99 99
214987      6.07   0.68   0.98 99.99 99.99    99.99 99.99    8.57  3    6.51  4    5.73  4    99.99 99    99.99 99
215114      6.45   0.12   0.17 99.99 99.99    99.99 99.99    7.74 12    6.45  3    6.49  2    99.99 99    99.99 99
215143      6.41  -0.11  -0.04 99.99 99.99    99.99 99.99    7.27 12    6.29  2    6.54  2    99.99 99    99.99 99
215191      6.43  -0.81  -0.09 99.99 99.99    99.99 99.99    6.32  4    6.29  2    6.59  2    99.99 99    99.99 99
215369      4.85   1.17   1.18 99.99 99.99    99.99 99.99    8.12  8    5.40  4    4.42  4    99.99 99    99.99 99
215405      5.51   1.42   1.32 99.99 99.99    99.99 99.99    9.20 10    6.14  4    5.01  4    99.99 99    99.99 99
215545      6.56   0.11   0.30 99.99 99.99    99.99 99.99    7.91  8    6.63  3    6.54  2    99.99 99    99.99 99
215631      6.73   0.10   0.11 99.99 99.99    99.99 99.99    7.96 13    6.70  3    6.79  2    99.99 99    99.99 99
215669      6.28 99.99   1.16 99.99 99.99    99.99 99.99    99.99 99    6.82  4    5.86  4    99.99 99    99.99 99
215682      6.37 99.99   1.06 99.99 99.99    99.99 99.99    99.99 99    6.86  4    5.99  4    99.99 99    99.99 99
215724      6.71 99.99   0.48 99.99 99.99    99.99 99.99    99.99 99    6.88  3    6.60  3    99.99 99    99.99 99
215729      6.34   0.10   0.07 99.99 99.99    99.99 99.99    7.55 14    6.28  3    6.42  2    99.99 99    99.99 99
215789      3.49   0.10   0.08 99.99 99.99    99.99 99.99    4.70 14    3.44  3    3.57  2    99.99 99    99.99 99
216042      6.33 99.99   0.31 99.99 99.99    99.99 99.99    99.99 99    6.41  3    6.30  2    99.99 99    99.99 99
216048      6.54   0.05   0.29 99.99 99.99    99.99 99.99    7.80  7    6.60  3    6.52  2    99.99 99    99.99 99
216385      5.16  -0.01   0.48 99.99 99.99    99.99 99.99    6.45  3    5.33  3    5.05  3    99.99 99    99.99 99
216435      6.04 99.99   0.62 99.99 99.99    99.99 99.99    99.99 99    6.29  3    5.87  3    99.99 99    99.99 99
216437      6.05   0.24   0.66 99.99 99.99    99.99 99.99    7.77  3    6.32  3    5.86  3    99.99 99    99.99 99
216637      6.19   1.41   1.28 99.99 99.99    99.99 99.99    9.84 11    6.80  4    5.71  4    99.99 99    99.99 99
216640      5.56   1.09   1.14 99.99 99.99    99.99 99.99    8.70  7    6.09  4    5.14  4    99.99 99    99.99 99
216701      6.11   0.15   0.20 99.99 99.99    99.99 99.99    7.46 11    6.13  3    6.13  2    99.99 99    99.99 99
216718      5.72   0.60   0.88 99.99 99.99    99.99 99.99    8.05  4    6.11  4    5.43  3    99.99 99    99.99 99
216953      6.31   0.72   0.94 99.99 99.99    99.99 99.99    8.84  5    6.73  4    5.99  3    99.99 99    99.99 99
217019      6.28   1.08   1.12 99.99 99.99    99.99 99.99    9.39  8    6.80  4    5.87  4    99.99 99    99.99 99
217096      6.13 99.99   0.58 99.99 99.99    99.99 99.99    99.99 99    6.35  3    5.98  3    99.99 99    99.99 99
217101      6.17  -0.80  -0.15 99.99 99.99    99.99 99.99    6.04  3    5.99  2    6.35  2    99.99 99    99.99 99
217107      6.16   0.43   0.74 99.99 99.99    99.99 99.99    8.18  4    6.47  3    5.93  3    99.99 99    99.99 99
217131      6.37   0.02   0.35 99.99 99.99    99.99 99.99    7.62  5    6.47  3    6.32  2    99.99 99    99.99 99
217186      6.33   0.02   0.06 99.99 99.99    99.99 99.99    7.43 12    6.27  3    6.42  2    99.99 99    99.99 99
217264      5.43   0.83   0.98 99.99 99.99    99.99 99.99    8.13  6    5.87  4    5.09  4    99.99 99    99.99 99
217364      4.12   0.70   0.98 99.99 99.99    99.99 99.99    6.64  4    4.56  4    3.78  4    99.99 99    99.99 99
217403      5.68 99.99   1.42 99.99 99.99    99.99 99.99    99.99 99    6.37  4    5.13  5    99.99 99    99.99 99
217428      6.21   0.60   0.89 99.99 99.99    99.99 99.99    8.55  4    6.60  4    5.91  3    99.99 99    99.99 99
217459      5.83   1.57   1.34 99.99 99.99    99.99 99.99    9.73 12    6.47  4    5.32  4    99.99 99    99.99 99
217531      6.21   1.77   1.41 99.99 99.99    99.99 99.99   10.42 15    6.89  4    5.67  4    99.99 99    99.99 99
217563      5.94   0.75   1.00 99.99 99.99    99.99 99.99    8.54  4    6.40  4    5.59  4    99.99 99    99.99 99
217701      6.15   1.90   1.58 99.99 99.99    99.99 99.99   10.63 13    6.92  5    5.53  5    99.99 99    99.99 99
217831      5.52   0.12   0.36 99.99 99.99    99.99 99.99    6.92  7    5.62  3    5.47  2    99.99 99    99.99 99
217877      6.68   0.06   0.58 99.99 99.99    99.99 99.99    8.12  3    6.90  3    6.53  3    99.99 99    99.99 99
```

```
217926      6.41  0.03  0.39 99.99 99.99     99.99 99.99      7.70  4    6.53  3    6.34  2    99.99 99    99.99 99
218060      5.43  0.07  0.30 99.99 99.99     99.99 99.99      6.72  7    5.50  3    5.41  2    99.99 99    99.99 99
218101      6.44  0.55  0.83 99.99 99.99     99.99 99.99      8.68  4    6.80  4    6.17  3    99.99 99    99.99 99
218103      6.39  0.72  0.94 99.99 99.99     99.99 99.99      8.92  5    6.81  4    6.07  3    99.99 99    99.99 99
218108      6.12  0.10  0.14 99.99 99.99     99.99 99.99      7.37 12    6.10  3    6.17  2    99.99 99    99.99 99
218227      4.28  0.16  0.42 99.99 99.99     99.99 99.99      5.76  6    4.42  3    4.20  3    99.99 99    99.99 99
218242      5.61 99.99  0.01 99.99 99.99     99.99 99.99     99.99 99    5.52  3    5.72  2    99.99 99    99.99 99
218261      5.83 99.99  0.48 99.99 99.99     99.99 99.99     99.99 99    6.00  3    5.72  3    99.99 99    99.99 99
218288      6.15  1.68  1.42 99.99 99.99     99.99 99.99     10.24 13    6.84  4    5.60  5    99.99 99    99.99 99
218407      6.66 -0.68 -0.05 99.99 99.99     99.99 99.99      6.75  3    6.54  2    6.80  2    99.99 99    99.99 99
218527      5.40  0.56  0.91 99.99 99.99     99.99 99.99      7.69  3    5.81  4    5.09  3    99.99 99    99.99 99
218558      6.47  0.71  0.95 99.99 99.99     99.99 99.99      8.99  4    6.90  4    6.14  4    99.99 99    99.99 99
218559      6.41  1.84  1.50 99.99 99.99     99.99 99.99     10.76 14    7.14  5    5.83  5    99.99 99    99.99 99
218630      5.81 99.99  0.48 99.99 99.99     99.99 99.99     99.99 99    5.98  3    5.70  3    99.99 99    99.99 99
218700      5.39 -0.29 -0.08 99.99 99.99     99.99 99.99      5.99  9    5.25  2    5.54  2    99.99 99    99.99 99
218759      6.51 99.99  0.27 99.99 99.99     99.99 99.99     99.99 99    6.56  3    6.50  2    99.99 99    99.99 99
218804      5.95 -0.05  0.44 99.99 99.99     99.99 99.99      7.16  3    6.10  3    5.86  3    99.99 99    99.99 99
218918      5.16  0.08  0.13 99.99 99.99     99.99 99.99      6.37 12    5.14  3    5.21  2    99.99 99    99.99 99
219077      6.12 99.99  0.78 99.99 99.99     99.99 99.99     99.99 99    6.45  3    5.87  3    99.99 99    99.99 99
219263      5.77 99.99  1.18 99.99 99.99     99.99 99.99     99.99 99    6.32  4    5.34  4    99.99 99    99.99 99
219402      5.55  0.06  0.06 99.99 99.99     99.99 99.99      6.70 13    5.49  3    5.64  2    99.99 99    99.99 99
219482      5.66 99.99  0.51 99.99 99.99     99.99 99.99     99.99 99    5.85  3    5.54  3    99.99 99    99.99 99
219531      6.47 99.99  1.08 99.99 99.99     99.99 99.99     99.99 99    6.97  4    6.08  4    99.99 99    99.99 99
219571      3.99 -0.02  0.40 99.99 99.99     99.99 99.99      5.22  3    4.12  3    3.92  3    99.99 99    99.99 99
219572      6.33  0.66  0.91 99.99 99.99     99.99 99.99      8.76  4    6.74  4    6.02  3    99.99 99    99.99 99
219644      6.13  1.58  1.35 99.99 99.99     99.99 99.99     10.05 12    6.78  4    5.62  4    99.99 99    99.99 99
219693      5.54 99.99  0.44 99.99 99.99     99.99 99.99     99.99 99    5.69  3    5.45  3    99.99 99    99.99 99
219765      5.49  1.43  1.27 99.99 99.99     99.99 99.99      9.16 11    6.09  4    5.01  4    99.99 99    99.99 99
219832      4.98 99.99 -0.02 99.99 99.99     99.99 99.99     99.99 99    4.87  2    5.10  2    99.99 99    99.99 99
219834      5.08 99.99  0.80 99.99 99.99     99.99 99.99     99.99 99    5.43  3    4.82  3    99.99 99    99.99 99
219877      5.55  0.00  0.39 99.99 99.99     99.99 99.99      6.80  4    5.67  3    5.48  2    99.99 99    99.99 99
219912      6.37 99.99  1.30 99.99 99.99     99.99 99.99     99.99 99    6.99  4    5.88  4    99.99 99    99.99 99
219962      6.32  1.05  1.12 99.99 99.99     99.99 99.99      9.39  7    6.84  4    5.91  4    99.99 99    99.99 99
220009      5.05  1.12  1.20 99.99 99.99     99.99 99.99      8.26  6    5.62  4    4.61  4    99.99 99    99.99 99
220035      6.17  0.97  1.07 99.99 99.99     99.99 99.99      9.11  7    6.66  4    5.79  4    99.99 99    99.99 99
220096      5.65 99.99  0.82 99.99 99.99     99.99 99.99     99.99 99    6.01  4    5.38  3    99.99 99    99.99 99
220278      5.20 99.99  0.20 99.99 99.99     99.99 99.99     99.99 99    5.22  3    5.22  2    99.99 99    99.99 99
220401      6.10 99.99  1.46 99.99 99.99     99.99 99.99     99.99 99    6.81  4    5.54  5    99.99 99    99.99 99
220406      6.31  1.95  1.61 99.99 99.99     99.99 99.99     10.87 14    7.10  5    5.68  5    99.99 99    99.99 99
220465      6.19 99.99  1.02 99.99 99.99     99.99 99.99     99.99 99    6.66  4    5.83  4    99.99 99    99.99 99
220759      6.45  1.77  1.47 99.99 99.99     99.99 99.99     10.69 13    7.16  4    5.88  5    99.99 99    99.99 99
```

```
220790      5.63 99.99   0.98 99.99 99.99     99.99 99.99    99.99 99   6.07  4   5.29  4   99.99 99   99.99 99
220802      6.20 99.99  -0.08 99.99 99.99     99.99 99.99    99.99 99   6.06  2   6.35  2   99.99 99   99.99 99
220858      6.25  0.80   1.02 99.99 99.99     99.99 99.99     8.93  5   6.72  4   5.89  4   99.99 99   99.99 99
220929      6.32 99.99   1.20 99.99 99.99     99.99 99.99    99.99 99   6.89  4   5.88  4   99.99 99   99.99 99
221051      6.43 99.99   1.18 99.99 99.99     99.99 99.99    99.99 99   6.98  4   6.00  4   99.99 99   99.99 99
221081      6.18  1.70   1.44 99.99 99.99     99.99 99.99    10.31 13   6.88  4   5.63  5   99.99 99   99.99 99
221148      6.26  1.14   1.08 99.99 99.99     99.99 99.99     9.43 10   6.76  4   5.87  4   99.99 99   99.99 99
221308      6.39  1.29   1.26 99.99 99.99     99.99 99.99     9.87  8   6.99  4   5.92  4   99.99 99   99.99 99
221323      6.02 99.99   1.02 99.99 99.99     99.99 99.99    99.99 99   6.49  4   5.66  4   99.99 99   99.99 99
221345      5.22  0.86   1.02 99.99 99.99     99.99 99.99     7.98  6   5.69  4   4.86  4   99.99 99   99.99 99
221356      6.49 -0.01   0.54 99.99 99.99     99.99 99.99     7.81  4   6.69  3   6.35  3   99.99 99   99.99 99
221409      6.38  1.10   1.18 99.99 99.99     99.99 99.99     9.55  7   6.93  4   5.95  4   99.99 99   99.99 99
221420      5.81  0.31   0.68 99.99 99.99     99.99 99.99     7.64  3   6.09  3   5.61  3   99.99 99   99.99 99
221525      5.58  0.13   0.23 99.99 99.99     99.99 99.99     6.92 10   5.61  3   5.59  2   99.99 99   99.99 99
221675      5.87  0.16   0.30 99.99 99.99     99.99 99.99     7.29  9   5.94  3   5.85  2   99.99 99   99.99 99
221835      6.39  0.57   0.88 99.99 99.99     99.99 99.99     8.68  4   6.78  4   6.10  3   99.99 99   99.99 99
221950      5.67 -0.08   0.44 99.99 99.99     99.99 99.99     6.84  3   5.82  3   5.58  3   99.99 99   99.99 99
222060      6.00  0.65   0.90 99.99 99.99     99.99 99.99     8.41  4   6.40  4   5.70  3   99.99 99   99.99 99
222377      5.97  0.13   0.20 99.99 99.99     99.99 99.99     7.29 11   5.99  3   5.99  2   99.99 99   99.99 99
222433      5.31 99.99   0.97 99.99 99.99     99.99 99.99    99.99 99   5.75  4   4.97  4   99.99 99   99.99 99
222485      6.60 99.99   1.57 99.99 99.99     99.99 99.99    99.99 99   7.37  5   5.98  5   99.99 99   99.99 99
222493      5.89 99.99   1.00 99.99 99.99     99.99 99.99    99.99 99   6.35  4   5.54  4   99.99 99   99.99 99
222602      5.89  0.08   0.10 99.99 99.99     99.99 99.99     7.09 13   5.85  3   5.96  2   99.99 99   99.99 99
222764      5.06  2.03   1.68 99.99 99.99     99.99 99.99     9.77 14   5.89  5   4.39  5   99.99 99   99.99 99
222803      6.09 99.99   0.98 99.99 99.99     99.99 99.99    99.99 99   6.53  4   5.75  4   99.99 99   99.99 99
222805      6.07  0.62   0.91 99.99 99.99     99.99 99.99     8.45  4   6.48  4   5.76  3   99.99 99   99.99 99
222806      5.75  1.06   1.11 99.99 99.99     99.99 99.99     8.83  7   6.27  4   5.35  4   99.99 99   99.99 99
222820      5.72  1.63   1.40 99.99 99.99     99.99 99.99     9.73 12   6.40  4   5.18  4   99.99 99   99.99 99
222872      6.17 99.99   0.50 99.99 99.99     99.99 99.99    99.99 99   6.35  3   6.05  3   99.99 99   99.99 99
223011      6.31 99.99   0.20 99.99 99.99     99.99 99.99    99.99 99   6.33  3   6.33  2   99.99 99   99.99 99
223024      5.29 99.99   0.28 99.99 99.99     99.99 99.99    99.99 99   5.35  3   5.27  2   99.99 99   99.99 99
223145      5.18 99.99  -0.19 99.99 99.99     99.99 99.99    99.99 99   4.98  2   5.38  2   99.99 99   99.99 99
223148      6.89  0.06   0.46 99.99 99.99     99.99 99.99     8.26  4   7.05  3   6.79  3   99.99 99   99.99 99
223252      5.49  0.70   0.94 99.99 99.99     99.99 99.99     7.99  4   5.91  4   5.17  3   99.99 99   99.99 99
223311      6.07  1.71   1.45 99.99 99.99     99.99 99.99    10.22 13   6.77  4   5.51  5   99.99 99   99.99 99
223346      6.46 -0.02   0.44 99.99 99.99     99.99 99.99     7.71  3   6.61  3   6.37  3   99.99 99   99.99 99
223438      5.77  0.11   0.16 99.99 99.99     99.99 99.99     7.04 12   5.76  3   5.81  2   99.99 99   99.99 99
223444      6.59 99.99   1.48 99.99 99.99     99.99 99.99    99.99 99   7.31  4   6.02  5   99.99 99   99.99 99
223466      6.42 99.99   0.12 99.99 99.99     99.99 99.99    99.99 99   6.39  3   6.48  2   99.99 99   99.99 99
223647      5.11  0.60   0.92 99.99 99.99     99.99 99.99     7.46  3   5.52  4   4.80  3   99.99 99   99.99 99
223719      5.54  1.85   1.53 99.99 99.99     99.99 99.99     9.92 14   6.29  5   4.94  5   99.99 99   99.99 99
```

```
223778      6.40  0.71  0.98 99.99 99.99    99.99 99.99      8.94  4    6.84  4    6.06  4   99.99 99   99.99 99
223807      5.75  1.15  1.17 99.99 99.99    99.99 99.99      8.99  8    6.30  4    5.32  4   99.99 99   99.99 99
223825      5.93  0.96  1.07 99.99 99.99    99.99 99.99      8.85  6    6.42  4    5.55  4   99.99 99   99.99 99
223855      6.28 -0.01 -0.01 99.99 99.99    99.99 99.99      7.30 14    6.18  3    6.40  2   99.99 99   99.99 99
223991      6.35 99.99  0.20 99.99 99.99    99.99 99.99     99.99 99    6.37  3    6.37  2   99.99 99   99.99 99
224022      6.03 99.99  0.57 99.99 99.99    99.99 99.99     99.99 99    6.25  3    5.88  3   99.99 99   99.99 99
224103      6.21 -0.19 -0.07 99.99 99.99    99.99 99.99      6.95 11    6.08  2    6.36  2   99.99 99   99.99 99
224112      6.83 99.99 -0.08 99.99 99.99    99.99 99.99     99.99 99    6.69  2    6.98  2   99.99 99   99.99 99
224361      5.97 99.99  0.11 99.99 99.99    99.99 99.99     99.99 99    5.94  3    6.03  2   99.99 99   99.99 99
224362      5.73  0.92  1.05 99.99 99.99    99.99 99.99      8.59  6    6.21  4    5.36  4   99.99 99   99.99 99
224392      5.00  0.08  0.06 99.99 99.99    99.99 99.99      6.18 14    4.94  3    5.09  2   99.99 99   99.99 99
224481      6.26  1.03  1.08 99.99 99.99    99.99 99.99      9.28  8    6.76  4    5.87  4   99.99 99   99.99 99
224533      4.86  0.70  0.93 99.99 99.99    99.99 99.99      7.36  5    5.28  4    4.54  3   99.99 99   99.99 99
224686      4.50 -0.28 -0.08 99.99 99.99    99.99 99.99      5.11  9    4.36  2    4.65  2   99.99 99   99.99 99
224750      6.29 99.99  0.76 99.99 99.99    99.99 99.99     99.99 99    6.61  3    6.05  3   99.99 99   99.99 99
224834      5.71 99.99  0.91 99.99 99.99    99.99 99.99     99.99 99    6.12  4    5.40  3   99.99 99   99.99 99
224889      4.78  1.41  1.27 99.99 99.99    99.99 99.99      8.42 11    5.38  4    4.30  4   99.99 99   99.99 99
224926      5.11 -0.51 -0.12 99.99 99.99    99.99 99.99      5.39  6    4.95  2    5.28  2   99.99 99   99.99 99
224930      5.75  0.05  0.67 99.99 99.99    99.99 99.99      7.22  5    6.02  3    5.55  3   99.99 99   99.99 99
224995      6.32  0.14  0.18 99.99 99.99    99.99 99.99      7.64 12    6.32  3    6.35  2   99.99 99   99.99 99
225003      5.69  0.01  0.31 99.99 99.99    99.99 99.99      6.91  6    5.77  3    5.66  2   99.99 99   99.99 99
225009      5.86  0.91  1.09 99.99 99.99    99.99 99.99      8.73  5    6.36  4    5.47  4   99.99 99   99.99 99
225045      6.25  0.08  0.53 99.99 99.99    99.99 99.99      7.68  3    6.45  3    6.12  3   99.99 99   99.99 99
225233      7.31  0.01  0.44 99.99 99.99    99.99 99.99      8.60  3    7.46  3    7.22  3   99.99 99   99.99 99
225239      6.12  0.09  0.62 99.99 99.99    99.99 99.99      7.62  3    6.37  3    5.95  3   99.99 99   99.99 99
225253      5.59 -0.42 -0.12 99.99 99.99    99.99 99.99      5.99  8    5.43  2    5.76  2   99.99 99   99.99 99
 22001      4.71 -0.04  0.39 99.99 99.99    99.99 99.99      5.91  3    4.83  3    4.64  2   99.99 99   99.99 99
 23466      5.35 -0.61 -0.11 99.99 99.99    99.99 99.99      5.50  4    5.19  2    5.52  2   99.99 99   99.99 99
 26677      6.51  0.12  0.16 99.99 99.99    99.99 99.99      7.80 12    6.50  3    6.55  2   99.99 99   99.99 99
 43745      6.04 99.99  0.59 99.99 99.99    99.99 99.99     99.99 99    6.27  3    5.88  3   99.99 99   99.99 99
 83104      6.31 99.99  0.06 99.99 99.99    99.99 99.99     99.99 99    6.25  3    6.40  2   99.99 99   99.99 99
 94363      6.13  0.61  0.91 99.99 99.99    99.99 99.99      8.49  4    6.54  4    5.82  3   99.99 99   99.99 99
135101      6.68  0.25  0.68 99.99 99.99    99.99 99.99      8.43  3    6.96  3    6.48  3   99.99 99   99.99 99
210884      5.50 -0.04  0.38 99.99 99.99    99.99 99.99      6.69  4    5.61  3    5.44  2   99.99 99   99.99 99
  2626      5.94 -0.36  0.01 99.99 99.99    99.99 99.99      6.49  6    5.85  3    6.05  2   99.99 99   99.99 99
  7927      4.99  0.49  0.68 99.99 99.99    99.99 99.99      7.06  6    5.27  3    4.79  3   99.99 99   99.99 99
 27084      5.45  0.12  0.22 99.99 99.99    99.99 99.99      6.77 10    5.48  3    5.46  2   99.99 99   99.99 99
 35519      6.15  1.68  1.45 99.99 99.99    99.99 99.99     10.26 12    6.85  4    5.59  5   99.99 99   99.99 99
 45546      5.05 -0.76 -0.18 99.99 99.99    99.99 99.99      4.96  3    4.86  2    5.25  2   99.99 99   99.99 99
 45321      6.15 -0.62 -0.14 99.99 99.99    99.99 99.99      6.27  4    5.98  2    6.33  2   99.99 99   99.99 99
 46241      5.83  0.78  1.00 99.99 99.99    99.99 99.99      8.47  5    6.29  4    5.48  4   99.99 99   99.99 99
```

```
 49643       5.75 -0.47 -0.10 99.99 99.99    99.99 99.99    6.09  6    5.60 2    5.91 2    99.99 99    99.99 99
 55879       6.04 -0.96 -0.18 99.99 99.99    99.99 99.99    5.68  5    5.85 2    6.24 2    99.99 99    99.99 99
 61224       6.53 -0.30 -0.02 99.99 99.99    99.99 99.99    7.15  8    6.42 2    6.65 2    99.99 99    99.99 99
 61642       6.21 99.99  1.02 99.99 99.99    99.99 99.99   99.99 99    6.68 4    5.85 4    99.99 99    99.99 99
 61831       4.85 -0.65 -0.19 99.99 99.99    99.99 99.99    4.90  5    4.65 2    5.05 2    99.99 99    99.99 99
 61899       5.76 99.99 -0.07 99.99 99.99    99.99 99.99   99.99 99    5.63 2    5.91 2    99.99 99    99.99 99
 62226       5.43 99.99 -0.15 99.99 99.99    99.99 99.99   99.99 99    5.25 2    5.61 2    99.99 99    99.99 99
 62893       5.89 99.99 -0.11 99.99 99.99    99.99 99.99   99.99 99    5.73 2    6.06 2    99.99 99    99.99 99
 62991       6.55 99.99 -0.10 99.99 99.99    99.99 99.99   99.99 99    6.40 2    6.71 2    99.99 99    99.99 99
 63032       3.61  1.71  1.73 99.99 99.99    99.99 99.99    7.91  5    4.47 5    2.92 5    99.99 99    99.99 99
 63465       5.09 99.99 -0.10 99.99 99.99    99.99 99.99   99.99 99    4.94 2    5.25 2    99.99 99    99.99 99
 65662       5.74 99.99  1.55 99.99 99.99    99.99 99.99   99.99 99    6.50 5    5.13 5    99.99 99    99.99 99
 65907       5.60 99.99  0.57 99.99 99.99    99.99 99.99   99.99 99    5.82 3    5.45 3    99.99 99    99.99 99
 66341       6.33 99.99 -0.06 99.99 99.99    99.99 99.99   99.99 99    6.20 2    6.47 2    99.99 99    99.99 99
 68862       6.43 99.99  0.10 99.99 99.99    99.99 99.99   99.99 99    6.39 3    6.50 2    99.99 99    99.99 99
 68450       6.44 99.99 -0.01 99.99 99.99    99.99 99.99   99.99 99    6.34 3    6.56 2    99.99 99    99.99 99
 73665       6.39  0.83  0.98 99.99 99.99    99.99 99.99    9.09  6    6.83 4    6.05 4    99.99 99    99.99 99
 73710       6.44  0.90  1.02 99.99 99.99    99.99 99.99    9.26  6    6.91 4    6.08 4    99.99 99    99.99 99
 73731       6.30  0.16  0.17 99.99 99.99    99.99 99.99    7.65 12    6.30 3    6.34 2    99.99 99    99.99 99
 72779       6.58 99.99  0.68 99.99 99.99    99.99 99.99   99.99 99    6.86 3    6.38 3    99.99 99    99.99 99
 85250       6.06 99.99  0.94 99.99 99.99    99.99 99.99   99.99 99    6.48 4    5.74 3    99.99 99    99.99 99
 87283       5.94 99.99  0.26 99.99 99.99    99.99 99.99   99.99 99    5.99 3    5.93 2    99.99 99    99.99 99
 87436       6.19 99.99  0.17 99.99 99.99    99.99 99.99   99.99 99    6.19 3    6.23 2    99.99 99    99.99 99
 96544       6.01 99.99  1.20 99.99 99.99    99.99 99.99   99.99 99    6.58 4    5.57 4    99.99 99    99.99 99
101189       5.15 99.99 -0.02 99.99 99.99    99.99 99.99   99.99 99    5.04 2    5.27 2    99.99 99    99.99 99
101021       5.15 99.99  1.12 99.99 99.99    99.99 99.99   99.99 99    5.67 4    4.74 4    99.99 99    99.99 99
111904       5.76 99.99  0.34 99.99 99.99    99.99 99.99   99.99 99    5.85 3    5.72 2    99.99 99    99.99 99
162515       6.45 -0.07  0.02 99.99 99.99    99.99 99.99    7.40 11    6.37 3    6.56 2    99.99 99    99.99 99
162586       6.17 -0.35 -0.03 99.99 99.99    99.99 99.99    6.71  7    6.06 2    6.30 2    99.99 99    99.99 99
162587       5.60  1.01  1.14 99.99 99.99    99.99 99.99    8.63  6    6.13 4    5.18 4    99.99 99    99.99 99
162817       5.96 -0.10 -0.01 99.99 99.99    99.99 99.99    6.85 12    5.86 3    6.08 2    99.99 99    99.99 99
163245       6.52  0.05  0.05 99.99 99.99    99.99 99.99    7.65 13    6.45 3    6.61 2    99.99 99    99.99 99
164794       5.97 -0.89  0.00 99.99 99.99    99.99 99.99    5.80  8    5.88 3    6.08 2    99.99 99    99.99 99
170200       5.72 -0.36 -0.03 99.99 99.99    99.99 99.99    6.25  7    5.61 2    5.85 2    99.99 99    99.99 99
192004       5.49  1.50  1.41 99.99 99.99    99.99 99.99    9.33  9    6.17 4    4.95 4    99.99 99    99.99 99
191747       5.53  0.13  0.08 99.99 99.99    99.99 99.99    6.79 14    5.48 3    5.61 2    99.99 99    99.99 99
192044       5.92 -0.43 -0.11 99.99 99.99    99.99 99.99    6.31  7    5.76 2    6.09 2    99.99 99    99.99 99
221246       6.17  1.71  1.46 99.99 99.99    99.99 99.99   10.33 12    6.88 4    5.61 5    99.99 99    99.99 99
 75466       6.29 99.99 -0.10 99.99 99.99    99.99 99.99   99.99 99    6.14 2    6.45 2    99.99 99    99.99 99
 93030       2.76 -1.00 -0.22 99.99 99.99    99.99 99.99    2.32  5    2.54 2    2.98 2    99.99 99    99.99 99
 93163       5.77 99.99  0.01 99.99 99.99    99.99 99.99   99.99 99    5.68 3    5.88 2    99.99 99    99.99 99
```

```
 93194      4.82 99.99 -0.13 99.99 99.99      99.99 99.99     99.99 99     4.65  2     5.00  2     99.99 99     99.99 99
 93540      5.34 -0.45 -0.10 99.99 99.99      99.99 99.99      5.71  6     5.19  2     5.50  2     99.99 99     99.99 99
 93549      5.23 -0.47 -0.07 99.99 99.99      99.99 99.99      5.59  5     5.10  2     5.38  2     99.99 99     99.99 99
 93607      4.85 -0.64 -0.15 99.99 99.99      99.99 99.99      4.94  4     4.67  2     5.03  2     99.99 99     99.99 99
224990      5.00 -0.55 -0.16 99.99 99.99      99.99 99.99      5.20  6     4.82  2     5.19  2     99.99 99     99.99 99
225200      6.40 99.99  0.00 99.99 99.99      99.99 99.99     99.99 99     6.31  3     6.51  2     99.99 99     99.99 99
 74772      4.07  0.52  0.87 99.99 99.99      99.99 99.99      6.29  3     4.45  4     3.78  3     99.99 99     99.99 99
 75387      6.43 -0.77 -0.20 99.99 99.99      99.99 99.99      6.31  3     6.22  2     6.64  2     99.99 99     99.99 99
 51283      5.29 99.99 -0.18 99.99 99.99      99.99 99.99     99.99 99     5.10  2     5.49  2     99.99 99     99.99 99
 53244      4.12 -0.47 -0.11 99.99 99.99      99.99 99.99      4.46  6     3.96  2     4.29  2     99.99 99     99.99 99
 54605      1.84  0.55  0.67 99.99 99.99      99.99 99.99      3.99  8     2.11  3     1.64  3     99.99 99     99.99 99
 49028      6.53 99.99 -0.12 99.99 99.99      99.99 99.99     99.99 99     6.37  2     6.70  2     99.99 99     99.99 99
 52018      5.59 -0.70 -0.16 99.99 99.99      99.99 99.99      5.59  3     5.41  2     5.78  2     99.99 99     99.99 99
 52140      6.43 99.99 -0.13 99.99 99.99      99.99 99.99     99.99 99     6.26  2     6.61  2     99.99 99     99.99 99
 56618      4.60  1.89  1.60 99.99 99.99      99.99 99.99      9.08 13     5.39  5     3.97  5     99.99 99     99.99 99
 58325      6.60 99.99 -0.19 99.99 99.99      99.99 99.99     99.99 99     6.40  2     6.80  2     99.99 99     99.99 99
 58346      6.21 99.99 -0.10 99.99 99.99      99.99 99.99     99.99 99     6.06  2     6.37  2     99.99 99     99.99 99
 58612      5.78 99.99 -0.10 99.99 99.99      99.99 99.99     99.99 99     5.63  2     5.94  2     99.99 99     99.99 99
 58286      5.39 99.99 -0.18 99.99 99.99      99.99 99.99     99.99 99     5.20  2     5.59  2     99.99 99     99.99 99
 58535      5.35 99.99  1.08 99.99 99.99      99.99 99.99     99.99 99     5.85  4     4.96  4     99.99 99     99.99 99
 58766      6.31 99.99 -0.17 99.99 99.99      99.99 99.99     99.99 99     6.12  2     6.50  2     99.99 99     99.99 99
 19373      4.05  0.12  0.60 99.99 99.99      99.99 99.99      5.58  3     4.29  3     3.89  3     99.99 99     99.99 99
 19735      6.33  1.64  1.43 99.99 99.99      99.99 99.99     10.37 12     7.02  4     5.78  5     99.99 99     99.99 99
 20365      5.15 -0.56 -0.06 99.99 99.99      99.99 99.99      5.39  4     5.02  2     5.29  2     99.99 99     99.99 99
 20418      5.03 -0.53 -0.06 99.99 99.99      99.99 99.99      5.31  4     4.90  2     5.17  2     99.99 99     99.99 99
 20902      1.79  0.38  0.48 99.99 99.99      99.99 99.99      3.60  9     1.96  3     1.68  3     99.99 99     99.99 99
 21278      4.98 -0.56 -0.10 99.99 99.99      99.99 99.99      5.20  4     4.83  2     5.14  2     99.99 99     99.99 99
 21362      5.58 -0.44 -0.04 99.99 99.99      99.99 99.99      6.00  5     5.46  2     5.71  2     99.99 99     99.99 99
 21428      4.67 -0.57 -0.09 99.99 99.99      99.99 99.99      4.88  4     4.53  2     4.83  2     99.99 99     99.99 99
 21455      6.24 -0.25  0.13 99.99 99.99      99.99 99.99      7.01  5     6.22  3     6.29  2     99.99 99     99.99 99
 21552      4.38  1.53  1.34 99.99 99.99      99.99 99.99      8.23 12     5.02  4     3.87  4     99.99 99     99.99 99
 21551      5.82 -0.31 -0.04 99.99 99.99      99.99 99.99      6.41  8     5.70  2     5.95  2     99.99 99     99.99 99
 23288      5.46 -0.33 -0.04 99.99 99.99      99.99 99.99      6.03  7     5.34  2     5.59  2     99.99 99     99.99 99
 23302      3.70 -0.40 -0.12 99.99 99.99      99.99 99.99      4.13  8     3.54  2     3.87  2     99.99 99     99.99 99
 23324      5.65 -0.36 -0.07 99.99 99.99      99.99 99.99      6.16  8     5.52  2     5.80  2     99.99 99     99.99 99
 23338      4.30 -0.46 -0.11 99.99 99.99      99.99 99.99      4.65  6     4.14  2     4.47  2     99.99 99     99.99 99
 23408      3.87 -0.40 -0.07 99.99 99.99      99.99 99.99      4.32  7     3.74  2     4.02  2     99.99 99     99.99 99
 23432      5.76 -0.23 -0.04 99.99 99.99      99.99 99.99      6.46  9     5.64  2     5.89  2     99.99 99     99.99 99
 23441      6.43 -0.15 -0.02 99.99 99.99      99.99 99.99      7.25 11     6.32  2     6.55  2     99.99 99     99.99 99
 23630      2.87 -0.35 -0.09 99.99 99.99      99.99 99.99      3.38  8     2.73  2     3.03  2     99.99 99     99.99 99
 23753      5.45 -0.32 -0.07 99.99 99.99      99.99 99.99      6.01  8     5.32  2     5.60  2     99.99 99     99.99 99
```

```
23850     3.62 -0.36 -0.09 99.99 99.99    99.99 99.99    4.12  8   3.48  2   3.78  2   99.99 99   99.99 99
23923     6.17 -0.19 -0.05 99.99 99.99    99.99 99.99    6.92 11   6.05  2   6.31  2   99.99 99   99.99 99
23950     6.07 -0.32 -0.01 99.99 99.99    99.99 99.99    6.67  7   5.97  3   6.19  2   99.99 99   99.99 99
24357     5.97  0.00  0.34 99.99 99.99    99.99 99.99    7.19  5   6.06  3   5.93  2   99.99 99   99.99 99
25102     6.37  0.00  0.42 99.99 99.99    99.99 99.99    7.64  3   6.51  3   6.29  3   99.99 99   99.99 99
26015     6.01  0.02  0.40 99.99 99.99    99.99 99.99    7.29  4   6.14  3   5.94  3   99.99 99   99.99 99
26462     5.73  0.00  0.36 99.99 99.99    99.99 99.99    6.96  4   5.83  3   5.68  2   99.99 99   99.99 99
27176     5.65  0.08  0.28 99.99 99.99    99.99 99.99    6.95  8   5.71  3   5.63  2   99.99 99   99.99 99
27371     3.65  0.81  0.99 99.99 99.99    99.99 99.99    6.33  5   4.10  4   3.30  4   99.99 99   99.99 99
27429     6.11  0.03  0.37 99.99 99.99    99.99 99.99    7.39  5   6.22  3   6.05  2   99.99 99   99.99 99
27483     6.17  0.02  0.46 99.99 99.99    99.99 99.99    7.49  3   6.33  3   6.07  3   99.99 99   99.99 99
27697     3.76  0.82  0.99 99.99 99.99    99.99 99.99    6.45  6   4.21  4   3.41  4   99.99 99   99.99 99
27749     5.64  0.14  0.30 99.99 99.99    99.99 99.99    7.03  9   5.71  3   5.62  2   99.99 99   99.99 99
27819     4.80  0.12  0.15 99.99 99.99    99.99 99.99    6.08 12   4.79  3   4.85  2   99.99 99   99.99 99
27901     5.97  0.05  0.37 99.99 99.99    99.99 99.99    7.28  5   6.08  3   5.91  2   99.99 99   99.99 99
27934     4.22  0.14  0.13 99.99 99.99    99.99 99.99    5.52 13   4.20  3   4.27  2   99.99 99   99.99 99
27946     5.28  0.10  0.25 99.99 99.99    99.99 99.99    6.59  9   5.32  3   5.28  2   99.99 99   99.99 99
27991     6.46  0.02  0.49 99.99 99.99    99.99 99.99    7.79  3   6.63  3   6.35  3   99.99 99   99.99 99
28226     5.72  0.10  0.27 99.99 99.99    99.99 99.99    7.04  9   5.77  3   5.71  2   99.99 99   99.99 99
28294     5.90  0.06  0.32 99.99 99.99    99.99 99.99    7.19  6   5.98  3   5.87  2   99.99 99   99.99 99
28305     3.54  0.87  1.01 99.99 99.99    99.99 99.99    6.31  6   4.00  4   3.19  4   99.99 99   99.99 99
28307     3.83  0.71  0.95 99.99 99.99    99.99 99.99    6.35  4   4.26  4   3.50  4   99.99 99   99.99 99
28355     5.03  0.12  0.23 99.99 99.99    99.99 99.99    6.35 10   5.06  3   5.04  2   99.99 99   99.99 99
28485     5.58  0.10  0.32 99.99 99.99    99.99 99.99    6.93  7   5.66  3   5.55  2   99.99 99   99.99 99
28527     4.78  0.13  0.17 99.99 99.99    99.99 99.99    6.08 12   4.78  3   4.82  2   99.99 99   99.99 99
28546     5.48  0.10  0.26 99.99 99.99    99.99 99.99    6.79  9   5.53  3   5.47  2   99.99 99   99.99 99
28556     5.41  0.10  0.26 99.99 99.99    99.99 99.99    6.72  9   5.46  3   5.40  2   99.99 99   99.99 99
28677     6.02  0.04  0.34 99.99 99.99    99.99 99.99    7.30  6   6.11  3   5.98  2   99.99 99   99.99 99
28736     6.40  0.00  0.41 99.99 99.99    99.99 99.99    7.66  4   6.53  3   6.32  3   99.99 99   99.99 99
29375     5.79  0.06  0.31 99.99 99.99    99.99 99.99    7.08  7   5.87  3   5.76  2   99.99 99   99.99 99
29388     4.27  0.13  0.12 99.99 99.99    99.99 99.99    5.55 13   4.24  3   4.33  2   99.99 99   99.99 99
29499     5.39  0.12  0.25 99.99 99.99    99.99 99.99    6.73 10   5.43  3   5.39  2   99.99 99   99.99 99
29488     4.70  0.13  0.14 99.99 99.99    99.99 99.99    5.99 13   4.68  3   4.75  2   99.99 99   99.99 99
30034     5.40  0.08  0.25 99.99 99.99    99.99 99.99    6.68  9   5.44  3   5.40  2   99.99 99   99.99 99
30210     5.37  0.13  0.19 99.99 99.99    99.99 99.99    6.69 11   5.38  3   5.40  2   99.99 99   99.99 99
31236     6.37  0.06  0.29 99.99 99.99    99.99 99.99    7.65  7   6.43  3   6.35  2   99.99 99   99.99 99
32301     4.64  0.15  0.16 99.99 99.99    99.99 99.99    5.97 13   4.63  3   4.68  2   99.99 99   99.99 99
33254     5.43  0.14  0.24 99.99 99.99    99.99 99.99    6.79 10   5.47  3   5.43  2   99.99 99   99.99 99
25202     5.89  0.04  0.32 99.99 99.99    99.99 99.99    7.16  6   5.97  3   5.86  2   99.99 99   99.99 99
18404     5.80  0.01  0.41 99.99 99.99    99.99 99.99    7.07  4   5.93  3   5.72  3   99.99 99   99.99 99
13871     5.79  0.04  0.44 99.99 99.99    99.99 99.99    7.12  4   5.94  3   5.70  3   99.99 99   99.99 99
```

```
 25570      5.46  0.00  0.36 99.99 99.99    99.99 99.99    6.69  4    5.56  3    5.41  2    99.99 99    99.99 99
 30197      6.01  1.32  1.21 99.99 99.99    99.99 99.99    9.50 10    6.58  4    5.56  4    99.99 99    99.99 99
 37147      5.54  0.10  0.22 99.99 99.99    99.99 99.99    6.83 10    5.57  3    5.55  2    99.99 99    99.99 99
 40932      4.13  0.11  0.16 99.99 99.99    99.99 99.99    5.40 12    4.12  3    4.17  2    99.99 99    99.99 99
105805      6.04  0.08  0.11 99.99 99.99    99.99 99.99    7.24 12    6.01  3    6.10  2    99.99 99    99.99 99
107168      6.27  0.15  0.17 99.99 99.99    99.99 99.99    7.60 12    6.27  3    6.31  2    99.99 99    99.99 99
107326      6.15  0.08  0.30 99.99 99.99    99.99 99.99    7.46  7    6.22  3    6.13  2    99.99 99    99.99 99
107655      6.20 -0.05  0.00 99.99 99.99    99.99 99.99    7.17 12    6.11  3    6.31  2    99.99 99    99.99 99
107700      4.81  0.26  0.49 99.99 99.99    99.99 99.99    6.47  6    4.98  3    4.70  3    99.99 99    99.99 99
108007      6.42  0.08  0.27 99.99 99.99    99.99 99.99    7.71  8    6.47  3    6.41  2    99.99 99    99.99 99
108123      6.03  1.00  1.10 99.99 99.99    99.99 99.99    9.03  6    6.54  4    5.63  4    99.99 99    99.99 99
108283      4.95  0.18  0.27 99.99 99.99    99.99 99.99    6.38 10    5.00  3    4.94  2    99.99 99    99.99 99
108381      4.37  1.15  1.13 99.99 99.99    99.99 99.99    7.58  9    4.90  4    3.96  4    99.99 99    99.99 99
108382      5.00  0.13  0.08 99.99 99.99    99.99 99.99    6.26 14    4.95  3    5.08  2    99.99 99    99.99 99
108642      6.54  0.11  0.18 99.99 99.99    99.99 99.99    7.82 11    6.54  3    6.57  2    99.99 99    99.99 99
108651      6.65  0.08  0.22 99.99 99.99    99.99 99.99    7.91  9    6.68  3    6.66  2    99.99 99    99.99 99
108722      5.48  0.09  0.43 99.99 99.99    99.99 99.99    6.87  5    5.62  3    5.40  3    99.99 99    99.99 99
109307      6.29  0.10  0.11 99.99 99.99    99.99 99.99    7.52 13    6.26  3    6.35  2    99.99 99    99.99 99
 36959      5.67 -0.91 -0.24 99.99 99.99    99.99 99.99    5.34  3    5.44  2    5.90  2    99.99 99    99.99 99
 36960      4.79 -1.02 -0.25 99.99 99.99    99.99 99.99    4.31  4    4.56  2    5.02  2    99.99 99    99.99 99
 37016      6.25 -0.70 -0.15 99.99 99.99    99.99 99.99    6.25  3    6.07  2    6.43  2    99.99 99    99.99 99
 37018      4.59 -0.94 -0.19 99.99 99.99    99.99 99.99    4.25  4    4.39  2    4.79  2    99.99 99    99.99 99
 37040      6.31 -0.71 -0.13 99.99 99.99    99.99 99.99    6.31  3    6.14  2    6.49  2    99.99 99    99.99 99
 37077      5.27  0.17  0.23 99.99 99.99    99.99 99.99    6.66 11    5.30  3    5.28  2    99.99 99    99.99 99
 37150      6.50 -0.82 -0.16 99.99 99.99    99.99 99.99    6.34  3    6.32  2    6.69  2    99.99 99    99.99 99
 37209      5.74 -0.91 -0.22 99.99 99.99    99.99 99.99    5.42  3    5.52  2    5.96  2    99.99 99    99.99 99
 37303      6.06 -0.92 -0.21 99.99 99.99    99.99 99.99    5.73  4    5.85  2    6.27  2    99.99 99    99.99 99
218537      6.25 -0.60 -0.01 99.99 99.99    99.99 99.99    6.47  3    6.15  3    6.37  2    99.99 99    99.99 99
 36591      5.35 -0.93 -0.19 99.99 99.99    99.99 99.99    5.02  4    5.15  2    5.55  2    99.99 99    99.99 99
 36646      6.53 -0.65 -0.10 99.99 99.99    99.99 99.99    6.63  3    6.38  2    6.69  2    99.99 99    99.99 99
 36779      6.20 -0.81 -0.16 99.99 99.99    99.99 99.99    6.05  3    6.02  2    6.39  2    99.99 99    99.99 99
 88318      6.27 -0.05 -0.02 99.99 99.99    99.99 99.99    7.23 13    6.16  2    6.39  2    99.99 99    99.99 99
 87808      5.60 99.99  1.50 99.99 99.99    99.99 99.99   99.99 99    6.33  5    5.02  5    99.99 99    99.99 99
 88437      6.13 99.99  0.03 99.99 99.99    99.99 99.99   99.99 99    6.05  3    6.23  2    99.99 99    99.99 99
 90432      3.79  1.81  1.48 99.99 99.99    99.99 99.99    8.09 14    4.51  4    3.22  5    99.99 99    99.99 99
```